\documentclass[a4paper,11pt]{article}
\pdfoutput=1
\usepackage{jheppub}
\usepackage{hyperref}
\usepackage{mathdots}
\usepackage{MnSymbol}
\usepackage{amsmath}
\usepackage{amsfonts}
\usepackage{graphicx}
\usepackage{caption}
\usepackage{subcaption}
\usepackage{placeins}
\usepackage{tikz}
\usepackage{svg}
\usepackage{subfiles}

\usepackage[english]{babel}

\usepackage{CJK}

\usepackage{array}
\usepackage{mathtools}
\def\be{\begin{equation}}
\def\ee{\end{equation}}
\def\ba{\begin{eqnarray}}
\def\ea{\end{eqnarray}}

\newcommand*\diff{\mathop{}\!\mathrm{d}}
\title{
Scalar-Scaffolded Gluons and the Combinatorial Origins of Yang-Mills Theory} 

\author[a]{Nima Arkani-Hamed,}\author[b,c]{Qu Cao (曹趣),}\author[b,d]{Jin Dong (董晋),}\author[e]{Carolina Figueiredo,} \author[b,f,g]{Song He (何颂)}

\affiliation[a]{School of Natural Sciences, Institute for Advanced Study, Princeton, NJ, 08540, USA}
\affiliation[b]{CAS Key Laboratory of Theoretical Physics, Institute of Theoretical Physics, Chinese Academy of Sciences, Beijing 100190, China}
\affiliation[c]{Zhejiang Institute of Modern Physics, Department of Physics, Zhejiang University, Hangzhou, 310027, China}
\affiliation[d]{School of Physical Sciences, University of Chinese Academy of Sciences, No.19A Yuquan Road, Beijing 100049, China}
\affiliation[e]{Jadwin Hall, Princeton University, Princeton, NJ 08540, USA}
\affiliation[f]{School of Fundamental Physics and Mathematical Sciences, Hangzhou Institute for Advanced Study \& ICTP-AP, UCAS, Hangzhou 310024, China}
\affiliation[g]{Peng Huanwu Center for Fundamental Theory, Hefei, Anhui 230026, P. R. China}
\emailAdd{arkani@ias.edu}
\emailAdd{qucao@zju.edu.cn}
\emailAdd{dongjin@itp.ac.cn}
\emailAdd{cfigueiredo@princeton.edu}
\emailAdd{songhe@itp.ac.cn}

\abstract{We present a new formulation for Yang-Mills scattering amplitudes in any number of dimensions and at any loop order, based on the same combinatorial and binary-geometric ideas in kinematic space recently used to give an all-order description of Tr $\phi^3$ theory. We propose that in a precise sense the amplitudes for a suitably ``stringy" form of these two theories are identical, up to a simple shift of kinematic variables. This connection is made possible by describing the amplitudes for $n$ gluons via a ``scalar scaffolding", arising from the scattering of $2n$ colored scalars coming in $n$ distinct pairs of flavors fusing to produce the gluons. Fundamental properties of the ``$u$-variables", describing the ``binary geometry" for surfaces appearing in the topological expansion, magically guarantee that the kinematically shifted Tr $\phi^3$ amplitudes satisfy the physical properties needed to be interpreted as scaffolded gluons. These include multilinearity, gauge invariance, and factorization on tree- and loop- level gluon cuts. Our ``stringy'' scaffolded gluon amplitudes coincide with amplitudes in the bosonic string for extra-dimensional gluon polarizations at tree-level, but differ (and are simpler) at loop-level. We provide many checks on our proposal, including matching non-trivial leading singularities through two loops. The simple counting problem underlying the $u$ variables autonomously ``knows" about everything needed to convert colored scalar to gluon amplitudes, exposing a striking ``discovery" of Yang-Mills amplitudes from elementary combinatorial ideas in kinematic space.}

\begin{document}
\begin{CJK*}{UTF8}{}
\CJKfamily{gbsn}
\maketitle
\end{CJK*}
  \section{Introduction and Summary}

Scattering amplitudes are the ``boundary observables'' in flat space. The observations of the asymptotic states are made at infinity, while the interior of the spacetime serves as an auxiliary arena allowing us to incrementally build up the final amplitude in terms of interactions at and propagation between spacetime points. But given that the observable is anchored at infinity, it is natural to take the holographic point of view and ask: is there a way of calculating scattering amplitudes that jettisons the interior of the spacetime entirely, formulated purely in the kinematic space describing the scattering amplitude?

Tackling this question has turned out to involve ideas very far from the much more familiar (and so far more successful) instantiations of holography in Anti-de-Sitter space~. The reason is simple: the ``kinematic space'' that labels boundary correlation functions in AdS space is simply all the points in the one lower-dimensional spacetime on the boundary, with a conventional notion of locality, time, and quantum mechanical evolution. By contrast, the kinematic space labelling scattering processes has none of this familiar structure: for $n$ particle scattering we have the space of $n$ on-shell momenta, without any of the familiar properties of the arenas in which physics is usually described. 

But over the past decade, it has become apparent that in a number of cases, there is just enough structure in this seemingly barren kinematic space to discover entirely new mathematical questions to which the scattering amplitudes, with all their richness and complexity, are the answer. The emerging picture is that kinematic space contains elementary but deep combinatorial structures that in turn specify a tower of related mathematical objects above them, ultimately giving a completely autonomous definition of amplitudes. In this way, the usual foundational principles of locality and unitarity can be seen as derivative notions, ultimately arising from this fundamentally combinatorial rubric. 

The first example of this sort of structure was seen with the discovery of the amplituhedron for ${\cal N}=4$ SYM in the planar limit~\cite{Arkani-Hamed:2013jha}. The integrand for the planar amplitude, at any loop order, is given as the ``canonical form"~\cite{Arkani-Hamed:2017tmz} of a certain region in the kinematic space of momentum(-twistor) data~\cite{Hodges:2009hk}, entirely specified by topological winding number and positivity properties~\cite{Arkani-Hamed:2017vfh}. There is no mention of diagrams, locality, or unitarity anywhere in sight, instead in this example, these very simple but rather alien and more abstract combinatorial and geometric concepts bring the physics of spacetime and quantum mechanics to life, along the way manifesting the hidden infinite Yangian symmetry enjoyed by the scattering amplitudes. However, the very special nature of planar ${\cal N}=4$ SYM--its maximal supersymmetry and integrability--might be thought of as a dissuasion from the thought that such structures generalize to more realistic theories, instead perhaps being an artifact of studying the most special toy model we know of.  
\begin{figure}[t]
    \centering
    \includegraphics[width=\linewidth]{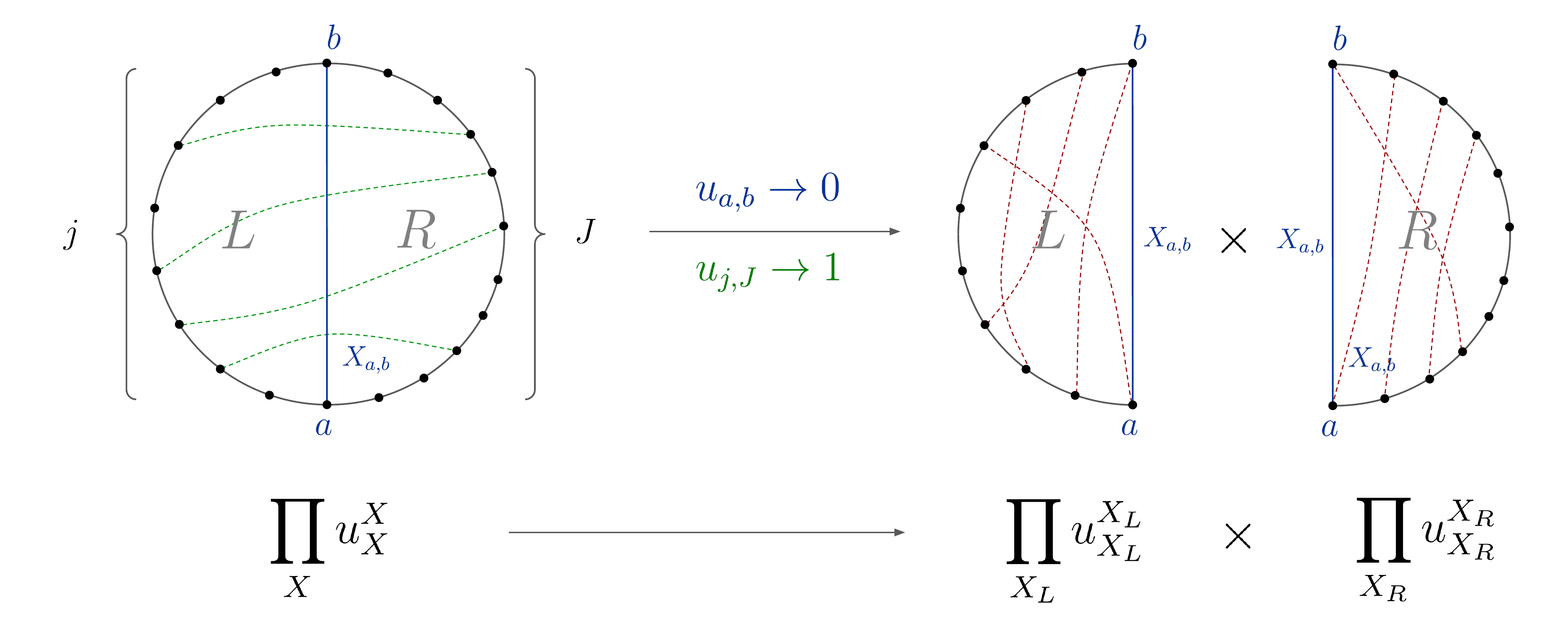}
    \caption{Binary behaviour of $u$-variables and factorizations.}
    \label{fig:ueqn}
\end{figure}

In the past few years, however, morally similar but technically quite different structures have been seen in a completely different setting, for the simplest theory of colored scalars with cubic self-interactions--the Tr $\phi^3$ theory--at all loop orders and to all orders in the topological 't Hooft expansion~\cite{Arkani-Hamed:2017mur, Arkani-Hamed:2019vag}~\cite{Arkani-Hamed:2023lbd, Arkani-Hamed:2023mvg}. This ``curve-integral" formalism features a simple counting problem defined at any order in the topological expansion, which defines a set of variables, $u_C$, defined for all propagators/ curves $C$ on a surface, satisfying the remarkable non-linear equations $u_C + \prod_D u_D^{\#{\rm int}(C,D)} = 1$. These equations define a ``binary geometry''~\cite{Arkani-Hamed:2019plo, Arkani-Hamed:2019mrd}: if we set $u_{\cal C} \geq 0$ they also force $0 \leq u_{\cal C} \leq 1$, and as we approach a boundary where $u_{\cal C} \to 0$ all the ``incompatible" curves ${\cal D}$ that cross ${\cal C}$ have $u_{\cal D} \to 1$. The amplitudes are defined by  ${\cal A} = \int \omega \prod_C u_C^{P_X^2}$, where $\omega$ is a simple canonical form with logarithmic singularities on the boundaries where  $u_C \to 0$, and $P^\mu_C$ is the momentum associated with $C$. The binary character of the $u_{\cal C}$ variables tells us that the ``curve integral" for ${\cal A}$ deserves to be called an ``amplitude'', by guaranteeing the correct patterns of factorization near singularities as in figure \ref{fig:ueqn} at tree-level. The curve integral can give us both ``stringy" amplitudes, or, via ``tropicalization", directly amplitudes in the field theory limit. There is no trace of the idea of ``summing over all spacetime processes' in these formulas. Instead, the simple counting problem attached to curves on surfaces magically converts combinatorial geometry into local and unitary physics in this model. 

But in the same way that ${\cal N} = 4$ SYM theory seems ``too special" to offer a hope for connecting combinatorial geometry to real-world physics, the Tr $\phi^3$ theory might seem too simple. While the singularities/poles/``denominators" of Tr $\phi^3$ are universal and shared by all theories of colored particles, real-world amplitudes for e.g. colored gluons differ in a crucial way, in their numerator structure.  However, so far the connection between amplitudes and geometry has centered on long-distance singularities captured only by the ``denominators", which are simple poles associated in many settings with the simplest ``logarithmic" singularities. This is reflected in the ubiquity of ``canonical forms with dlog singularities on boundaries of positive geometries" in this world of ideas. Interesting numerators are instead associated with new singularities at infinite momentum, not captured by ``dlog'' singularities. This basic fact has been a thorn in the side of the ``amplitudes=combinatorial geometry" program for over a decade, going back to ``poles at infinity" seen in non-supersymmetric BCFW shifts,  and non-dlog forms in the Grassmannian required to describe leading singularities in non-supersymmetric gauge theory \cite{Arkani-Hamed:2012zlh}. Clearly the realistic non-supersymmetric Yang-Mills theory demands a move away from such ``dlog forms", but in what setting, and how? Until very recently, there was no plan of attack to make progress on this fundamental mystery. 

But in~\cite{Zeros} we reported a wonderful surprise that has transformed the situation. Far from the toy model it appears to be, the ``stringy'' Tr $\phi^3$ amplitudes secretly {\it contains} the scattering amplitudes for pions~\cite{NLSM}, as well as non-supersymmetric gluons, in any number of dimensions. The amplitudes for the different theories are given by one and the same function, related by a simple shift of the kinematics. Working at tree-level, the stringy Tr $\phi^3$ amplitude is given as
\begin{equation}
{\cal A}^{{\rm Tr} \phi^3} = \int_0^\infty \prod \frac{dy_i}{y_i} \prod_{i,j} u_{i,j}^{\alpha^\prime X_{i,j}},
\end{equation}
where $X_{i,j} = (p_i + p_{i+1} + \cdots +p_{j-1})^2$ are the natural planar variables occurring as poles in the amplitude. 
This same function remarkably determines Yang-Mills amplitudes by a simple kinematical shift: 
\begin{equation}
{\cal A}^{{\rm gluons}}(X) = {\cal A}^{{\rm Tr} \phi^3}(X_{e,e} \to X_{e,e} + 1/\alpha^\prime, X_{o,o} \to X_{o,o} - 1/\alpha^\prime, X_{e,o} \to X_{e,o}),
\end{equation}
where $X_{e,e},X_{o,o},X_{e,o}$ refer to $X_{i,j}$ where the pair of indices $(i,j)$ are both even, both odd, or one of each respectively. 

To be more precise, this kinematical shift gives us the amplitude for what we call ``scaffolded gluons": an amplitude for $2n$ colored scalars coming in $n$ pairs of distinct flavors. As we will explain in detail, completely general amplitudes for $n$ gluons, with arbitrary choices of (null) polarizations, are captured by these ``scaffolded" amplitudes the ``scaffolding residues" are taken, corresponding to factorizing this amplitude on the channel where the pairs of same-species scalar fuse into gluons. This will turn out to be a simple and powerful picture both for presenting gluon amplitudes in an entirely canonical way free of the familiar annoying redundancies, as well as seeing their hidden relationships with other theories. 
\begin{figure}[t]
    \centering
    \includegraphics[width=0.85\textwidth]{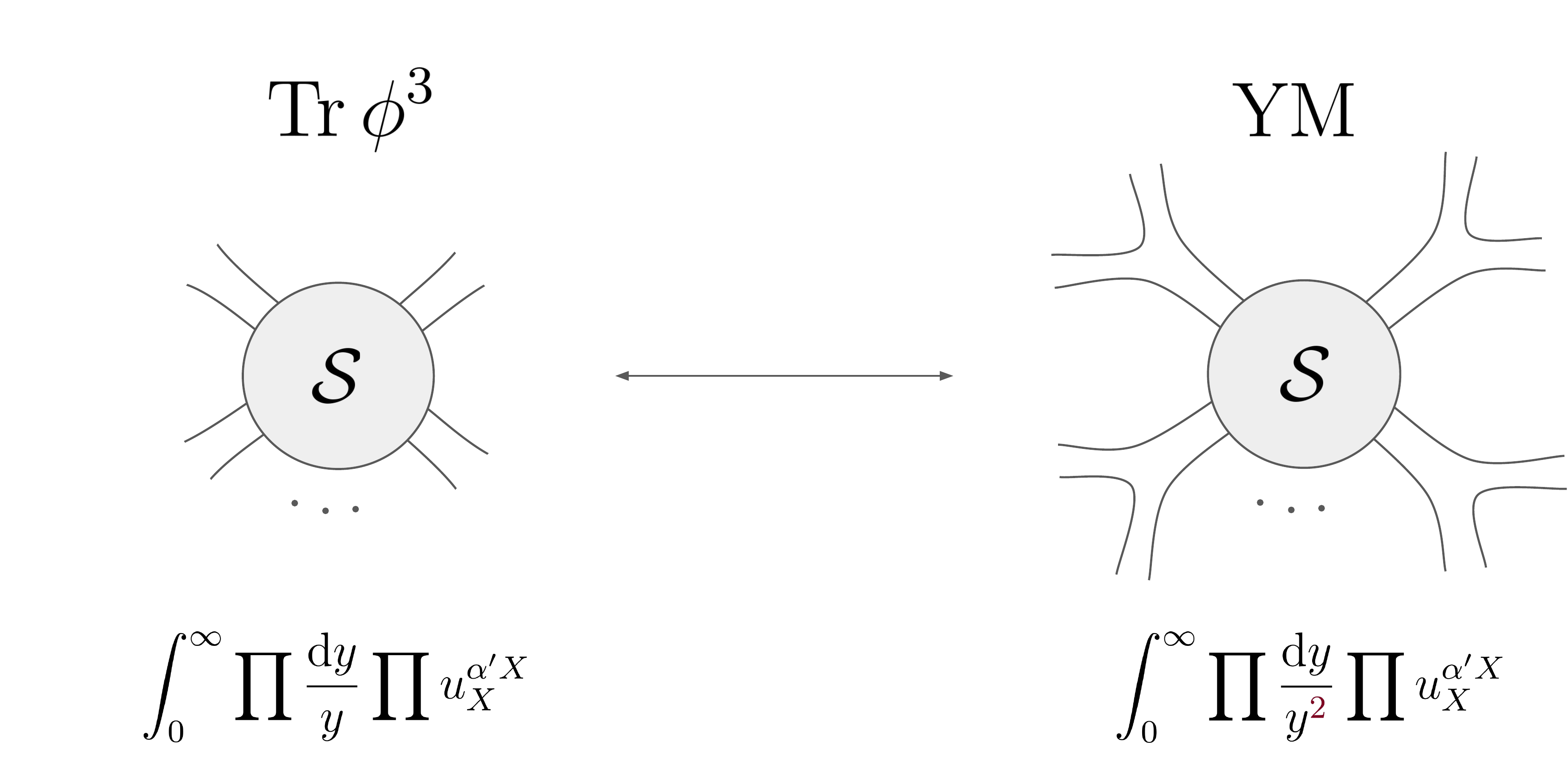}
    \caption{From colored scalars to scalar-scaffolded gluons.}
    \label{fig:ScalarToGluon}
\end{figure}

Our kinematic shift can equivalently be described by multiplying the Tr $\phi^3$ integrand by a factor $(\prod u_{e,e}^{+1} u_{o,o}^{-1})$. This factor has a beautiful interpretation directly in terms of the $y$ variables, if we choose a particularly natural triangulation for our $2n$-gon surface. We simply choose any triangulation/fat graph we like for the ``inner" $n$-gon, and turn it into a ``scaffolding triangulation'' by splitting each of the $n$ external legs in two, just as suggested from the scalar scaffolding picture. Quite beautifully, this product $(\prod u_{e,e}^{+1} u_{o,o}^{-1}) = \prod_i y_i^{-1}$. This means that the entire effect of the kinematical shift for the $2n$ scalar problem is to replace the dlog-form measure $\prod dy_i/y_i \to \prod dy_i/y_i^2$ (see figure \ref{fig:ScalarToGluon})! Note this is no longer a dlog form, as had long been anticipated was necessary for non-supersymmetric Yang-Mills theory. 

As shown in~\cite{Zeros} and discussed at further length in this paper, at tree level this kinematic shift precisely matches gluon amplitudes in the bosonic string, choosing the momenta to lie in $d$ dimensions and pairs of polarizations to live in $n$ extra dimensions. In this paper we will go much further, and propose that, appropriately interpreted, this expression gives us a definition for the amplitudes of non-supersymmetric Yang-Mills theory at all loop orders, in any number of spacetime dimensions. 
But where does this strange kinematic shift come from? And why on earth should it have had anything to do with describing gluon amplitudes? 

We will see striking answers to these questions throughout this paper. Clearly generic functions of $X$'s for a $2n$ problem can not be interpreted as amplitudes for scaffolded gluons. Instead as we will see these functions must have a symmetry under certain linear shifts of the $X_{i,j}$, guaranteeing both {\bf gauge invariance} and {\bf multilinearity} of the gluon amplitudes they are describing. These properties emerge magically from the shifted integral with the $dy/y^2$ measure, when the scaffolding residue to land on gluon amplitudes is taken. We have already remarked that the fundamental ``binary" property of the $u$ variables--that when $u_{\cal C} \to 0$ the $u_{\cal D} \to 1$ for ${\cal D}$ intersecting ${\cal C}$--is responsible for factorization and therefore the right to call the curve integral an ``amplitude'' for the scalar Tr $\phi^3$ theory. Shifting to the $dy/y^2$ form and taking the scaffolding residue ends up also probing the behavior of the $u$'s in the first-order neighborhood of where they vanish, which takes us beyond the most basic ``binary" properties. Amazingly, fundamental identities relating $u$ variables of the inner $n$-gon and outer $2n$-gon problems give exactly the needed first-order behavior to force the integral to have the symmetries needed for gauge invariance and multilinearity!

This is already strong evidence that we must be describing gluons, but to clinch the case we have to show consistent factorization on gluon poles, both for tree-like factorization, as well the loop-level  ``single cut''. As we will see, once again properties of the $u$ variables that follow transparently from their definition in terms of the counting problem, give precisely the needed identities for these ``tree-cut'' and ``loop-cut = forward limit'' factorizations to arise exactly as needed. All of these properties hold not just in the field theory limit but also at finite $\alpha^\prime$. 

We find it fascinating that elementary but deep properties of the $u$ variables, ultimately deriving from the simple counting problem associated with curves on surfaces that define them \cite{Arkani-Hamed:2023lbd, Arkani-Hamed:2023mvg}, contain exactly the magical properties needed to produce functions that can be interpreted as scattering amplitudes for gluons enjoying gauge invariance. This mirrors the famous ``discovery of gluons" arising from M{\"o}bius invariance of correlators for the spin-one vertex operators in string theory. But in this new picture there is no worldsheet, no conformal field theory living on it, and no vertex operators. Instead we have pure combinatorics, with the magical ability to produce ``gluon amplitudes" seemingly out of thin air!

In the course of these explorations we will also see that merely asking for our new integrals to have any kind of natural factorization properties on the poles, dictates how to more precisely interpret the $2n$ scaffolded formula. At loop level the product over curves ${\cal C}$ must include open curves with up to one self-intersection per loop. It must also include a product over closed curves $\Delta$ with exponents $u_\Delta^{1- D}$ where $D$ is the number of spacetime dimensions, which is where the dependence on spacetime dimensionality, absent in the Tr $\phi^3$ theory, makes an appearance for gluons. The need for including curves with one self-intersection per loop is very interesting, and shows a pronounced difference with what is expected of string theory at one-loop. For Tr $\phi^3$ theory it is already known that matching (bosonic) string theory at loop-level from the curve-integral formalism \cite{combstring} asks us to consider the product over {\it all} curves, including those with arbitrarily high intersection numbers. It is fascinating that in order to have consistent factorization on massless gluon poles--guaranteeing that we are describing Yang-Mills theory at low energies--we must make the curve-integral ``slightly more stringy" by including curves with a finite number of self-intersections. Of course, this is still vastly simpler than the infinite set of curves that string theory demands even to describe the scalar Tr $\phi^3$ theory. 

The consistent factorization of ``loop-cut = forward limit"~\cite{Feynman:1963ax, Caron-Huot:2010fvq} is especially non-trivial, where not only algebraic relations but total derivative identities are needed to establish the correspondence; nonetheless, they all fall out naturally from the counting problem formulae for the $u$ variables. As we will emphasize, this solves a long-standing conceptual problem afflicting all attempts to date to make sense of the notion of ``the Yang-Mills integrand" at loop-level for non-supersymmetric theories. The well-known problem is again a very basic one. The Feynman diagrams have diagrams with ``tadpoles" and hence $1/0$ propagators at the level of the integrand. But a general feature of the curve-integral formalism already for Tr $\phi^3$ theory allows us the flexibility to give any values we like to the $X$ variables--in particular, nothing mandates that the variables corresponding to tadpoles (or massless external bubbles) be set equal to zero. Leaving them free then lets us work with completely well-defined integrands. Remarkably, it is precisely these well-defined integrands that satisfy the exact ``loop-cut = forward limit" formula!  So the surface formalism {\it defines} objects for us that naturally enjoy a ``surface" notion of gauge-invariance, and also factorize precisely onto themselves. Amongst other things, this opens the door to finding all-loop recursion relations for pure Yang-Mills amplitudes in any number of dimensions~\cite{YMrec}.

We believe that the remarkable link between Tr $\phi^3$ theory, pions, and gluons reported in~\cite{Zeros}, extended to all loops in the present paper and~\cite{NLSM} represents a major step forward in the adventure of describing real-world scattering processes in the radically new language of combinatorial geometry. Our goal in this paper is to initiate an exploration of the new world of physical and mathematical ideas involved in this ``combinatorial discovery" of pure Yang-Mills theory, and present the most basic checks on our proposal, leaving many interesting and important questions for future work. 

We will begin in section \ref{sec:Surfaceology} with a review of ``surfaceology" and the curve integral formalism. This is intended as a crash-course introduction to all the formulas needed to understand the surface picture for kinematic space, a user manual for computing $u$ variables and using them to construct stringy amplitudes, and a compendium of important facts about $u$ variables we will use throughout this paper. This short section is meant as a self-contained summary of the facts we will need from the curve-integral formalism, whose origin and conceptual underpinning are presented in a different series of papers beginning with~\cite{Arkani-Hamed:2023lbd, Arkani-Hamed:2023mvg}; the review of $u$ variables will borrow results to appear in \cite{curvy}.  We next move on in section \ref{sec:scalarkin} to describing the picture of scalar-scaffolded gluons, establishing the simple dictionary between gluon and scalar variables, and understanding how the basic gluonic kinematic facts of on-shell gauge-invariance and linear dependence on polarization-vectors are given a unified description in scalar terms. We discuss gluon amplitudes at tree level in section \ref{sec:tree}, beginning by showing how our proposal precisely matches gluon amplitudes in the open bosonic string, with an appropriate choice of extra-dimensional polarizations.  We will see how the unified statements of gauge-invariance and multilinearity arise from simple properties of the $u$ variables. The same properties also imply consistent factorization of the amplitude on massless gluon poles. We present an explicit $n$-point formula for scaffolding residues, and see how the interpretation of the scalar-scaffolded amplitudes as those of extra-dimensional polarization gluons is naturally discovered in our language.  We also give simple examples of three- and four-point string/particle amplitudes. We move on to present our proposal at loop-level in section \ref{sec:loops}, where a number of essential new ingredients first make an appearance, and the difference with standard string amplitudes becomes stark.  This includes the need for keeping minimally self-intersecting curves as well as closed curves encoding the dependence on the number of spacetime dimensions we have alluded to above. Finally, we present a series of checks we have performed on our proposal. In section \ref{sec:LS}, we compute leading singularities corresponding to gluing together on-shell three-gluon amplitudes for arbitrary graphs: in our picture this turns into a simple residue computation in the $y$ variables, and we present non-trivial examples through two loops, and we find perfect matching with the gluing computation includes both the Yang-Mills and $F^3$ three-particle amplitudes. Of course this computation of leading singularities using residues in $y$ variables bears no resemblance at all to the direct gluing of three-particle amplitudes, and is also much more efficient. We then proceed to our most fundamental and important check in section \ref{sec:Loopcut}, showing how the single one loop-cut magically matches the tree-forward limit, at full ``stringy'' level, following from many further fundamental properties of $u$ variables hardwired into their combinatorial definition. We finally end with an outlook on obvious avenues for future exploration immediately following from the developments we discuss in this paper.

  %\newpage
  \section{Review of Surfaceology}\label{sec:Surfaceology}

A recent series of papers \cite{Arkani-Hamed:2023lbd} has introduced the curve integral formalism for describing both ``stringy" Tr $\phi^3$ amplitudes, as well as the closely related ``tropical" representation of these amplitudes in the field theory limit, at all orders in the `t Hooft topological expansion. The kinematic space for the amplitude at any order in the topological expansion is associated with a surface ${\cal S}$, which is specified by providing a triangulation or what is the same, a single ``fat graph" ${\cal F}$, whose vertices are associated with triangles of the triangulation, glued to each other as dictated by the edges. All the kinematic variables at this order in the topological expansion are naturally associated with curves that can be drawn on the surface or equivalently, paths that can be drawn on ${\cal F}$. This space of curves on ${\cal S}$ specifies the kinematic arena for Tr $\phi^3$ amplitudes. The heart of the curve integral formalism lies in the existence of a remarkable set of ``$u_{\cal C}$-variables" attached to any curve ${\cal C}$ on surface ${\cal S}$. As already mentioned in the introduction, these variables non-trivially satisfy the beautiful equations
\begin{equation}
u_{\cal C} + \prod_{{\cal D}} u_{{\cal D}}^{{\rm \# \, int}({\cal C},{\cal D})} = 1.
\end{equation}
These equations define a variety whose dimensionality is given by the number of internal edges of ${\cal F}$, or more fancily, the dimensionality of the Teichmuller space of ${\cal S}$. There are a variety of ways of presenting the solution space of these equations. One uses ideas from the hyperbolic geometry of ${\cal S}$. A more surprising and deeper presentation is fundamentally combinatorial, expressing $u_{\cal C}$'s as the solution to a certain extremely simple counting problem associated with the curve ${\cal C}$ as drawn on the fat-graph ${\cal F}$, giving $u_{\cal C}(y_i)$ as a ratio of polynomials in variables $y_i$ associated with all the internal edges of ${\cal F}$.

The $u$ variables are the stars of the Tr $\phi^3$ show in \cite{Arkani-Hamed:2023lbd}, playing many crucial roles. They manifest a ``binary" property, where asking $u_{\cal C} \geq 0$ also forces $0 \leq u_{\cal C} \leq 1$. This also implies that if some $u_{\cal C} \to 0$, all the $u_{\cal D} \to 1$ for the curves ${\cal D}$ that cross ${\cal C}$. Mathematically, this gives us a completely algebraic characterization of all possible degenerations of the surface. Said in more fancy (but not more contentful) language, the $u_{\cal C}$ variables give a new, practical, algebraic, and ``global'' definition of the compactification of the Teichmuller space of ${\cal S}$. This is extremely useful: all the usual ways of choosing coordinates for the surface are ``local"; as such they make some degenerations/singularities easy to see, but others are always more complicated, needing a detailed analysis and manual ``blow-ups" as the boundaries of the co-ordinate chart are approached. But the $u$ variables do all the blowups for us, once and for all. 

This ``binary'' property also has a crucial physical interpretation. If we consider a product of the form $\prod_X u_X^{P^2_X}$, then as any particular $u_{X^*} \to 0$, the product factorizes exactly to the same product on the smaller surface we get by cutting the big surface along $X^*$. This precisely reflects the factorization of scattering amplitudes when the propagator $P^2_{X^*} \to 0$ goes on-shell. This is what allows the curve integral formulas for both the ``stringy" and field-theoretic Tr $\phi^3$ theory to be interpreted as ``amplitudes" for these models. 

Amazingly, the $u/y$ variables turn out to have a number of further properties, well beyond the ``binary" property manifested in the $u$ equations, that allow simple kinematic shifts to transmute the seemingly ``dull" amplitudes of Tr $\phi^3$ theory into the amplitudes of gluons. The central goal of this paper is to show how this magic happens. 

But our goal in this section is to provide a self-contained introduction to all aspects of ``surfaceology" needed for the reader to follow all of our arguments, without needing to have previously read the series of papers associated with~\cite{Arkani-Hamed:2023lbd, Arkani-Hamed:2023mvg}. As such our purpose is to simply present the logical connection between various important concepts, and provide recipes for concrete computations, without explaining where the concepts and formulas come from. The reader interested in these conceptual underpinning as well as many concrete examples of full amplitudes computed using the curve integral formalism is encouraged to consult \cite{Arkani-Hamed:2023lbd, Arkani-Hamed:2023mvg}. Our review of the properties of the $u/y$ variables borrows from the complete account that will be given in \cite{curvy}.

\subsection{The Tr $\phi^3$ theory}

Let $\phi$ be an $N\times N$ matrix, the Lagrangian of the theory is:
\begin{equation}
	\mathcal{L} = \text{Tr} \left(\partial \phi\right)^2 + g \, \text{Tr} \, \phi^3.
	\label{eq:LTrphi}
\end{equation}
so $\phi$ is a massless field that interacts through cubic interactions with the coupling constant $g$ \footnote{It is trivial to add a mass term, but to keep things as simple as possible with our ultimate aim of describing gluons in mind, we will stick with the massless Tr $\phi^3$ theory.}. We use the double line notation to keep track of the color indices in the interactions.

Since this is a scalar theory, the data of a scattering process in this theory is simply the momentum of the particles, $p_i$. As for the amplitude, since it has to be a Lorentz-invariant function, it will only be a function of the \textbf{Mandelstam invariants}: $p_i \cdot p_j$. 

As we will explain in the next sections, the surface provides a very useful way of organizing the kinematic data of the problem. In addition, we can use it to read off the building blocks of string amplitudes -- the \textbf{u-variables}.

\subsection{Defining the surface}
\label{subsec:DefSurface}
\begin{figure}[t]
\begin{center}
\includegraphics[width=\textwidth]{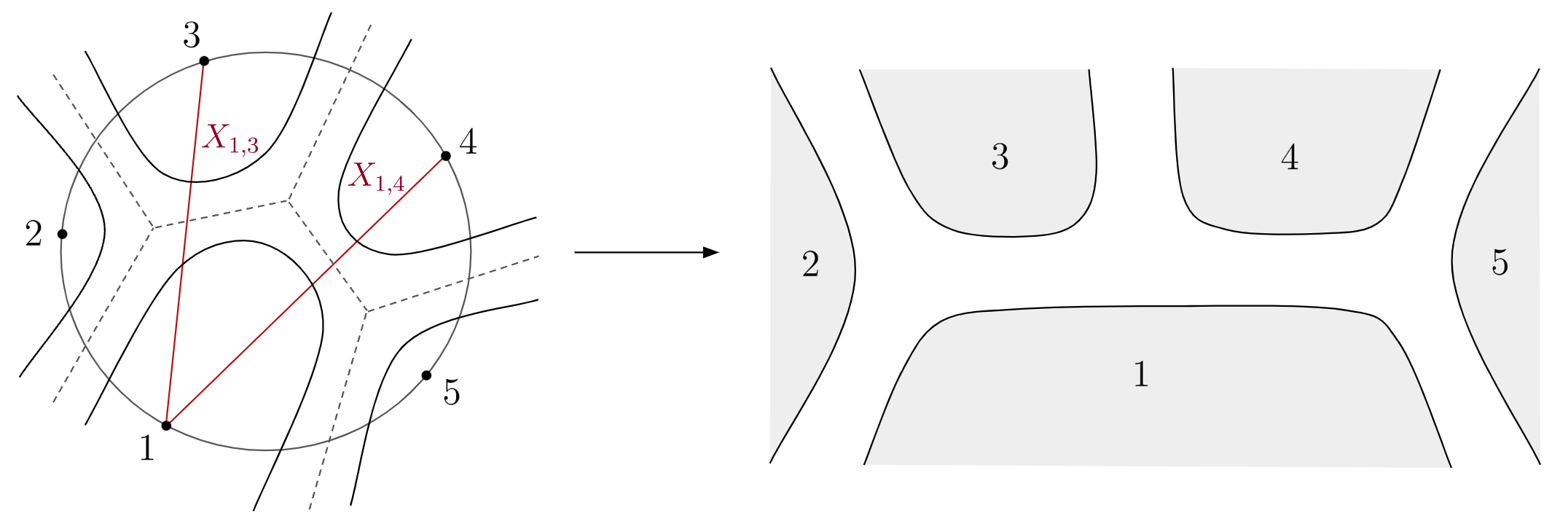}
\caption{Triangulation of the disk with 5 marked points on the boundary and corresponding dual fat graph.}
\label{fig:Triang5pt}
\end{center}
\end{figure}
The first step consists of defining the surface. By doing so, we specify both the scattering process and the order in the topological expansion we are considering.
  
A surface is defined as a collection of triangles together with a rule on how to glue the edges of the triangles together. So, one way of defining it is by providing a diagram, as it uniquely determines a triangulation via duality. 
  
For the most part in this paper, the surfaces we will be considering are disks with $n$ marked points on the boundary, and $l$ punctures -- corresponding to single-trace amplitudes of $n$ particles at $l$ loop order in the planar limit--though of course all surfaces corresponding to all orders in the `t Hooft topological expansion can be chosen. 

\subsubsection{Tree-level}

At tree level let us consider the case of a 5-pt amplitude, which corresponds to a disk with 5 marked points on the boundary. Taking into account the diagram with propagators $X_{1,3}$ and $X_{1,4}$, we obtain the triangulation presented in Figure \ref{fig:Triang5pt} (left).

The color graph, or ribbon graph, can be obtained from the Feynman diagram by fattening the edges of the graph and assigning arrows that keep track of the contraction of the color indices. However for our purposes, we only need the color graph without the color indices -- we call it the \textbf{fat graph}, as shown in figure \ref{fig:Triang5pt} (right). The fat graph is then dual to a triangulation of the surface and thus uniquely defines the surface. In fact, one can explicitly go from the fat graph to the surface: for example, in the 5-point case presented in figure \ref{fig:Triang5pt}, we can recover the disk with marked points on the boundary by shrinking the shaded regions to points. 

\subsubsection{Loop-level}

\begin{figure}[t]
\begin{center}
\includegraphics[width=\textwidth]{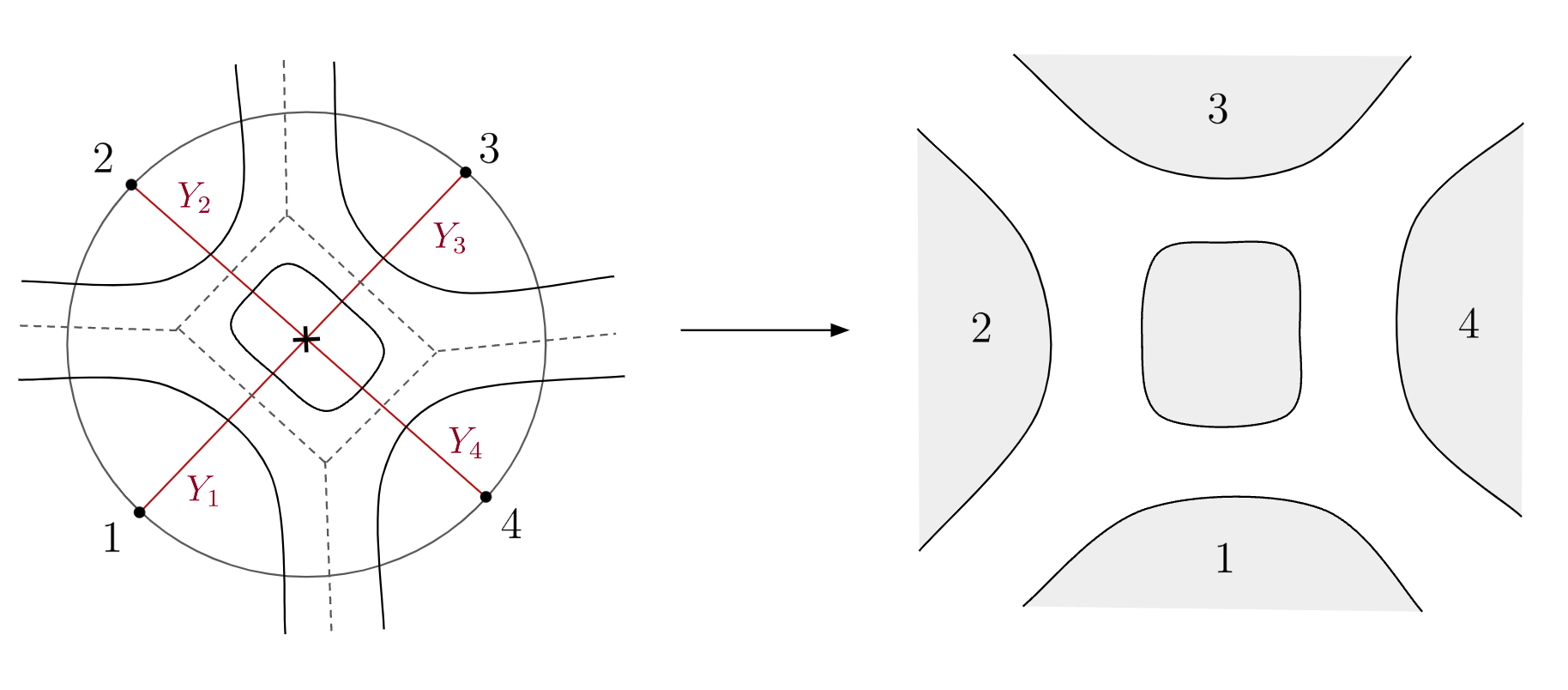}
\caption{Box triangulation of the punctured disk with 4 marked points on the boundary and corresponding dual fat graph.}
\label{fig:1loopbox}
\end{center}
\end{figure}

Let us now consider a loop example, a 1-loop amplitude at 4 points. The corresponding surface is now a disk with a puncture and four marked points on the boundary, for which we choose a box triangulation. We denote the four chords of this triangulation by $Y_i$, (instead of $X$) because they end on the puncture.  (see Figure \ref{fig:1loopbox} for triangulation (left) and corresponding fat graph (right)). 
Similarly, by shrinking the shaded regions of the fat graph, we can recover the punctured disk with 4 marked points on the boundary.
\subsection{Kinematical data and curves on the surface}

Next, we need to understand how the kinematics, the data of the scattering problem is encoded in the surface. 
  
The mapping can be summarized as follows:
\begin{center}
\begin{tabular}{c c c c c c }
 \hspace*{5mm} &\textbf{Kinematical Data } & \hspace*{5mm} & & \hspace*{5mm}&\textbf{Surface} \\ 
\hspace*{5mm} &\text{Propagators} & &$\xrightarrow{\hspace*{3cm}} $  & \hspace*{5mm} & \text{Chords/Curves},   [$X_{{\cal C}}$]\\  
\hspace*{5mm} &\text{Feynman Diagrams} & & $\xrightarrow{\hspace*{3cm}} $  & \hspace*{5mm}& \text{Triangulations}  $\{X_{i_1,j_1},...,X_{i_{n-3},j_{n-3}}\} $  \\
& & & & &
\end{tabular}
\end{center}

The scalar amplitudes we are considering are purely functions of the propagators, so, according to the mapping, all the data of the problem is encoded on the curves we can draw on the surface. Of course, naively, there seems to be an infinite number of curves one can draw on a surface, so we need to understand which ones we need to consider.
  
The answer is: we consider all curves that are not homologous to each other. Now we need to say how to consistently assign momenta to these curves. 

\subsubsection{Tree-level}
At tree level, we start by assigning momentum to the boundary curves of the disk, corresponding to the momentum of the external particles. Now the planar variables associated with the boundary curves are $X_{i,i+1}=p_i^2$, so for boundary curves, we have:  $X_{i,i+1}=0$, for massless particles. 
  
Now, at tree-level, we can read off the momentum assigned to a non-boundary curve,  $X_{i,j}$, by deforming it into the boundary, where it can be seen as the union of the boundary curves from $i$ to $i+1$,..., $j-1$ to $j$. So then the momentum of this curve is just the sum of the momentum of these boundary curves, and thus we get $X_{i,j} = (p_i + ... +p_{j-1})^2$, as we expect for the planar propagators. 

\begin{figure}[t]
\begin{center}
\includegraphics[width=\textwidth]{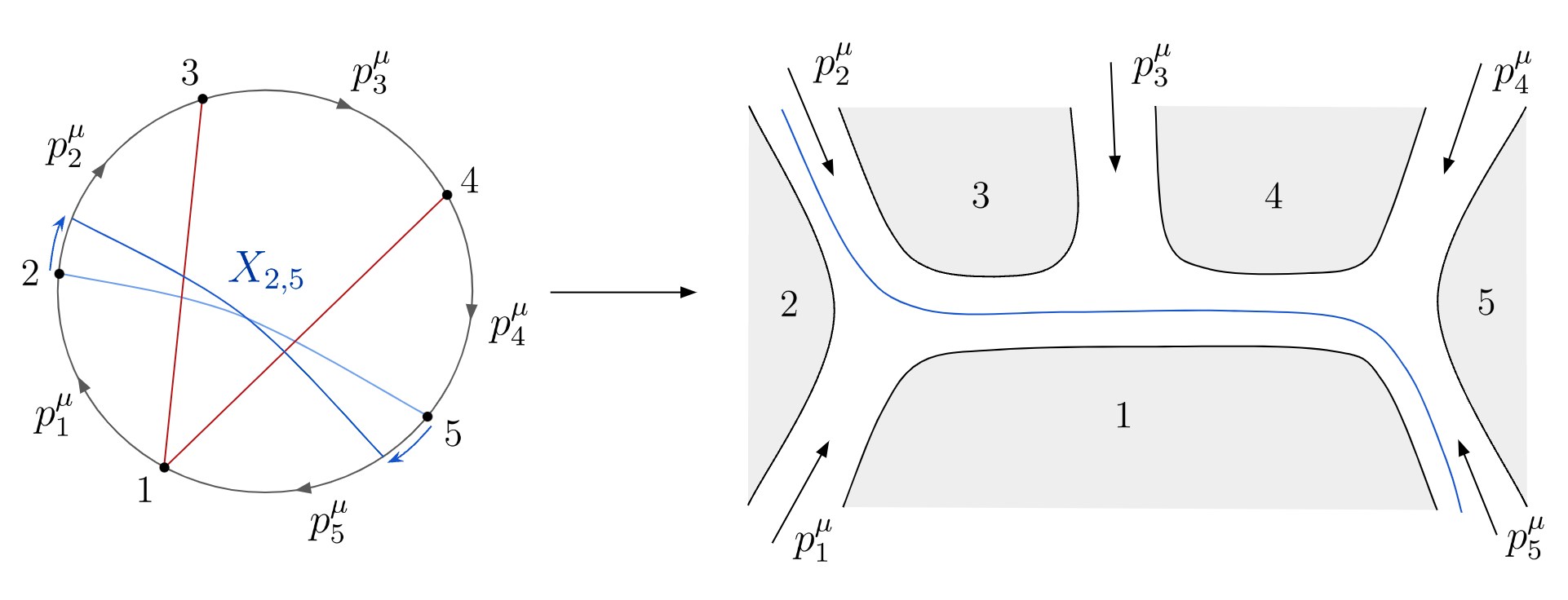}
\caption{Determining momentum of curve $X_{2,5}$.}
\label{fig:5ptMom}
\end{center}
\end{figure}
Alternatively, we can read off the momentum of a curve by using the fat graph. Let's go back to our 5-point example and consider the curve $X_{2,5}$, which on the surface side is connecting the marked points 2 and 5. In order to draw this curve on the fat graph, we first need to slightly rotate the ends of this curve \textbf{clock-wise} (see figure \ref{fig:5ptMom} (left)), this new curve goes by the name of a \textbf{lamination}. Laminations start and end in the boundaries of the disk, instead of marked points. These boundaries correspond to the edges of the fat graph, and this is why laminations are the objects that are naturally defined in the fat graph. From now on we will refer to the edges of the fat graph as \textbf{roads}, since, as opposed to 1-dimensional edges of graphs, the roads of the fat graph are crucially 2-dimensional. Therefore, the momentum of the boundaries of the disk, $p_i^{\mu}$, determines the momentum associated with the external roads of the fat graph. 

Thus to each curve on the surface, there is a unique lamination which is also defined in the fat graph. The distinction between a curve and a lamination will not be important for most of the matters we will encounter, so we will refer to curves and laminations interchangeably, and highlight the situations in which the difference is important.

So as a lamination, we can draw curve $X_{2,5}$ on the fat graph as shown in figure \ref{fig:5ptMom} (right). To read off the momentum, we follow the curve and start by writing down the momentum of the road in which the curve enters the graph -- so for $X_{2,5}$, we would write down $p_2^{\mu}$. Now each time the curve reaches an intersection and turns \textbf{right}, we add the momentum of the road coming from the \textbf{left} -- so for $X_{2,5}$, it turns right in the first intersection, so we add $p_3^{\mu}$, and right in the second intersection, so we also add $p_4^{\mu}$. If, instead, the curve turns left at an intersection, we don't add anything. In the end, the momentum of the curve is simply the sum of momentum obtained squared -- so for $X_{2,5}$ we get $(p_2+p_3+p_4)^2$, as expected. 

\subsubsection{Loop-level}
\begin{figure}[t]
\begin{center}
\includegraphics[width=\textwidth]{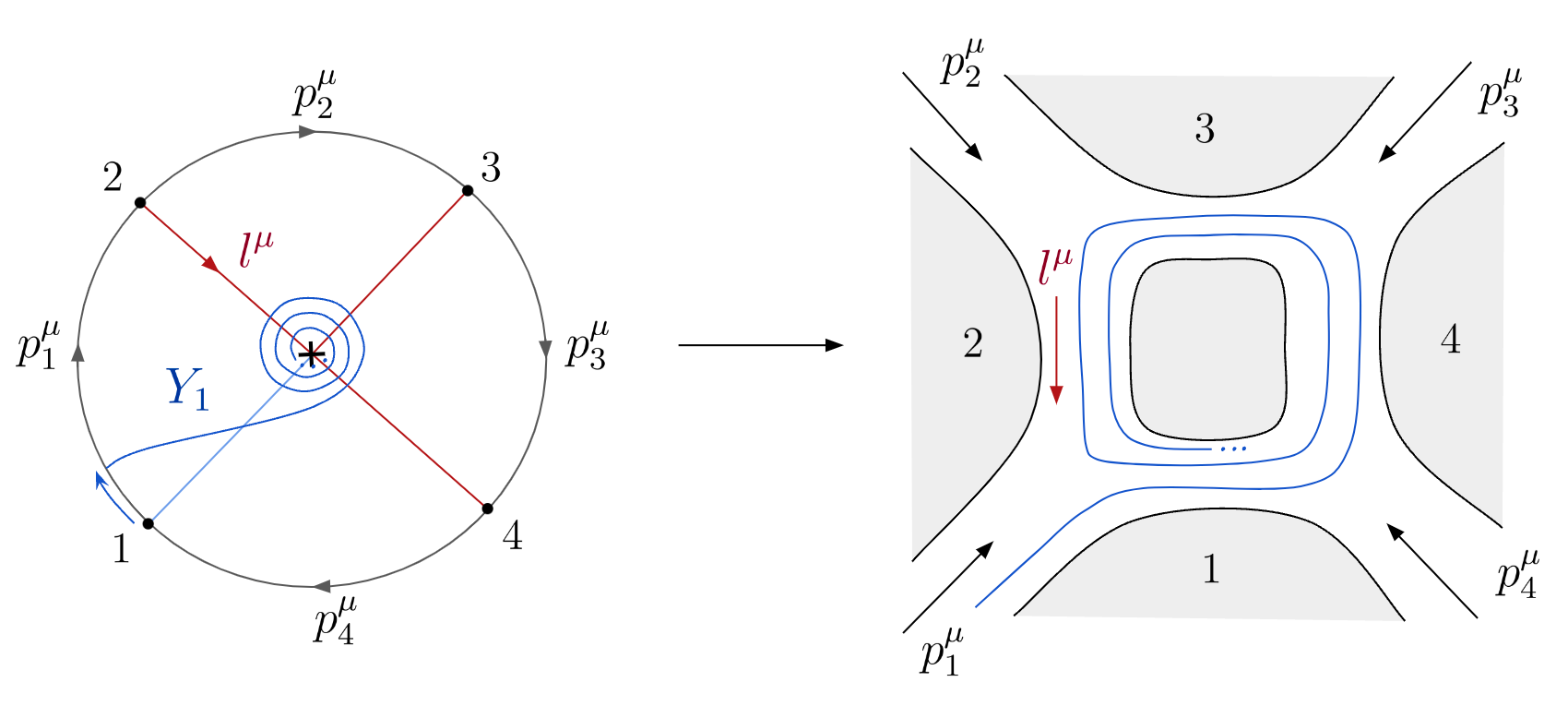}
\caption{Determining momentum of curve $Y_1$.}
\label{fig:LoopMom}
\end{center}
\end{figure}
At loop level, we also start by assigning momentum to the boundary curves. However, because of the presence of the puncture, we can now draw curves that are \textbf{not homologous} to any boundary, for example, any curve going from a boundary-marked point to the puncture. Therefore we need to assign a momentum to one of such curves -- this is exactly the loop momentum $l^{\mu}$ (see figure \ref{fig:LoopMom}, left).  
  
As opposed to the case of tree-level, at loop-level we now have a genuine infinite family of curves that are not homologous to each other. One way of seeing all such curves is to consider a curve starting and ending in some boundary-marked points. At tree-level, all such curves are unique, but in the presence of the puncture we can consider curves that start and end in the same points but that wind around the puncture $n$ times and thus self-intersecting $n$ times. In addition, another curve we can draw in the punctured disk is a \textbf{closed} curve. This closed curve is homologous to the full boundary and so has zero momentum. In any case, we will associate a variable, $\Delta$, to this curve and it will turn out to play an important role later. 
  
Just like in the tree-level case, we can read off the momentum of any curve using the fat graph. To do so we go from the curve to the lamination -- note that for curves that end in the puncture, the corresponding laminations spiral \textbf{counter-clockwise} around the puncture an infinite number of times (see figure \ref{fig:LoopMom}). Then we follow exactly the same rule as in the tree-level case: so for $Y_1$, we start by writing down $p_1^{\mu}$, then it turns right in the first intersection, so we add the momentum coming from the left which in this case is $l^{\mu}$, the loop momentum. Finally, it spirals around forever, and keeps turning left while doing so, which means we don't have to add any more momentum. Therefore we are left with $Y_1=(p_1+l)^2$. 

Having defined the mapping between the surface variables and the kinematic invariants, we now have to understand what operation we need to do on the surface that gives us the amplitude in terms of these variables. 

\subsection{From curves on the fat graph to words}

In this step, we want to find a way of encoding the information about the curves on the surface of some new objects. These objects should be such that using them one is fully able to reconstruct the surface as well as the curve in question. 
  
\begin{figure}[t]
\begin{center}
\includegraphics[width=\textwidth]{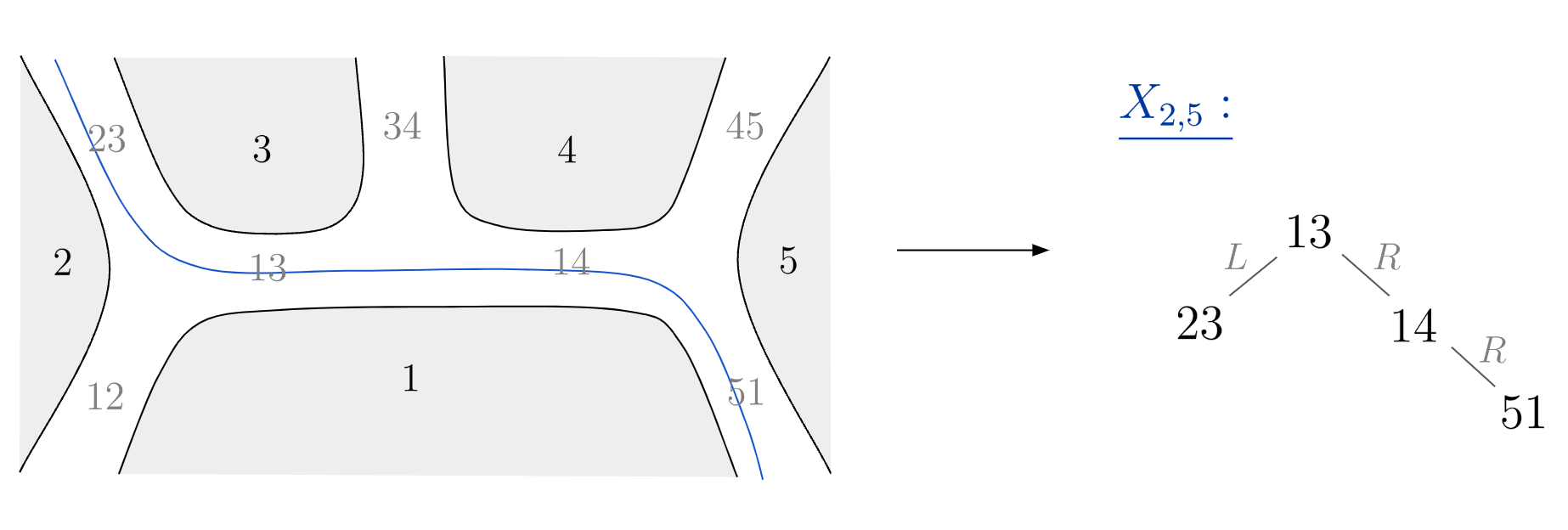}
\caption{From curve to word: curve $X_{2,5}$.}
\label{fig:CurveWord}
\end{center}
\end{figure}

To do this we use the fat graph defining the surface, and start by labelling its \textbf{roads} according to the sides they touch (see figure \ref{fig:CurveWord} (left)). 
   
Making use of this labelling, we can record the path of a given curve along the fat graph by keeping track of the roads it goes through as well as whether it turns right or left each time it reaches an intersection. So let's establish the rule that for each left-turn we go up, while for each right turn, we go down. By following this rule we build up a mountainscape uniquely associated with the curve we follow, we call this object the \textbf{word}.
  
As an example, let's consider a 5-point process at tree-level. Let's pick the fat graph with propagators $X_{1,3}$ and $X_{1,4}$, and look at the curve from 2 to 5 (see figure \ref{fig:CurveWord} (left)). Following the curve, we see it enters the graph in road 23, then it turns \textbf{left} to 13, then \textbf{right} to 14, and finally \textbf{right} to 45. Following the rule described in the previous paragraph, we get the word for curve $X_{2,5}$ (see figure \ref{fig:CurveWord} (right)).

The words are the most basic objects defined on the surface that encode all the kinematic data. The last step is to understand what manipulation we need to do on these words to obtain the amplitude. This involves building the \textbf{u-variables}.
\subsection{From words to $u$-variables}\label{sec:words to u}

At tree-level, the $u$-variables are well-studied objects in the literature, appearing already in the earliest days of the dual resonance model in the work of Koba and Nielsen~\cite{Koba:1969rw}, in the work of Francis Brown on the study of $\zeta$ values from stringy integrals~\cite{Brown:2009qja}, and directly in the context of binary geometries in~\cite{Arkani-Hamed:2019plo}. One of the central discoveries of the curve integral formalism \cite{Arkani-Hamed:2023lbd} is the generalization of $u$-variables to all surfaces. There are a number of ways of doing this, either framed in the language of hyperbolic geometry, or more abstractly, in terms of a simple counting problems associated with the words attached to curves on surfaces. This counting problem approach is both the more elementary and deeper one, and was summarized in \cite{Arkani-Hamed:2023lbd}. A more systematic exposition of all the different ways of thinking about $u$ variables from both the counting problem and hyperbolic geometry viewpoints, aimed at physicists, will be given in \cite{curvy,combstring}. 

To each curve, ${\cal C}$, on the surface we can associate a u-variable, $u_{{\cal C}}$, that satisfies the \textbf{u-equation}:
\begin{equation}
	u_{\cal C} + \prod_{\cal D}  u_{{\cal D}}^{{\rm \# int}({\cal C},{\cal D})}=1,
	\label{eq:u-eq}
\end{equation}
where the product is taken over all other curves  ${\cal D}$ and $\#{\rm int}({\cal C},{\cal D})$ the intersection number between the two curves. For positive $u$'s, we see that all $u$'s are bounded by 1. In addition, from equation \eqref{eq:u-eq}, when $u_{\cal C}$ goes to zero, all the $u_{\cal D}$, for the curves ${\cal D}$ that intersect ${\cal C}$, have to go to one. This \textbf{binary} behavior is crucial to ensure that amplitudes build out of the $u$'s satisfy \textbf{factorization} near poles. 
  
The $u$-equations form a non-linear set of equations that make it obvious that the $u$-variables are not independent of each other. So it is useful to find a good parametrization of the space of solutions of the $u$-equations. This is a non-trivial problem for which the surface gives a very nice answer.  
  
As we explained in \ref{subsec:DefSurface}, in order to define the surface one has to provide a fat graph, dual to a triangulation of the surface. Of course, there are multiple fat graphs that describe the same surface, in particular for an $n$-point tree, there are Catalan$_{n-2}$ such possibilities. Each choice of fat graph corresponds to a \textbf{positive parametrization} of the u-variables that automatically satisfies the u-equations. 
  
To obtain this parametrization we start by assigning variables to the internal roads of the fat graph -- the ones corresponding to the propagators, $y_i$. Now the $u$'s will be functions of the $y_i$'s, in particular, will be ratios of polynomials in the $y_i$'s. To get these functions we define two matrices:

\begin{equation}
	 M_{L}(y_{i}) = \begin{bmatrix}
		y_{i} & y_{i}  \\
		0 & 1 
	\end{bmatrix} \quad ;\quad M_{R}(y_{i}) = 
 \begin{bmatrix}
		y_{i} & 0  \\
		1 & 1 
	\end{bmatrix} ,
\end{equation}
where $L$ and $R$ stand for left and right. To build the u-variable from some word, $W_C$ of some curve $C$, all we need to do is multiply these matrices in the order the curve tells us to, $i.e.$ each time it turns left/right at some road $y_i$ we multiply by $M_L(y_i)/M_R(y_i)$,  respectively. Any non-boundary curve starts in some boundary road and then turns left or right into some internal road, since there is no $y$ associated with the starting boundary road, the first matrix appearing $M_L$ or $M_R$ should be evaluated at 1. By the end of this, we have a 2$\times$2 matrix, whose entries are polynomials in the $y$s, $m_{i,j} (y)$. The u-variable for the curve is simply:
\begin{equation}
	u_C = \frac{m_{1,2}\cdot m_{2,1}}{m_{1,1}\cdot m_{2,2}}.
	\label{eq:udefm}
\end{equation}
 
Let's go through an example to see how the rule works in practice. Consider the 5-pt tree level amplitude, so the disk with 5 punctures, and let's get the u-variable for curve $X_{2,5}$ (figure \ref{fig:CurveWord}). In this case, the sequence of $L/R$ matrices is:
\begin{equation}
	 M_{L}(1)  M_{R}(y_{1,3})  M_{R}(y_{1,4}) = \begin{bmatrix}
		1 & 1  \\
		0 & 1 
	\end{bmatrix} \begin{bmatrix}
		y_{1,3} & 0  \\
		1 & 1 
	\end{bmatrix} \begin{bmatrix}
		y_{1,4} & 0  \\
		1 & 1 
	\end{bmatrix} = \begin{bmatrix}
		1+y_{1,4}+y_{1,3}y_{1,4} & 1  \\
		1+y_{1,4} & 1 
	\end{bmatrix}.
\end{equation}

Hence from \eqref{eq:udefm}, we have:
\begin{equation}
	u_{2,5} = \frac{1+y_{1,4}}{1+y_{1,4}+y_{1,3}y_{1,4}}.
\end{equation}

The $u$ variables for the ``spiraling curves'' into punctures are interesting. Let us denote the chords in the triangulation hitting the puncture read counterclockwise as  $y^{(p)}_1,y^{(p)}_2,\cdots,y^{(p)}_r$. Then any word corresponding to a spiraling curve will end with the infinite repeating upward sequence: 

\begin{equation}
    \begin{matrix}
     & &  & & &  &  & &  \udots \\
    & &  & & &  &  & y^{(p)}_1 &\\
    & &  & & &  &  \diagup & &\\
       & &  & & &  y^{(p)}_r & & &\\
       & &  & & \udots  & && &\\
       & & &y^{(p)}_2 & & & && &\\
       & &\diagup & & & && &\\
       & y^{(p)}_1& & & & & && &\\
       \udots  & & & & & & && &\\
    \end{matrix}
\end{equation}

So at the end of the word, we encounter the matrix $M_{{\rm spiral}} = M_L(y^{(p)}_1) M_L(y^{(p)}_2) \cdots M_L(y^{(p)}_r)$, which is raised to the $N$'th power with $N$ sent to infinity. Now it is easy to see that
\begin{equation}
M^N_{{\rm spiral}} = \left(\begin{array}{cc} Y_{{\rm spiral}}^N  & F_{{\rm spiral}} ( 1 + Y_{{\rm spiral}} + \cdots + Y_{{\rm spiral}}^{N-1}) \\ 0 & 1 \end{array} \right),
\end{equation}
where 
\begin{equation}
Y_{{\rm spiral}} = (y^{(p)}_1 \cdots y^{(p)}_r), \quad  F_{{\rm spiral}} = y^{(p)}_1 + y^{(p)}_1 y^{(p)}_2 + \cdots + y^{(p)}_1 \cdots y^{(p)}_r.
\end{equation}

Not that in order for the geometric series to converge, we must have $Y_{\text{spiral}}<1 \Leftrightarrow \prod_i y_i^{(p)}<1$. This is a restriction we must impose on the $y_i^{(p)}$ above and beyond their positivity. 

If we call the matrix product for the part of the word before we begin spiraling as
\begin{equation}
    M_{{\rm before}} = \left(\begin{array}{cc} a & b \\ c & d \end{array} \right),
\end{equation}
then if we take the $u$ variable for the matrix $M_{{\rm before}} M_{{\rm spiral}}^N$ and take $N \to \infty$ we find 
\begin{equation}
    u_{{\rm spiral}} = \frac{c (a F_{{\rm spiral}} + b (1 -  Y_{{\rm spiral}}))}{a (c F_{{\rm spiral}} + d (1 - Y_{{\rm spiral}}))}.
\end{equation}

\subsubsection{u-variable for closed curve}
The rules described above allow us to obtain the $u$ variables for all curves defined on the surface. We will also have occasion to define $u$ variables for closed curves around the punctures, $\Delta$, for which we put:
\begin{equation}
	u_{\Delta_I} = \left(1 - \prod_{i} y_{i}^{(p_I)} \right),
\end{equation}
where $\Delta_I$ stands for the closed curve around puncture $I$, and $y_{i}^{(p_I)}$ are the $y$'s present in the underlying triangulation that go from one boundary point, $i$, and end in the puncture, $I$. For example, let's consider again the punctured disk with 4 marked points on the boundary -- the 4-point one-loop amplitude. Choosing the box triangulation of figure \ref{fig:1loopbox}, means we have four chords going from the boundary to the puncture, $Y_1,Y_2,Y_3$ and $Y_4$. So in this case, the u-variable for the closed curve is:
\begin{equation}
	u_{\Delta}= \left(1 - y_1^{(p)} y_2^{(p)} y_3^{(p)} y_4^{(p)}\right).
\end{equation} 

Note this $u_{\Delta} \to 0$ precisely when the product $Y_{{\rm spiral}}$ we encountered for spiraling $u$ variables hits the boundary $Y_{{\rm spiral}} \to 1$ we just saw above.  So it is natural to have $u_{\Delta}$ detect this boundary. 

Using the relation between the $y$ and $u$ variables we will review just below, it is easy to see that this definition is independent of the choice of triangulation; we have that $\prod_i y_i^{(p)} = \prod_Y u_{Y}$ where the $Y$ denote all the curves spiraling at the puncture.

\subsection{Summary of $u$ facts}

Here present a summary of a few facts about the $u$ variables that will be useful throughout the rest of the paper. These results are all borrowed from \cite{curvy} where a systematic understanding of these and many other properties of $u/y$ variables will be given. 

\subsubsection{$u$'s of sub-surfaces} 

Suppose we have a surface $\mathcal{S}$, and a subsurface $s$ contained inside it. (We can concretely define a subsurface $s$ simply by giving its fat-graph as a subgraph of the parent fat-graph for ${\cal S}$). It is natural to wonder whether we can build out of the $u$ variables $u^{(\mathcal{S})}_\mathcal{C}$ for curves $\mathcal{C}$ variables on the big surface $\mathcal{S}$, a set of objects $u^{(s)}_c$ that satisfy the $u$ equations labelled by curves $c$ on the smaller surface $s$. This question has a beautiful and natural answer: we can build $u$ variables for the small surface out of simple monomials in the $u$ variables for the big one. Given a curve $c$ in $s$, we simply consider {\it all} ways of extending $c$ into curves on the big surface, $\mathcal{C}_c$. Then we have 
\begin{equation}
u^{(s)}_c = \prod_{\text{ all curves } \mathcal{C}_c, \text{ extending }   c  \text{ into } \mathcal{S}} u^{(\mathcal{S})}_{\mathcal{C}_c}.
\end{equation}

\subsubsection{Generalized $u$-equations}

We have just seen that the $u$ variables for smaller surfaces $s$ can be expressed as monomials in the $u$ variables of the big surface $\mathcal{S}$. Thus the $u$ equations for the smaller surface $s$ turn into new equations when these are expressed as monomials in the original surface $\mathcal{S}$--these are the {\it generalized $u$ equations}. For instance for the disk at tree-level, we can consider the subsurface given by a 4-point fatgraph bounded by regions $i,a,j,b$ as in figure \ref{fig:subsurface}. The curve $(i,j)$ in this 4-point subsurface extends to the bigger tree into all the curves $(x,y)$ with $x$ in the set between $(i,a-1)$ and $y$ in the set between $(j,b-1)$. Then we have $u^{(4)}_{i,j} = \prod_{x \subset(i,a-1), y \subset(j, b-1)} u_{x,y}$. Similarly the curve $(a,b)$ in the subsurface extends to all the $(w,z)$ with $w$ between $(a,j-1)$ and $z$ between $(b,i-1)$, so $u^{(4)}_{a,b} = \prod_{w \subset (a,j-1), z \subset (b,i-1)} u_{w,z}$. 
Then from the trivial 4-point $u$ equations we get a non-trivial expression for sums of monomials in the parent $u$'s: 
\begin{figure}[t]
\begin{center}
\includegraphics[width=0.6\textwidth]{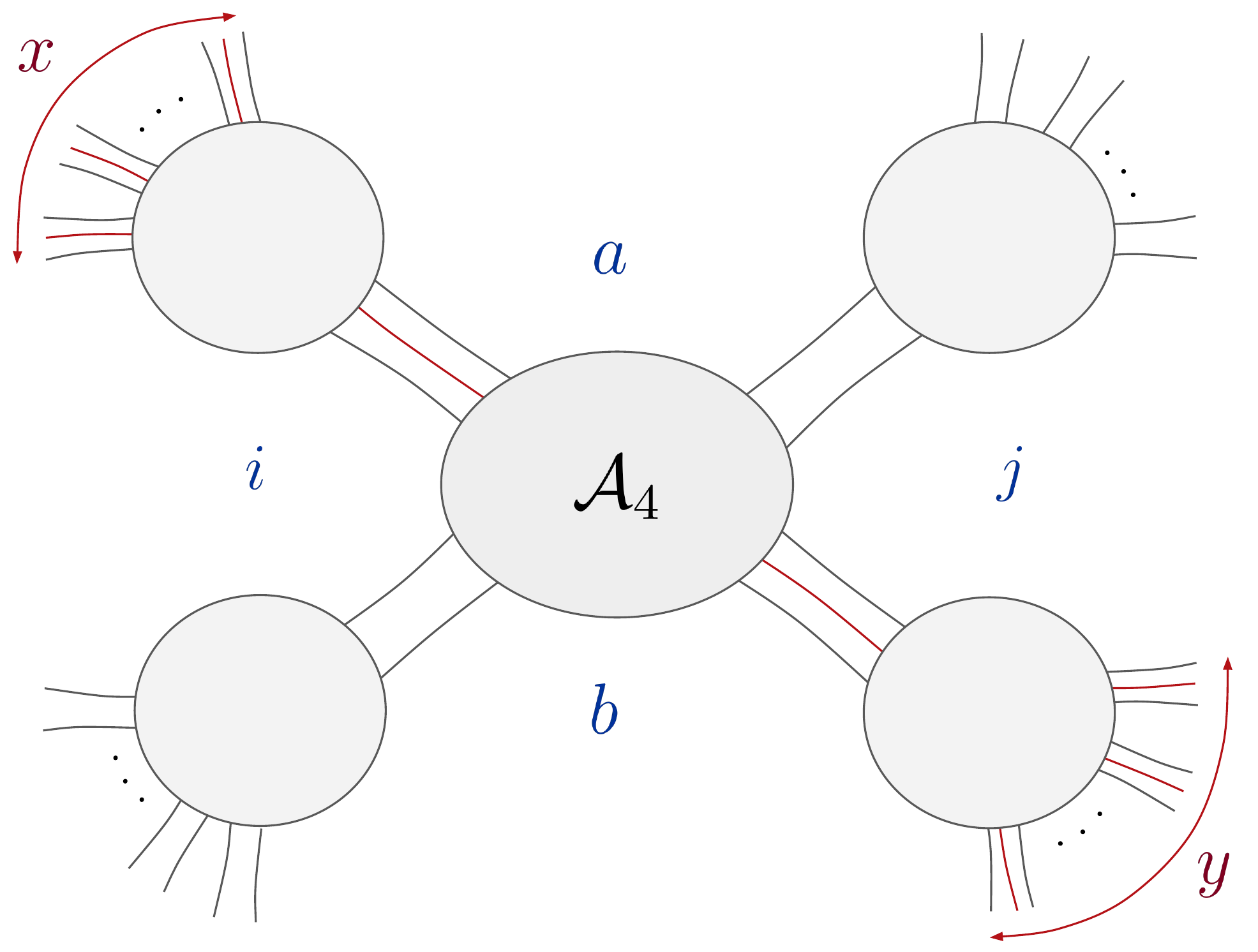}
\caption{Smaller 4-point inside bigger surface.}
\label{fig:subsurface}
\end{center}
\end{figure}
\begin{equation}
u^{(4)}_{i,j} + u^{(4)}_{a,b} = 1 \rightarrow \prod_{x \subset(i,a-1), y \subset(j, b-1)} u_{x,y} + \prod_{w \subset (a,j-1), z \subset (b,i-1)} u_{w,z} = 1.
\end{equation}

All the extended $u$ equations at tree-level can be summarized by a simple picture: we divide all the indices going around the disk into four consecutive sets $(A,B,C,D)$. Then the extended equations are $U_{A,C} + U_{B,D} = 1$, where $U_{A,C}$ is the product of all the $u_{i,j}$ with $i$ in $A$ and $j$ in $C$, and $U{B,D}$ is the product of all $u_{k,l}$ with $k$ in $B$ and $l$ in $D$. The special case where $A$ is the single $i$ and $C$ is the single $j$, with $B,D$ filling out the complementary indices, gives us the standard $u$ equation for $u_{i,j}$. As an example of a new generalized $u$ equations for $n=6$ we can take $A = (1,2), B = (3), C = (4), D = (5,6)$, then we have $u_{1,4} u_{2,4} + u_{3,5} u_{3,6} = 1$.

\subsubsection{$y$'s and $u$'s}
\label{sec:prody}
We now explain how to express the product of all the $y$'s in terms of $u$'s, for any given triangulation of the surface.
  
We can write each $y$ as a product of $u$'s in the following way:
\begin{equation}
	y_i = \prod_{X} \, u_X^{g_X^{(i)}},	\label{eq:yi}
\end{equation} 
where $g_X^{(i)}$ stands for the $i$-th component of the \textbf{g-vector} corresponding to curve $X$. We can get $g_X$ by looking at the corresponding word describing the curve. This word is a mountainscape with some roads appearing as peaks, $\mathcal{P}$, and others as valleys, $\mathcal{V}$. These peaks and valleys are always roads associated with the propagators of the underlying triangulation/ internal edges of the underlying fat-graph. To each such internal road we assign a vector, $\vec{g}_{i}$, then the $g$-vector of some curve, $X$, can be written as a linear combination of $\vec{g}_i$ as follows:
\begin{equation}
	\vec{g}_X = - \sum_{I\in\mathcal{P}} \vec{g}_{I} + \sum_{J\in\mathcal{V}} \vec{g}_{J}.
	\label{eq:peakValley}
\end{equation} 
\begin{figure}[t]
\begin{center}
\includegraphics[width=0.8\textwidth]{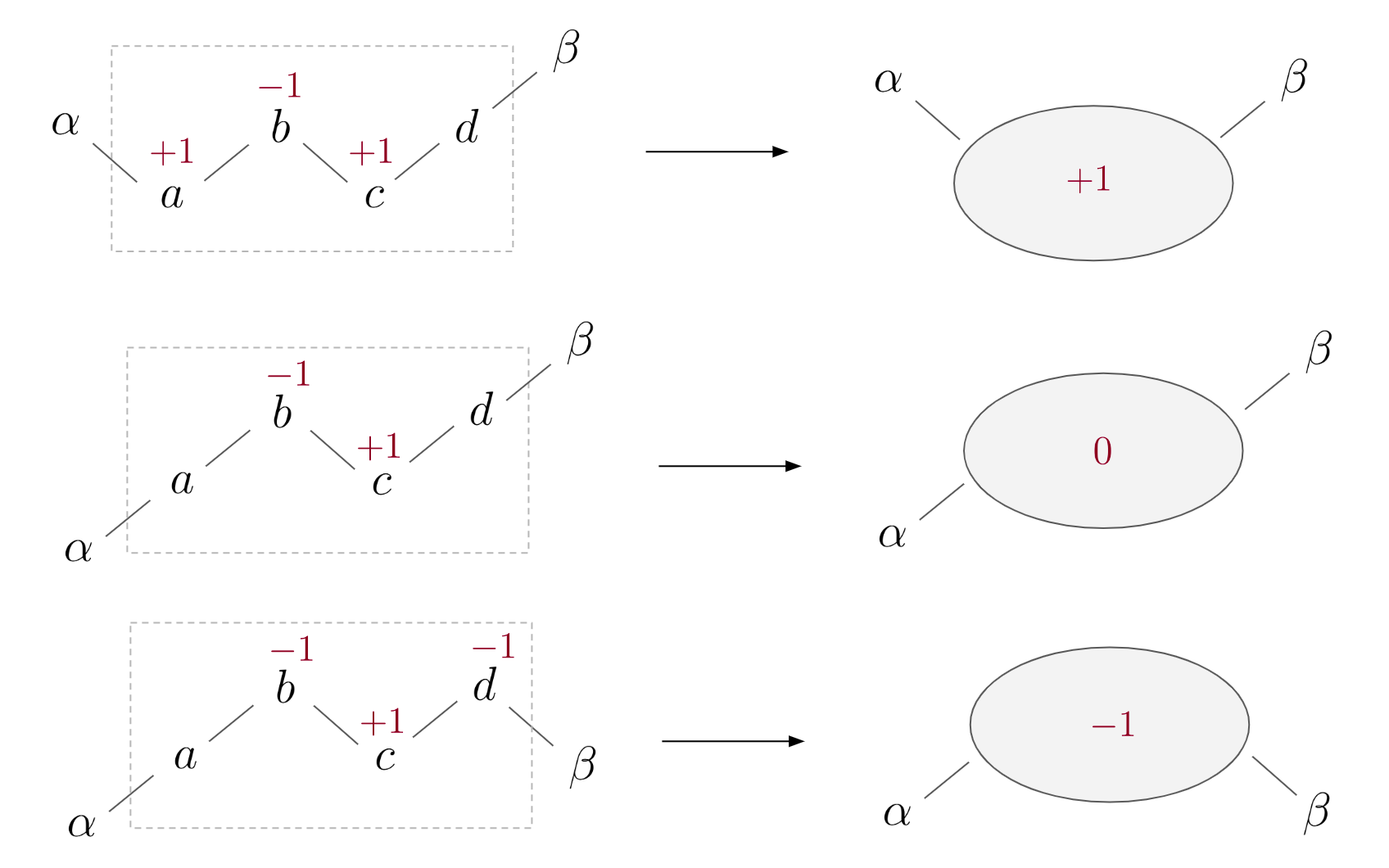}
\caption{$P_X$ for different words.}
\label{fig:Px}
\end{center}
\end{figure}

Therefore we have that $g_X^{(i)}$ will either be $\pm 1$ depending on whether $i$ appears as a valley or a peak in the word for $X$. 
Ultimately, the product of all the $y$'s is simply:
\begin{equation}
	\prod_i y_i = \prod_{X} \, u_X^{p_X}, \quad p_X \equiv \sum_i g_X^{(i)}.
\end{equation}

From \eqref{eq:peakValley}, we have that $p_X = \#$ (valleys) - $\#$ (peaks), appearing in the word for curve $X$. It turns out that $p_X$ is completely determined by what the word looks like at the boundaries since the net number of right/left turns in the middle must even out (see figure \ref{fig:Px}).In summary, we have:
\begin{equation}
	\prod_I y_I = \prod_X u_X^{P_X}, \quad \text{with }\, p_X = \begin{cases}
		+1, \quad &\text{if $X$ turns $R$ at the start and $L$ at the end,}\\
		-1, \quad &\text{if $X$ turns $L$ at the start and $R$ at the end,}\\
		\quad 0, \quad &\text{otherwise.}
	\end{cases}
 \label{eq:prody}
\end{equation}

\subsection{``Stringy'' amplitudes}
Now we want to explain how we can use the variables defined from the surface, such as the curves and respective $u$-variables, to build string amplitudes. 
\subsubsection{Tree-level} 
String amplitudes are given as integrals over possible configurations of the string worldsheet. An $n$-point string tree-level amplitude corresponds to an integral over the moduli-space of real points $z_1,\dots,z_n$ on the boundary of the disk~\cite{Veneziano:1968yb, Koba:1969rw}:
\begin{equation}
	\mathcal{A}^{\text{tree,tachyon}}_n(1,2,...,n) = \int \frac{\mathrm{d} z_1 \ldots \mathrm{d} z_n}{\mathrm{SL}(2, \mathbb{R})}\times \underbrace{ \prod_{i<j}\,	(ij)^{2 \alpha^\prime k_i \cdot k_j}}_{\text{Koba-Nielsen factor}}\label{eq:StringAmpZ},
\end{equation}
\\
with $z_1<z_2<\dots<z_n$, so that $(ij) = z_j -z_i >0$ for $i<j$, $k_i$ are the particle momenta and $\alpha^\prime$ is the string constant.  The worldsheet SL$(2,\mathbb{R})$ invariance demands that $\alpha^\prime k_i^2= -1$ and so the external states are tachyons with $m^2 = -(\alpha^\prime)^{-1}$. \footnote{We use the mostly-plus signature so that $k^2 = -m^2$. }. 
  
It turns out that the $z_i$'s also define a parametrization of the solution to the $u$ equations in which each $u$ variable is an  $\text{SL(2,}\mathbb{R})$ invariant cross-ratios of the $(ij)$'s defined as follows:
\begin{equation}
    u_{i,j} = \frac{(i{-}1,j)(i,j{-}1)}{(i,j)(i{-}1,j{-}1)}.
	\label{eq:udef}
\end{equation}

We can now express the integrand in terms of $u$'s, finding~\cite{Arkani-Hamed:2017mur, Arkani-Hamed:2019mrd, Arkani-Hamed:2019plo}:
\begin{equation}
	\mathcal{A}^{\text{tree,tachyon}}_n(1,2,...,n) = \int \frac{\diff^{n} z_i/\text{SL(2,}\mathbb{R})}{(12)(23)(34)...(n1)}\, \prod_{i<j} u_{i,j}^{\alpha^{\prime} (x_{i,j}-1)},
	\label{eq:tachyon}
\end{equation}
where the $x_{i,j} = (k_i + ...+k_{j-1})^2$ are the planar-variables made from the external tachyon momenta. We will find it convenient to instead interpret this expression not as scattering amplitudes for tachyons but for massless external scalars with momenta $p_i$ with $p_i^2=0$, simply by identifying the planar kinematic invariants for the massless problem $X_{i,j} = (p_i + \cdots + p_{j-1})^2$  with those of the tachyons $x_{i,j}$ via $X_{i,j} = x_{i,j} -1$. Thus we study finally 
\begin{equation}
	\mathcal{A}^{\text{tree}}_n(1,2,...,n) = \int \frac{\diff^{n} z_i/\text{SL(2,}\mathbb{R})}{(12)(23)(34)...(n1)}\, \prod_{i<j} u_{i,j}^{\alpha^{\prime} X_{i,j}},
	\label{eq:Trphi3U}
\end{equation}
This is useful for talking about a ``low-energy field theory limit", something which doesn't make sense for tachyons whose mass is set by the string scale, but does make sense for the massless $X_{i,j}$ kinematics (where we also set $X_{i,i+1} = 0$).  Indeed the low-energy limit of these ``Parke-Taylor'' or ``$Z$-theory'' amplitudes~\cite{Mafra:2011nw, Mafra:2016mcc, Carrasco:2016ldy}, where $\alpha^\prime X_{i,j} \ll 1$, are just those of massless Tr $\phi^3$ theory \eqref{eq:LTrphi}. For this reason we will refer to this as the bosonic string Tr $\phi^3$ amplitude in the rest of this paper.  
 
We can easily convert between the $z_i$ and $y_i$ variables that have both been used to give parametrizations of the same $u_{i,j}$ variables. Doing this we find that the ``Parke-Taylor'' measure written in terms of the $z_i$ variables simplifies to a ``dlog form'' in the $y_i$, and the string integral takes it's simplest form as: 

\begin{equation}
	\mathcal{A}^{\text{tree}}_n(1,2,...,n) = \int_{0}^{\infty} \prod_{i}\frac{\diff y_i}{y_i} \, \prod_{C} u_C^{\alpha^{\prime} X_C},
	\label{eq:treeStringSurface}
\end{equation}
where the product is taken over all the curves $C$ on the disk with $n$-marked points on the boundary, and $u_C$ are the corresponding $u$-variables parametrized by the $y_i$.

\subsubsection{Loop-level} 

The $n$-point 1-loop amplitude in terms of surface variables is given exactly by \eqref{eq:treeStringSurface} but where the product is taken over all the curves of the punctured disk with $n$ marked points on the boundary. As pointed out before, at 1-loop, there are now an infinite number of curves coming from curves that self-intersect infinitely often. So we can rewrite this product as follows:

\begin{equation}
		\mathcal{A}^{\text{1-loop}}_n(1,2,...,n) = \int_{0}^{\infty} \prod_{i}\frac{\diff y_i}{y_i} \, \prod_{C} u_C^{\alpha^{\prime} X_C} \times \prod_{C^\prime \in \, \text{ S.I.}} u_{C^\prime}^{\alpha^{\prime} X_{C^\prime}} \times u_{\Delta}^\Delta,
	\label{eq:loopStringSurface}
\end{equation}
where the first product includes a finite number of terms corresponding to curves that never self-intersect. The second product is an infinite product over all the self-intersecting curves with different numbers of self-intersections. At last, we have the closed curve contribution, $\Delta$. 

A precise matching between this form of the amplitude and the conventional presentation of tachyon bosonic string amplitudes through 1-loop will appear in \cite{combstring}. But as stressed in \cite{Arkani-Hamed:2023lbd}, at loop-level there are more options for ``stringy'' presentations of amplitudes at least so long as we are primarily interested in matching the correct field theoretic low energy limit. In particular, the contributions from the self-intersecting and closed curves are irrelevant to the field theory or ``tropical'' limit of the amplitudes; indeed these curves don't even occur dually as propagators in any Tr $\phi^3$ diagrams.  This means that if we are only interested in the field theory limit, we can effectively use a truncated integrand where we only keep the first product over non-self-intersecting curves. We emphasize again that this truncated object is \textbf{not} the full string amplitude at loop level, rather it is a simpler but still ``stringy''  object that correctly gives Tr $\phi^3$ amplitudes at low-energies.

\subsubsection{Higher loops}

Beyond single-trace amplitudes at one loop, an important novelty appears with the first non-planar amplitudes. For instance in the double-trace one-loop amplitudes, 
we have infinitely many curves that we can draw on this fat graph: they differ from one another only in how many times they wind around the graph. 
This is a consequence of the \emph{mapping class group} of the surface, which acts by increasing the winding of curves. In fact, this infinity of windings is the heart of the well-known difficulty in defining a loop integrand for non-planar amplitudes. Fortunately as described explicitly in \cite{Arkani-Hamed:2023lbd,Arkani-Hamed:2023mvg}, it is easy to \emph{mod out} by the action of the mapping class group, using the ``\emph{Mirzakhani trick}" \cite{Mirzakhani:2006fta}. This is a version of the familiar Fadeev-Popov idea for modding out by gauge redundancies in the path integral, and is accomplished by inserting a ``\emph{Mirzakhani Kernel}" $K(u)$ into the integration over the $y$ variables. We will not display these factors explicitly in the rest of this paper, and they will not be relevant for the explicit two-loop computations we will perform for leading singularities.

\subsubsection{Integration Contour}

The stringy Tr $\phi^3$ integral is defined naively on the integration contour where $0< y_i < \infty$. But this form only converges when the $X_{i,j}>0$; continuing to the physically interesting region where we see resonances and the amplitude has poles, where $X_{i,j} < 0$, is carried out by analytic continuation. 

This is a frustrating state of affairs. For four-point, the explicit expression in terms of the Beta function $(\Gamma(X_{1,3}) \Gamma(X_{2,4}))/\Gamma(X_{1,3} + X_{2,4})$ gives us the analytic continuation and allows us to compute the amplitude everywhere, but for general $n$-point scattering no analytic formula is known and so the analytic continuation is not handed to us. 

More conceptually, we would like to find the correct integration contour on which to define this integral. The contour should agree with the naive one for $0<y_i<\infty$ when the integral is convergent, but give us the correct analytic continuation everywhere. 

This is essentially the subject of defining the $i \epsilon$ prescription for string amplitudes, and a prescription for this contour was given by Witten in \cite{Witten:2013pra}. This prescription is essentially ``local" in character, describing a deformation of the contour in the neighborhood of all possible singularities of the amplitude, ultimately associated with all the possible degenerations of the surface equivalent to Feynman diagrams, and so involves a local analysis of the integration domain with an amount of work that grows exponentially with the number of particles. 

Now we know that one of the purposes in life of the $u$ variables and their $y$ parametrization is to explicitly ``blow up" all possible degenerations of the surface, and thereby generate these exponentially many degenerations from objects that are described only by at most $O(n^2)$ variables. As will be described in much greater length \cite{contour}, for this reason, the $u, y$ formalism is ideally suited to define the $i \epsilon$ contour in a one-shot ``global" way. In practice, this gives us a concrete contour prescription that can be written down once and for all, for all $n$, on which all the integrals in our paper are convergent returning well-defined amplitudes. We defer a detailed discussion of this contour to \cite{contour}, contenting ourselves here only with illustrating how things work in the simplest case of four-particle scattering. Setting $u_{1,3} = y_{1,3}/(1 + y_{1,3})$ and $u_{2,4} = 1/(1 + y_{1,3})$, we have one of the classic forms of the beta function integral 
\begin{equation}
{\cal A}_4 = \int_0^\infty \frac{\diff y_{1,3}}{y_{1,3}} y_{1,3}^{X_{1,3}} (1 + y_{1,3})^{-(X_{1,3} + X_{2,4})}.
\end{equation}

To define the analytic continuation we imagine giving $X_{1,3},X_{2,4}$ small imaginary parts, i.e. setting $X_{1,3} \to X^*_{1,3} + i \epsilon_{1,3}$ and $X_{2,4} \to X^*_{2,4} + i \epsilon_{2,4}$ where $X^*_{1,3},X^*_{2,4}$ are real.  
It is useful to change from the $y$ variables to the effective Schwinger parameter $y_{1,3}=e^{-t_{1,3}}$. The naive contour $0<y_{1,3}<
\infty$ is then given by an integration over the entire real $t_{1,3}$ axis. It is important that the contour goes to $t_{1,3} \to + \infty$ to capture the pole in the amplitude as $X_{1,3} \to 0$, and similarly that it goes to $t_{1,3} \to -\infty$ to capture the pole in $X_{2,4}$. These are of course also the degenerations where $u_{1,3} \to 0$ and $u_{2,4} \to 0$, respectively. 

Now, the $i \epsilon$ contour is very natural in this Schwinger parameterization language. After we proceed on the real axis for a little while, we tilt the contour up till it's nearly vertical in the complex $t_{1,3}$ plane for Re$(t_{1,3}) > 0$, but with a small tilt towards the right so it does reach Re$(t_{1,3}) \to \infty$. Similarly, we tilt it down in the complex plane for Re$(t_{1,3}) < 0$. In other words for some $t_{1,3} > t_{1,3}^{\star}$, our contour can be parametrized as $t_{1,3}(\rho)= t_{1,3}^* + \rho(i + \delta)$ with $0<\rho<\infty$, for small $\delta$. The contour for negative values of Re$(t_{1,3})$ can be defined in a similar way. This contour is a smooth deformation of the original one, but now gives us a well-defined integral, provided $\delta$ is chosen to be small enough. For instance on the right, as $\rho$ becomes large, the integrand scales as exp$[(X^*_{1,3} + i \epsilon_{1,3})(\rho(i + \delta))]$. The real part of the argument of the exponential is $(X^*_{1,3} \delta - \epsilon_{1,3}) \rho$. So no matter how negative $X^*_{1,3}$ becomes, we can make $\delta$ small enough for the $- \epsilon_{1,3}$ term to suppress the integrand as $\rho \to \infty$. The same thing holds for the left side of the contour. We can readily check that numerically integrating along this contour gives us the correct result. 
\begin{figure}[t]
    \centering
    \includegraphics[width=0.9\textwidth]{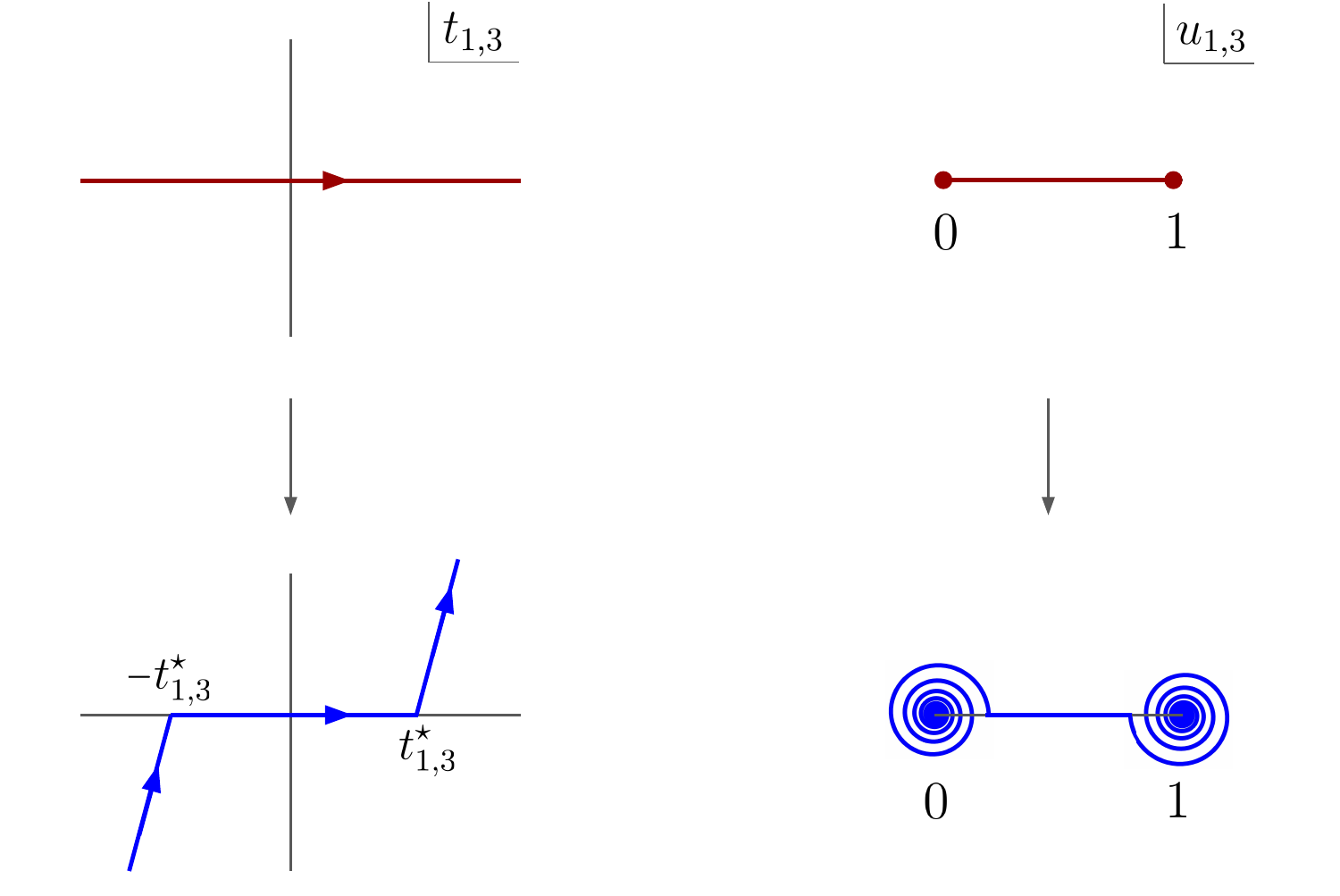}
    \caption{$i\epsilon$ contour for the 4-point amplitude.}
    \label{fig:contours}
\end{figure}
Note that this natural contour in $t_{1,3}$ translates in a pretty contour in $u_{1,3} = e^{-t_{1,3}}/(1 + e^{-t_{1,3}})$. The usual contour has $u_{1,3}$ covering the entire real line $0<u_{1,3} < 1$. The new contour instead spirals around $u_{1,3}=0,1$, exactly as in Witten's prescription~\cite{Witten:2013pra}. The original and deformed contours in both $t_{1,3}$ and $u_{1,3}$ are shown in figure \ref{fig:contours}. 

This example is too simple to illustrate the real power of the $u,y$ variables in defining the $i \epsilon$ contour, which we will explain in more detail in~\cite{contour}. But briefly, the point is that essentially the same contour, only with some care taken in where the ``turns" into the complex plane are taken, can be chosen for each $y_i$ independently. The magic of the $u, y$ variables then correctly implements the needed shifts to make the integrals convergent in all degenerate limits, for precisely the same reason that the $u_{i,j}$'s ``blow up" all the singularities as $u_{i,j} \to 0$ and the $y_{i,j}$ furnish a positive parametrization of the $u_{i,j}$. For the purposes of this paper, this is enough to concretely specify a contour on which our scaffolded gluon amplitudes can be concretely evaluated. 

\subsubsection{Low Energy Limit}

The $u/y$ variable formalism makes it especially easy to take the low-energy limit $\alpha^\prime X_{i,j} \ll 1$ of the stringy Tr $\phi^3$ integrals and extract the field theory limit of the amplitudes. The reason is simply that the singularities in the $dy/y$ measure are logarithmic, regulated by the $u_X^X$ factors. 

Putting $y_i = e^{-t_i}$, the naive integration runs over all real $t_i$, and for 
$X \to 0$ we have divergences at infinity in $t$ space. Thus for $\alpha^\prime X \ll 1$ the integral is dominated by  $|\vec{t}| \to \infty$. In this limit, the $u$ variables simplify. Each $u$ is a ratio of polynomials in the $y$, and the polynomials in turn simplify. In different cones in $\vec{t}$ space, corresponding to a triangulation/diagram ${\cal T}$ when a collection of compatible $u$ variables all vanish together, exactly because the incompatible $u$'s $\to 1$, the integral in $t$ space becomes nothing than the Schwinger parametrization for ${\cal T}$ and the integral is dominated by $\prod_{{\cal X} \subset {\cal T}} (1/X)$ which is just the contribution to the amplitude from the diagram ${\cal T}$.  

This shows that the field-theory limit of the stringy integrals matches the correct sum over diagrams. The fact that we have an analytic representation of $u$ variables actually gives us much more: an integral expression for the amplitude directly in the field-theory limit, as an integral over a single object, just as the full stringy integral, but now interpreted as giving a ``global Schwinger parametrization" for the sum over all diagrams at once. Returning to examining the stringy integrand as $|\vec{t}| \to \infty$, in different cones, one monomial in the polynomials defining the $u$'s will dominate over others, and so $u_X \to \exp^{\alpha_X(t)}$, where $\alpha_X$ is a {\bf piece-wise linear} function of the $t_i$ that is called the ``tropicalization" of $u_X$, which serves as the global Schwinger parameter for $X$. This variable was also called the ``headlight function" in \cite{Arkani-Hamed:2023lbd}, to emphasize that it ``lights up" in all the cones in $t$ space compatible with the chord $X$.  Replacing the full stringy integrand by this ``tropicalized" version gives the correct field-theory limit. This gives a much more striking representation of the amplitude without any reference to sum over diagrams.

As an example at 5-points, consider the polynomial $(1 + y_{1,4} + y_{1,4} y_{1,3})$ appearing in the expression for the $u$ variables. Seeing $y_{1,3} = e^{-t_{1,3}}, y_{1,4} = e^{-t_{1,4}}$, then as $|\vec{t}| \to \infty$ we have
\begin{equation}
(1 + e^{-t_{14}} + e^{-t_{14}} e^{-t_{13}}) \xrightarrow{|\vec{t}| \to \infty} e^{{\rm max}(0, -t_{14},-t_{14} - t_{13})},
\end{equation}
and similarly for the other polynomials. In this way we can read off the $\alpha_X$, for instance 
\begin{equation}
u_{2,4} = \frac{1 + e^{-t_{1,4}} + e^{-t_{1,4}} e^{-t_{1,3}}}{(1 + e^{-t_{1,3}})(1 + e^{-t_{1,4}})} \xrightarrow{|\vec{t}| \to \infty} e^{\alpha_{2,4}(t_{13},t_{1,4})},
\end{equation}
where 
\begin{equation}
\alpha_{2,4}(\vec{t}) = {\rm max}(0,-t_{1,4},-t_{1,4} - t_{1,3}) - {\rm max}(0,-t_{1,3}) - {\rm max}(0,-t_{1,4}),
\end{equation}
is the piece-wise linear global Schwinger parameter or ``headlight function" associated with the curve $(2,4)$. In this way the full 5-point tree amplitude can be written as a single ``tropical" integral, 
\begin{equation}
{\cal A}_5[X_{i,j}] = \int_{-\infty}^{+\infty} \diff t_{1,3} \diff t_{1,4}\,  e^{\sum_{i,j} X_{i,j} \alpha_{i,j}(t)} .
\end{equation}

Furthermore, at loop-level, as familiar from Schwinger parametrization, the loop integrations can be performed as Gaussian integrals, so we are left with a global Schwinger parametrization of the sum of all loop diagrams at any given order in the topological expansion.

  %\newpage
  \section{Scalar-Scaffolded Gluons}
\label{sec:scalarkin}
In this paper, we are unveiling a hidden direct connection between the seemingly ``boring" Tr $\phi^3$ theory and the much more exciting physics of gluon scattering amplitudes in Yang-Mills theory. Obviously the first order of business in making this connection is to answer the simplest possible question: how can the basic kinematics for these theories possibly be the same? After all, gluons have spin described by polarization vectors, which are obviously missing in Tr $\phi^3$ theory. So how can gluons and scalars possibly be related? 

This has been a constant source of tension in attempting to connect the world of polytopes and curves on surfaces to real-world scattering amplitudes.  The purely scalar $X_{i,j}$ variable description of kinematics played a crucial role from the very outset of the ABHY associahedron story, but there was seemingly no natural way to manually ``add polarization vectors'' to include gluons. For one thing, due to gauge-redundancy, there was no invariant way of doing this. And besides, adding any source of redundancy ran very much counter to the spirit of the $X_{i,j}$, one of whose virtues was in giving a non-redundant description of the momentum kinematics! 

In this section we will describe an extremely simple but beautiful resolution to this basic problem. We will describe $n$ gluon amplitudes in a purely scalar language using kinematic $X$ variables, by imagining that we are scattering $2n$ colored scalars. There is always a pole of the amplitude where the scalars $(1,2)$ fuse into an on-shell gluon$_1$, scalars $(3,4)$ fuse to gluon$_2$, and so on. The residue on this pole then gives us the $n-$gluon amplitude \footnote{Note that in order for the massless residue on $(p_i + p_{i+1})^2 \to 0$ to {\it only} be given by factorization on the gluon pole, we further have to assume that there are no cubic interactions between these scalars. This can be trivially arranged with the $Z_2$ symmetry flipping the sign of every scalar. It also automatically occurs from one of the natural origins for these scalars from higher-dimensional gluons}. For obvious reasons, we will refer to this as   ``the scaffolding residue". 

We will show explicitly how to describe gluon scattering amplitudes for any \footnote{Strictly speaking we will describe {\it null} polarization vectors $\epsilon^2 = 0$. Of course, these provide a basis for all possible polarizations.} polarizations in this scalar language. 
We will also discuss the constraints on functions of the $2n$-gon $X_{ij}$ variables, that allow us to interpret scaffolding residues as gluon amplitude, satisfying both the constraints of on-shell gauge-invariance and multilinearity in the polarization vectors. Despite the fact that these two properties are physically very different, we will see that they are both naturally unified in a single pretty mathematical statement. 

We remark that the ``scaffolding" method is a very general one, and can be used to describe particles of any mass and spin, since we can always imagine producing the spinning particles with a pair of scalars. We can for instance also ``scaffold" gravitons, only the scaffolded avatar of the gauge-invariance/multilinearity conditions will change. Similarly we can scaffold massive string states, with some extra care to be taken to disentangle mass degeneracy in the string spectrum. 

But while the basic strategy is essentially trivial, the scaffolding picture allows us to talk about seemingly very different amplitudes in the same kinematical language. It is this ability that allows us to bring the highly non-trivial relations between the amplitudes for very different theories to light. 

\subsection{From $2n$ scalars to $n$ gluons}

\begin{figure}[t]
\begin{center}
\includegraphics[width=0.5\textwidth]{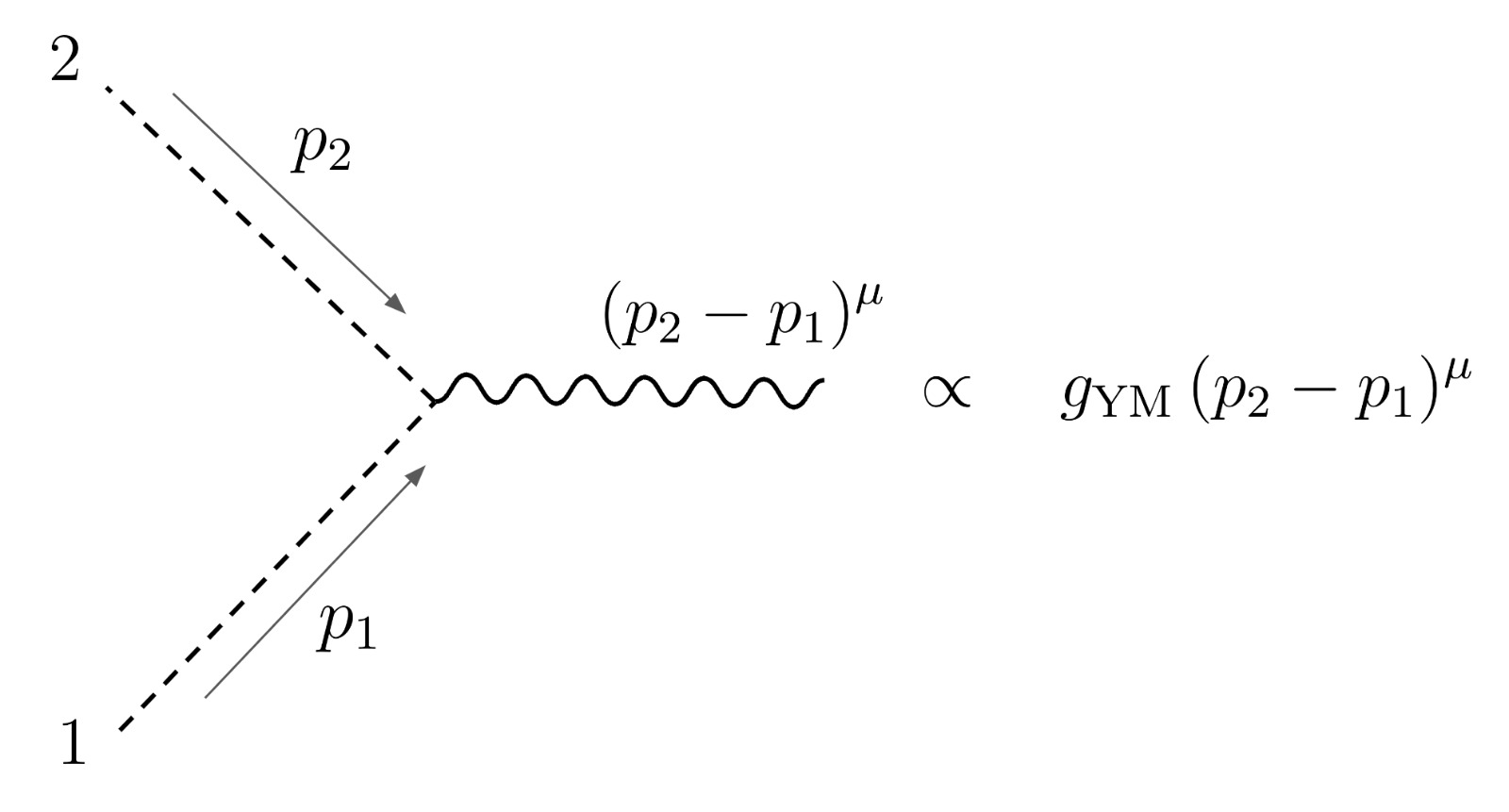}
\caption{Three particle amplitude between two colored scalars and a gluon.}
\label{fig:FeynRules}
\end{center}
\end{figure}

The scalars we consider interact with gluons via the unique three-point coupling: $g_{\text{YM}}(p_{2} -p_1)^\mu$, as depicted in figure \ref{fig:FeynRules}.
If we start with a $2n$-scalar scattering process and take $n$ residues to make all the gluons (produced by adjacent scalars) on-shell, we land on the $n$-gluon amplitude. 
  
\begin{figure}[t]
\begin{center}
\includegraphics[width=0.9\textwidth]{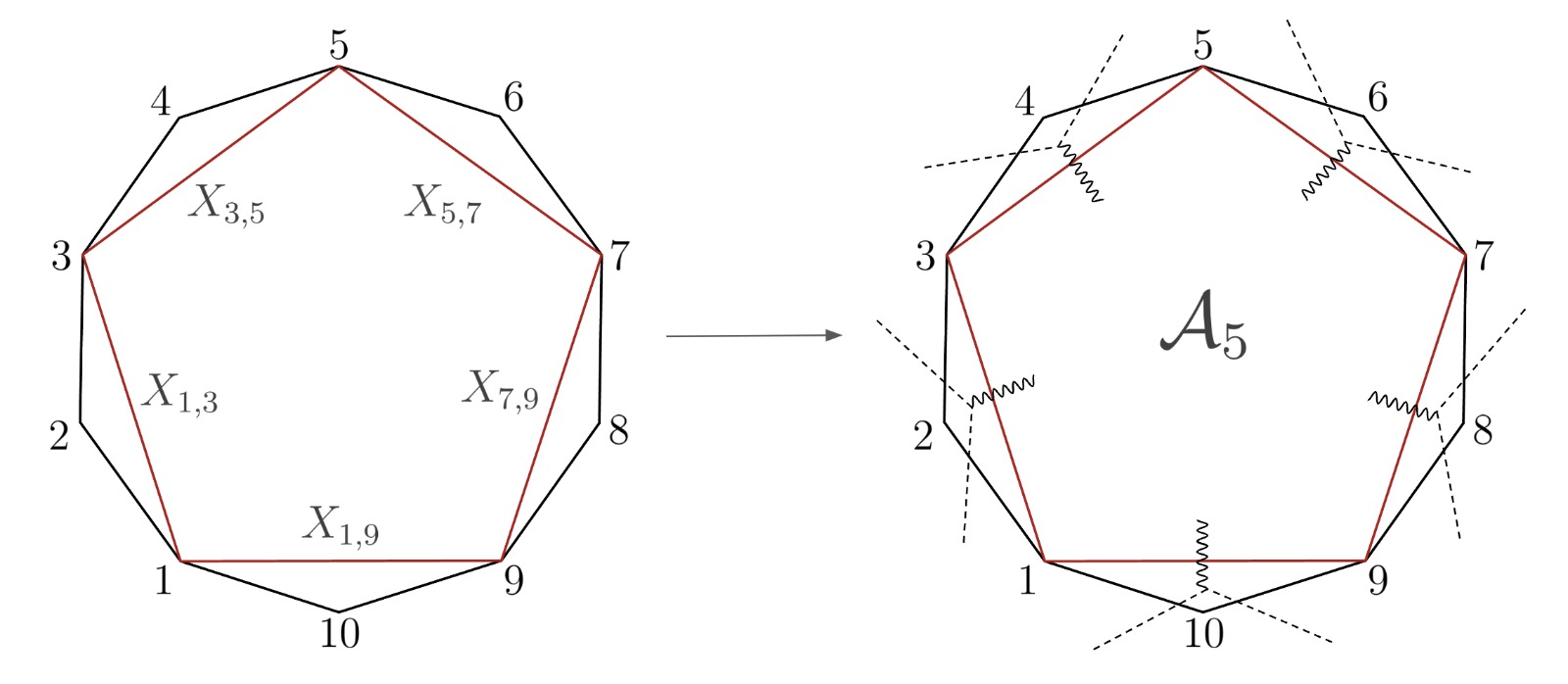}
\caption{Scaffolding residue from 10-scalar amplitude to 5-gluon.}
\label{fig:Scaff5pt}
\end{center}
\end{figure}

Explicitly, to get the $n$-point gluon amplitude we take residues on the simple poles corresponding to $X_{1,3} = X_{3,5} = X_{5,7}= ... = X_{2 n -1,1}=0$, or equivalently, in terms of the momentum, $(p_1+p_2)^2 = (p_3+p_4)^2 = (p_5+p_6)^2 = ... = (p_{2n-1}+p_{2n})^2 =0$, which is exactly the condition to have the intermediate particles, in this case the gluons, on-shell. This procedure is illustrated in figure \ref{fig:Scaff5pt} for the case of a 5-point gluon amplitude. 

At tree-level, making the external gluons on the shell is enough to ensure that all the internal particles are gluons as well. For this reason, by taking this residue we are automatically left with the \textbf{tree-level pure gluon amplitude}. From this point onward, we will denote this residue that takes us from the scalar 2$n$-gon to the gluon $n$-gon, as the \textbf{scaffolding residue}, for purely pictorial reasons inspired by figure \ref{fig:Scaff5pt}.
  
However, exactly because we are still starting from a scalar scattering, all the gluon information in the n-point amplitude left after taking the scaffolding residue is now encoded in scalar language, and so all the explicit dependence on polarizations characteristic of gluon amplitudes is less manifest. So we need to understand how we can extract this information from the starting scalar process. 

For the rest of this section, we will explain how the information about the gluons is encoded in the $X_{i,j}$'s as well as understand how the properties of gluon amplitudes such as gauge invariance and multilinearity in each polarization are manifest in what we get from the scaffolding residue of the $2n$-scalar amplitude.

\subsubsection{Counting number of degrees of freedom}

As a sanity check, let's start by checking that after taking the 2$n$ scaffolding residue the number of degrees of freedom left in the scalar problem, matches that of an n-point gluon problem. Originally, for the scattering of these 2$n$-scalars, we have a total of $2n(2n-3)/2$ independent planar Mandelstams, $X_{i,j}$. After taking the scaffolding residue, we fix $n$ of these variables to zero, so we have $2n(2n-3)/2 - n = 2n(2n-4)/2$ degrees of freedom left. For an n-point gluon problem, we have $n(n-3)/2$ independent planar Mandelstams, $n(n-1)/2$ possible $\epsilon_i \cdot \epsilon_j$, and $n(n-1)-n= n(n-2)= n(2n-2)/2$ different $\epsilon_i \cdot p_j$. So summing these three: $(4n^2 -8n)/2 = 2n(2n-4)/2$, which agrees with the scaffolding residue, as expected. 
  
Let's now proceed and see in detail how to extract the gluon amplitude from the 2$n$-scalar amplitude, via this residue.

\subsection{Polarizations and momenta of the gluons}

\begin{figure}[t]
\begin{center}
\includegraphics[width=0.6\textwidth]{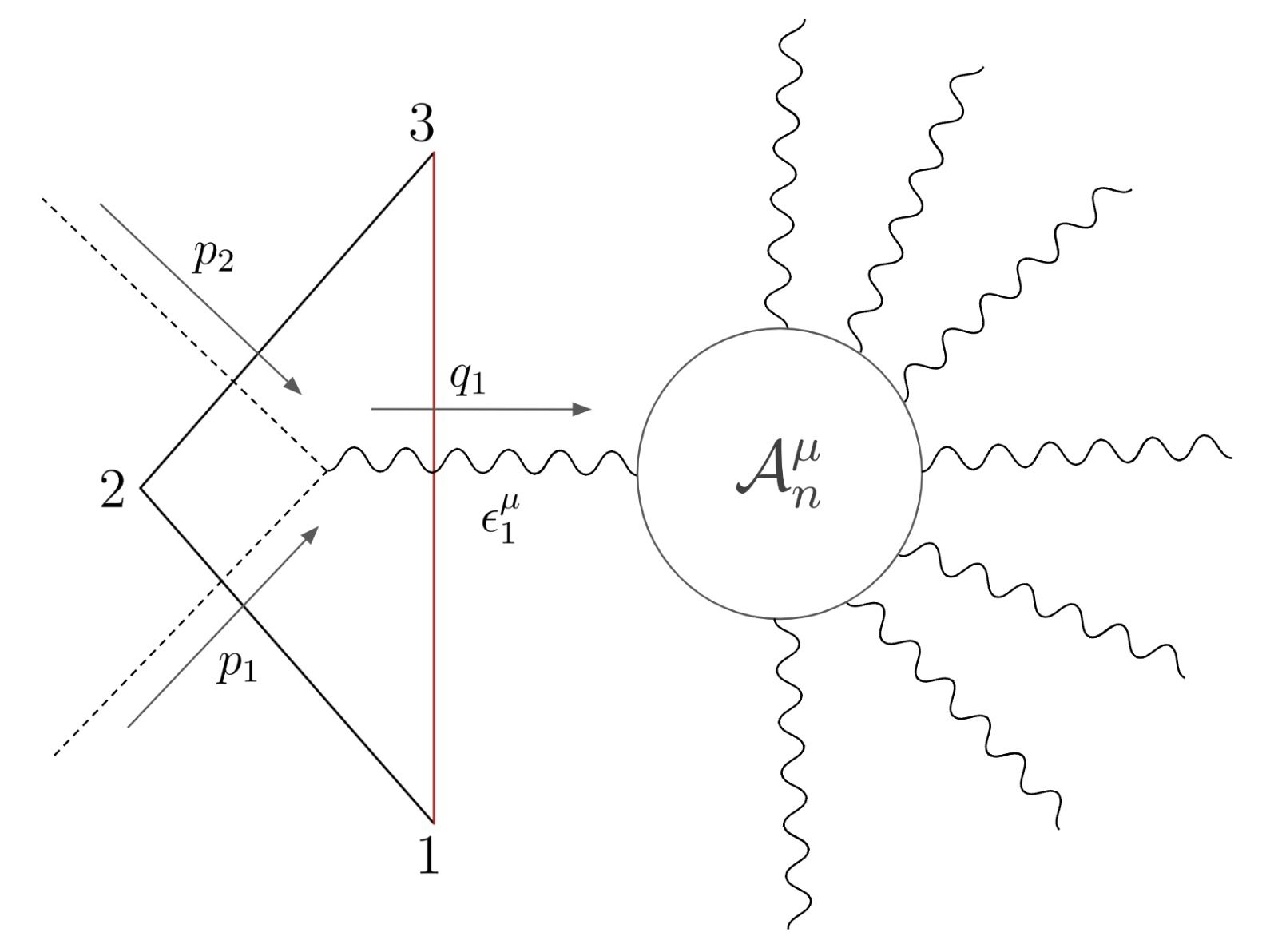}
\caption{Polarization and momentum of scaffolded gluon.}
\label{fig:Pol1}
\end{center}
\end{figure}

Let's now understand how the polarizations and momentum of the gluons are encoded in the planar variables $X_{i,j}$ of the scaffolding scalars. 
To do so, we focus on a single gluon of the full process, see figure \ref{fig:Pol1}.
By momentum conservation, we can read off directly that the momentum of the incoming gluon is going to be $q_i^{\mu} = (p_{2i-1}+p_{2i})^{\mu}$, which satisfies $q_i^2 =0$ on the scaffolding residue. So we have $q_1^\mu = (p_1+p_2)^\mu$. By convention, we will always use $q$ to denote the momentum of the gluons and $p$ the momentum of the scaffolding scalars.
  
As for the polarization vector, $\epsilon_i$, of course it isn't uniquely fixed but can be shifted by gauge transformations, whose geometric interpretation we will give in a moment. But one symmetrical choice for $\epsilon_i$ is the one familiar from the Feynman rule for the 3-point vertex between the scalars and the gluon, which gives us $\epsilon_{i}^{\mu} \propto (p_{2i}- p_{2i-1})^{\mu}$. On the scaffolding residue, it satisfies $\epsilon_i \cdot p_i=0$, as necessary for on-shell gluons. It also satisfies  $\epsilon_i^2=0$, so we are describing generalized ``circular polarizations''; this is not a restriction since these furnish a basis for all possible polarization states.

To better understand the geometrical meaning of the polarizations, it is useful to label the edges of the momentum polygon by vectors $X_i^{\mu}$, such that $p_i^{\mu} = (X_{i+1}-X_{i})^{\mu}$. In terms of these variables we have:
\begin{equation}
	\epsilon_1 ^{\mu} \propto (X_{3}-X_2)^{\mu}-(X_2-X_1)^{\mu} \propto X_2^{\mu} - \frac{(X_3 +X_1)^{\mu}}{2}.
\end{equation}

\begin{figure}[t]
\begin{center}
\includegraphics[width=0.4\textwidth]{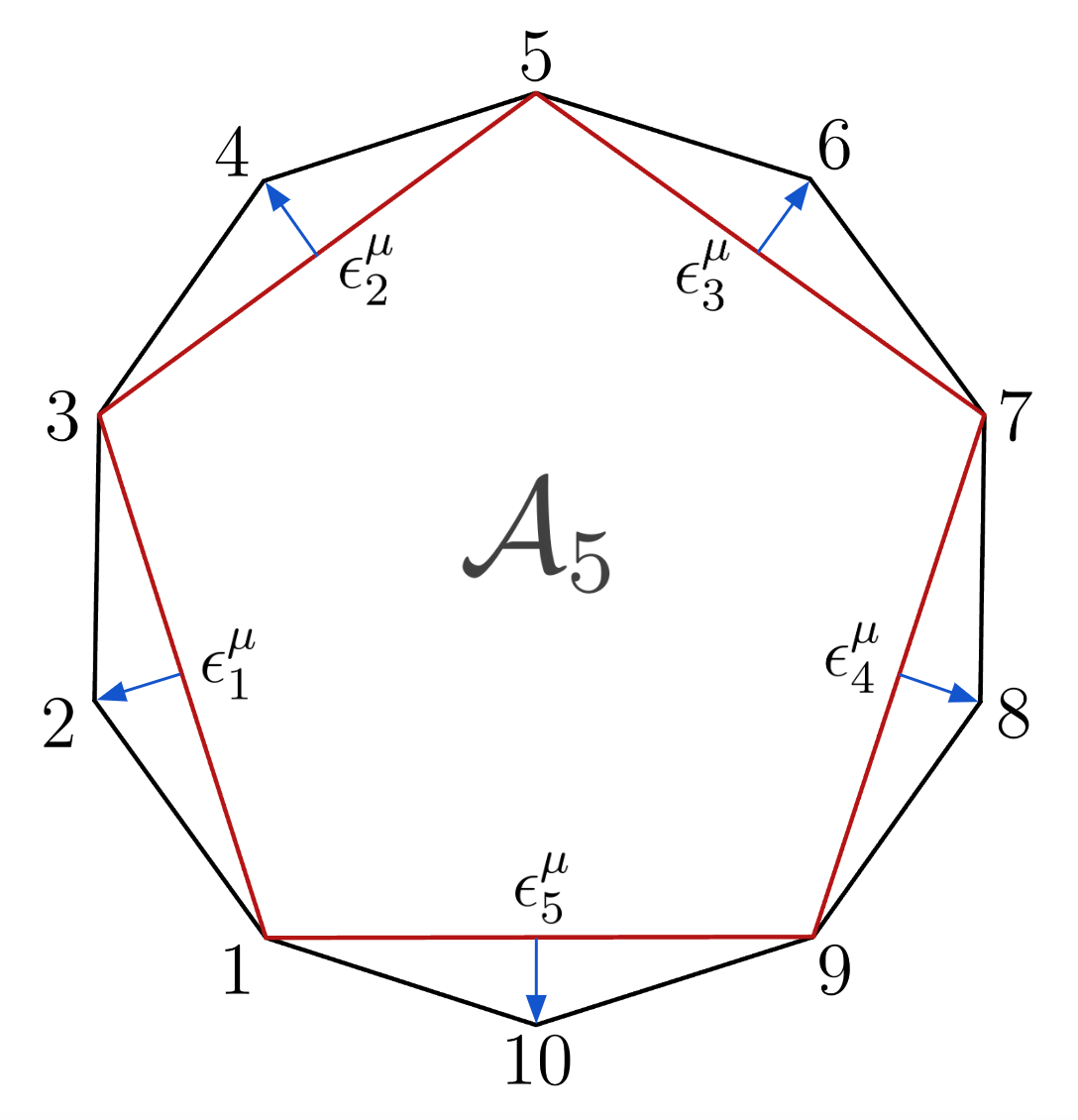}
\caption{Polarizations of scaffolded gluons.}
\label{fig:Polarizations}
\end{center}
\end{figure}

So we can see that the information about the polarizations is also included in the momentum polygon: they correspond to the vectors going from the midpoint of the triangle to the extra edge in the scaffolding (see figure \ref{fig:Polarizations}). Picking this particular representation of the polarization vectors gives a misleading significance to the midpoint, however, as we will see, invariance under gauge transformations tells us precisely that the polarization vector can be attached anywhere on this edge. 

\subsection{Unified form of gauge invariance and multi-linearity}
\label{sec:ginvMultLin}
We now want to see what the statements about gauge invariance and multi-linearity translate to in the scalar degrees of freedom, $X_{i,j}$. Let's start with \textbf{gauge invariance}. To avoid cumbersome indices, we will study gauge invariance for gluon $1$, but all translate to the $i$th gluon by replacing $1$ by $2i-1$, 2 by $2i$, etc. 
  
Invariance under gauge transformations tells us: 
\begin{equation}
	\epsilon_1 ^{\mu} \rightarrow \epsilon_1 ^{\mu} + \alpha q_1^{\mu} \quad \Rightarrow \quad  \delta \mathcal{A} = 0 \quad \forall \alpha,
\end{equation} 
which can be phrased in terms of the momentum of the 2n-scalars as 
\begin{equation}
	(p_{2}-p_{1}) ^{\mu} \rightarrow (p_{2}-p_{1}) ^{\mu}  + \alpha (p_{1}+p_{2})^{\mu} \quad \Rightarrow \quad  \delta \mathcal{A} = 0 \quad \forall \alpha.
	\label{eq:GInv_1}
\end{equation}

Now we want to understand what this shift means in terms of the $X_{i,j}$ since these are the variables the amplitude depends on. To do this, let's express the gauge transformations in terms of the variables $X_i^{\mu}$: 
\begin{equation}
	\left(X_2 - \frac{X_1+X_3}{2}\right) ^{\mu} \rightarrow \left(X_2 - \frac{X_1+X_3}{2}\right)^{\mu} + \alpha (X_3-X_1)^{\mu},
\end{equation}
which is equivalent to simply shifting $X_2^{\mu}$ as follows:
\begin{equation}
	X_2^{\mu} \rightarrow X_2^{\mu} + \alpha (X_3-X_1)^{\mu}. 
\end{equation}

\begin{figure}[t]
\begin{center}
\includegraphics[width=0.95\textwidth]{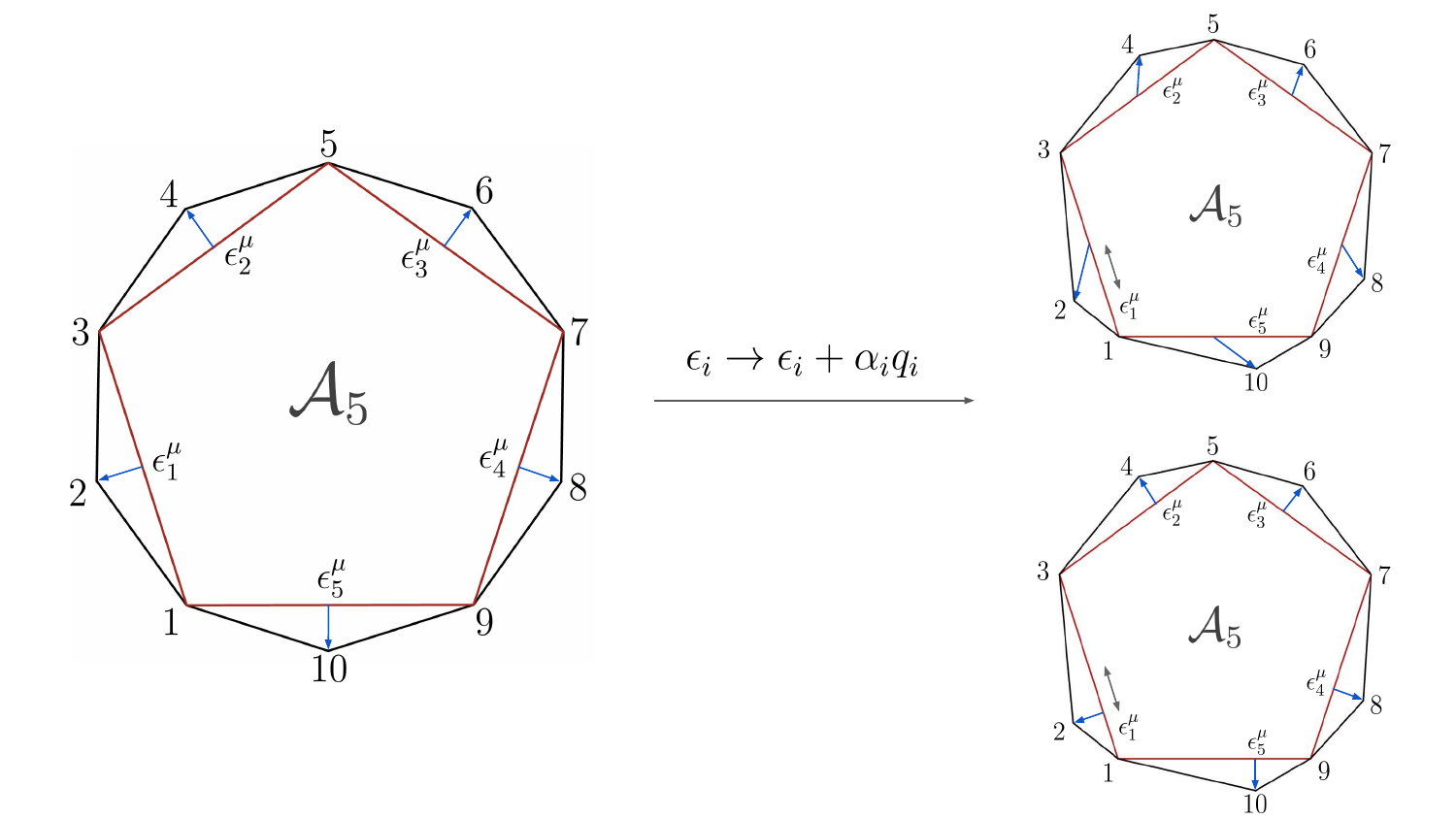}
\caption{Gauge Transformations.}
\label{fig:GaugeTransf}
\end{center}
\end{figure}

Geometrically, gauge invariance is then equivalent to the statement that we are allowed to move the base point of the polarization of gluon $i$ along $X^\mu_{2i{-}1,2i{+}1}$, see figure \ref{fig:GaugeTransf}. 

Indeed, while we have emphasized a ``top-down" picture of the scalar scaffolding, starting with the idea of using pairs of colored scalars to fuse into gluons, we can also very nicely motivate the scaffolding idea from the bottom-up. So let's start from the beginning just with the data for $n$ gluons--their momenta $p_i$ and polarization vectors $\epsilon_i$. We can use the $p_i$ to draw the momentum polygon as usual. But it is also natural to record the information about the polarization vectors simply by placing the end of the vector $\epsilon_i$ somewhere on the edge $p_i$ of the momentum polygon. This immediately defines our $2n$-gon, on the locus where $X_{1,3}, X_{3,5}, \cdots = 0$ reflecting $p_i^2 = 0$. Since there is no invariant place on the edge $p_i$ to place $\epsilon_i$, it is natural to demand that any question we ask of this $2n$-gon should be invariant under shifting where we place the endpoint of $\epsilon_i$, and of course very nicely shifting this endpoint is precisely gauge-invariance $\epsilon^\mu_i \to \epsilon^\mu_i + \alpha p^\mu_i$. As an aside, the on-shell field-strength $f^{\mu \nu}_i = p_i^\mu \epsilon_i^\nu - p_i^\nu \epsilon_i^\mu$ has an obvious and pretty geometric interpretation as well--it is simply the antisymmetric tensor determined by the plane formed, $(X_{2i -1}, X_{2i}, X_{2i + 1})$.

Ultimately this transformation will produce a shift in the planar variables involving $X_2^{\mu}$ which can be derived as follows:
\begin{equation}
\begin{aligned}
	(X_2^{\mu} -  X_j^{\mu})&\rightarrow (X_2^{\mu} - X_j^{\mu}) + \alpha (X_3^{\mu}-X_1^{\mu})\\
	\Rightarrow (X_2^{\mu} -  X_j^{\mu})(X_{2\mu} -  X_{j\mu})&\rightarrow (X_2^{\mu} - X_j^{\mu})(X_{2\mu} -  X_{j\mu}) + \alpha (X_3^{\mu}-X_1^{\mu})(X_{2\mu} -  X_{j\mu}) \\
	 \Rightarrow  X_{2,j} &\rightarrow X_{2,j} + \alpha (X_{3,j}-X_{1,j}).
\end{aligned}
\label{eq:GInvX2j}
\end{equation}

So we can state gauge invariance in gluon $i$th as
\begin{equation}
	\mathcal{A}\left[X_{2i,j}\rightarrow X_{2i,j} - \alpha (X_{2i+1,j}-X_{2i-1,j}) \right] - \mathcal{A}[X_{2i,j}] = 0.
\end{equation}

In addition, we know the amplitude, $A_{2n}$, should be \textbf{multi-linear} in each polarization:
\begin{equation}
	\epsilon_i^{\mu} \rightarrow \epsilon_i^{\mu} + \beta \epsilon_i^{\mu} \quad \Rightarrow \quad \delta \mathcal{A} = \beta \mathcal{A}.
	\label{eq:MultiLin}
\end{equation}

Again, let's see what this translates to in scalar language for gluon 1. Rewriting this in terms of $X_i^{\mu}$ yields 
\begin{equation}
	\left(X_2 - \frac{X_1+X_3}{2}\right) ^{\mu} \rightarrow \left(X_2 - \frac{X_1+X_3}{2}\right)^{\mu} + \beta \left(X_2 - \frac{X_1+X_3}{2}\right)^{\mu},
\end{equation} 
which is equivalent to 
\begin{equation}
	X_2^{\mu} \rightarrow X_2^{\mu} + \frac{\beta (X_2^{\mu} - X_1^{\mu})}{2} + \frac{\beta (X_2^{\mu} - X_3^{\mu})}{2}.
\end{equation}

Proceeding as we did in \eqref{eq:GInvX2j}, we get that the $X_{2,j}$ transform as
\begin{equation}
	X_{2,j} \rightarrow X_{2,j} + \frac{\beta (X_{2,j}-X_{1,j})}{2} +\frac{\beta (X_{2,j}-X_{3,j})}{2},
\end{equation}
and so linearity in polarization $i$th can be restated as:
\begin{equation}
	\mathcal{A}\left[X_{2i,j} \rightarrow X_{2i,j} + \frac{\beta}{2} (X_{2i,j}-X_{2i-1,j}) +\frac{\beta}{2} (X_{2i,j}-X_{2i+1,j})\right] - \mathcal{A}[X_{2i,j}] = \beta \mathcal{A}[X_{2i,j}] .
\end{equation}

Combining \textbf{gauge invariance} and \textbf{multi-linearity} for gluon 1, we get:
\begin{equation}
\begin{aligned}
	&\mathcal{A}\left[X_{2,j} \rightarrow X_{2,j} + \frac{\beta (X_{2,j}-X_{1,j})}{2} +\frac{\beta (X_{2,j}-X_{3,j})}{2} + \alpha (X_{3,j}-X_{1,j})\right] - \mathcal{A}[X_{2,j}] = \beta \mathcal{A}[X_{2,j}] \\
	\Leftrightarrow \quad  &\mathcal{A}\left[X_{2,j} \rightarrow X_{2,j} + \left(\frac{\beta}{2} +\alpha\right) (X_{2,j}-X_{1,j}) +\left(\frac{\beta}{2} -\alpha \right) (X_{2,j}-X_{3,j}) \right] - \mathcal{A}[X_{2,j}]  = \beta \mathcal{A}[X_{2,j}].
\end{aligned}
\end{equation}

Picking $\alpha= \pm \beta/2$, we get:
\begin{eqnarray}
	\mathcal{A}\left[X_{2,j} \rightarrow X_{2,j} + \beta (X_{2,j}-X_{1,j})  \right] - \mathcal{A}[X_{2,j}]  &=& \beta \mathcal{A}[X_{2,j}], \nonumber \\
\mathcal{A}\left[X_{2,j} \rightarrow X_{2,j} + \beta (X_{2,j}-X_{3,j})  \right] - \mathcal{A}[X_{2,j}]  &=& \beta \mathcal{A}[X_{2,j}].
\end{eqnarray}

This is then the unified form of the statements for both multilinearity and gauge invariance: we have a rescaling symmetry under shifting $X_{2,j}$ by either of its neighbors $X_{1,j}$ or $X_{3,j}$. The shift by $X_{1,j}$ tells us that the amplitude must be expressible as 
\begin{eqnarray}
	\mathcal{A} &=& \sum_{j\neq\{1,2,3\}} (X_{2,j}-X_{1,j}) \times \mathcal{Q}_{2,j} \nonumber, \\
\mathcal{A} &=& \sum_{j\neq\{1,2,3\}} (X_{2,j}-X_{3,j}) \times \tilde{\mathcal{Q}}_{2,j},
\end{eqnarray}
where $\mathcal{Q}_{2,j}, \tilde{\mathcal{Q}}_{2,j}$ some function of the kinematics, $X_{i,j}$, but \textbf{independent} of $X_{2,j}$. This tells us they are equal: 
\begin{equation}
\mathcal{Q}_{2,j} = \tilde{\mathcal{Q}}_{2,j} = \partial_{X_{2,j}}\mathcal{A}.
\end{equation}

In summary, translating this property for a general gluon $i$, we conclude that the amplitude \textbf{must} be of the form:
\begin{equation}
\begin{aligned}
	\mathcal{A}_{n}^{\text{gluon}} &= \sum_{j\neq\{2i-1,2i,2i+1\}} (X_{2i,j}-X_{2i-1,j}) \times \mathcal{Q}_{2i,j} \\
	&= \sum_{j\neq\{2i-1,2i,2i+1\}} (X_{2i,j}-X_{2i+1,j}) \times \mathcal{Q}_{2i,j}. \\
\end{aligned}
\quad,
\label{eq:GInv_Mult}
\end{equation}
where $\mathcal{Q}_{2i,j}$ is function of the remaining kinematics, \textbf{independent} of $X_{2i,j}$. As pointed out before and is made clear in this section, gluon amplitudes expressed in terms of the scalar data have a \textbf{unique} form, that makes gauge invariance manifest, without introducing any redundancies in the representation. 

\subsection{Dictionary: From scalar to gluon kinematics}
\label{sec:Dict}
In this section we present a systematic way of mapping from the scalar kinematic invariants of the scaffolding $2n$-gon into the momenta, $q_{i}$, and polarizations, $\epsilon_i$, of the gluons we obtain after extracting the scaffolding residue. 

Note that all the $X_{i,j}$ with $i$ and $j$ \textbf{odd}, $X_{o,o}$, correspond to chords living inside the $n$-gon, and thus are the planar variables of the $n$ gluon problem:
\begin{equation}
X_{2i{-}1, 2j{-}1}=\left(q_i+q_{i+1} + \dots + q_{j-1}\right)^2\equiv X^{\text{g}}_{i,j},
\end{equation}
where $X_{2i{-}1, 2i{+}1}=(p_{2i-1}+p_{2i})^2 = q_i^2=0$ on the scaffolding residue. As usual, these $n(n{-}3)/2$ planar variables are related to the same number of linearly independent Mandelstam variables of $n$ gluons via
\begin{equation}
\begin{aligned}
   &-2 q_i \cdot q_j= X^{\text{g}}_{i,j} + X^{\text{g}}_{i+1,j+1}- X^{\text{g}}_{i,j+1} - X^{\text{g}}_{i+1,j} \Leftrightarrow \\
   &-2 (p_{2i-1}+p_{2i}) \cdot (p_{2j-1}+p_{2j})=  X_{2 i-1,2 j-1} + X_{2 i+1,2 j+1}- X_{2 i-1,2 j+1}-X_{2 i+1,2 j-1}.
\end{aligned}
\label{qq}
\end{equation}

Let's now work out the relations between the remaining planar variables: $X_{e,e}$ and $X_{e,o}$, and the gluon data, $i.e.$ (linearly independent) products of the form $\epsilon_i \cdot \epsilon_j$, $\epsilon_i \cdot q_j$. In the previous section, we saw that the polarization of each gluon could be associated with a vector starting at the midpoint of the edge of the gluon $n$-gon and ending on the vertex of the scalar $2n$-gon. However, gauge invariance tells us that the polarization vector does \textbf{not} have to lie at the midpoint, and can lie in any point along the gluon edge, allowing us to write:
\begin{equation}
\epsilon_i=(1-\alpha) p_{2i}-\alpha p_{2i-1},  
\end{equation} 
for an arbitrary parameter $\alpha$. Recall that $q_i^\mu= (p_{2i}+p_{2i-1})^\mu$, thus $\epsilon_i^2=\epsilon_i    \cdot q_i=0$ exactly on the scaffolding residue. The choice of midpoint corresponds to $\alpha=1/2$, but let us keep $\alpha$ generic. Both the momentum and polarization of the $i$th gluon can be expressed in terms of the vertices of the $2n$-gon as
\begin{align}
    \epsilon_i^\mu& \propto (1-\alpha) (X_{2i+1}-X_{2i})^\mu-\alpha (X_{2i}-X_{2i-1})^\mu, \label{eq:eps} \\
    q_i^\mu& \propto (X_{2i+1}-X_{2i})^\mu+(X_{2i}-X_{2i-1})^\mu=(X_{2i+1}-X_{2i-1})^\mu\,.\label{eq:q}
\end{align}

Note that $q_i^\mu$ only depends on ``odd" vertices $X_1, X_3, \cdots, X_{2n{-}1}$, thus $\epsilon_i \cdot q_j$ must depend on $X_{k,l}$ with at least one of $k,l$ being odd. This means that the only dependence on $X_{e,e}$ is through $\epsilon_i \cdot \epsilon_j$ --  this will be important shortly. 
The remaining Lorentz products can be expressed in terms of $X$'s as follows: Using \eqref{eq:eps} and \eqref{eq:q} we can write the remaining gluon kinematics as 
\begin{equation}\label{ee}
\begin{aligned}
2 \epsilon_i \cdot \epsilon_j=&(1-\alpha)\left(-(1-\alpha ) X_{2 i+1,2 j+1}-\alpha  X_{2 i-1,2 j+1}-\alpha  X_{2 i+1,2 j-1}+X_{2 i,2 j+1}+X_{2 i+1,2 j}\right)\\
   & -\alpha ^2 X_{2 i-1,2 j-1}+\alpha  X_{2 i,2 j-1}+\alpha  X_{2 i-1,2 j}-X_{2 i,2 j}\,,
\end{aligned}
\end{equation}
\begin{equation}
\begin{aligned}\label{eq}
2 \epsilon_i \cdot q_j=&(1-\alpha) X_{2 i+1,2 j-1}-(1-\alpha ) X_{2 i+1,2 j+1}+\alpha  X_{2 i-1,2 j-1}-\alpha  X_{2 i-1,2 j+1}\\
&+X_{2 i,2 j+1}-X_{2 i,2 j-1}\,,
\end{aligned}
\end{equation}
with $X_{1,3}=X_{3,5}=\ldots=X_{2n-1,1}=0$ and $X_{i,i}=X_{i,i+1}=0$. Again by plugging in $\alpha=\frac 1 2$ we recover the special midpoint choice above. If, instead, we choose $\alpha=0$ or $1$ both equations simplify, {\it e.g.}, for $\alpha=1$ we have
\begin{align}
&2 \epsilon_i \cdot \epsilon_j=X_{2 i,2 j-1}+ X_{2 i-1,2 j}- X_{2 i-1,2 j-1}-X_{2 i,2 j}=-c_{2i{-}1, 2j{-}1}\,,\\\nonumber
&2 \epsilon_i \cdot q_j=X_{2i, 2j{+}1}-X_{2i, 2j{-}1}+X_{2i{-}1, 2j{-}1}-X_{2i{-}1, 2j{+}1}\,.
\end{align}

Note that if and only if an expression of Lorentz products is {\bf gauge invariant} and {\bf multilinear} in all $\epsilon$'s, the corresponding expression in $X$'s becomes independent of the parameter $\alpha$. On the other hand, by solving \eqref{ee} and \eqref{eq} we can express all $X_{e, e}$ and $X_{e, o}$ in terms of a basis of Lorentz products, and vice-versa. 

\subsection{3-point gluon amplitude in scalar language}\label{sec:3pt}

To finish this section, we want to provide an example of a gluon amplitude written in this scalar language. So will consider the $3-$point gluon amplitude that is extracted from a $6-$point scalar process. 

Using \eqref{ee} and \eqref{eq} for $n=3$, {\it e.g.} with $\alpha=1$ we have 
\begin{equation}
\begin{aligned}
&2\epsilon_1 \cdot \epsilon_2=X_{1,4}-X_{2,4}, \quad 2 \epsilon_2 \cdot \epsilon_3=X_{3,6}-X_{4,6}, \quad
2 \epsilon_3 \cdot \epsilon_1=X_{2,5}-X_{2,6},\\
&2 \epsilon_1 \cdot q_2=X_{2,5}, \quad 2 \epsilon_2 \cdot q_3=X_{1,4}, \quad 2 \epsilon_3\cdot q_1=X_{3,6}.
\end{aligned}
\label{eq:Dic3}
\end{equation}

At 3-points we have two different types of interactions: the Yang-Mills (YM) vertex, $\mathcal{A}_3^{\rm YM}$,  and the $F^3$ vertex, $\mathcal{A}_3^{F^3}$.
Using \eqref{eq:Dic3}, we can write $\mathcal{A}_3^{\rm YM}$ in terms of $X$'s as 
\begin{equation} \label{eq:3ptYM}
A^{\mathrm{YM}}_3(1,2,3)=-(X_{2,5}-X_{2,6})X_{1,4}-(X_{3,6}-X_{4,6})X_{2,5}-(X_{1,4}-X_{2,4})X_{3,6}\,,
\end{equation}
which become nicely
\begin{equation} A^{\mathrm{YM}}_3(1,2,3)=-4(\epsilon_1 \cdot \epsilon_3\; \epsilon_2 \cdot q_3 + \epsilon_2 \cdot \epsilon_3\; \epsilon_1 \cdot q_2+\epsilon_1 \cdot \epsilon_2\; \epsilon_3\cdot q_1).
\end{equation} 

Similarly, for the 3-point $F^3$ amplitude, $-X_{1,4}X_{2,5}X_{3,6}$, we find 
\begin{equation}
    A^{F^3}_3(1,2,3)= -8\; \epsilon_1 \cdot q_2
    \; \epsilon_2 \cdot q_3\; \epsilon_3 \cdot q_1\,.
\end{equation}

  %\newpage
  \section{Tree-level Gluon Amplitudes}
\label{sec:tree}
Having learned how to describe the kinematics of gluon scattering amplitudes in a scalar language--thinking of each gluon as being ``scaffolded" by a pair of colored scalars--we now turn to dynamics, and make our proposal for gluon scattering amplitudes. 

We start at tree-level, and define a stringy integral representation of scaffolded gluon amplitudes as a very simple deformation of the stringy  Tr $\phi^3$ as~\cite{Zeros}:
\begin{equation}
\mathcal{A}_{2n}(1,2,...,2n) [X_{i,j}] =\int_0^\infty \underbrace{\prod_{\mathcal{C} \in \mathcal{T}} \frac{\diff y_{\mathcal{C}}}{y_{\mathcal{C}}} \,   \prod_{i<j} u_{ij}^{\alpha^{\prime} X_{i,j}} }_{\text{Integrand of \eqref{eq:Trphi3U}}} \, \left( \frac{\prod_{(i,j)\in\text{even}} u_{i,j}}{\prod_{(i,j)\in\text{odd}} u_{i,j}} \right),
\label{eq:2nScalarUs}
\end{equation}
where $y_\mathcal{C}$ corresponds to the positive parametrization associated with triangulation $\mathcal{T}$ of the disk with $2n$ marked points in the boundary. Note that this deformation simply amounts to shifting the kinematics entering the stringy Tr $\phi^3$: $X_{o,o} \rightarrow X_{o,o} - 1/\alpha^\prime$ and $X_{e,e} \rightarrow X_{e,e} + 1/\alpha^\prime$. 

In the rest of this section, we will explain {\it why} \eqref{eq:2nScalarUs} describes the scaffolded gluons. 
At tree-level, we can ``cheat'' to be assured that we are describing gluons, by showing that this proposal precisely matches the amplitude for gluons in the open bosonic string, and we will begin by doing this. 

We will then proceed to see why these expressions deserve to be called ``gluon amplitudes" more intrinsically, by showing that the amplitudes defined
in this way satisfy all the physical properties expected of gluons: on-shell gauge invariance,  linearity in polarizations, as well as consistent factorization on massless gluon poles. These will all follow from simple but fundamental facts about the $u$ variables summarized in our discussion of ``surfaceology". This more intrinsic understanding of the amplitudes directly in $u$ variable language will be crucial in the extension of our proposal to loop-level we undertake in the next section, where (welcome!) stark differences with string loop amplitudes appear and we can no longer ``cheat" by borrowing the (in)consistency of the bosonic string.

\subsection{Connection to the bosonic string}

At tree-level, gluon amplitudes in the bosonic string are given as a worldsheet integral~\cite{green1988superstringv1}:

\begin{equation}	\mathcal{A}^{\text{tree}}_n(1,2,\dots,n) = \int \frac{\diff^{n} z_i}{\text{SL(2,}\mathbb{R})} \,  \left( \prod_{i,j} (ij)^{2\alpha^{\prime} p_i \cdot p_j} \right) \, \left.\exp{\left( \sum_{i\neq j} 2 \frac{\epsilon_i \cdot \epsilon_j }{(ij)^2} - \frac{\sqrt{\alpha^{\prime}}\epsilon_i \cdot p_j}{(ij)}\right)}\right\vert_{\text{multi-linear in }\epsilon_i},
\label{eq:BosStringTree}
\end{equation}
where $\epsilon_i$ are the polarizations and $p_i$ the respective momentum, and $(ij)=z_i-z_j$. Since we are describing the scattering of gluons, in addition to the Koba-Nielsen factor~\cite{Koba:1969rw}, we also have a part encoding the information about the polarizations of the gluons, which is compactly represented in the exponential. Since gluon amplitudes are \textbf{linear} in each polarization, we are meant to keep only the term in the expansion of the exponential which is multilinear in all the polarizations.
  
This integral representation of the bosonic string amplitudes is fairly compact, however, it is less interesting for practical purposes such as computing the amplitudes. As it is usually the case for stringy integrals, the integrals in \eqref{eq:BosStringTree} do not converge and so if one wants to evaluate it for some physical kinematics, one has to first resort to analytic continuation to properly define it (we come back to this in section \ref{sec:FurtherCom}.) 
  
In addition, expanding the exponential we can see that to each term proportional to $\epsilon_i \cdot \epsilon_j$ we get a factor of $(ij)^2$ in the denominator. The presence of such terms tells us that the integral has double poles when $(ij) \rightarrow 0$. Of course, such double poles are not physical and can be eliminated using integration by parts (IBPs). 
  
Even forgetting about the convergence issues, one key difference between the \eqref{eq:BosStringTree} and the tachyon integral, \eqref{eq:StringAmpZ}, is that in the first, \eqref{eq:BosStringTree}, we have a sum of different stringy integrals, each multiplied by a factor of $p_i \cdot \epsilon_j$, $\epsilon_i \cdot \epsilon_j$, while in the latter, \eqref{eq:StringAmpZ}, there is a single integral that generates everything. So a natural question is whether there is a different starting point, that still describes gluons, but in which the amplitude, like in the scalar case, is represented as single term integral -- this can be achieved exactly by \textbf{scaffolding the gluons in a scalar problem}. 

To reach the $2n$-scalar problem,  define these scalars as the remaining degrees of freedom after fixing some configuration for the polarizations: 
  
Concretely, we consider the scattering to happen in a sufficiently high-dimensional space, with $\tilde{D}$ dimensions, so that we can achieve the following kinematical configuration: 

\begin{equation} 
\begin{aligned}
	p_i \cdot \epsilon_j &= 0, \quad \forall \, \, (i,j) \in (1,...,2n) , \\
	\epsilon_i \cdot \epsilon_j &= \begin{cases}
		1 \quad \text{if } (i,j) \in \{(1,2);(3,4);(5,6);...;(2n-1,2n)\} ,\\
		0 \quad \text{otherwise}.
	\end{cases}
\end{aligned}
\label{eq:kinConfig}	
\end{equation}

This kinematics can be achieved by demanding that the momentum, $p_i$ live in the first $D$ dimensions of spacetime, and the polarizations live in the extra dimensions, of this big dimensional, $\tilde{D}$ spacetime:
\begin{equation}
\begin{aligned}
    \begin{aligned}
    \begin{tikzpicture}
    \node at (0,0) {$p_i =\left(
\begin{array}{c}
|\\
p^{\mu}_i\\
| \\
0\\
0\\
\vdots \\
0\\
0\\
\end{array}
\right)$};
\draw[<->] (1.3,2.1) -- (1.3,0.7); 
\node at (1.6,1.4) {$D$};
\draw[<->] (1.3,0.4) -- (1.3,-2.1); 
\node at (2,-0.65) {Extra};
\node at (2,-1.05) {dims.};
    \end{tikzpicture}
    \end{aligned} ; \quad \quad
\end{aligned}
\begin{aligned}
\begin{tikzpicture}
    \node at (0,0) {$\epsilon_{2j/2j-1} =\left(
\begin{array}{c}
0\\
\vdots\\
0\\
0\\
\vdots \\
1\\ 
\vdots \\
0\\
\end{array}
\right)$ };
\draw[<->] (1.6,2.1) -- (1.6,0.7); 
\node at (1.9,1.5) {$D$};
\node at (1.95,-0.8) {$\leftarrow \, j \,$th};
\end{tikzpicture} 
\end{aligned}.
\end{equation}

By making this kinematical choice, we fixed all the polarizations, and all the data left in the problem are the momenta. Therefore we are left only with scalar degrees of freedom, whose kinematics is described by the Mandelstam invariants, $p_i \cdot p_j$ living in $D$ spacetime dimensions -- these are exactly $2n$ \textbf{scaffolding scalars}.
  
Plugging this kinematical choice in the bosonic string integral \eqref{eq:BosStringTree}, most of the terms in the exponential vanish, and the big sum collapses into a single term:
\begin{equation}
	\mathcal{A}_{2n}(1,2,...,2n) 
	\xrightarrow{\text{2n-scalar}}\int \frac{\diff^{2n} z_i}{\text{SL(2,}\mathbb{R})} \,   \prod_{i,j\in(1,2,...,2n)} (ij)^{2\alpha^{\prime} p_i \cdot p_j} \, \frac{1}{(12)^2(34)^2(56)^2...(2n-1,2n)^2}.
	\label{eq:2nScalarTree}
\end{equation}

Recalling the definition of the $u$ variables in terms of $(ij)$ \eqref{eq:udef}, we can rewrite \eqref{eq:2nScalarTree} in terms of $u$'s as follows:
\begin{equation}
\begin{aligned}
	\mathcal{A}_{2n}(1,2,...,2n) [X_{i,j}] &= \int  \frac{\diff^{n} z_i/\text{SL(2,}\mathbb{R})}{(12)(23)(34)...(2n,1)} \,   \prod_{i,j} u_{ij}^{\alpha^{\prime} X_{i,j}} \, \frac{(23)(45)(67)...(2n,1)}{(12)(34)(56)...(2n-1,2n)} \\
	&=\int \underbrace{\frac{\diff^{n} z_i/\text{SL(2,}\mathbb{R})}{(12)(23)(34)...(2n,1)} \,   \prod_{i,j} u_{ij}^{\alpha^{\prime} X_{i,j}} }_{\text{Integrand of \eqref{eq:Trphi3U}}} \, \left( \frac{\prod_{(i,j)\in\text{even}} u_{i,j}}{\prod_{(i,j)\in\text{odd}} u_{i,j}} \right).
\end{aligned}
\label{eq:2nScalarU}
\end{equation}

This is exactly the form of equation \eqref{eq:2nScalarUs} -- the integrand of \eqref{eq:Trphi3U}, times the extra multiplicative factor of the ratio of the product of all the $u_{i,j}$ for $i$ and $j$ even and the product of all the $u_{i,j}$ for $i$ and $j$ odd. 

So, from this perspective, we learn that the external scaffolding scalars we are scattering are secretly \textbf{gluons in higher dimensions}, where we are expanding around a kinematic point where the gluons effectively become colored scalars. Exactly because of the gluon $3$
-point interaction, we have that these colored scalars interact with gluons through a cubic vertex whose Feynman rule is, once again, the expected one in figure \ref{fig:FeynRules}. Note that, however, since we picked $\epsilon_i \cdot \epsilon_j =0$, except for $(i,j) = (2k-1,2k)$, this cubic interaction is only present between \textbf{adjacent} pairs of scalars, $(2k-1,2k)$. This means that for a $2n$-scalar process we have $n$-different species, $\{A_1,A_2,\dots,A_n\}$, of these colored scalars and the scattering process we are describing is $(1^{A_1},2^{A_1},3^{A_2},4^{A_2},  \cdots, (2n-1)^{A_n},(2n)^{A_n})$. Such $2n$-scalar bosonic string amplitudes were also studied in~\cite{He:2018pol, He:2019drm}, where a systematic IBP algorithm was developed and the field-theory limit produces $2n$-scalar amplitudes of Yang-Mills-scalar theory~\cite{Cachazo:2014xea} naturally given in CHY formulas~\cite{Cachazo:2013hca, Cachazo:2013iea}.

Therefore, to extract the gluon amplitude from \eqref{eq:2nScalarUs} all we need to do is compute the scaffolding residue: $X_{2k-1,2k+1}\rightarrow0$. In this kinematical locus, each pair of adjacent scalars, $(2k-1,2k)$, produces an on-shell gluon, uncovering the n-point gluon amplitude. Of course, the answer will be in terms of the scalar Mandelstam invariants, $X_{i,j}$, and so to recover the usual expression for the amplitude in terms of polarizations we can use directly the dictionary presented in section \ref{sec:Dict}.

\subsection{The scaffolding residue}

\begin{figure}[t]
\begin{center}
\includegraphics[width=0.5\textwidth]{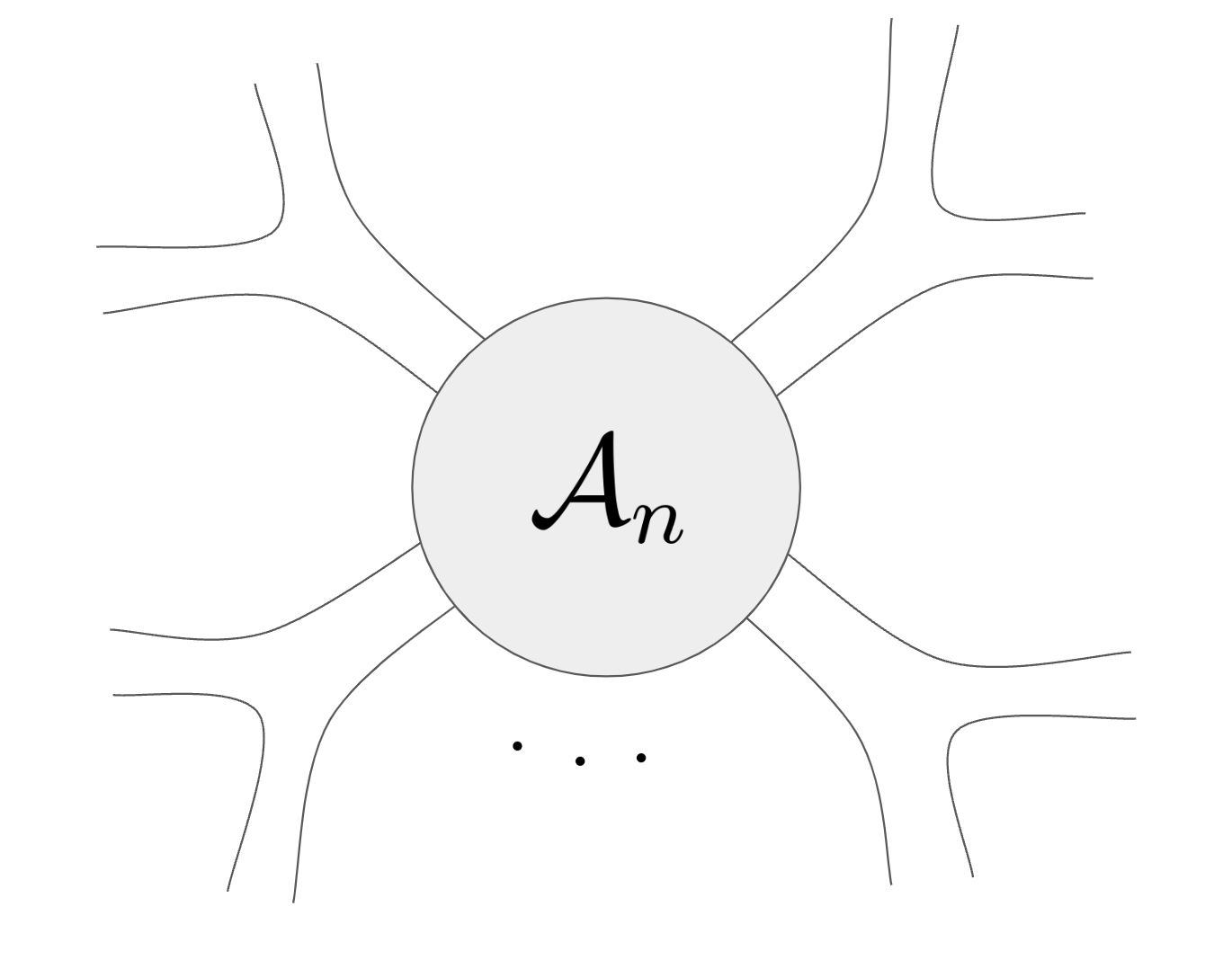}
\caption{Scaffolding triangulation: including all the $X_{2i-1,2i+1}$ propagators.}
\label{fig:ScaffTriang}
\end{center}
\end{figure}

Let us now proceed and see in detail how to extract the scaffolding residue from the 2n-scalar amplitude \eqref{eq:2nScalarUs}. Following the procedure laid out in section \ref{sec:Surfaceology}, each a positive parametrization, $u_{i,j}[y]$, is associated with a triangulation of the disk with $2n$ marked points on the boundary, $\mathcal{T}$. As pointed out before, the final answer is \textbf{independent} of this choice, but it is simpler to access the singularities of a particular diagram by choosing the parametrization associated with its dual triangulation. Therefore, since we want to access the residue $X_{2i-1,2i+1}\rightarrow0$, it is helpful to choose a triangulation that includes the scaffolding chords -- we call such triangulations as \textbf{scaffolding triangulation} (see figure \ref{fig:ScaffTriang}). 
\subsubsection{The $2n$-scalar form in the scaffolding triangulation}

For such a choice of triangulation, the ratio of $u_{i,j}$ in the 2n-scalar form \eqref{eq:2nScalarUs} reduces to something very simple. Recall from section \ref{sec:prody}, that the product of all the $y_{\mathcal{C}}$ can be written in terms of $u$'s as described in eq. \eqref{eq:prody}. 

Now something especially nice happens in the case of a scaffolding triangulation: by drawing the curves on the fat graph we can see that all the $X_{e,e}$ will be such that $P_{X_{e,e}} = -1 $ and  $X_{o,o}$ will be such that $P_{X_{o,o}} = +1 $, while all the remaining will be zero\footnote{Note that this is \textbf{independent} of the internal details of the triangulation, and so it is true for any surface, which means it holds both tree-level as well as loop-level}. Therefore, as long as we are working in the scaffolding triangulation we have:

\begin{equation}
	\prod_\mathcal{C} y_\mathcal{C} = \prod_s y_s \prod_i y_i =  \left( \frac{\prod_{(i,j)\in\text{odd}} u_{i,j}}{\prod_{(i,j)\in\text{even}} u_{i,j}} \right),
\end{equation}
where $y_s$ stands for the y's corresponding to propagators in the scaffolding, $X_{2k-1,2k+1}$, while $y_i$ are those associated with propagators of the internal gluon problem, $X_{o,o}$. 

Therefore picking a positive parametrization of the stringy integral corresponding to a scaffolding triangulation, $\mathcal{T}_s$,  \eqref{eq:2nScalarU} reduces to:

\begin{equation}
\begin{aligned}
	\mathcal{A}_{2n}(1,2,...,2n) [X_{i,j}] &=\int_0^{\infty} \prod_{s} \frac{\diff y_{s}}{y_{s}}\times \prod_{i} \frac{\diff y_{i}}{y_{i}} \,   \prod_{i,j} u_{ij}^{ X_{i,j}}[y_s,y_i] \, \left( \frac{1}{\prod_s y_s \prod_i y_i} \right) \\
	&= \int_0^{\infty} \underbrace{\prod_{s} \frac{\diff y_{s}}{y_{s}^2} \, \prod_{i} \frac{\diff y_{i}}{y_{i}^2} \,\prod_{i,j} u_{ij}^{ X_{i,j}}[y_s,y_i]}_{\Omega_{2n}},
\end{aligned}
\label{eq:2nScaffTriang}
\end{equation}
where we have set $\alpha^{\prime}=1$. Note that this is exactly the scalar integral \eqref{eq:Trphi3U}, but where instead of having a $\diff \log(y) = \diff y/y$ form, we have $\diff y/y^2$. We call this non-dlog form that we are integrating the 2n-scalar form, $\Omega_{2n}$. 

\subsubsection{Extracting the residue}
\label{sec:ExtRes}
Since we have picked the scaffolding chords to be part of the triangulation, we can factor out of the Koba-Nielsen factor a piece that looks like the $\prod_s y_s^{X_s}$, with $X_s$ denoting the scaffolding chords, $X_{2k-1,2k+1}$. Therefore the integral over the scaffolding $y_s$ reduces schematically to:
\begin{equation}
	\int_0^{\infty} \prod_{s} \frac{\diff y_{s}}{y_{s}^2}y_s^{X_s} \times  \mathcal{F}\,[y_{s},y_i,X_{i,j}],
\end{equation}
where $\mathcal{F}$ represents some function of the remaining variables which is regular and non-zero when the $y_s$ and $X_s$ go to zero. 
   
Looking at the integral we can see that if we set one of the scaffolding variables $X_{s_i}$ to zero, then the integral over $y_{s_i}$ is divergent for $y_{s_i}$ near $0$. Therefore, for $X_{s_i} \rightarrow 0$, the integral is dominated by the region $y_{s_i}$ near $0$. Focusing on this region, we can rewrite $\mathcal{F}$ using the appropriate Taylor expansion:
\begin{equation}
	\int_0^{\infty} \prod_{s\neq s_i} \frac{\diff y_{s}}{y_{s}^2}y_s^{X_s} \times  \int_0^{\epsilon} \frac{\diff y_{s_i}}{y_{s_i}^2}y_{s_i}^{X_{s_i}} \left( \mathcal{F}\,[y_{s_i}=0,y_s,y_i,X_{i,j}]+y_{s_i} \frac{\diff \mathcal{F}\,[y_s,y_i,X_{i,j}] }{\diff y_{s_i}}\bigg \vert_{y_{s_i}=0}  + ... \right),
\end{equation}
where the integration in $y_{s_i}$ is from $0$ to $\epsilon$, and $\epsilon$ is some number arbitrarily close to zero, but non-zero. 
  
One can now perform the integral, term by term in the expansion, using analytic continuation:
\begin{equation}
	\int_0^{\epsilon} \frac{\diff y}{y} y^A =  \frac{\epsilon^{A}}{A} , 
\end{equation}
which gives:
\begin{equation}
	\int_0^{\infty} \prod_{s\neq s_i} \frac{\diff y_{s}}{y_{s}^2}y_s^{X_s} \, \times \left( \frac{\epsilon^{X_{s_i}-1}}{X_{s_i}-1}\mathcal{F}\,[y_{s_i}=0,y_s,y_i,X_{i,j}]+\frac{\epsilon^{X_{s_i}}}{X_{s_i}} \cdot \frac{\diff \mathcal{F}\,[y_s,y_i,X_{i,j}] }{\diff y_{s_i}}\bigg \vert_{y_{s_i}=0}  + ... \right).
\end{equation}

From here we can read off directly the residue on $X_{s_i} =0$ to be
\begin{equation}
\begin{aligned}
	\text{Res}_{X_{s_i}=0}(\mathcal{A}_{2n}(1,2,...,2n))&= \int_0^{\infty} \prod_{i} \frac{\diff y_{i}}{y_{i}^2}\int_0^{\infty} \prod_{s\neq s_i} \frac{\diff y_{s}}{y_{s}^2}y_s^{X_s} \, \times \left(  \frac{\diff \mathcal{F}\,[y_s,y_i,X_{i,j}] }{\diff y_{s_i}}\bigg \vert_{y_{s_i}=0}  \right)\\
	&= \int_0^{\infty} \text{Res}_{y_{s_i}=0}\left(\Omega_{2n}\right).
\end{aligned}
\end{equation}

Repeating this procedure for all the scaffolding variables $y_s$, we obtain:
\begin{equation}
	\mathcal{A}_n^{\text{gluon}}= \int_0^{\infty} \, \text{Res}_{y_{s_1}=0} \left(\text{Res}_{y_{s_2}=0}\left(...\left(\text{Res}_{y_{s_n}=0}\Omega_{2n}\right)...\right)\right) ,
\end{equation}
and we conclude that this consecutive residue is simply given by the part of $\mathcal{F}\,[y_{s},y_i,X_{i,j}]$ which is \textbf{linear in all the scaffolding} $y$'s, $\{y_{s_1},$...$,y_{s_n}\}$.

\subsection{Scaffolding residue $\Rightarrow$ Gauge invariant and multi-linear amplitude}
\label{sec:scaffGinv}

Let's now understand how the scaffolding residue of the 2n-scalar amplitude, automatically gives something of the form of \eqref{eq:GInv_Mult}, $i.e.$ an object which is automatically gauge-invariant and multilinear in the corresponding polarizations. Without loss of generality, let us focus on what happens after taking the residue corresponding to putting gluon 1 on-shell, $i.e.$ $X_{1,3} \rightarrow 0$, which is given by the residue of $\Omega_{2n}$ at $y_{1,3}=0$:
\begin{equation}
\begin{aligned}
	\text{Res}_{y_{1,3}=0} \Omega_{2n}(1,2,...,2n) [X_{i,j}] &=\prod_{s\neq (1,3)} \frac{\diff y_{s}}{y_{s}^2} \, \prod_{i} \frac{\diff y_{i}}{y_{i}^2} \, \sum_{(IJ)} \left.\left( X_{I,J}\frac{\partial \log ( u_{IJ})}{\partial y_{1,3}}\prod_{(km) } u_{km}^{ X_{k,m}}[y] \right) \right\vert_{y_{1,3}=0}.
\end{aligned}
\end{equation}

However, not all the $u_{i,j}$ depend on $y_{1,3}$. In fact, only the $u$'s corresponding to curves that go through road 13, can depend on $y_{1,3}$. Therefore the only curves for which this derivative is non-zero are curves  $X_{1,j}$ and $X_{2,j}$ (see figure \ref{fig:X1JX2J1}):
\begin{equation}
\begin{aligned}
	\text{Res}_{y_{1,3}=0} \Omega_{2n}[X_{i,j}] =\prod_{s\neq (1,3)} \frac{\diff y_{s}}{y_{s}^2} \, \prod_{i} \frac{\diff y_{i}}{y_{i}^2} \, \sum_{J}& \left( X_{1,J}\frac{\partial \log ( u_{1J})}{\partial y_{1,3}}\prod_{(km) } u_{km}^{ X_{k,m}}[y] \right. \\ 
	& \quad \quad \quad   \left. \left. + X_{2,J}\frac{\partial \log ( u_{2J})}{\partial y_{1,3}}\prod_{(km) } u_{km}^{ X_{k,m}}[y] \right)  \right\vert_{y_{1,3}=0}.
\end{aligned}
\end{equation}

Now we claim that $\partial_{y_{1,3}} \log ( u_{1J})=-\partial_{y_{1,3}}\log (u_{2J})$. This follows immediately from one of the fundamental facts about the $u$ variables we summarized in our surfaceology review. We can consider the smaller surface obtained simply by cutting just before the forking into $(1,2)$. Let's consider the curve $[1 J]$ on the smaller surface that starts on the road $(13)$ and exits at $J$; we use square brackets to remind us this is a curve on the smaller surface. This manifestly does not depend on $y_{1,3}$ which is a boundary in the smaller surface. This curve can be extended to a curve on the full surface in two ways, into $(1J)$ and $(2J)$. So we have that $U_{[1,J]} = u_{1,J} u_{2,J}$ is independent of $y_{1,3}$ and hence $\partial_{y_{1,3}} u_{1,J} = - \partial_{y_{1,3}} u_{2,J}$ as desired. 
We can also see this very simply directly, from the matrix formula for the $u$ variables. Let's consider the words corresponding to curves $X_{2,J}$ and $X_{1,J}$. From figure \ref{fig:X1JX2J1}, we see that these curves agree everywhere except for the last intersection, at the end of road $13$. So we can write their $u$-variables as follows:
\begin{figure}[t]
\begin{center}
\includegraphics[width=0.5\textwidth]{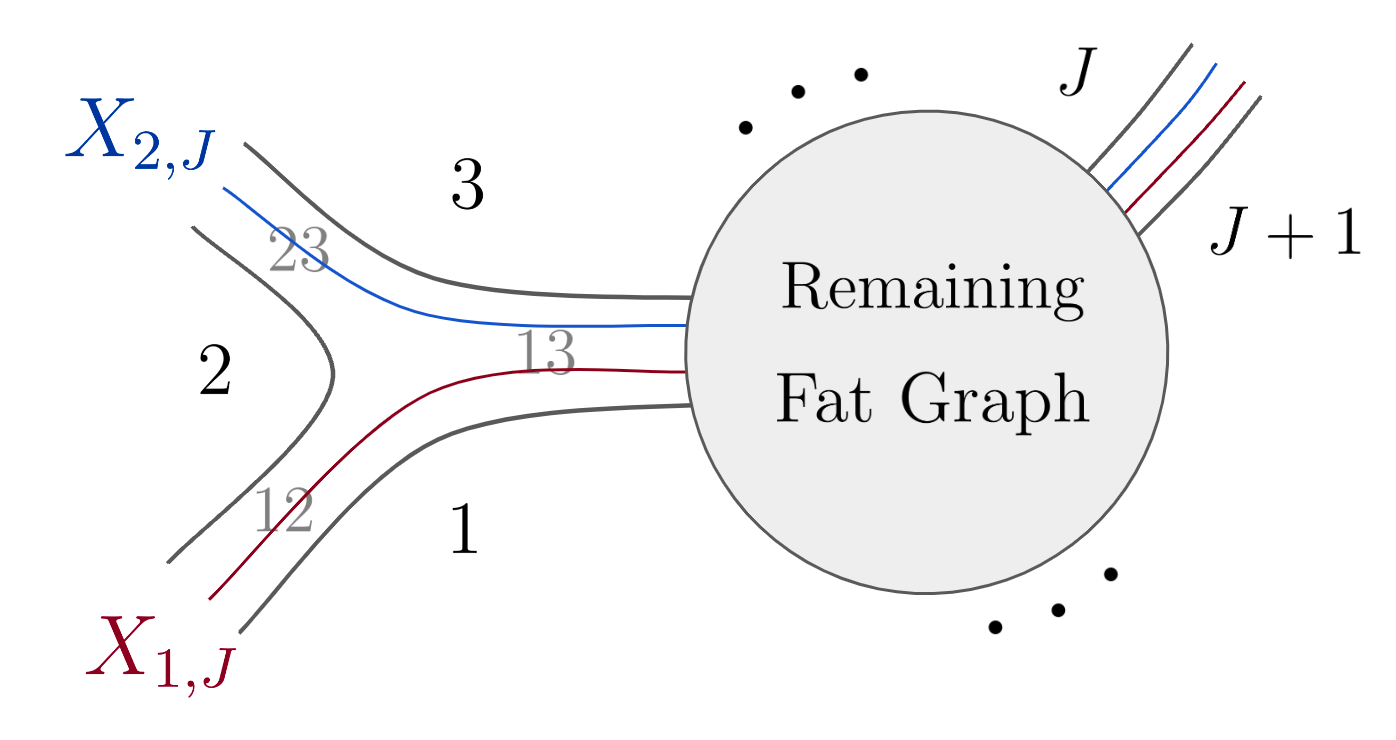}
\caption{Curves $X_{1,J}$ and $X_{2,J}$.}
\label{fig:X1JX2J1}
\end{center}
\end{figure}
\begin{equation}
\begin{aligned}
	& \underline{1J} \,: M_{\text{remaining}} \times M_{L}(y_{1,3}) = 
	\begin{bmatrix}
		A & B \\
		C & D 
	\end{bmatrix} \times \begin{bmatrix}
		y_{1,3} & y_{1,3}  \\
		0 & 1 
	\end{bmatrix} = \begin{bmatrix}
		A\,y_{1,3} &  A\,y_{1,3}+B  \\
		C\,y_{1,3}  & C\,y_{1,3}+D  
	\end{bmatrix} \\ 
	&\Rightarrow u_{1J} = \frac{C\, (A\,y_{1,3}+B) }{A\,(C\,y_{1,3}+D)},	\\
	& \underline{2J} \,: M_{\text{remaining}} \times M_{R}(y_{1,3}) = 
	\begin{bmatrix}
		A & B \\
		C & D 
	\end{bmatrix} \times \begin{bmatrix}
		y_{1,3} & 0  \\
		1 & 1 
	\end{bmatrix} = \begin{bmatrix}
		A\,y_{1,3}+B  &  B  \\
		C\,y_{1,3} + D  & D  
	\end{bmatrix} \\
	&\Rightarrow u_{2J} = \frac{B\, (C\,y_{1,3}+D) }{D\,(A\,y_{1,3}+B)}	,
\end{aligned}
 \end{equation}
where $A,B,C$ and $D$ are functions of the remaining $y$'s. Even without knowing the details of the rest of the curves, we can check that the product $u_{1,J} u_{2,J}$ does \textbf{not} depend on $y_{1,3}$. Thus we have $ \diff \log u_{1J} = - \diff \log{u_{2J}}$. This relation allows us to write the residue as:
\begin{equation}	\text{Res}_{y_{1,3}=0} \Omega_{2n}[X_{i,j}] = \sum_{J} \left(X_{2,J} - X_{1,J} \right) \underbrace{\prod_{s\neq (1,3)} \frac{\diff y_{s}}{y_{s}^2} \, \prod_{i} \frac{\diff y_{i}}{y_{i}^2} \times \frac{\partial \log ( u_{2J})}{\partial y_{1,3}}\prod_{(km) \neq (2J)} u_{km}^{ X_{k,m}}[y] }_{\mathcal{Q}_{2,J}} ,
\end{equation}
\begin{figure}[t]
    \centering
\includegraphics[width=0.4\textwidth]{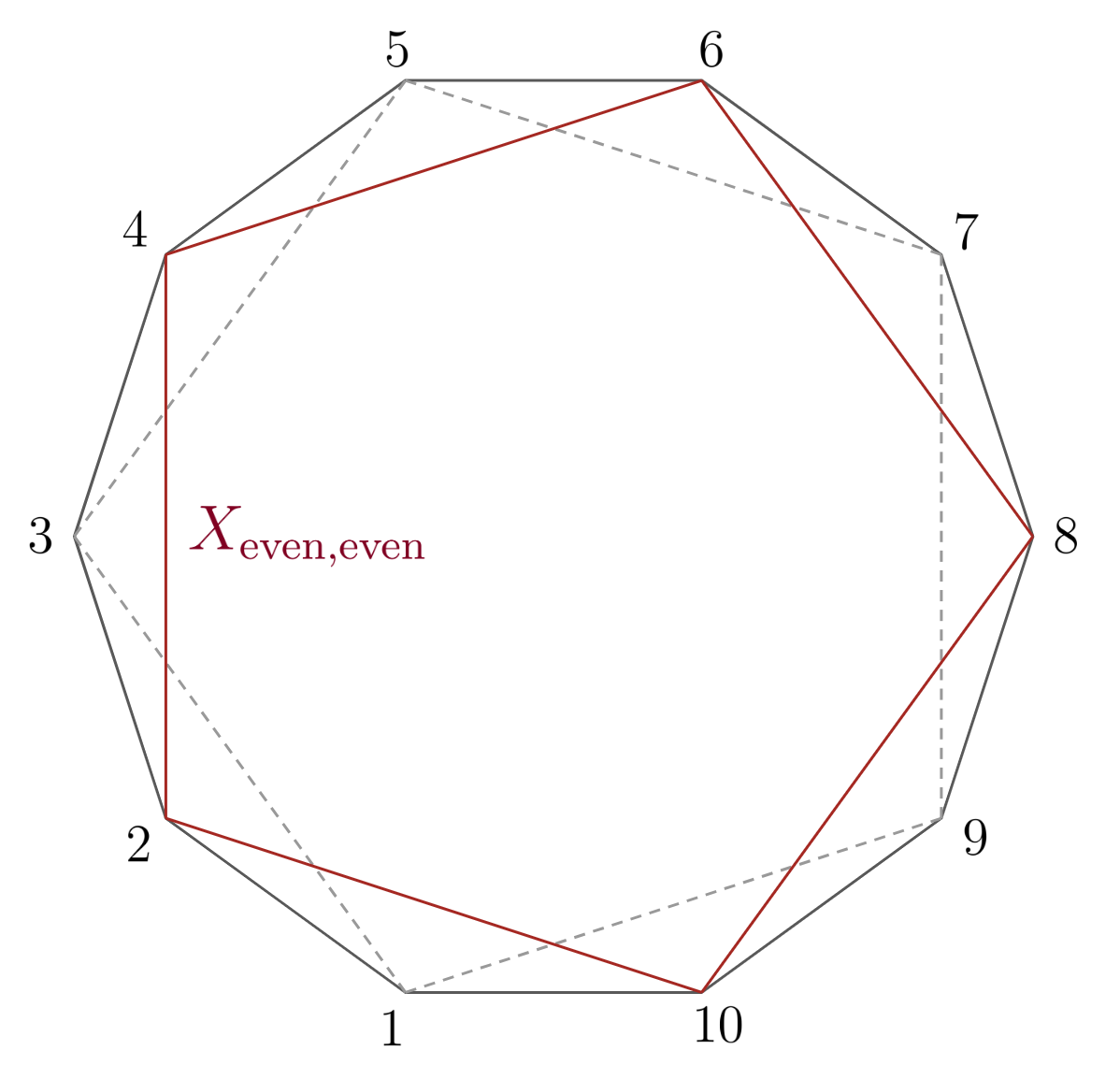}
    \caption{Triangulation containing $\{X_{2i,2i+2}\}$.}
    \label{fig:NewTriang}
\end{figure}
which is exactly of the form required by gauge-invariance and multi-linearity \eqref{eq:GInvX2j}.  Here the product over $(km)$ excludes $(2J)$ since $u_{2J}|_{y_{1,3}=0}=1$. From \eqref{eq:GInvX2j}, we should also be able to express this residue as a sum of terms proportional to $(X_{2,J}-X_{3,J})$. This is not as obvious and, to prove that this is the case, it is useful to triangulate the surface in a different way -- using a triangulation containing $X_{2i,2i+2}$, with $i \in \{1,..,n\}$ (see figure \ref{fig:NewTriang}).

In this case, the $u$'s are functions of different $y$'s, that we denote by $\tilde{y}$. In this parametrization, the residue corresponding to the original scaffolding residue, $i.e.$ $X_{1,3}=X_{3,5}=...=X_{1,2n-1}=0$, is obtained by taking the residue of $\Omega_{2n}$ at $y_{2i,2i+2}\rightarrow \infty$. Now in this new triangulation, by exactly the same argument presented for the original triangulation, we have that  $\frac{\partial \log ( u_{3J})}{\partial \tilde{y}_{2,4}}=-\frac{\partial \log ( u_{2J})}{\partial \tilde{y}_{2,4}}$, therefore we also have:
\begin{equation}
\text{Res}_{\tilde{y}_{2,4}=\infty} \Omega_{2n}[X_{i,j}] = \sum_{J} \left(X_{2,J} - X_{3,J} \right) \underbrace{\prod_{s\neq (2,4)} \frac{\diff \tilde{y}_{s}}{\tilde{y}_{s}^2} \, \prod_{i} \frac{\diff \tilde{y}_{i}}{\tilde{y}_{i}^2} \times \frac{\partial \log ( u_{2J})}{\partial \tilde{y}_{2,4}}\prod_{(km) \neq (3J)} u_{km}^{ X_{k,m}}[\tilde{y}] }_{\mathcal{\tilde{Q}}_{2,J}}.
\end{equation}

Repeating the same argument, one can show that such a form holds for any triple of indices $(2i-1,2i,2i+1)$, and thus for each gluon in the problem. Just in the same way as before, (after taking the remaining scaffolding residues) we have $\int \tilde{Q}_{2,J} = \int Q_{2,J} = \partial_{X_{2,J}} \mathcal{A}_n$.
Hence, we have proved that taking the scaffolding residue automatically gives us something that is manifestly \textbf{gauge invariant} as well as \textbf{multi-linear} in all polarizations. 

\subsection{Tree-level factorization}
\label{sec:factTree}

Our goal now is to understand what factorization looks like in this scalar language. As we know, for the usual way we write the amplitudes directly in terms of the polarizations of the gluons, when we extract the residue of the amplitude on a pole, $X_{a,b}=0$,  we obtain the gluing of the two lower point amplitudes, $\mathcal{A}_L$ and $\mathcal{A}_R$:
\begin{equation}
    \mathcal{A}_{\text{glued}} = \mathcal{A}_L^\mu \left( -\eta_{\mu,\nu} + \frac{p_\mu q_\nu + p_\nu q_\mu}{p \cdot q} \right) \mathcal{A}_R^\nu = -  \mathcal{A}_L^\mu  \eta_{\mu,\nu}  \mathcal{A}_R^\nu,
\end{equation}
where the term between brackets is doing the polarization sum for the intermediate gluon, $I,I^\prime$ we are producing. We define
\begin{equation}
	\mathcal{A}_L^{\mu} = \partial_{\epsilon_{I,\mu}} \mathcal{A}_L \quad , \quad \mathcal{A}_R^\nu = \partial_{\epsilon_{I^\prime,\nu}} \mathcal{A}_R,
\end{equation}
and $p_\mu= p_{I,\mu} = -p_{I^\prime,\mu}$ and $q^{\mu}$ is some reference vector for which $q \cdot p \neq 0$. Note that the final answer does not depend on the arbitrary choice of $q^\mu$. In the second equality, we use the fact that $M_{L}^\mu \,p_{I,\mu} = M_R^\nu \,p_{I^\prime,\nu}=0$, as required by gauge invariance.

\begin{figure}[t]
\begin{center}
\includegraphics[width=\linewidth]{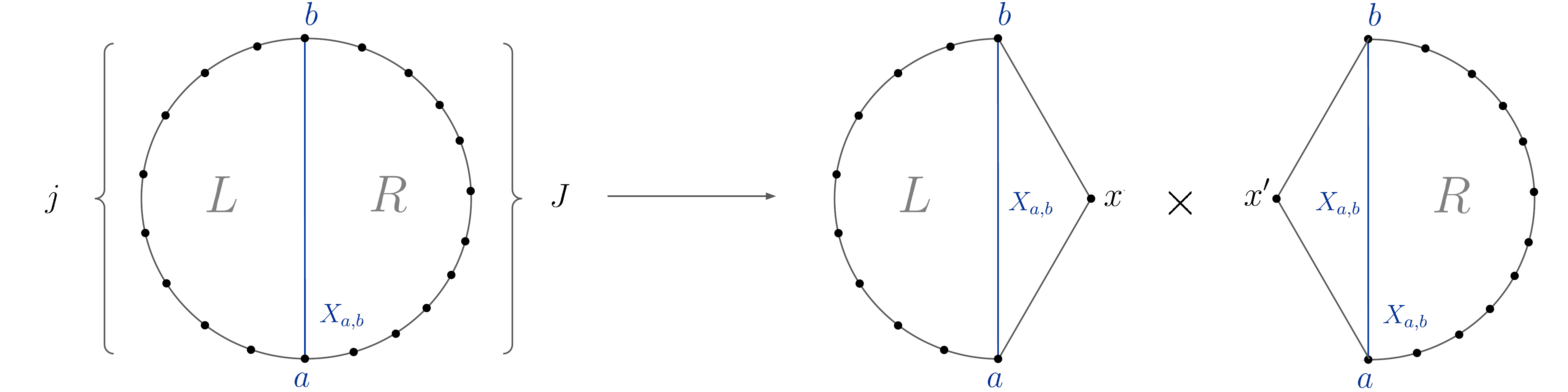}
\caption{Tree-level factorization.}
\label{fig:TreeFact}
\end{center}
\end{figure}

We now want to phrase this rule for summing over polarizations (``gluing'' rule) in terms of the scalar variables of the scaffolded problem. To do so, we think of gluons, $I$ and $I^\prime$ as being scaffolded by a pair of scalars, such that in the momentum polygon this interaction is associated with a triangle whose vertices are $(a,b,x)$ and $(a,b,x^\prime)$, respectively  (see figure \ref{fig:TreeFact} ). Therefore, the polarizations of $I$ and $I^\prime$ are, respectively, encoded in the scalar variables, $X_x^\mu$, and $X_{x^\prime}^\nu$. Now using our scalar definition of gauge-invariance and multi-linearity of $\mathcal{A}_L$ and $\mathcal{A}_L$ in gluons, $I$ and $I^\prime$, respectively, we have:
\begin{equation}
\begin{aligned}
    \mathcal{A}_L^{\mu} = \partial_{X_{x}^\mu} \mathcal{A}_L =& \frac{\partial}{\partial X_{x}^\mu} \left( \sum_{j} (X_{x,j}-X_{b,j})\mathcal{Q}_{x,j} \right) = 2   \left( \sum_{j} (X_{x}^{\mu}-X_{j}^{\mu})\frac{\partial \mathcal{A}_L}{\partial X_{x,j}} \right), \\
    \mathcal{A}_R^{\nu} = \partial_{X_{x^\prime}^\nu} \mathcal{A}_R =& \frac{\partial}{\partial X_{x^\prime}^\nu} \left( \sum_{J} (X_{x^\prime,J}-X_{a,J})\mathcal{Q}_{x^\prime,J} \right) = 2   \left( \sum_{J} (X_{x^\prime}^{\nu}-X_{J}^{\nu})\frac{\partial \mathcal{A}_R}{\partial X_{x^\prime,J}} \right) .
\end{aligned}
\end{equation}

Therefore contracting with the flat metric, we obtain:
\begin{equation}
    \mathcal{A}_{\text{glued}} = -4 \sum_{j,J} (X_{x}^{\mu}-X_{j}^{\mu})(X_{x^\prime,\mu}-X_{J,\mu}) \frac{\partial \mathcal{A}_L}{\partial X_{x,j}} \frac{\partial \mathcal{A}_R}{\partial X_{x^\prime,J}},
\end{equation}
using a gauge transformation and a rescaling, we can choose $X_x^{\mu}= X_b^\mu$ and $X_{x^\prime}^{\nu}= X_a^\nu$ which yields:
\begin{equation}
\begin{aligned}
    \mathcal{A}_{\text{glued}} & \propto \sum_{j,J} (X_{b}^{\mu}-X_{j}^{\mu})(X_{a,\mu}-X_{J,\mu}) \partial_ {X_{x,j}}\mathcal{A}_L \partial_ {X_{x^\prime,J}} \mathcal{A}_R \\
    & \propto \sum_{j,J} (-X_{b,J} -X_{a,j}+X_{j,J}) \partial_ {X_{x,j}}\mathcal{A}_L \partial_ {X_{x^\prime,J}} \mathcal{A}_R.
\end{aligned}
\end{equation}

To check factorization we want to extract the massless residues and interpret the result in terms of lower point gluon amplitudes. Let's study the residue corresponding to having $X_{a,b}$ on-shell, so $X_{a,b}=0$. Choosing an appropriate scaffolding triangulation that contains $X_{a,b}$, this residue is given by the coefficient multiplying $y_{a,b}$ in the expansion of $\prod u_X^X$. 
Similarly to what was argued in \ref{sec:scaffGinv}, the only $u$'s that depend on $y_{a,b}$ are those that correspond to curves that intersect $X_{a,b}$ as laminations, i.e. all the curves that go through road $ab$ in the fat graph. There are three types of such curves: $X_{j,J}$, $X_{a,J}$ and $X_{j,b}$, with $j \in \{a+1, ..., b-1\}$ and $J \in \{b+1,...,a-1\}$. So expanding the Koba-Nielsen factor up to linear order in $y_{a,b}$ yields:
\begin{equation}
\begin{aligned}
	\prod_X u_X ^{X} = 1 &+ y_{a,b} \left. \left[\sum_{j,J} X_{j,J} \frac{\partial \log(u_{j,J})}{\partial y_{a,b}}  + \sum_{j} X_{j,b} \frac{\partial \log(u_{j,b})}{\partial y_{a,b}}  + \sum_{J} X_{a,J} \frac{\partial \log(u_{a,J})}{\partial y_{a,b}} \right] \right\rvert_{y_{a,b}=0}  \\
	& + \mathcal{O}(y_{a,b}^2) .
\end{aligned}
\label{eq:Factu}
\end{equation}

\begin{figure}[t]
\begin{center}
\includegraphics[width=0.6\textwidth]{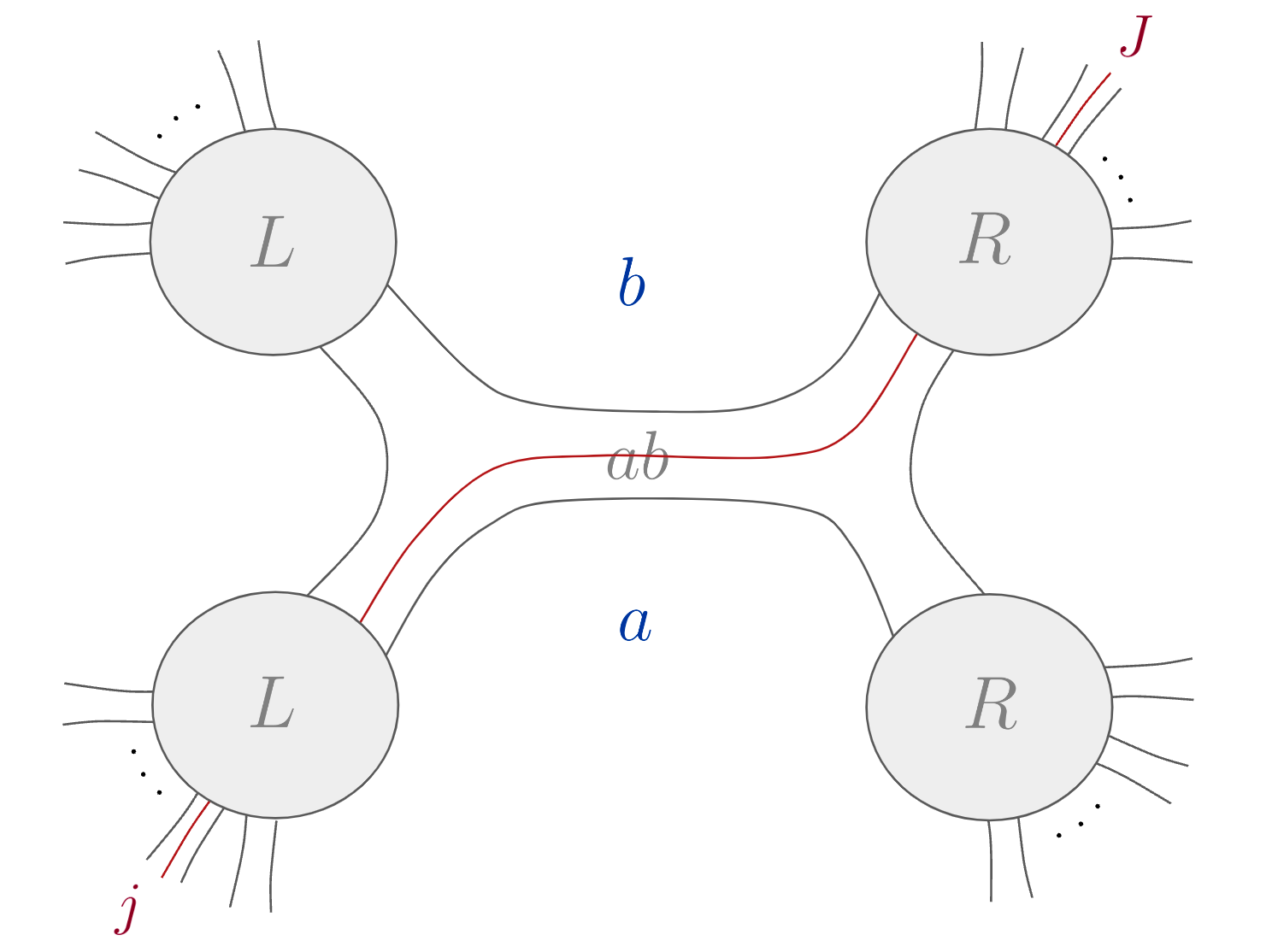}
\caption{Fat graph around road $ab$.}
\label{fig:factab}
\end{center}
\end{figure}

Now we note that any triangulation containing road $ab$ will look schematically like figure \ref{fig:factab}, and therefore, similarly to what we proved in \ref{sec:scaffGinv}, we have:
\begin{align}
	&\prod_j u_{j,J}\times u_{a,J} \equiv \mathcal{U}_{ab, J}^{\text{right}}, \\
	&\prod_J u_{j,J}\times u_{j,b} \equiv \mathcal{U}_{j, ab}^{\text{left}},
\end{align}
where $\mathcal{U}_{ab, J}^{\text{right}}$ and $\mathcal{U}_{j, ab}^{\text{left}}$ are the $u$'s of the smaller surfaces on the right/left of $ab$ respectively, and thus \textbf{do not depend} on $y_{a,b}$. Therefore we have:
\begin{equation}
\begin{aligned}
	\frac{\partial \log(u_{a,J})}{\partial y_{a,b}} &= - \sum_j \frac{\partial \log(u_{j,J})}{\partial y_{a,b}}, \\
	\frac{\partial \log(u_{j,b})}{\partial y_{a,b}} &= - \sum_j \frac{\partial \log(u_{j,J})}{\partial y_{a,b}}.
\end{aligned}
\end{equation}

Plugging these relations into \eqref{eq:Factu}, we get:
\begin{equation}
	\prod_X u_X ^{X} = 1 + y_{a,b} \sum_{j,J} (X_{j,J} -X_{j,b} -X_{a,J}) \frac{\partial \log(u_{j,J})}{\partial y_{a,b}}  \bigg \rvert_{y_{a,b}=0}  + \, \mathcal{O}(y_{a,b}^2) .
\label{eq:Factu2}
\end{equation}

Finally, we can get  $\frac{\partial \log(u_{j,J})}{\partial y_{a,b}}  \bigg \rvert_{y_{a,b}=0}$ using the $u$-equations:
\begin{equation}
	u_{j,J} + u_{a,b} \prod_{X_L \in L_j} u_{X_L} \prod_{X_R \in R_J} u_{X_R} \prod_{X \in \mathcal{I}} u_X =1, 
\end{equation}
where $R_J$ and $ L_j$ stand for the set of curves that intersects $X_{j,J}$ but that live on the right/left on $X_{a,b}$, respectively, and so do \textbf{not} intersect $X_{a,b}$; $ \mathcal{I}$ stands for the set of curves that are simultaneously incompatible with $X_{a,b}$ and $X_{j,J}$. Since $u_{a,b}=0$, when $y_{a,b}=0$, we have:
\begin{equation}
	\frac{\partial \log(u_{j,J})}{\partial y_{a,b}}  \bigg \rvert_{y_{a,b}=0} = - \frac{\partial u_{a,b}}{\partial y_{a,b}}  \bigg \rvert_{y_{a,b}=0} \prod_{ X_L \in L_j } \tilde{u}_{X_L} \prod_{ X_R \in R_J } \tilde{u}_{X_R} ,
\end{equation}
where $\tilde{u}_X = u_X \vert_{y_{a,b}=0}$, are the $u$'s of the smaller right and left surfaces, obtained when cut through curve $X_{a,b}$.  In addition, we don't have any $u_X$ for $X \in \mathcal{I}$ since these all go to 1 when $u_{a,b} \rightarrow0$. 

We now need to compute $\partial_{y_{a,b}} u_{a,b}\vert_{y_{a,b}=0}$. It is easy to see that this expression nicely factorizes in precisely the way needed to ensure factorization of the amplitude. Since $(a,b)$ is road present in our triangulation, the word for $(a,b)$ is a ``$V$'', with a series of right turns ending on $(a,b)$ then a series of left turns thereafter, of the form $\alpha R a_1 \cdots a_l R y_{a,b} L b_1 L \cdots b_r L \beta$ where $\alpha, \beta$ are boundaries and the $a$'s and $b$'s denote the roads on the left and right of $(a,b)$. Then from the matrix product formula for $u_{a,b}$ it is very easy to see that 
\begin{equation}
u_{a,b} = y_{a,b} F_L F_R + O(y_{a,b}^2),
\end{equation}
where
\begin{equation}
F_L = ( 1 + a_l + a_l a_{l-1} + \cdots+ a_{l} a_{l-1} \cdots a_1), \, F_R = (1 + b_r + b_{r} b_{r-1} + \cdots + b_r b_{r-1} \cdots b_1).
\end{equation}

So we have that $\partial_{y_{a,b}} u_{a,b}\vert_{y_{a,b}=0} = F_L F_R$ nicely factorizes. But each of these factors $F_L, F_R$ is also directly interpreted as the coefficient of the leading term in $y_{a,b}$ for $(a,b)$ curve on each of the obvious smaller $L,R$ surfaces. The reason is simply that on each of the $L,R$ surfaces, the $(a,b)^{L,R}$ curves is still a road in the triangulation, the word is still a $V$, but e.g. on the L surface, the word for $(a,b)^L$ immediately runs into a boundary after turning left on $(a,b)$, similarly for $(a,b)^R$. Thus $\partial_{y_{a,b}} u^L_{a,b}\vert_{y_{a,b}=0} = F_L, \partial_{y_{a,b}} u^R_{a,b}\vert_{y_{a,b}=0} = F_R$. Thus we have that  
\begin{equation}
\frac{\partial u_{a,b}}{\partial y_{a,b}}\vert_{y_{a,b} = 0} = \frac{\partial u^{L}_{a,b}}{\partial y_{a,b}}\vert_{y_{a,b} = 0} \times \frac{\partial u^{R}_{a,b}}{\partial y_{a,b}}\vert_{y_{a,b} = 0} .
\end{equation}

So then we have that the residue of the amplitude at $X_{a,b}=0$ is:
\begin{equation}\label{eq-fac}
	\text{Res}_{X_{a,b}=0} \, \mathcal{A}_n = - \sum_{j,J} (X_{j,J} -X_{j,b} -X_{a,J}) \underbrace{ \frac{\partial u^{L}_{a,b}}{\partial y_{a,b}}\vert_{y_{a,b} = 0} 
 \prod_{X_L\in \mathcal{L}_j} \tilde{u}_{X_L} }_{Q_j^L} \quad \cdot \underbrace{\frac{\partial u^{R}_{a,b}}{\partial y_{a,b}}\vert_{y_{a,b} = 0} \prod_{X_R \in \mathcal{R}_J} \tilde{u}_{X_R}   }_{Q_J^R}.
\end{equation}

But $Q_j^L$ and $Q_J^R$ are precisely the objects that show up in the lower-point amplitudes when we take the scaffolding residue on $y_{a,b}$! Consider two lower-point amplitudes: $\mathcal{A}_L (a, a+1, ...,b-1, b, x)$ and $\mathcal{A}_R (b, b+1, ...,a-1, a, x^\prime)$. As we saw in our discussion of gauge invariance and multilinearity above, when we take the scaffolding residue as $X_{a,b} \to 0$ we have ${\cal A}_L = \sum_j (X_{x,j} - X_{b,j}) \partial_{y_{a,b}} u_{x,j} \vert_{y_{a,b} = 0}$ and similarly for ${\cal A}_R$. But by precisely the same steps above, by looking at the $u$ equation for $u_{x,j}/u_{x^\prime,J}$ and differentiating with respect to $y_{a,b}$, we find that 
\begin{equation}
\begin{aligned}
	\text{Res}_{X_{a,b}=0} \, \mathcal{A}_L(a, ...,b, x)=&\sum_j (X_{x,j} -X_{b,j}) \, Q_{x,j}^L, \\
	\text{Res}_{X_{a,b}=0} \, \mathcal{A}_R(b, ...,a, x^\prime)=&\sum_J (X_{x^\prime,J} -X_{a,J}) \, Q_{x^\prime,J}^R, \
\end{aligned}
\end{equation}
where the $Q_{x,j}^L, Q_{x^\prime,J}^R$ are precisely the factors that appeared in the residue of the full amplitude on the $X_{a,b} \to 0$ pole above, $i.e.$ $Q_{x,j}^L \equiv Q_{j}^L$ and $ Q_{x^\prime,J}^R \equiv Q_{J}^R$. Since $Q_j^L = \partial_{X_{x,j}} {\cal A}_L$ and similarly for $Q_J^R$, we finally have that 
\begin{equation}\label{eq-fac2}
	\text{Res}_{X_{a,b}=0} \, \mathcal{A}_n = - \sum_{j,J} (X_{j,J} -X_{j,b} -X_{a,J}) \partial_{X_{x,j}} {\cal A}_L \partial_{X_{x^\prime,J}} {\cal A}_R,
\end{equation}
which is precisely the statement of factorization on the $X_{a,b} \to 0$ pole. %Note that in this argument no use was made of the ``left'' and ``right'' surfaces being trees; we have proven general ``tree exchange'' factorizations for general loop amplitudes on the left and right. 

\subsection{Computing scaffolding residues for general $n$}

In this section, we present explicit formulas for scaffolding residues for all $n$. We will see that the result becomes exactly the $n$-gluon bosonic string amplitude written in our new scalar language, whose integrand is a sum of products of surprisingly simple building blocks we call $V$'s and $W$'s, corresponding to first- and second-derivatives with respect to the scaffolding $y$'s. We will also see that the $V$'s and $W$'s can be expressed purely in terms of products of $u$ variables of the inner $n$-gon problem in a nice graphical way. Our purpose here is to show how the bosonic string gluons amplitude naturally falls out of taking the scaffolding residue in the world of $u$ and $y$ variables; in an appendix~\ref{sec:bosonicString} we show how to see this directly in terms of the conventional worldsheet $z$ variables. 

Of course, the consistent factorization of the scaffolded bosonic string amplitude at tree-level guarantees that taking the scaffolding residue will correctly land on the gluon amplitude, but we are performing this computation here for two reasons. First, in order to illustrate how the residues can be taken and subsequently interpreted directly in terms of integrands involving the $u$ variables on the smaller $n$-gon surface. We expect this exercise will be useful for higher loops, where our picture for gluon amplitudes differs significantly (and is much simpler) than the string amplitude. Second, the final expressions written in terms of $X$ and $u$ variables have a very pretty direct combinatorial characterization, which is not manifested in the conventional bosonic string gluon amplitudes written in terms of polarization vectors and momenta. 

We begin by considering a scaffolding triangulation, denoting by $y_{k} \equiv y_{2k{-}1, 2k{+}1} $ for $k\leq n$ to represent the $y$'s associated with the ``outer" scaffolding propagators $X_{2k-1,2k+1}$, and $y_i$ with $n<i\leq 2n-3$ for the ones corresponding to the ``inner" propagators of the gluon problem. For example, for $n=4$, we have denoted $y_{1}=y_{1,3}, y_{2}=y_{3,5}, y_{3}=y_{5,7}, y_{4}=y_{1,7}$ and $y_5=y_{1,5}$ or $y_5=y_{3,7}$.

As we've discussed already taking the scaffolding residue amounts to extracting the coefficient of $y_1 \cdots y_n$ of the factor $\prod u^{\alpha' X}$, which is a function of $y_{n{+}1}, \cdots, y_{2n{-}3}$ and the $X$ variables in the exponents. As we have seen in sec.~\ref{sec:scaffGinv}, only $u_{1,j}$ and $u_{2,j}$ depend on $y_1$ since these are the only two chords that pass the road $13$; they differ only by taking different turns after hitting $(13)$ and the product $u_{1,j} u_{2,j}$ is independent of $y_1$. Thus, the derivative with respect to $y_1$ brings down a factor
\begin{equation}
V_1:=\alpha' \sum_j (X_{2,j} -X_{1,j}) \left. \frac{\partial u_{2,j}}{\partial y_1}\right\vert_{y_1=0}\,,
\end{equation}
where the sum is over $j=1,\cdots, 2n$, but the terms with $j=1,2,3$ vanish ($X_{1,3}=0$), and we have used the fact that $u_{2,j}=1$ on the support of $y_1=0$ and so we omit it in the denominator. 

Similarly only $u_{2a-1,j}$, $u_{2a,j}$ depend on $y_a$ and their product is independent of it, thus we define for $a=1,2, \cdots, n$
\begin{equation}
V_a:=\alpha' \sum_j (X_{2a,j}-X_{2a-1,j}) \left. \frac{\partial u_{2a,j}}{\partial y_a}\right\vert_{y_a=0},
\end{equation}
and the simplest term of the residue contains the product of $n$ such factors $V_1 V_2 \cdots V_n$, which is at order $\alpha'^n$. 

More complicated terms appear when the derivative with respect to $y_b$ hits on $V_a$ from previous steps. However, the higher-than-second-order derivative with respect to $y_1, \cdots, y_n$ of any $u$ variable must vanish, since any chord has exactly one entrance and exit. Moreover, exactly four $u$ variables, namely $u_{2a-1, 2b-1}$, $u_{2a-1, 2b}$ , $u_{2a, 2b-1}$ and $u_{2a, 2b}$ have non-vanishing second derivatives with respect to $y_a, y_b$. Thus, in the summation over $j$ of $\partial V_a/\partial y_b$, only two terms with $j=2b$ and $j=2b-1$ are non-vanishing:
\begin{equation}
\alpha' \left. \frac{\partial V_a}{\partial y_b}\right\vert_{y_b=0}=(X_{2a, 2b}-X_{2a-1, 2b})\frac{\partial^2 u_{2a,2b}}{\partial y_a \partial y_b}+ \alpha' (X_{2a, 2b-1}-X_{2a-1, 2b-1})\frac{\partial^2 u_{2a,2b-1}}{\partial y_a \partial y_b}
,
\end{equation}
where we have suppressed the conditions $y_a=y_b=0$. Note that similarly for $\partial V_b/\partial y_a$ we have only two terms with $j=2a$ and $2a-1$ non-vanishing. An explicit computation shows that these two derivatives are very nicely equal to each other, so we can write the result in a more symmetric way and define
\begin{equation}
W_{a,b}:=\alpha' c_{2a-1, 2b-1} \left. \frac{\partial^2 u_{2a, 2b}}{\partial y_a \partial y_b}\right\vert_{y_a=y_b=0},
\end{equation}
where we have used $X_{2a,2b}+X_{2a-1,2b-1}-X_{2a-1, 2b}-X_{2a, 2b-1}=c_{2a-1, 2b-1}$. 
Putting everything together, we have a pretty formula for the scaffolding residue
\begin{equation}
\int \prod_{a=n{+}1}^{2n{-}3} \frac{dy_a}{y_a^2}\left(\sum_{r=0}^{\lfloor n/2 \rfloor{+}1} \sum_{\{i,j\}, \{k\}} \prod_{a}^r W_{i_a, j_a} \prod_{b}^{n-2r} V_{k_b}\right) \prod_{(i j)} u_{i,j}^{\alpha' X_{i,j}}, \label{VWformula}
\end{equation}
where we have the $dy/y^2$ form for the remaining $n{-}3$ variables within the $n$-gon, and the $n$-point Koba-Nielsen factor $\prod u^X$; inside the bracket, we have a summation over all partitions of $\{1,2,\cdots, n\}$ into $r$ pairs $\{i_a, j_a\}$ and $n-2r$ singlets $k_b$, each given by the product of $W$'s and $V$'s. For example, the $n=3$ case reads $V_1 V_2 V_3+ W_{1,2} V_3 + W_{2,3} V_1+ W_{1,3} V_2$, and for $n=4$ we have
\begin{equation}\label{4ptVW}
V_1 V_2 V_3 V_4+ \left(W_{1,2} V_3 V_4+ {\rm perm.} \right)+ \left( W_{1,2} W_{3,4} + {\rm perm.}\right),
\end{equation}
with $6$ terms of the form $W V V$ and $3$ terms of the form $W W$. Note that in the $\alpha' \to 0$ limit, only some of these terms contribute to the lowest-order, or the pure Yang-Mills amplitude. For example, for $n=3$ only $V W$ terms contribute and for $n=4$ only $W W$ and $W V V$ terms contribute, while $V^n$ terms only contribute to amplitude with pure higher-dimensional operators such as $F^3$ and $F^4$. 

We remark that we can already learn many interesting facts about bosonic string gluon amplitudes and their field-theory limit (Yang-Mills plus higher-dimensional operators starting with $F^3$) from these expressions.  For example, if we are interested in helicity amplitudes in $D=4$, there is a prescription to directly translate $V$ and $W$ into spinor-helicity expressions, which lead to formulas for $n$-gluon string and field-theory amplitudes in spinor-helicity variables~\cite{cexpansion}. Soft theorems for gluons can be derived using \eqref{VWformula} which take the form of ``double-soft theorems" for scaffolded scalars~\cite{Arkani-Hamed:2008owk, Cachazo:2015ksa}.  We also discover hidden properties of gluon amplitudes, for instance, the field-theory amplitudes admit an interesting expansion associated with a sum over all facets of the associahedron~\cite{cexpansion}.

\subsubsection{Triangulation-independent formula}

One apparent drawback of our result \eqref{VWformula} is that it seems to depend on the triangulation, namely both the form $dy/y^2$ and those first and second derivatives in $V$ and $W$ depend on the triangulation of the $n$-gon. Here we will show that by absorbing some $y$ factors into $V$ and $W$ we can make both of them triangulation independent. Such a rewriting also makes it manifest that the new $V$ and $W$ can be expressed in terms of products of $u$'s, which have nice graphical interpretations. 

To begin with, note that in the scaffolding triangulation, each $u_{2a{-}1,{2a{+}1}}$ is proportional to $y_{a}$, and we define \begin{equation}
u_{2a{-}1,{2a{+}1}}\equiv y_{a}\hat{u}_{2a{-}1,{2a{+}1}}, 
\end{equation} 
and it is crucial to use the $u$-equations
\begin{equation}\label{eq-u-2aj}
u_{2a,j}+u_{2a-1,2a+1}\prod_{i\in D_j} u_i=1\,,
\end{equation}
where subset $D_j$ includes the curves intersecting with $(2a,j)$ except  $(2a{-}1,2a{+}1)$. In the limit $y_{a}\to 0$, we expand $u$-equations~\eqref{eq-u-2aj} as
\begin{equation}
        u_{2a,j}\vert_{y_{a}=0}=1,
        \end{equation}
as well as 
\begin{equation}
\left. \frac{\partial u_{2a,j}}{\partial y_{a}}\right\vert_{y_{a}=0}=-\hat{u}_{2a{-}1,{2a{+}1}}\prod_{i\in D_j}u_i|_{y_{a}=0},
        \end{equation}
where we have expressed the first derivative in terms of $\hat{u}$ and product of those $u$'s crossing $(2a{-}1, 2a{+}1)$.  Similarly from the $u$-equation of $u_{2a,2b}$, we can derive that the second derivative is given by
\begin{equation}
\frac{\partial^2 u_{2a, 2b}}{\partial y_a \partial y_b}=-\hat{u}_{2a{-}1,{2a{+}1}}\hat{u}_{2b{-}1,{2b{+}1}}\prod_{i\in E}u_i|_{y_{a}=0,y_{b}=0},   
\end{equation} 
where subset $E$ includes curves intersecting with $(2a,2b)$ except for $(2a{-}1,2a{+}1)$ and $(2b{-}1,2b{+}1)$. 

Now it becomes obvious that for any term of the summation in \eqref{VWformula}, there is a common factor $\prod_{a=1}^{n}\hat{u}_{2a{-}1,2a{+}1}$ from the $\hat{u}$'s in each $V$ and $W$'s, which cancels the ratio $\left( \prod_{(i,j)\in\text{even}} u_{i,j}/\left(\prod_{(i,j)\in\text{odd}} u_{i,j} \right)\right)|_{y_{a}=0}$ to give:
\begin{equation}
\prod_{a=1}^n \hat{u}_{2a{-}1, 2a{+}1} \left.\left( \frac{\prod_{(i,j)\in\text{even}} u_{i,j}}{\prod_{(i,j)\in\text{odd}} u_{i,j}} \right)\right\vert_{y_{a}=0}=\prod_{i,j\in n-{\rm gon}}\frac{1}{u_{i,j}}\,,
\end{equation}
\begin{figure}[t]
    \centering
    \includegraphics[width=0.8\textwidth]{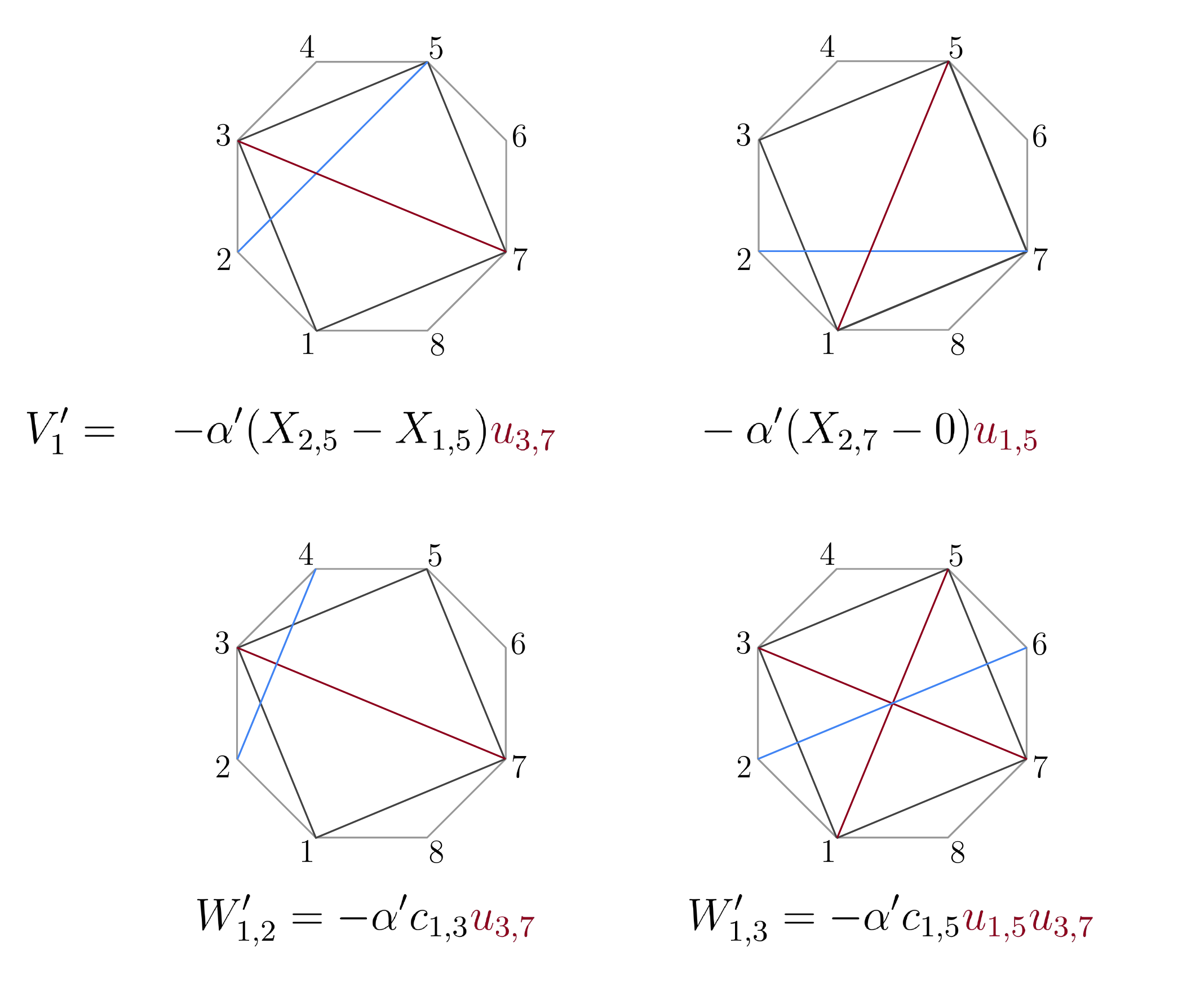}
    \caption{$V'$ and $W'$ in terms $u$'s for $n=4$.}
    \label{fig:VWinu4pt}
\end{figure}

This gives us a triangulation-independent formula for the scaffolding residue, which contains the invariant $(n{-}3)$-form $dy/y$ and the Koba-Nielsen factor $\prod u^{X-1}$: 
\begin{equation}\label{VW2}
    \begin{aligned}
     \int_{0}^{\infty} \prod_{a}\frac{\diff y_a}{y_a} \left(\sum_{r=0}^{\lfloor n/2 \rfloor{+}1} \sum_{\{i,j\}, \{k\}} \prod_{a}^r W'_{i_a, j_a} \prod_{b}^{n-2r} V'_{k_b}\right)\prod_{(i j)} 
     u_{i,j}^{\alpha' X_{i,j}-1},     \end{aligned}
\end{equation}
where we have redefined $V$ and $W$ to be
\begin{equation}
    V'_{a}=-\alpha' \sum_j (X_{2a,j}-X_{2a-1,j})\prod_{i\in D_j}u_i |_{y_{a}=0}, \qquad  W'_{a,b}=-\alpha' c_{2a-1, 2b-1} \prod_{i\in E}u_i|_{y_{a}=0,y_{b}=0},
\end{equation}
which are written in terms of a product of $u$'s without referring to any specific triangulation. 

The product of the $u$'s appearing in the definitions of $V'_{a}$ and $W'_{a,b}$ have a very simple graphical interpretation. For the coefficient of $(X_{2a,j} - X_{2a-1,j})$ in $V'_{a}$ we simply take the product over all the $u$'s in the inner $n$-gon that intersects $(2a,j)$. For $W'_{a,b}$ we similarly have the product over all the $u$'s of the inner $n$-gon that intersect $(2a,2b)$. 

For example, the $n=3$ amplitude is given by $V'_1 V'_2 V'_3+ V'_1 W'_{2,3} + {\rm perm.}$ with
\begin{equation}
    V'_1=-\alpha^\prime X_{2,5}, \qquad W'_{1,2}=-\alpha^\prime c_{1,3},
\end{equation}
and cyclic images {\it e.g.} $V_2=-\alpha^\prime X_{1,4}$. In this case, there is no integral left in \eqref{VW2} and we reproduce the familiar result~\eqref{eq:3ptCs}.

For $n=4$ case, we have the following $V'$ and $W'$, where the product of $u$'s are shown in Fig.~\ref{fig:VWinu4pt}:
\begin{equation}
    V'_1= -\alpha^\prime( u_{3,7} (X_{2,5}-X_{1,5})+u_{1,5} X_{2,7}), \quad
    W'_{1,2}=-\alpha^\prime c_{1,3} u_{3,7},\quad
    W'_{1,3}=-\alpha^\prime c_{1,5} u_{1,5} u_{3,7},
\end{equation}
as well as their cyclic images. Similar to \eqref{4ptVW}, we have $3+6+1$ terms of the form $W' W'$, $W' V' V'$ and $V' V' V' V'$ respectively, and \eqref{VW2} becomes: 
\begin{equation}\label{4ptVW2}
\int_0^\infty \frac{dy}{y} \left((W'_{1,2}W'_{3,4}+ {\rm perms.}) + (W'_{1,2} V'_3 V'_4+ {\rm perm.})+V'_1 V'_2 V'_3 V'_4\right) u_{1,5}^{\alpha' X_{1,5}-1} u_{3,7}^{\alpha' X_{3,7}-1}\,,
\end{equation}
where only one integration variable is left, which can be chosen as $y=y_{1,5}$ or $y_{3,7}$ depending on the inner triangulation. For the former we have $u_{1,5}=y_{1,5}/(1+y_{1,5})$ and $u_{3,7}=1-u_{1,5}=1/(1+y_{1,5})$, and the other way around for the latter. 
\begin{figure}[t]
    \centering
    \includegraphics[width=0.9\textwidth]{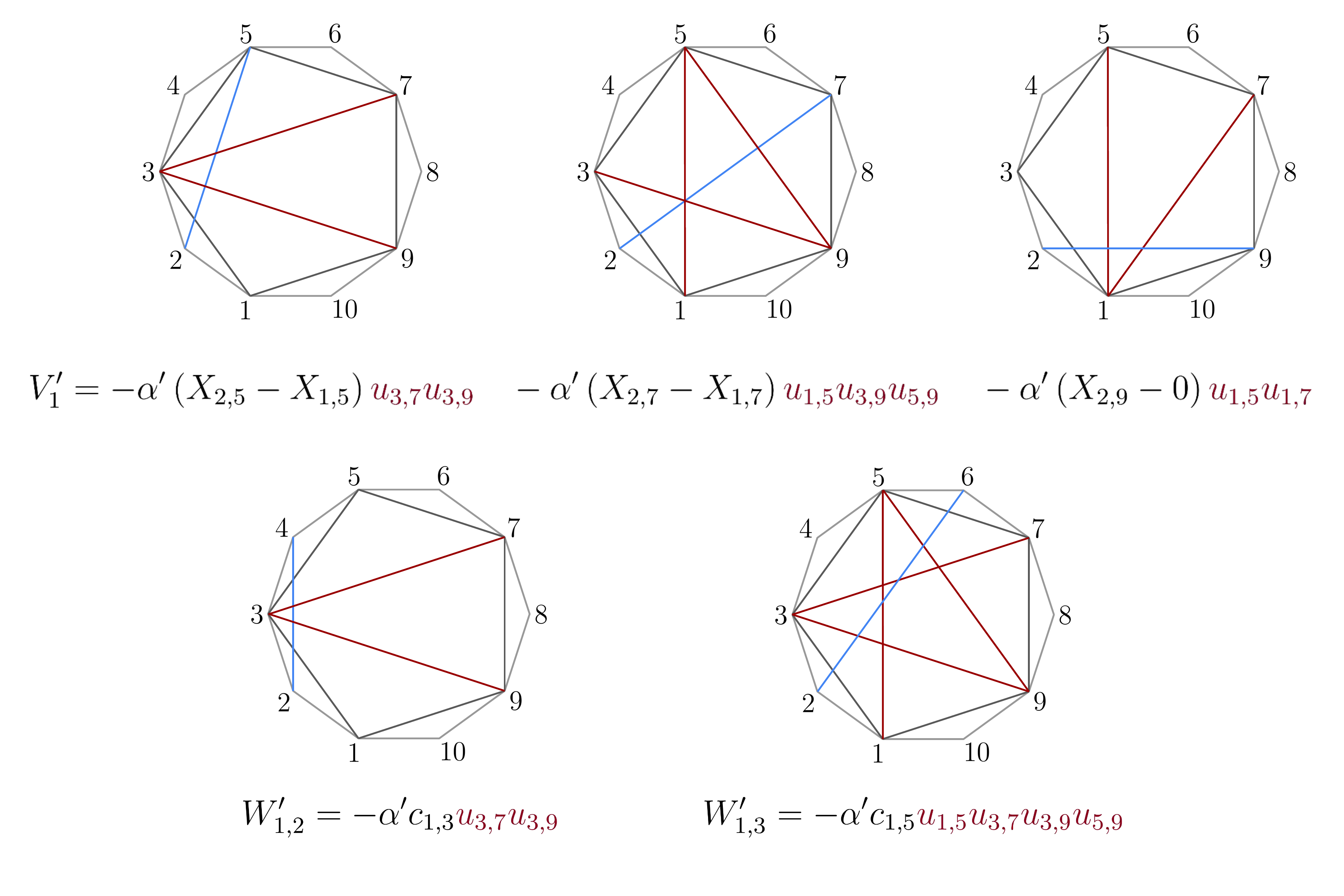}
    \caption{$V'$ and $W'$ in terms $u$'s for $n=5$.}
    \label{fig:VWinu5pt}
\end{figure}

Finally, we present these building blocks for $n=5$ amplitude which can be nicely summarized in figure~\ref{fig:VWinu5pt}. All cyclically inequivalent ones are:
\begin{equation}
    V'_1=-\alpha^\prime (u_{3,7} u_{3,9} \left(X_{2,5}-X_{1,5}\right)+u_{1,5} u_{3,9} u_{5,9} \left(X_{2,7}-X_{1,7}\right)+u_{1,5} u_{1,7} X_{2,9}),
\end{equation}
\begin{equation}
    W'_{1,2}=-\alpha^\prime c_{1,3} u_{3,7} u_{3,9}, \quad W'_{1,3}=-\alpha^\prime c_{1,5} u_{1,5} u_{3,7} u_{3,9} u_{5,9}.
\end{equation}
The $n=5$ string amplitude is a two-fold integral as in \eqref{VW2}, where in the summation we have $V'_1 \cdots V'_5$, $10$ permutations of $V'_1 V'_2 V'_3 W'_{4,5}$, and $15$ permutations of $V'_1 W'_{2,3} W'_{4,5}$.

\subsection{Consistency check: From 4$n$-scalars to 2$n$-scalars}

Now that we have an explicit form for the gluon amplitude obtained after taking the scaffolding residue, we can perform a nice consistency check. Suppose we did not know the bosonic string answer, and so we did not know a priori that the scalars we are scattering can be thought of as higher-dimensional polarizations of gluons.  How could we discover this fact inherently within our world?

One way of doing this is to start with a $4n$-scalar string amplitude. In this case, taking the scaffolding residue leads to a $2n$-gluon string amplitude; if we then choose the special polarizations in \eqref{eq:kinConfig}, the result must simplify dramatically to land on the $2n$-scalar string amplitude again! 

We will now see that very nicely this is indeed the case. 
Let us consider the $4n$-scalar scattering labelled by $\{1,1^\prime,2,2^\prime,\ldots,2n,2n^\prime\}$, which results in the $2n$-gluon scattering by taking the scaffolding residues $\{X_{1,2},X_{2,3},\ldots,X_{2n,1}\}$. Recalling the dimensional reduction that turns the gluons into the $2n$-scalars, eq. \eqref{eq:kinConfig}, we see that this is equivalent to take the derivative $\partial_{\epsilon_1 \cdot \epsilon_2}\partial_{\epsilon_3 \cdot \epsilon_4}\ldots \partial_{\epsilon_{2n-1} \cdot \epsilon_{2n}}$ on the conventional string correlator. On the other hand, in the $4n$-scalar kinematic configuration, we have:
\begin{align}
    \epsilon_i^\mu& \propto (1-\alpha) (X_{i+1}-X_{i^\prime})^\mu-\alpha (X_{i^\prime}-X_{i})^\mu, \\
    q_i^\mu& \propto (X_{i+1}-X_{i})^\mu\,.
\end{align}

Importantly, as we have mentioned before, the dependence on $X_{i^\prime,j^\prime}$ only appears in $\epsilon_i \cdot \epsilon_j$. Therefore, the above derivative on Lorentz product of $\epsilon$ is equivalent to the following derivative in our kinematic configuration:
\begin{equation} \label{eq:derivX}
    \partial_{\epsilon_1 \cdot \epsilon_2}\partial_{\epsilon_3 \cdot \epsilon_4}\ldots \partial_{\epsilon_{2n-1} \cdot \epsilon_{2n}} \sim  (-2)^n \partial_{X_{1^\prime,2^\prime}} \partial_{X_{3^\prime,4^\prime}} \ldots \partial_{X_{2n-1^\prime,2n^\prime}},
\end{equation}
where in the RHS we have included the normalization factor. Now recall the $X$'s dependence in the building blocks of our ``stringy correlator'': $V_i^\prime$ depends on $X_{i^\prime,j},X_{i,j}$ for some $j$(with no prime), $W_{i,j}^\prime$ depends on $c_{i,j}=X_{i,j}+X_{i^\prime,j^\prime}-X_{i^\prime,j}-X_{i,j^\prime}$.
\begin{figure}[t]
    \centering
    \includegraphics[width=\linewidth]{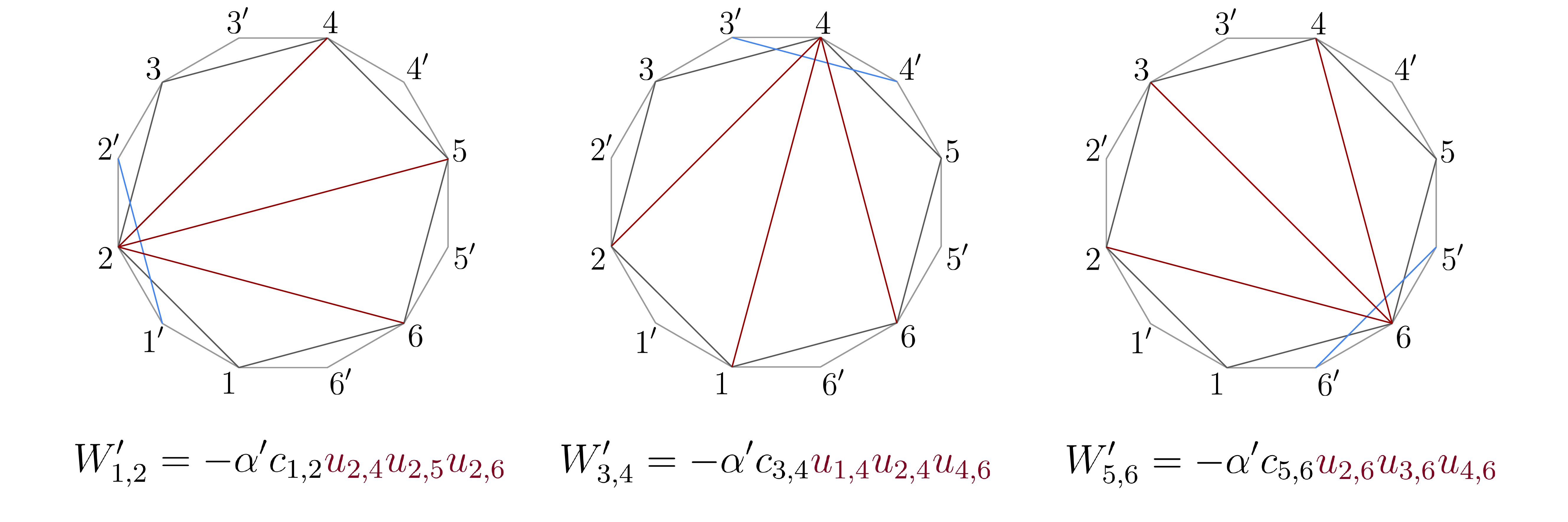}
    \caption{$W^\prime$'s that contribute to the derivative from $6$-scaffolding-gluon to $6$-scalar.}
    \label{fig:consistency}
\end{figure}
Therefore, it is clear that the derivative~\eqref{eq:derivX} resulting on only one term:
\begin{equation}
\begin{aligned}
    W_{1,2}^\prime W_{3,4}^\prime \ldots W_{2n-1,2n}^\prime&\to(-1)^n (-2)^n \prod_{i\in \text{even}} \prod_{j\neq i-1,i,i+1} u_{i,j} \\
    &=2^n \prod_{(i,j)\in \text{even}} u_{i,j}^2 \prod_{k\in \text{even},l\in \text{odd}} u_{k,l},
\end{aligned}
\end{equation}
where all the indices in the subscript only involve labels with no prime. For example, consider $(4n=12)$-scalar to $(2n=6)$-scalar, the one term that contributes to the derivative reads(see Fig.~\ref{fig:consistency}):
\begin{equation}
    W_{1,2}^\prime W_{3,4}^\prime W_{5,6}^\prime \to 8 u_{2,4}^2 u_{2,6}^2 u_{4,6}^2 u_{2,5}u_{1,4}u_{3,6},
\end{equation}
where we have set $\alpha^\prime=1$. Recall in~\eqref{VW2} we have an overall factor $1/(\prod_{(i,j)}u_{i,j})$, therefore the dimensional reduction transforms the $2n$-scaffolding-gluon to the $2n$-scalar correctly with the factor $2^n \prod_{(i,j)\in\text{even}} u_{i,j}/\prod_{(i,j)\in\text{odd}} u_{i,j}$, which in the above example reads $8(u_{2,4}u_{2,6}u_{4,6})/(u_{1,3}u_{1,5}u_{3,5})$ as expected.

\subsection{Examples}

We end our tree-level discussion by presenting some explicit examples of how to use \eqref{eq:2nScalarUs} to extract gluon amplitudes. For simplicity, we will be focusing on the 3- and 4-point amplitudes. In section \ref{sec:FurtherCom}, we make some comments on how to systematically extract the field theory gluon amplitude at higher points. 
\subsubsection{3-point example}

To obtain the 3-point gluon interaction we start with the 6-scalar amplitude. Picking the scaffolding triangulation $\{X_{1,3},X_{3,5},X_{1,5}\}$ we obtain 
\begin{equation}
	\mathcal{A}_6=\int_0^\infty \frac{\diff y_{1,3}\diff y_{3,5}\diff y_{1,5}}{y_{1,3}^2y_{3,5}^2y_{1,5}^2}  \,   \prod_{i,j} u_{ij}^{\alpha^{\prime} X_{i,j}}  \, \left( \frac{u_{2,4}u_{2,6}u_{4,6}}{u_{1,3}u_{3,5}u_{5,7}} \right).
	\label{eq:6ptInt}
\end{equation}

Plugging in the explicit parametrization, $u_{i,j}(y_{1,3},y_{3,5},y_{1,5})$, yields:
\begin{equation}
\begin{aligned}
	\mathcal{A}_6 = \int_0^{\infty} &\frac{\diff y_{1,3}\diff y_{3,5}\diff y_{1,5}}{y_{1,3}^2y_{3,5}^2y_{1,5}^2}  \, y_{1,3}^{\alpha^{\prime}X_{1,3}} y_{3,5}^{\alpha^{\prime}X_{3,5}} y_{1,5}^{\alpha^{\prime}X_{1,5}}(1+y_{1,3})^{-\alpha^{\prime}c_{1,4}} (1+y_{1,5})^{-\alpha^{\prime}c_{2,5}}  (1+y_{3,5})^{-\alpha^{\prime}c_{3,6}} \\
	& \times (1+y_{1,3}+y_{1,3}\, y_{3,5})^{-\alpha^{\prime}c_{1,3}}(1+y_{1,5}+y_{1,5}\, y_{1,3})^{-\alpha^{\prime}c_{1,5}} (1+y_{3,5}+y_{3,5}\, y_{1,5})^{-\alpha^{\prime}c_{3,5}}.
\end{aligned} 
\end{equation}

As discussed in \ref{sec:ExtRes}, taking the scaffolding residue amounts to setting $X_{1,3}=X_{3,5}=X_{1,5}=0$, and collecting the piece of the integrand that is linear in $y_{1,3},y_{3,5}$ and $y_{1,5}$. In this case, after taking this residue there are no integrations left and we get:
\begin{equation}
\begin{aligned}
	A_3^{\text{gluon}}=& \,  \alpha^{\prime \,2}\left(c_{1,3} c_{1,5} + c_{1,3} c_{2,5} + c_{1,3} c_{3,5} + c_{1,4} c_{3,5} + c_{1,5} c_{3,5} + c_{1,5} c_{3,6} \right)\quad + \\
	& -\alpha^{\prime \, 3}\left(X_{1,4} X_{2,5} X_{3,6} \right) \quad \\
 =&\, \alpha^{\prime \,2} \left(-(X_{2,5}-X_{2,6})X_{1,4}-(X_{3,6}-X_{4,6})X_{2,5}-(X_{1,4}-X_{2,4})X_{3,6}\right)\\
 & -\alpha^{\prime \, 3}\left(X_{1,4} X_{2,5} X_{3,6} \right). \quad \\
\end{aligned} 
\label{eq:3ptCs}
\end{equation}

The leading piece in $\alpha^\prime$ is exactly the Yang-Mills amplitude, and we can check it precisely matches \eqref{eq:3ptYM}. The sub-leading term corresponds to the $F^3$ vertex, which has higher units. 
  
Before proceeding, let's use this simple example to highlight two non-trivial features of this scaffolded gluon amplitude: First, since we extract the answer via this scaffolding residue, the amplitude can be written as a function of \textbf{solely non-planar variables}, $c_{i,j} = -p_i \cdot p_j$, with $i$ and $j$ non-adjacent. This is a non-trivial general fact that will be explored further in \cite{cexpansion}. Second, as we have mentioned already, in this scalar language the amplitude takes a unique form, there is no gauge-redundancy ambiguity in how it is expressed.

\subsubsection{4-point example: string and field-theory amplitudes}

For $n=4$, after taking the scaffolding residue we are left with one integration -- which we can trivially perform -- so we now present explicitly the integrated $4$-gluon string amplitude. To do this, we use the result we get after taking the scaffolding residue in terms of $V$'s and $W$'s presented in \eqref{4ptVW2}, but now we explicitly evaluate the last integration.

Let us define 
\begin{equation}
    I(a,b):=\frac{\Gamma(
    X_{1,5}+a)\Gamma( 
    X_{3,7}+b)}{\Gamma(X_{1,5}+X_{3,7}+a+b)},
\end{equation}
with integer shifts $a,b$. The first term reads \begin{equation}
    \int_{0}^\infty \frac{dy_{1,5}}{y_{1,5}} u_{1,5}^{X_{1,5}-1}u_{3,7}^{X_{3,7}-1} W_{1,2}^\prime W_{3,4}^\prime=c_{1,3} c_{5,7} I(1,-1).
\end{equation}

The second term, {\it i.e.} the integral of $W_{2,3}^\prime W_{1,4}^\prime$ is given by the cyclic rotation for the $4$-gon: $X_{i,j}\to X_{i+2,j+2}$, which acts on $I(a,b)$ as $I(a,b)\to I(b,a)$. The third term reads:
\begin{equation}
    \int_{0}^\infty \frac{dy_{1,5}}{y_{1,5}} u_{1,5}^{X_{1,5}-1}u_{3,7}^{X_{3,7}-1} W_{1,3}^\prime W_{2,4}^\prime=c_{1,3} c_{3,7} I(1,1).
\end{equation}
For combinations of the type $W^\prime V^\prime V^\prime$, we have
\begin{equation}
\begin{aligned}
&\int_{0}^\infty \frac{dy_{1,5}}{y_{1,5}} u_{1,5}^{X_{1,5}-1}u_{3,7}^{X_{3,7}-1} W_{1,2}^\prime V_3^\prime V_4^\prime \\
=&\frac{c_{1,3}I(1,0)}{\left(X_{1,5}-1\right) X_{1,5}} \left(\left(X_{1,5}-1\right) \left(X_{1,5} \left(X_{3,6}-X_{3,7}\right)+X_{1,6} X_{3,7}\right) \left(X_{3,7}-X_{3,8}\right)\right.\\
& \left. +X_{3,7} \left(X_{3,6}-X_{1,6} \left(X_{3,7}+1\right)+X_{1,5} \left(-X_{3,6}+X_{3,7}+1\right)\right) X_{5,8}\right)\,,
\end{aligned}
\end{equation}
and three more terms with $W'_{2,3} V'_4 V'_1, W'_{3,4} V'_1 V'_2$, and $W'_{1,4} V'_2 V'_3$ are obtained from cyclic rotations (proportional to $I_{0,1}, I_{1,0}$ and $I_{0,1}$). Similarly, for $W'_{1,3} V'_2 V'_4$ we have
\begin{equation}
\begin{aligned}
&\int_{0}^\infty \frac{dy_{1,5}}{y_{1,5}} u_{1,5}^{X_{1,5}-1}u_{3,7}^{X_{3,7}-1} W_{1,3}^\prime V_2^\prime V_4^\prime \\
=&\frac{c_{1,5}I(1,1)}{X_{1,5} X_{3,7}} (X_{1,5} \left(X_{3,7}-X_{3,8}\right) \left(\left(X_{1,4}-X_{1,5}-1\right) X_{3,7}+\left(X_{1,5}+1\right) X_{4,7}\right)\\
&-X_{3,7} \left(X_{1,4} \left(X_{3,7}+1\right)+X_{1,5} \left(X_{4,7}-X_{3,7}\right)\right) X_{5,8})\,,
\end{aligned}
\end{equation}
and another term with $W'_{2,4} V'_1 V'_3$ by cyclic rotation.
The last case, the integral with $V_1^\prime V_2^\prime V_3^\prime V_4^\prime$, is given by
\begin{equation}
P_0 I(-1,3)+P_1 I(0,2)+P_2 I(1,1)+P_3 I(2,0)+P_4 I(3,-1),
\end{equation}
where we have degree-$4$ polynomials in $X$'s,  $P_i$ with $i=0,1, \cdots, 4$, as follows
\begin{equation}
P_0=X_{1,4} X_{5,8} \left(X_{2,5}-X_{1,5}\right) \left(X_{1,6}-X_{1,5}\right),
\end{equation}
\begin{equation}
\begin{aligned}
P_1=&X_{1,4} \left(X_{1,6}-X_{1,5}\right) \left(X_{2,5}-X_{1,5}\right) \left(X_{3,8}-X_{3,7}\right)+X_{1,4} \left(X_{1,6}-X_{1,5}\right) X_{2,7} X_{5,8}\\
&+X_{1,4} \left(X_{2,5}-X_{1,5}\right) X_{3,6} X_{5,8}+\left(X_{1,6}-X_{1,5}\right) \left(X_{2,5}-X_{1,5}\right) \left(X_{4,7}-X_{3,7}\right) X_{5,8}\,,
\end{aligned}
\end{equation}
\begin{equation}
\begin{aligned}
P_2=&X_{1,4} \left(X_{1,6}-X_{1,5}\right) X_{2,7} \left(X_{3,8}-X_{3,7}\right)+X_{1,4} \left(X_{2,5}-X_{1,5}\right) X_{3,6} \left(X_{3,8}-X_{3,7}\right)\\
&+\left(X_{1,6}-X_{1,5}\right) \left(X_{2,5}-X_{1,5}\right) \left(X_{4,7}-X_{3,7}\right) \left(X_{3,8}-X_{3,7}\right)+X_{1,4} X_{2,7} X_{3,6} X_{5,8}\\
&+\left(X_{1,6}-X_{1,5}\right) X_{2,7} \left(X_{4,7}-X_{3,7}\right) X_{5,8}+\left(X_{2,5}-X_{1,5}\right) X_{3,6} \left(X_{4,7}-X_{3,7}\right) X_{5,8}\,;
\end{aligned}
\end{equation}
and $P_3$ and $P_4$ are obtained from $P_1$ and $P_0$ by cyclic rotation, respectively. Adding these contributions we obtain the explicit result of $n=4$ string amplitude.  

Finally, we present the $4$-point Yang-Mills amplitude, which is given by the $\alpha'\to 0$ limit of the full stringy result. Just as we saw at $3$-points case, we find that the result takes a suggestive ``c-expansion" form:
\begin{equation}
A_4^{\rm YM}=\alpha'^2 \left(\frac{N_{1,5} (\{c_{i,j}\})}{X_{1,5}} +  \frac{N_{3,7}(\{c_{i,j}\})}{X_{3,7}}+ c_{1,5} c_{3,7}\right)\,,    
\end{equation}
where we have two poles, $X_{1,5}$ and $X_{3,7}$, with numerators $N_{1,5}$ and $N_{3,7}=N_{1,5}|_{i \to i+2}$, as well as a ``contact term" $c_{1,5} c_{3,7}$. Both the contact term and these two numerators can be expressed solely in terms of $c_{i,j}$'s. This is not at all guaranteed a priori but reflects a general phenomenon we will explore for all $n$ in forthcoming work~\cite{cexpansion}. The numerator $N_{1,5}$ is given by:
\begin{equation}
\begin{aligned}
N_{1,5}=&-c_{1,5} c_{3,7}^2-c_{1,3} c_{1,5} c_{3,7}-c_{1,5} c_{1,7} c_{3,7}-c_{1,3} c_{2,5} c_{3,7}-c_{1,5} c_{2,7} c_{3,7}-c_{1,3} c_{3,5} c_{3,7}\\
&-c_{1,5} c_{3,6} c_{3,7}-c_{1,5} c_{3,8} c_{3,7}-c_{1,5} c_{4,7} c_{3,7}-c_{1,3} c_{5,7} c_{3,7}-c_{1,4} c_{5,7} c_{3,7}-c_{1,5} c_{5,7} c_{3,7}\\
&-c_{1,3} c_{5,8} c_{3,7}-c_{1,4} c_{5,8} c_{3,7}-c_{1,7} c_{5,8} c_{3,7}-c_{1,3} c_{1,5} c_{1,7}-c_{1,3} c_{1,7} c_{2,5}-c_{1,3} c_{1,5} c_{2,7}\\
&-c_{1,4} c_{1,7} c_{3,5}-c_{1,5} c_{1,7} c_{3,5}-c_{1,3} c_{2,7} c_{3,5}-c_{1,4} c_{2,7} c_{3,5}-c_{1,5} c_{2,7} c_{3,5}-c_{1,5} c_{1,7} c_{3,6}\\
&-c_{1,5} c_{2,7} c_{3,8}-c_{1,3} c_{1,5} c_{4,7}-c_{1,3} c_{2,5} c_{4,7}-c_{1,3} c_{3,5} c_{4,7}-c_{1,4} c_{3,5} c_{4,7}
-c_{1,5} c_{3,5} c_{4,7}\\
&-c_{1,3} c_{1,5} c_{5,7}-c_{1,3} c_{1,6} c_{5,7}
-c_{1,3} c_{1,7} c_{5,7}-c_{1,3} c_{2,5} c_{5,7}-c_{1,3} c_{2,6} c_{5,7}-c_{1,3} c_{2,7} c_{5,7}\\
&-c_{1,5} c_{3,5} c_{5,7}-c_{1,6} c_{3,5} c_{5,7}-c_{1,7} c_{3,5} c_{5,7}-c_{1,3} c_{3,6} c_{5,7}-c_{1,4} c_{3,6} c_{5,7}-c_{1,5} c_{3,6} c_{5,7}\\
&-c_{1,5} c_{3,8} c_{5,7}-c_{1,6} c_{3,8} c_{5,7}-c_{1,7} c_{3,8} c_{5,7}-c_{1,3} c_{1,7} c_{5,8}-c_{1,3} c_{2,7} c_{5,8}-c_{1,7} c_{3,5} c_{5,8}\\
&-c_{1,5} c_{3,5} c_{3,7}-c_{1,7} c_{5,7} c_{3,7}-c_{1,3} c_{1,7} c_{3,5}-c_{1,5} c_{1,7} c_{3,8}-c_{1,5} c_{3,8} c_{4,7}-c_{1,4} c_{3,5} c_{5,7}\\
&-c_{1,4} c_{3,5} c_{3,7}-c_{1,6} c_{5,7} c_{3,7}-c_{1,3} c_{2,5} c_{2,7}-c_{1,5} c_{2,7} c_{3,6}-c_{1,5} c_{3,6} c_{4,7}-c_{1,3} c_{3,5} c_{5,7}\\
&-c_{1,7} c_{3,6} c_{5,7}-c_{1,7} c_{3,8} c_{5,8}-c_{1,6} c_{3,6} c_{5,7}-c_{1,7} c_{3,6} c_{5,8}.
\end{aligned}
\end{equation}

This reduces to the ``leading singularity" given in  ~\eqref{eq-ls-4pt-ym} upon sending $X_{1,5} \to 0$.  
It is possible to simplify this formula somewhat by introducing the notation $c_{A,B} = \sum_{k \in A, l \in B} c_{k,l}$, so that e.g. $c_{12,34} = c_{1,3} + c_{1,4} + c_{2,3} + c_{2,4}$. Then we have
\begin{equation}
    \begin{aligned}
      N_{1,5}= &-c_{3,5678}\left(c_{1,67}c_{5,7}+c_{1,5} c_{12345,7}+c_{1,7} c_{5,8}\right)-c_{1,4} \left(c_{12345,7} c_{3,5}+c_{3,67} c_{5,7}+c_{3,7} c_{5,8}\right)\\
        &-c_{1,3} \left(c_{123,5}c_{1234,7}+c_{123,567}c_{5,7}+c_{123,7}c_{5,8}\right).
    \end{aligned}
\end{equation}

Note that already the 4-point gluon amplitude looks more complicated than we are used to from the familiar Parke-Taylor formulas in four dimensions. This reflects the basic fact that $D$-dimensional gluon amplitudes simply are richer than in four dimensions, and that we are non-redundantly describing all possible gluon polarizations in a single expression. 

\subsection{Further comments}
\label{sec:FurtherCom}

\paragraph{Simplicity of residue extraction}

A fascinating general phenomenon has been encountered repeatedly in ``surfaceology" for amplitudes, which also appears in the context of the computation of scaffolding residues we began with in this section that we would like to highlight here. 

After taking the scaffolding residue, we end up (as with the bosonic string gluon expression) with a sum over exponentially many terms. What then is the point of beginning with the ``one-term'' expression for the $2n$ scalar problem, if the final gluon expression looks exponentially complicated in any case? Of course the scalar starting point is much more compact and crucially allows us to see the deep connection between scalar, pions, and gluons. But does it have a practical, computational advantage of any sort? 

Naively the answer would seem to be ``no", but remarkably, while the analytic expression for the residue indeed has exponentially many terms, the actual computation of the residue can be done in polynomial time! This is a famous occurrence in various branches of mathematics and physics. For instance, the analytic expression for the determinant of an $n \times n$ matrix has $n!$ terms but can be trivially computed in $n^3$ time using Gaussian elimination. Similarly, if we look at the $10$-th Newton polynomial of $10^3$ variables $\sum x_{i_1} \cdots x_{i_{10}}$ where the $i$'s range from $1, \cdots, 10^3$, there are roughly $10^{30}$ terms in the sum, but for any numerical values of the $x_i$ we can get the answer in a nanosecond on a computer, simply by looking at to the coefficient of $t^{10}$ in the expansion of $(1 + t x_1)(1 + t x_2) \cdots (1 + t x_{1000})$. 

The same thing occurs for the computation of our residue: the ``one-term" starting point from the $2n$ scalar problem allows us to compute a result naively needing exponentially many terms in polynomial time. As mentioned this is a special case of a general fact we have encountered repeatedly in the connection between amplitudes and the combinatorial geometry of surfaces. At the most basic level, the $u_X$ variables and their tropicalizations to the ``headlight functions" $\alpha_X$ generate the exponentially many cones associated with Feynman diagrams from polynomially many expressions involving polynomially many terms. Extensions of the tropical formalism for computing amplitudes for very general Lagrangians~\cite{troamp} see the same feature for generating numerator factors. It would be fascinating to better understand the origins of this huge reduction in apparent complexity for the sorts of polynomial expressions we encounter in scattering amplitudes. 

\paragraph{Integration Contour for Gluons}
Our integral form for the $2n$ scalar scaffolded amplitude is given as an integral over positive $y_i>0$, but of a form $\prod_i d y_i/y_i^2$ with power-law singularities as $y_i \to 0$. As such even in the field theory limit, where the $X_{ij}$ close to zero, the integral is not well defined on this naive contour and the result must be appropriately analytically continued from the domain where the $X$'s are positive enough to make the integral converge. Given that these amplitudes arise from a simple kinematic shift of the Tr $\phi^3$, this can be rephrased as the familiar fact discussed in our review of stringy Tr $\phi^3$ integrals, that the integral form of the Koba-Nielsen formula only converges when the $X_{i,j}>0$, and continuing to the physically interesting region where we see resonances and the amplitude has poles, where $X_{i,j} < 0$, is accomplished by analytic continuation. As discussed there and at greater length in \cite{contour}, the $u/y$ variables make it easy to define an explicit contour on which the integral converges for {\it any} $X_{i,j}$. For the purposes of this paper, this is enough to concretely specify a contour on which our scaffolded gluon amplitudes can be concretely evaluated. 

\paragraph{Field Theory Limit For Gluons}
Extracting the field theory limit of our scalar-scaffolded gluon amplitude is more challenging than for Tr $\phi^3$ theory. As we discussed earlier, the $dy/y$ singularity makes it easy to extract the low-energy limit of the amplitudes for $\alpha^\prime X \ll 1$, localizing to logarithmic divergences as $|{\rm log} y| \to \infty$. This procedure does not work immediately for our scaffolded gluons, where the $dy/y^2$ form has power-law divergences and the integral must be defined by analytic continuation, even at tree-level. In \cite{cexpansion} will report on an elementary and direct way of extracting the field theory limit, via a ``cone-by-cone" analysis of the singular behavior of the integrand. This will be analyzed systematically at tree and loop level in \cite{cexpansion}, but we can illustrate with the simplest but still non-trivial example of the four-point amplitude. 

We begin by arbitrarily dividing the integral over $y$ into two regions, for $y$ between (0,1) and $(1,\infty)$.
Then we have 
\begin{equation}
    {\cal A}^{{\rm gluon}}_4 = \left(\int_0^1 + \int_1^\infty \right) \frac{dy_{1,3}}{y_{1,3}^2} y_{1,3}^{X_{1,3}} (1 + y_{1,3})^{-c_{1,3}} ,
\end{equation}
where $c_{1,3} = X_{1,3} + X_{2,4}$ is a ``mesh constant" in the language of~\cite{Arkani-Hamed:2019vag,Zeros}. 

We will now extract the leading singular behavior from each piece in a standard way. For $y \subset (0,1)$, we can expand 
\begin{equation}
\begin{aligned}
\int_0^1 \frac{dy_{1,3}}{y_{1,3}^2} y_{1,3}^{X_{1,3}} &\left(1 + (-c_{1,3}) y_{1,3} +  \frac{(-c_{1,3})(-c_{1,3}-1)}{2} y_{1,3}^2 + \cdots \right)= \\ 
&= \frac{1}{X_{1,3} - 1} -\frac{c_{1,3}}{X_{1,3}} + \frac{(-c_{1,3}) (-c_{1,3} -1)}{2} + \cdots = -1 + \cdots -\frac{c_{1,3}}{X_{1,3}} + \cdots \,\, \, ,
\end{aligned}
\end{equation}
where in the second line we have kept the leading terms at low energies, which here scale as $X^0$ as opposed to $X^{-1}$ we are familiar with for the Tr $\phi^3$ amplitude. Note that in performing this integral we are assuming that $X_{1,3} > 1$ for convergence, and indeed the leading singular term $1/(X_{1,3} - 1)$ has a pole at $X_{1,3} =1$, but we are then free to analytically continue and expand in $X_{1,3}$ around zero. Doing this converts the naive power-law divergence we would get from the $\diff y_{1,3}/y_{1,3}^2$ behavior near $y_{1,3} \to 0$, to simply the constant term $-1$. Performing a similar analysis for $y_{1,3} \subset (1,\infty)$, we find that the leading behavior simply gives us $(+1)$: 
\begin{eqnarray}
\int_{1}^\infty \frac{dy_{1,3}}{y_{1,3}^2} y_{1,3}^{X_{1,3}} (1 + y_{1,3})^{-X_{1,3} - X_{2,4}} & =&  \int_1^\infty \frac{dy_{1,3}}{y_{1,3}^2} \left(\frac{1}{y_{1,3}}\right)^{X_{2,4}} (1 + \frac{-c_{1,3}}{y_{1,3}} + \frac{(-c_{1,3})(-c_{1,3} - 1)}{2 y_{1,3}^2} + \cdots \nonumber \\ 
 &=&  1 +\cdots \, \, \, .
\end{eqnarray}

This $(+1)$ cancels the $(-1)$ remnant from the naive power divergence in the first regions, leaving us with  
\begin{equation}
{\cal A}^{{\rm gluon}}_4 \rightarrow \frac{-c_{1,3}}{X_{1,3}}.
\end{equation}

The cancellation of the ``-1" and ``1" pieces from the two regions reflects the basic fact used repeatedly in the zeros analysis of \cite{Zeros} that the classic ``scale-less integral" $\int_0^\infty dy/y\, y^\alpha = 0$. 

The same strategy works in general. It is more convenient to set the $y_i = e^{-t_i}$ and work in $t$ space, which is divided into cones corresponding to the different degenerations of the surface/diagrams. In each cone, there is a leading behavior at infinity, that can be nicely controlled by understanding the ``$g$-vector fan", which we can factor out and then expand around. The resulting functions of the kinematic variables $X$ can then in turn be expanded about $X=0$, keeping the leading low-energy behavior. 

  %\newpage
  \section{Loop-level Gluon Amplitudes}
\label{sec:loops}
Having just learned how to describe tree-level gluon scattering amplitudes using scalar-scaffolding, we now turn to making our proposal for gluon amplitudes at loop level. At tree-level, we had the luxury of being able to easily compare with the corresponding expressions in the bosonic string, but at loop-level we will be proceeding more adventurously. 

At tree level, we understood that the simple kinematic shift -- which is equivalent to inserting the magic monomials in $u$'s $(\prod u_{e,e}/\prod u_{o,o})$, which is equal to the product of all $\prod_i y_i$ in the scaffolding triangulation -- is responsible for transmuting the ``boring" amplitudes of Tr $\phi^3$ theory into ones that enjoy the gauge invariance, multilinearity and factorization of gluon amplitudes. 

But in fact, as mentioned in our review of surfaceology, this monomial has exactly the same property for {\it any} surface. This motivates an extension of our proposal to any surface and hence all loop orders/orders in the 't Hooft expansion. 

Our goal in this section is to present this all-loop generalization, discussing a number of novelties that arise along the way: we will see that the product over curves ${\cal C}$ must include open curves with up to one self-intersection per loop, as well as a product over closed curves $\Delta$ with exponents that introduce the dependence on the spacetime dimension, $D$. The treatment of self-intersecting curves marks a major difference from the conventional bosonic string theory. For Tr $\phi^3$ theory matching (bosonic) string theory at loop-level from the curve-integral formalism asks us to consider the product over {\it all} curves \cite{combstring}, including those with arbitrarily high intersection numbers. Meanwhile, if we want a ``stringy'' integral that reduces to Tr $\phi^3$ at low energies it suffices to only keep non-self-intersecting curves, which are the only ones dually interpretable as propagators in Feynman diagrams. It is intriguing that in order to describe gluons we must make the curve integral ``slightly more stringy'' by including curves with a finite number of self-intersections, but still distinct from the infinite number of curves needed to describe the ``real bosonic string'' in this language.

\subsection{Loop scaffolding form}

In the tree-level case, we derived the scaffolding form $\Omega_{2n}$ starting from the bosonic string $2n$ gluon amplitude and choosing a kinematic configuration in which the gluons become effectively scalars -- $2n$ scalars. Following the same procedure at loop-level is much less trivial, as bosonic string loop amplitudes are significantly more complicated. 
  
Instead, we will follow a different strategy that once more relies on how the tree-level amplitudes are formulated in terms of the surface problem. As we saw in the previous section, the resulting scaffolding form, in terms of $u$'s, was simply that of the bosonic string Tr $\phi^3$ amplitude, times the factor corresponding to the ratio of $u$'s, \eqref{eq:2nScalarU}. In particular, choosing the scaffolding triangulation, we can simplify this to the bosonic string Tr $\phi^3$ amplitude where instead of a logarithmic form in the $y$'s, we have a $dy/y^2$ form \eqref{eq:2nScaffTriang}.  
   
In order to generalize this scaffolding form to the loop level, we consider exactly the same integrand but instead of being defined on the disk, it is now defined on the punctured disk. So, at one-loop, for example, we start with \eqref{eq:loopStringSurface} and replace the integration measure:
\begin{equation}
	\mathcal{A}^{\text{1-loop}}_{2n}(1,2,...,2n) = \int_{0}^{\infty} \prod_{i}\frac{\diff y_i}{y_i^2} \, \prod_{C} u_C^{\alpha^{\prime} X_C} \times \prod_{C^\prime \in \, \text{ s.i.}} u_{C^\prime}^{\alpha^{\prime} X_{C^\prime}} \times u_{\Delta}^\Delta.
	\label{eq:loopScaffForm}
\end{equation}

This is our candidate for the loop surface integral. The version for higher loops is exactly the same except that at higher loops there are more closed curves, so instead we have a product over $u_{\Delta_i}$. 
  
As pointed out previously, since there are an infinite number of self-intersecting curves, this integrand contains an infinite number of terms. In the case of Tr $\phi^3$, we can truncate this infinite product, since all self-intersection curves had no impact on the field-theory limit: their $u$ variables can never go to zero since they occur in both terms of the $u$ equations, and hence they don't affect the leading singularities that dominate the field theory limit.  We now want to understand how this generalizes to gluons, {\it i.e.} what is the minimal set of curves we need to include in the integrand that gives us gluons at low energies. 
  
Another feature of gluon loop amplitudes is the dependence on the spacetime dimension, $D$. If tree amplitudes are written either in terms of dot products of momenta and polarization vectors, or in our scalar scaffolding language, they are $D$ independent. However, loop amplitudes/integrands are certainly $D$ dependent, mechanically arising from the trace of the metric $\eta^{\mu \nu} \eta_{\mu \nu} = D$. However, from \eqref{eq:loopScaffForm} we do not see any explicit dependence on $D$, so where is the spacetime dimension dependence coming from?

\subsubsection{$\Delta$}
At the outset of our discussion we mentioned that in order to describe gluons at loop level, we have to consider closed curves that naturally live on the surface. In the planar limit, these are closed curves surrounding punctures. These are not needed for ``stringy'' Tr $\phi^3$ integrals, at least if we only want to match the low-energy field theory limit (they are needed for the full bosonic string amplitude), but together with the self-intersecting curves we will mention in a moment, they will be important to describe gluons. 

By homology closed curves should have zero momentum, but we will give the corresponding $u_\Delta$ variables an exponent $\Delta$. We will see that to describe gluons in $D$ spacetime dimensions at one loop we must take 
\begin{equation}
	\Delta = 1-D.
\end{equation}
Note again the importance of using the scalar scaffolding picture, which allows our description of momentum kinematics to be dimension agnostic, and now allows us to introduce the dependence on spacetime dimension $D$ as a continuous parameter we can choose at will. 

At general loop level we conjecture that the $D$-dependence comes in purely through the choice of the exponents for closed curves in the form $\Delta = n_{\Delta} - D$, where $n_{\Delta}$ is an integer. All the checks of our proposal from leading singularities we will discuss later are consistent with $n_{\Delta} =1$ just as at 1-loop. 

\subsubsection{Self-Intersecting Curves}

As we mentioned previously, in the case of Tr $\phi^3$ the contribution from self-intersecting curves does not affect the low energy limit. For gluons, however, the minimal set of curves we need includes curves that self-intersect at most once (around each puncture, in case we are considering a higher loop process). Thus the proposal for the gluon loop integrand is:

\begin{equation}
	\mathcal{A}^{\text{m-loop}}_{2n}(1,2,...,2n) = \int_{0}^{\infty} \underbrace{\prod_{i}\frac{\diff y_i}{y_i^2} \, \prod_{C} u_C^{\alpha^{\prime} X_C} \times \prod_{C^\prime \in \, \#\text{ s.i.}=1} u_{C^\prime}^{\alpha^{\prime} X_{C^\prime}} \times\prod_{k \in \{1,...,m\} } u_{\Delta_k}^{\Delta_k}}_{\Omega_{2n}^{\text{m-loop}}},
	\label{eq:loopScaffFormTrunc}
\end{equation}
where $\Delta_k$ denotes the closed curve going around puncture $k$.
  
Before proceeding, let's highlight the following: at tree-level, the $2n$-scaffolding form was derived directly from the bosonic string answer, and thus the low energy limit had to be gluons. For the loop integrand \eqref{eq:loopScaffForm}, this is \textbf{not} the case, so there is no guarantee a priori that we are describing gluons, and in particular that \eqref{eq:loopScaffFormTrunc} is the correct truncation.  
  
One encouraging feature is that the object we get after taking the scaffolding residue satisfies a notion of gauge-invariance/multi-linearity, just like in the tree-level case. We will explore this notion in section \ref{sec:gauginvLoop}.
  
Another check on this object is to compute the loop-leading singularities and check that they agree with gluon-leading singularities. Some nontrivial examples of this matching are presented in \ref{sec:LeadSingLoop}. These computations make manifest how the self-intersecting curves are important at low energies. 
  
However to prove the full consistency of this formulation what one has to do is show that starting with \eqref{eq:loopScaffFormTrunc} and going on the cut of the loop propagator, we land on the tree-level answer \eqref{eq:2nScaffTriang} of the corresponding cut surface. This turns out to be quite non-trivial, but we present the full proof of this matching in section \ref{sec:Loopcut}. As a byproduct, we will understand why we need to keep all the curves up to self-intersection one.

\subsection{Surface Kinematics}
\label{sec:surfaceKin}

Let us now explain how we can read off the $X_{\mathcal{C}}$, $i.e.$ the kinematic invariant $X$ associated to a given curve $\mathcal{C}$ on the surface, appearing in the loop surface integral \eqref{eq:loopScaffFormTrunc}.

Just like at tree-level, for the non-self-intersecting curves, each curve $\mathcal{C}$ is dual to a propagator entering in the $2n$ scalar amplitude, and, in particular, those that starts and ends on an odd marked point/puncture are dual to gluon propagators (post-scaffolding residue). Regardless, like we did at tree-level, we can read off the momentum flowing through the corresponding propagator by \textit{homology}, or equivalently by the left-right rule presented in \ref{sec:Surfaceology}. 

\begin{figure}[t]
    \centering
    \includegraphics[width=\linewidth]{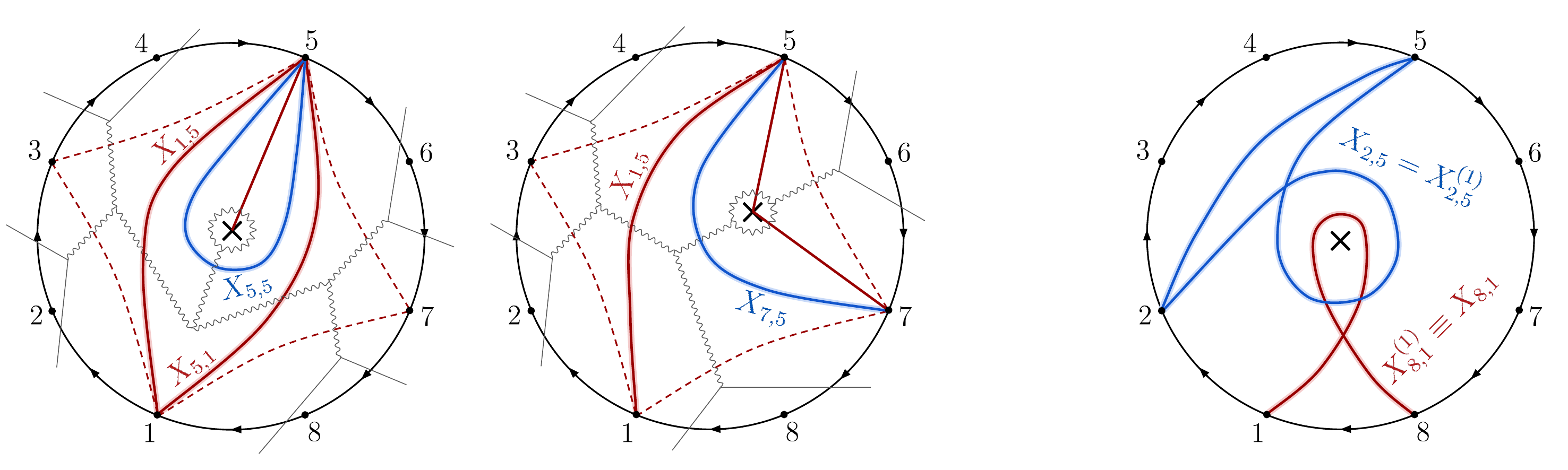}
    \caption{On the left, we have the tadpole and external bubble triangulations of the 4-point gluon one-loop surface, where each curve is assigned a different variable $X$, even when they are the same homology/have the same momentum. The scaffolding curves are dashed while the gluon internal propagators are solid lines. On the right, we have some examples of curves and the respective kinematic assignments defined by surface kinematics. }
    \label{fig:TadpoleBubble}
\end{figure}

So we could certainly follow this procedure at loop-level as well. Doing so and assuming the puncture carries no momenta, we would get that different curves on the surface are assigned the same kinematic invariant $X$. For example, already at one-loop, there are two non-self-intersecting going from boundary marked points $i$ and $j$: $X_{i,j}$, that goes through the left of the puncture, and $X_{j,i}$ going through the right. If we read off the momentum of these curves by homology we would have $X_{i,j} = X_{j,i}$. In particular, this would force certain curves to be assigned zero kinematic invariant, such as $X_{i+1,i}=X_{i,i}=0$, as well as $X_{j+2,j}=0$ for $j$ odd, as these are homologous to the scaffolding curves that are set to zero after the scaffolding residue. In particular, for $i$ odd, $X_{i,i}$ and $X_{i+2,i}$ are precisely the curves that are dual to gluon tadpoles and external bubbles propagators, respectively, which in momentum space are indeed assigned zero momentum (see figure \ref{fig:TadpoleBubble}, left).

However, the integrand that has all the canonical features of a gluon integrand, $i.e.$ that is gauge-invariant and where all the cuts are given in terms of gluing of lower objects, is naturally defined in terms of an extension of these kinematics -- this is what we will call \textit{surface kinematics}. Instead of reading the $X_{\mathcal{C}}$'s by homology, we assign a different variable $X_{\mathcal{C}}$ to each \textit{different} curve on the surface, so in particular, already at one-loop we have $X_{i,j} \neq X_{j,i}$. However we assign the same variable to curves that differ by self-intersection, $X_{\mathcal{C}^{(0)}}=X_{\mathcal{C}^{(1)}}$ except for the case of the boundary curves $X_{i,i+1}$ for which the only existing curve has self-intersection one (see figure \ref{fig:TadpoleBubble}, right). 

By reading off the kinematics in this way, the $n$-point gluon one-loop integrand depends on $(2n)^2- 2n$ $X_{i,j}$'s~\footnote{We have $X_{i,i+2}=0$ for $i$ odd from the scaffolding residue, and no $X_{a,a}$ with $a$ even can appear in the answer.}, and $2n$ $X_{i,p}\equiv Y_i$, {\it i.e.} curves ending on the puncture. As we will see momentarily, keeping \textit{all} these distinct kinematic variables is crucial to allow us to correctly interpret all the cuts of the integrand, as well as provide a simple generalization of gauge-invariance satisfied by this object. Let us now proceed to explaining how the statement of tree-level gauge-invariance that we found earlier nicely generalizes to loop-level for integrands defined in terms of surface kinematics.

\subsection{Surface gauge invariance}
\label{sec:gauginvLoop}
It is well known that the notion of gauge invariance for loop integrands is afflicted with difficulties associated with the presence of the tadpole (as well as massless external bubble) propagators. But we have already seen, at tree level, that simple properties of the $u's$ directly hand us a certain linear expansion of the amplitude in terms of the scaffolding kinematic variables, that automatically encodes gauge invariance and multilinearity. We will now see how this extends to loop level, to determine the general form of ``gluon amplitudes'' predicted by the binary geometry of the surface. As we will see these will give a natural home for incorporating tadpoles/massless external bubbles with no ambiguities, and we will use this to {\bf define} what we mean by gauge invariance and multilinearity. To avoid this mouthful of a phrase in what follows we will refer to this simply as ``surface gauge invariance''. 
  
To do this let us focus on gluon $1$, and look at the scaffolding residue, $y_{1,3} \to 0$. Similarly to the tree-level example, the $u$'s that depend on $y_{1,3}$ are: $u_{1,J}$, $u_{2,J}$, $u_{J,1}$, $u_{J,2}$, as well as the loop variables, $u_{Y_1}$ and $u_{Y_2}$.
  
Now for each curve $u^{(q)}_{2,J}/u^{(q)}_{J,2}$ (where q denotes the number of self-intersections, so $q=0$ or $1$), we have, just like at tree-level:
\begin{equation}
	\frac{\partial \log u^{(q)}_{1,J}}{\partial y_{1,3}} = - \frac{\partial \log u^{(q)}_{2,J}}{\partial y_{1,3}}, \quad \frac{\partial \log u^{(q)}_{J,1}}{\partial y_{1,3}} = - \frac{\partial \log u^{(q)}_{J,2}}{\partial y_{1,3}}, \quad 
	\frac{\partial \log u_{Y_{1}}}{\partial y_{1,3}} = - \frac{\partial \log u_{Y_2}}{\partial y_{1,3}}.
\end{equation} 

Note however that when $J=2n$, $u_{2n,1}^{(0)}=0$, as this in this case it is just a boundary curve, so the equality above only holds for $u_{2n,2}^{(1)}$ and $u_{2n,1}^{(1)}$. Bearing this in mind we get that after taking the scaffolding residue in $X_{1,3}$ we can write our answer as follows:

\begin{equation}
\begin{aligned}
	\mathop{\mathrm{Res}}_{y_{1,3}=0}\Omega^{\text{loop}}_{2n}[X_{i,j}] = &\sum_{J} \left(X_{2,J} - X_{1,J} \right) \underbrace{\prod_{s\neq (1,3)} \frac{\diff y_{s}}{y_{s}^2} \, \prod_{i} \frac{\diff y_{i}}{y_{i}^2} \times \frac{\partial \log ( \prod_{q=0,1} u^{(q)}_{2,J})}{\partial y_{1,3}}\prod_{(km)} u_{km}^{ X_{k,m}}[y] \bigg \vert_{y_{1,3}=0} }_{\mathcal{Q}_{2,J}} \\
 &+\sum_{J} \left(X_{J,2} - X_{J,1} \right) \underbrace{\prod_{s\neq (1,3)} \frac{\diff y_{s}}{y_{s}^2} \, \prod_{i} \frac{\diff y_{i}}{y_{i}^2} \times \frac{\partial \log ( \prod_{q=0,1} u^{(q)}_{J,2})}{\partial y_{1,3}}\prod_{(km)} u_{km}^{ X_{k,m}}[y] \bigg \vert_{y_{1,3}=0} }_{\mathcal{Q}_{J,2}} \\
	& +(Y_{2} - Y_{1}) \underbrace{\prod_{s\neq (1,3)} \frac{\diff y_{s}}{y_{s}^2} \, \prod_{i} \frac{\diff y_{i}}{y_{i}^2} \times \frac{\partial \log ( u_{Y_2})}{\partial y_{1,3}}\prod_{(km)} u_{km}^{ X_{k,m}}[y]  \bigg \vert_{y_{1,3}=0}}_{\mathcal{Q}_0} \\
 &+X_{2n,1}\times  \underbrace{\prod_{s\neq (1,3)} \frac{\diff y_{s}}{y_{s}^2} \, \prod_{i} \frac{\diff y_{i}}{y_{i}^2} \times \frac{\partial \log (  u^{(0)}_{2n,2})}{\partial y_{1,3}}\prod_{(km)} u_{km}^{ X_{k,m}}[y] \bigg \vert_{y_{1,3}=0} }_{\mathcal{Q}^{(0)}_{2n,2}},
\end{aligned}
\label{eq:GaugeInvLoop}
\end{equation} 
where $J \in \{1,3,4,\cdots,2n\}$. Note that there is the correction term in the last line proportional to $X_{2n,1}$ precisely because of the reason mentioned above. 

Now just like at tree-level, we have the $\mathcal{Q}_{2,J} = \partial_{X_{2,J}} \mathcal{I}_n$, $\mathcal{Q}_{J,2} = \partial_{X_{J,2}} \mathcal{I}_n$ and $\mathcal{Q}_{0} = \partial_{Y_2} \mathcal{I}_n$, where $\mathcal{I}_n$ stands for the object we get after taking the scaffolding residue, which at low energies gives us the gluon loop-integrand. Similarly we would like to be able to express the correction term, in terms of some operation on the integrand. To do this we start by observing that taking a derivative on \eqref{eq:GaugeInvLoop} with respect to $X_{2n,1}$ and observe that we can isolate $\mathcal{Q}^{(0)}_{2n,2}$ by evaluating the result at $X_{2,J}\to X_{1,J}$, $X_{J,2}\to X_{J,1}$ and $Y_2 \to Y_1$:
\begin{equation}
\begin{aligned}
	\frac{\partial \mathop{\mathrm{Res}}_{y_{1,3}=0}\Omega^{\text{loop}}_{2n}[X_{i,j}]}{\partial X_{2n,1}} \bigg \vert_{2 \to 1} = &\left[\sum_{J} \left(X_{2,J} - X_{1,J} \right) \frac{\partial \mathcal{Q}_{2,J}}{\partial X_{2n,1}} +\sum_{J} \left(X_{J,2} - X_{J,1} \right) \frac{\partial \mathcal{Q}_{J,2}}{\partial X_{2n,1}} \right.\\
	&\left. +(Y_{2} - Y_{1}) \frac{\partial \mathcal{Q}_{0}}{\partial X_{2n,1}} +\mathcal{Q}^{(0)}_{2n,2}  - \mathcal{Q}_{2n,2} \right] \bigg \vert_{2\to 1}\\ 
     =& -\mathcal{Q}^{(1)}_{2n,2}\vert_{2\to 1} = -\mathcal{Q}^{(1)}_{2n,2},
\end{aligned}
\end{equation} 
where we used $\mathcal{Q}_{2n,2} = \mathcal{Q}^{(1)}_{2n,2} + \mathcal{Q}^{(0)}_{2n,2}$ as well as that the $\mathcal{Q}$'s are independent of $X_{2,J},X_{J,2}$ and $Y_2$. Therefore we have that:
\begin{equation}
    \mathcal{Q}^{(0)}_{2n,2} =  \mathcal{Q}_{2n,2} - \mathcal{Q}^{(1)}_{2n,2} =  \frac{\partial  \mathcal{I}_n}{\partial X_{2n,2}} + \frac{\partial   \mathcal{I}_n}{\partial X_{2n,1}} \bigg \vert_{2\rightarrow 1}.
\end{equation}

In summary, the full gauge-invariance as well as multi-linearity of the integrand in gluon one can be written as follows: 

\begin{equation}
\begin{aligned}
   \mathcal{I}_n &= \sum_{j\neq 2} \left[ \left(X_{2,j} - X_{1,j}\right) \frac{\partial  \mathcal{I}_n}{\partial X_{2,j}} + \left(X_{j,2} - X_{j,1}\right) \frac{\partial  \mathcal{I}_n}{\partial X_{j,2}} \right]+ X_{2n,1} \times \left[ \frac{\partial  \mathcal{I}_n}{\partial X_{2n,2}} + \frac{\partial   \mathcal{I}_n}{\partial X_{2n,1}} \bigg \vert_{2\rightarrow 1} \right] \\
    &= \sum_{j\neq 2} \left[ \left(X_{2,j} - X_{3,j}\right) \frac{\partial   \mathcal{I}_n}{\partial X_{2,j}} + \left(X_{j,2} - X_{j,3}\right) \frac{\partial  \mathcal{I}_n}{\partial X_{j,2}} \right] + X_{3,4} \times \left[ \frac{\partial  \mathcal{I}_n}{\partial X_{2,4}} + \frac{\partial   \mathcal{I}_n}{\partial X_{3,4}} \bigg \vert_{2\rightarrow 3} \right].
\end{aligned}
\label{eq:SurfaceGaugeInv}
\end{equation}
where once again the same statement goes through for $X_{2,J},X_{3,J}$, just like at tree-level. 

%where we have that $\mathcal{Q}_J$ is independent of $X_{2,J}$ and $\mathcal{Q}_0$ is independent of $Y_2$. 
%Note importantly that $\mathcal{Q}_{1,1}, \mathcal{Q}_{2,2}$ {\bf can} still depend on $X_{1,1},X_{2,2}$, and indeed can (and do) have poles $1/X_{1,1}, 1/X_{2,2}$. This is the reflection of the familiar difficulty in defining non-supersymmetric Yang-Mills integrands where the tadpoles/massless external bubbles are set to zero by fiat, and this is also why we cannot simply set $e.g.$ the $X_{i,i}$ variables to zero. But the expansion of equation \ref{eq:surfgauge} is completely well-defined, and our curve integral for gluons gives a well-defined integrand with its own notion of gauge invariance and multilinearity. Of course this notion differs from what is standard in ordinary gauge theory, but the difference arises precisely from terms with tadpole/massless external gluon bubble poles, that multiply scale-less integrands integrating to zero. We will make further comments about this well-defined integrand after we see a concrete example below. 

\subsection{Examples: one-loop $1$- and $2$-gluon integrands}
In this section, we present simplest examples at one loop, namely $1$-gluon and $2$-gluon integrands, which are obtained from scaffolding residues of \eqref{eq:loopScaffFormTrunc} in the low energy limit with $2$ and $4$ scalars, respectively. We will see explicitly how the surface kinematics and surface gauge invariance work in these simple examples. 

\subsubsection{$1$-gluon integrand}
Let us begin with the $1$-point integrand, and we choose the underlying triangulation to be the tadpole triangulation for the gluon problem, $i.e.$ $\{X_{1,1},Y_1\}$, where $X_{1,1}$ is the scaffolding chord and $Y_1$ is the loop propagator of the gluon amplitude. Keeping only curves with self-intersection at most one as prescribed, we get that the stringy gluon integrand is:
\begin{equation}
	\begin{aligned}
		\mathcal{A}_1^{\text{gluon}} = \int_{0}^{\infty} \frac{\diff y_1 }{y_1^2 } \text{Res}_{y_{1,1}=0} \left( \frac{1}{y_{1,1}^2 } \right. & \left.  \prod_{i=1,2} u_{i,i}^{X_{i,i}}  u_{Y_i}^{Y_i} \times u_\Delta^\Delta \right. 
		& \left. (u_{1,2}^{(1)})^{X_{1,2}} (u_{2,1}^{(1)})^{ X_{2,1}}\prod_{i=1,2}(u_{i,i}^{(1)})^{ X_{i,i}}\right),
	\end{aligned}
\end{equation}
where we have used the algorithm of section~\ref{sec:words to u} to obtain $u$ variables in this triangulation as follows,
\begin{equation}
	\begin{aligned}
		u_{1,1}=&\frac{y_{1,1}(1+y_{1}+y_{1}y_{1,1})}{1+y_{1}+y_{1,1}+y_{1}y_{1,1}},\quad
    		u_{2,2}=\frac{1+y_{1,1}+y_{1}y_{1,1}}{1+y_{1,1}+y_{1}y_{1,1}+y_{1}y_{1,1}^2},\\
		u_{Y_{1}}=&\frac{y_1 \left(y_{1,1}+1\right)}{y_1 y_{1,1}+1},\quad
		u_{Y_{2}}=\frac{y_1 y_{1,1}+1}{y_{1,1}+1},\quad u_\Delta=1-y_{1},\\
		u_{1,2}^{(1)} =&\frac{\left(y_1+1\right) \left(y_{1,1}+1\right) \left(y_1 y_{1,1}+1\right)}{\left(\left(y_1+1\right) y_{1,1}+1\right) \left(y_1 \left(y_{1,1}+1\right)+1\right)},u_{2,1}^{(1)} =\frac{\left(y_1+1\right) \left(y_{1,1}+1\right) \left(y_1 y_{1,1}+1\right)}{\left(\left(y_1+1\right) y_{1,1}+1\right) \left(y_1 \left(y_{1,1}+1\right)+1\right)}\,,
	\end{aligned}
\end{equation}
where in the low-energy limit those $u_{i,i}^{(1)}$ do not contribute to the gluon integrand, so we do not list them here. Then, we take the scaffolding residue on $y_{1,1}=0$ and set $X_{1,1}=0$ as we have discussed in section~\ref{sec:ExtRes}, which gives the nice result: 
\begin{equation}
	\begin{aligned}
		\mathcal{A}_1^{\text{gluon}} = \int_{0}^{\infty} \frac{\diff y_1 }{y_1^2 }  \left( -\alpha' \frac{\left(1-y_1\right){}^{\Delta} y_1^{\alpha' Y_1}  \left(y_1 \left(X_{1,2}+X_{2,1}\right)+\left(y_1-1\right) \left(y_1+1\right) (Y_1-Y_{2})\right)}{y_1+1}\right)\,.
	\end{aligned}
\end{equation}

We can now directly integrate over $y_{1}$ and obtain the ``stringy" integrand, but all we need is the low-energy limit. It is straightforward to take $\alpha^{\prime}\to 0$ and obtain
\begin{equation}\label{1gluon}
		\mathcal{A}_1^{\text{gluon}} =-\frac{X_{1,2}+X_{2,1}+(\Delta+1)(Y_1- Y_2)}{Y_1}\,,
\end{equation}
which as expected involves the $1$-point tadpole graph only. 
\subsubsection{$2$-gluon integrand}

Next we present the result for one-loop $2$-gluon integrand, and we obtain it by starting with a 4-scalar process described by \eqref{eq:loopScaffFormTrunc}, taking the scaffolding residue, reducing us to the 2-point 1-loop stringy gluon integrand from which we ultimately extract the low-energy limit to obtain the field theory integrand.
 
We choose the underlying triangulation to be the bubble triangulation for the gluon problem, $i.e.$ $\{X_{1,3},X_{3,1},Y_1,Y_3\}$, where $X_{1,3}$ and $X_{3,1}$ are the scaffolding chords and $Y_1,Y_3$ are the loop propagators of the gluon amplitude. Keeping only curves with self-intersection at most one as prescribed, we get that the stringy gluon integrand is:
\begin{equation}
\begin{aligned}
    \mathcal{A}_2^{\text{gluon}} = \int \frac{\diff y_1 \diff y_3}{y_1^2 y_3^2} & \text{Res}_{y_{1,3}=y_{3,1}=0} \left( \frac{1}{y_{1,3}^2 y_{3,1}^2} \right.  \left. u_{2,4}^{X_{2,4}}u_{4,2}^{X_{4,2}} \prod_{i=1,\dots,4} u_{i,i}^{X_{i,i}} u_{i,i-1}^{X_{i,i-1}} u_{Y_i}^{Y_i} \times u_\Delta^\Delta \right.  \\
    & \left. (u_{2,4}^{(1)})^{X_{2,4}} (u_{4,2}^{(1)})^{ X_{4,2}}\prod_{i=1,\dots,4}(u_{i,i}^{(1)})^{ X_{i,i}} (u_{i,i-1}^{(1)})^{ X_{i,i-1}} (u_{i-1,i}^{(1)})^{ X_{i-1,i}} \right),
\end{aligned}
\label{eq:bubbleInt}
\end{equation}
where we have set $X_{1,3}=X_{3,1}=0$. Now since we picked the triangulation corresponding to the bubble diagram for the gluon problem, containing chords $\{Y_1, Y_3\}$, the singularity corresponding to this diagram is located near $y_1,y_3 \rightarrow0$. There are two other diagrams, corresponding to the tadpole diagrams with chords, $\{Y_1,X_{1,1}\}$ and $\{Y_3,X_{3,3}\}$, whose singularities are located in different boundaries of the integration domain. To extract the field theory limit of \eqref{eq:bubbleInt}, we proceed as outlined in section \ref{sec:FurtherCom}, fully explained in \cite{cexpansion}, and extract the contribution to the integrand from each of these 3 regions. Summing them gives the full low-energy integrand: 

\begin{equation}\label{2gluon}
\begin{aligned}
    \mathcal{A}_2^{\text{gluon}}=   &-\frac{X_{1,2} X_{1,4}+X_{2,1} X_{4,1}-(\Delta+1)(X_{2,1}X_{1,4}-Y_2 X_{1,4}-Y_4 X_{2,1}+Y_3\left(X_{1,4}+X_{2,1}-X_{2,4}\right))}{Y_1 X_{1,1}}\\
    &+\frac{X_{1,2}-X_{2,1}+X_{4,1}-X_{1,4}+2 X_{2,4}-X_{4,2}-(\Delta+1)(Y_2-Y_3+Y_4-X_{2,4})}{Y_1}+\frac{X_{2,4}}{X_{1,1}}
    \\&+\frac{Y_2 \left(X_{1,4}-X_{4,1}+X_{4,3}-X_{3,4}\right)+X_{1,1} \left(X_{3,2}-X_{4,2}+X_{4,3}\right)}{Y_1 Y_3}+(i\to i+2)\\
    &-\frac{X_{1,4} X_{3,2}+X_{2,1} X_{4,3}+X_{1,1} X_{3,3}-(\Delta+1)Y_2 Y_4}{Y_1 Y_3}+(\Delta+1)\,,
\end{aligned}
\end{equation}
where $(i\to i+2)$ denotes the cyclic permutation $\{1,2,3,4\}\to \{3,4,1,2\}$ for all preceding terms, and the terms on the last line are manifestly invariant under $(i\to i{+}2)$. This same integrand can be obtained from a recursion relation we present in \cite{YMrec}, which leverages our understanding of the loop-cuts as well as good behvaior at infinity under a uniform shift of the loop-variables.

We can as usual translate from our scalar description of the kinematics to standard presentations depending on dot products of momenta and polarization vectors if we like. The polarization vectors are still given by~\eqref{eq:eps}; we can make the gauge choice $\alpha=1$ so that
\begin{equation}
    \epsilon_i^\mu \propto - (X_{2i}-X_{2i-1})^\mu.
\end{equation}

In addition, we have the freedom to choose the loop momentum so that $Y_1=l^2$. Then the $X,Y$ variables are expressed in terms of the usual Lorentz invariants as
\begin{equation}
\begin{aligned}
Y_1=l^2,\;\; Y_2=Y_1-2 \epsilon _1\cdot l,\;\; Y_3=l^2+2 q_1\cdot l,\;\; Y_4=Y_3-2 \epsilon _2\cdot l,\;\; X_{2,4}=-2 \epsilon _1\cdot \epsilon _2.
\end{aligned}
\end{equation}

Note that we do {\it not} put $X_{1,1}, X_{3,3} = 0$ at the level of the integrand; we will make further comments on this point in a moment. We obviously have similar expressions for higher points and loops, once we specify some choice for polarizations and loop momenta. Of course, the integrated result is independent of such choices, and post-integration we are also free to put tadpoles to zero. 

\subsubsection{Gauge Invariance check}

In the previous section, we explained how this surface definition of the integrand gives us a generalized notion of gauge invariance for the loop integrand. Now that we have derived an explicit example of $1$- and $2$-gluon integrands, we can check that the form predicted in \eqref{eq:GaugeInvLoop} holds in these cases.

For \eqref{1gluon}, it is easy to check the surface gauge invariance (in gluon $1$):
\begin{equation}
\mathcal{A}_1^{\text{gluon}}=(X_{2,1}){\cal Q}_{X_{2,1}} +(Y_{2}-Y_{1}){\cal Q}_{Y_{2}}+ (X_{1,2}){\cal Q}_{X_{1,2}} \,,
\end{equation}
where ${\cal Q}_X:=\frac{\partial\mathcal{A}_1^{\text{gluon}}}{\partial X}$, and they read,
\begin{equation}
    {\cal Q}_{X_{2,1}}={\cal Q}_{X_{1,2}}=-\frac{1}{Y_{1}},{\cal Q}_{Y_{2}}=-\frac{(\Delta+1)}{Y_1}\,.
\end{equation}

% Let us present the explicit check of this form for gluon 1 (the same holds for gluon 2 by cyclic invariance). According to \eqref{eq:GaugeInvLoop}, we should be able to write the integrand as: $\mathcal{A}_2^{\text{gluon}} = X_{2,4} \mathcal{Q}_{2,4}+(Y_2-Y_1)\mathcal{Q}_{Y_2}+X_{1,1}\mathcal{Q}_{X_{1,1}}$, where $\mathcal{Q}_{2,4}$ and $\mathcal{Q}_{Y_2}$ are independent of $X_{2,4}$ and $Y_{2}$, respectively. Indeed one can explicitly check this is the case with
% \begin{equation}
% \begin{aligned}
%     &\mathcal{Q}_{2,4} =  \frac{(1+\Delta)(-X_{1,1} + Y_1  - Y_3)}{
%  Y_1 X_{1,1}} + \frac{X_{1,1} + X_{3,3}-Y_1-Y_3}{
%  Y_1 Y_3} + \frac{(1+\Delta)(-X_{3,3}  - Y_1  + Y_3)}{Y_3X_{3,3}}, \\
%   &\mathcal{Q}_{Y_2} = \frac{(1+\Delta)}{Y_3} + \frac{(1+\Delta) (Y_3 - Y_4)}{Y_1 Y_3}, \\
%   &\mathcal{Q}_{X_{1,1}} = \frac{X_{3,3}}{Y_1 Y_3} - \frac{(1 + \Delta) (Y_3 - Y_4)}{X_{1,1} Y_1 }.
% \end{aligned}
% \end{equation}

Similarly, we can write the $2$-gluon integrand~\eqref{2gluon} in the form as prescribed in~\eqref{eq:GaugeInvLoop}:
\begin{equation}
\mathcal{A}_2^{\text{gluon}}=\sum_{j=1,3,4,p}(X_{2,j}-X_{1,j}){\cal Q}_{X_{2,j}} + \sum_{j=1,3}(X_{j,2}-X_{j,1}){\cal Q}_{X_{j,2}} %+(X_{2,4}-X_{1,4}){\cal Q}_{X_{2,4}} 
+ X_{4,2} {\cal Q}_{X_{4,2}}+X_{4,1} {\cal Q}'_{X_{4,1}}\,,
\end{equation}
where ${\cal Q}_X:=\frac{\partial\mathcal{A}_2^{\text{gluon}}}{\partial X}$ except for the last boundary term ${\cal Q}'_{X_{4,1}}:=\frac{\partial\mathcal{A}_2^{\text{gluon}}}{\partial X_{4,1}}|_{2\to 1}$, and they read
\begin{equation}
\begin{aligned}
&{\cal Q}_{X_{2,1}}=\frac{X_{3,3}-X_{4,3}+Y_4}{Y_1 Y_3}-\frac{X_{4,1}}{Y_1 X_{1,1}}-\frac{1}{Y_1}+(D-2)\frac{X_{1,4}+Y_3-Y_4}{Y_1 X_{1,1}}\,,\\
&{\cal Q}_{X_{1,2}}=-\frac{X_{1,4}}{Y_1 X_{1,1}}-\frac{Y_4}{Y_1 Y_3}+\frac{1}{Y_1}\,,\\
&{\cal Q}_{X_{2,3}}=-\frac{X_{4,3}}{Y_3 X_{3,3}}-\frac{Y_4}{Y_1 Y_3}+\frac{1}{Y_3}\,,\\
&{\cal Q}_{X_{3,2}}=\frac{X_{1,1}-X_{1,4}+Y_4}{Y_1 Y_3}-\frac{X_{3,4}}{Y_3 X_{3,3}}-\frac{1}{Y_3}+(D-2)\frac{X_{4,3}+Y_1-Y_4}{Y_3 X_{3,3}}\,,\\
&{\cal Q}_{X_{2,4}}=-\frac{X_{3,3}+Y_1-2 Y_3}{Y_1 Y_3}+(D-2)\frac{-X_{1,1}+Y_1-Y_3}{Y_1 X_{1,1}}\,,\\
&{\cal Q}_{X_{4,2}}=\frac{-X_{1,1}+2 Y_1-Y_3}{Y_1 Y_3}+(D-2)\frac{-X_{3,3}-Y_1+Y_3}{Y_3 X_{3,3}}\,,\\
&{\cal Q}_{Y_2}=\frac{X_{1,4}-X_{4,1}+X_{4,3}-X_{3,4}}{Y_1 Y_3}+(D-2)\left(-\frac{X_{1,4}}{Y_1 X_{1,1}}-\frac{X_{4,3}}{Y_3 X_{3,3}}-\frac{Y_4}{Y_1 Y_3}+\frac{1}{Y_1}+\frac{1}{Y_3}\right)\,,\\
&{\cal Q}'_{X_{4,1}}=-\frac{1}{Y_3}\,.
\end{aligned}
\end{equation}

\subsection{A well-defined Yang-Mills integrand} 

It is interesting to compare the integrand we have obtained in this way with the conventional Yang-Mill bubble integrand. Here we run into the familiar difficulty that we do not have a well-defined conventional Yang-Mills integrand, precisely due to the existence of the tadpole/massless external bubble terms and the infamous ``1/0'' problem associated with them, as well as massless external bubbles. We are usually instructed to throw out the tadpoles/massless external bubbles, on the grounds that they are ``scale-less integrals that integrate to zero'', but clearly something nice about the integrand must be lost in doing so, and indeed it now fails to be gauge invariant. 

By contrast, the bubble integrand we have just produced is perfectly well defined. We encounter the familiar 1/0 difficulties if we were to set the tadpole $X_{1,1} \to 0$, but nothing forces us to do this, and at finite $X_{1,1}$ we have an object that is both well-defined and is consistent with its own internal-to-the-world-of-surfaces notion of gauge invariance. If we were to set $X_{1,1} \to 0$, we can only get singularities in terms with $1/X_{1,1}$ poles. These are of course precisely tadpole factorizations, and therefore must obviously multiply scale-less integrals that integrate to zero. So, we get to have our integrand cake and eat our amplitude too: we have a well-defined integrand at finite values of tadpole/massless external bubble variables, when these are set to zero as naively needed to match normal kinematics we do encounter divergences which are however guaranteed to be proportional to objects that integrate to zero. 

We can also check that our integrand has the correct cuts to match field theory gluing. The most straightforward check is the usual unitarity double-cut. But it also has the correct single-cuts, which again needs a proper discussion of $X_{1,1} \neq 0$ for the ``forward limit" term. We will discuss at length in our discussion of the loop-cut/forward limit connection, even at finite $\alpha^\prime$, in the penultimate section of this paper. 

The elementary mechanism by which tadpoles/massless external bubbles are naturally dealt with suggests a recursive procedure for computing integrands for Yang-Mills theory \cite{YMrec} in any number of dimensions, in the planar limit where integrands are well-defined. We have computed integrands determined by this recursion through a number of one- and two-loop examples, which have passed all obvious consistency checks \cite{YMrec}. Optimistically an all-loop recursion for non-supersymmetric Yang-Mills theory in the planar limit may be at hand, as a counterpart to the very different recursion for planar ${\cal N}=4$ SYM~\cite{Arkani-Hamed:2010zjl}.

  %\newpage
  \section{Leading Singularities}
\label{sec:LS}
We have given an integral representation for scalar-scaffolded gluon scattering amplitudes. In principle, with a correct contour prescription for integrating over the $y$ variables,  we have a ``curve integral'' representation for these amplitudes. Just as for Tr $\phi^3$ theory, at loop level we can also integrate over the loop momenta as Gaussian integrals, to represent the integrated amplitudes as an integral over $y$ space. We can also attempt to understand the field theory limit directly, which as we have indicated a number of times, is more interesting than the trivial tropicalization seen in Tr $\phi^3$ theory, due to the non-logarithmic character of the $dy/y^2$ measure. 

We will take up both these lines of investigation in future works, defining the contour for the amplitude~\cite{contour} as well as understanding a number of aspects of the field theory limit~\cite{cexpansion}. 
In this section we will instead content ourselves with performing the most basic but fundamental check that we are correctly describing gluon amplitudes at multi-loop level: we will show that our curve integral form correctly produces the leading singularities of gluon amplitudes in a number of non-trivial examples, from trees through to two loops.  

Unlike the computation of the full amplitude/integrand, which needs care in defining the contour in $y$ space, computing leading singularities is much simpler. We are interested in computing a leading singularity corresponding to gluing together three-particle amplitudes according to any particular trivalent diagram. 
Now, if we choose the underlying triangulation of the surface to be dual to the diagram of the leading singularity, it is easy to see that the residue on the pole setting all the $X \to 0$ for the glued internal lines, is computed simply by setting all these $X$'s to zero in the $y$ integrand and simply computing the {\bf residue} of the integrand around all the $y \to 0$. Since the form is a $dy/y^2$ form, this residue ultimately gives us simply the part of the expansion of the product of the $u_X^X$ for small $y$, that is linear in all the $y$'s, and so we need the expansion of each $u_X$ around $y \to 0$ up to first order in all the $y_i$.  

Note that leading singularities for the Tr $\phi^3$ theory are completely trivial--they are all equal to 1, since all we are doing in that case is taking residues of the $dy/y$ form at $y=0$ which sets the remaining $u \to 1$. This reflects the obvious fact that the Tr $\phi^3$ theory has only poles and no numerators. The leading singularities are (essentially by definition!) a clean probe of the ``purely numerator'' part of gluon amplitudes, and we see directly that this is associated with the $dy/y^2$ form. This already also alerts us to the possibility that curves with small self-intersection numbers can be important. Even though their $u$'s can never go to zero, expanding around $1$ they can have terms first-order in $y$. It is intuitive that curves with higher self-intersection will begin with higher powers of $y$ and so up to linear order we will only need curves with low self-intersection number. Indeed in a moment we will prove that curves with two or more self-intersections around a given puncture begin at quadratic order or higher in some $y$ variable and hence never contribute to leading singularities. But we also see that curves with up to one self-intersection around each loop puncture will indeed be needed to correctly match leading singularities. 

\subsection{Linearized $u$'s}

The computation of leading singularities needs an expansion for the $u$'s up to first order in the $y$'s. Quite beautifully, there is a direct expression for $u_X$ to first order in the $y$'s that we can simply read off by looking at the pattern of intersections of the curve $X$ and the base triangulation of the surface, which will greatly simplify our analysis. 
We are borrowing here from \cite{curvy} where this and more general results will be explained and derived.  
  \begin{figure}[t]
\begin{center}
\includegraphics[width=0.6\textwidth]{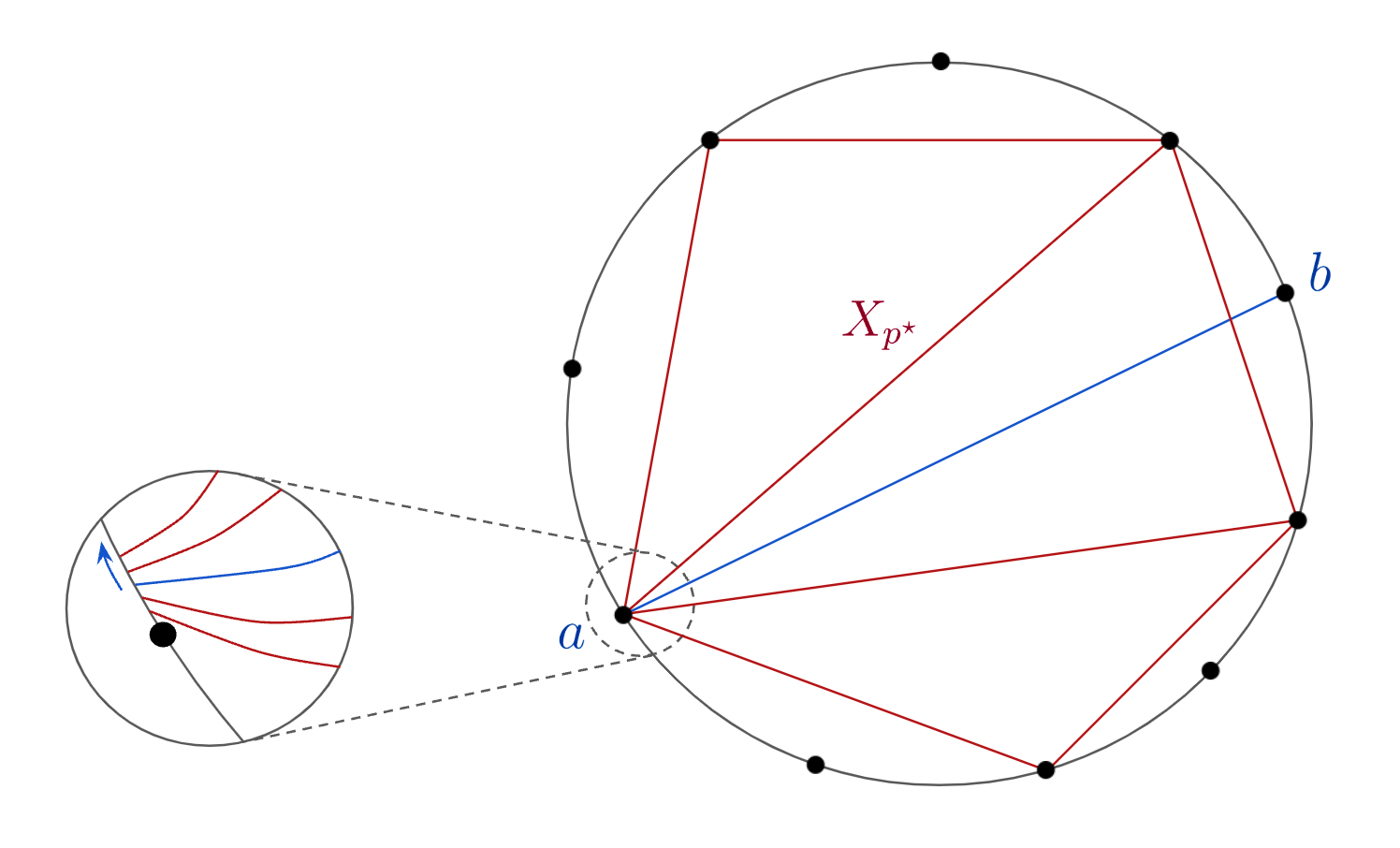}
\caption{Linearized $u$'s.}
\label{fig:LinUs}
\end{center}
\end{figure}
Let us start by picking the underlying triangulation, corresponding to the leading singularity we are interested in.  Now consider some curve, $X_{a,b}$, not in the triangulation, and its respective $u$ variable, $u_{X_{a,b}}$ (see figure \ref{fig:LinUs}).  
  
Let's denote by $\mathcal{I}_{a,b}$ the set of curves in the triangulation that $X_{a,b}$ intersects. Next, we look at each marked point $a$ and $b$, and check whether there are triangulation chords ending on these points. Let us say that this is the case for point $a$, and list these chords together with chord $X_{a,b}$ and twist them clockwise to obtain the respective laminations. Then let us list all these chords in clockwise order: $\{X_{p_1},\dots,X_{a,b}, X_{p^\star},\dots, X_{p_k}\}$ (see figure \ref{fig:LinUs}). We do the same for vertex $b$ and we get a set $\{X_{q_1},\dots,X_{a,b}, X_{q^\star},\dots, X_{q_l}\}$. Then the linearized $u_{a,b}$ is:
\begin{equation} \label{eq:linearu}
\begin{aligned}
	&u_{{a,b}}^{X_{a,b}} = 1 - X_{a,b} \left[\prod_{\mathcal{C} \in \mathcal{I}_{a,b}} y_C^{\text{Int}[\{\mathcal{C},X_{a,b}\}]} - \prod_{\mathcal{C} \in \mathcal{I}_{a,b}}y_C^{\text{Int}[\{\mathcal{C},X_{a,b}\}]} \left( y_{p^\star} + y_{p^\star} y_{p^\star_{+1}} +\dots +y_{p^\star} y_{p^\star_{+1}}\dots y_{p_k}
	\right) \right.\\
	& \left.\quad \quad - \prod_{\mathcal{C} \in \mathcal{I}_{a,b}}y_C^{\text{Int}[\{\mathcal{C},X_{a,b}\}]} \left( y_{q^\star} + y_{q^\star} y_{q^\star_{+1}} +\dots +y_{q^\star} y_{q^\star_{+1}}\dots y_{q_l}
	\right) \right.\\
	& \left.\quad \quad +\prod_{\mathcal{C} \in \mathcal{I}_{a,b}}y_C^{\text{Int}[\{\mathcal{C},X_{a,b}\}]} \left( y_{p^\star} + y_{p^\star} y_{p^\star_{+1}} +\dots +y_{p^\star} y_{p^\star_{+1}}\dots y_{p_k}
	\right)\left( y_{q^\star} + y_{q^\star} y_{q^\star_{+1}} +\dots +y_{q^\star} y_{q^\star_{+1}}\dots y_{q_l}
	\right) \right]\\
	& \quad \quad +\mathcal{O}^2(y).
\end{aligned}
\end{equation}

This makes manifest that any curve that intersects one of the triangulation chords more than once will \textbf{not} contribute to the leading singularity, as it will be automatically quadratic in that $y$. Thus as alluded to above, any curve that self-intersects twice or more going around any loop puncture will not contribute to leading singularities. But curves with up to one self-intersection per puncture can (and will) be relevant. 
  
\begin{figure}[t]
\begin{center}
\includegraphics[width=0.9\textwidth]{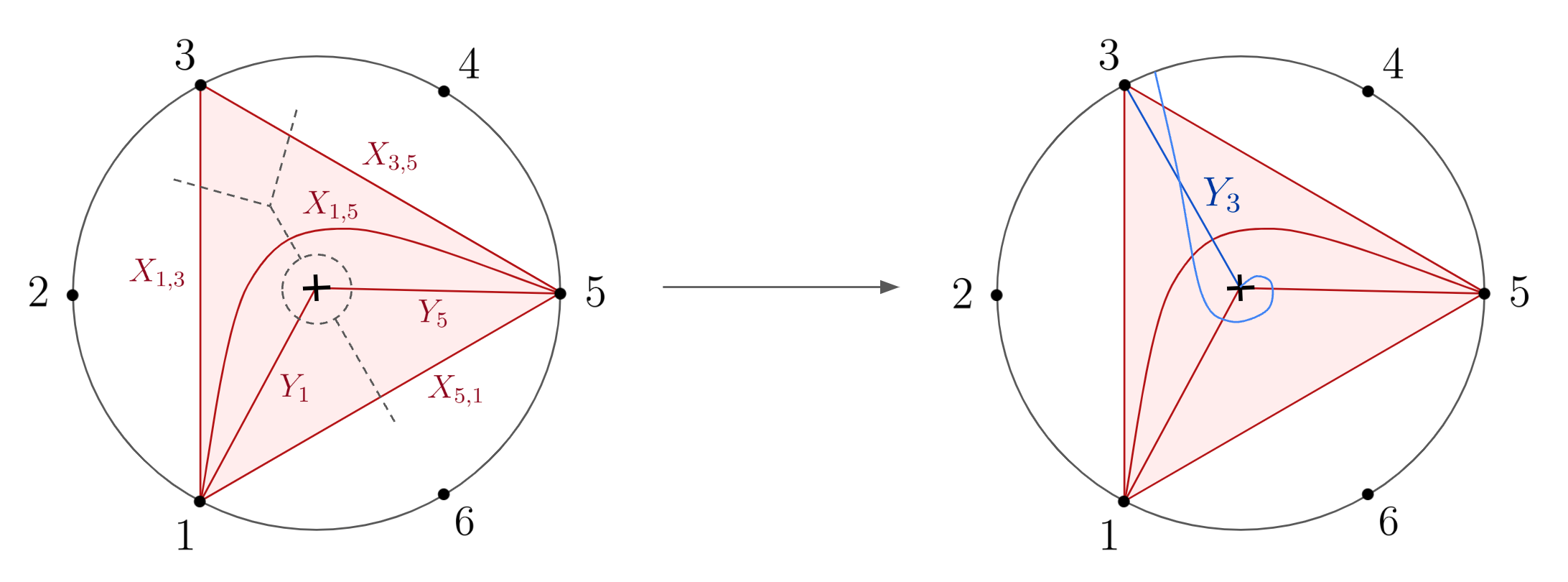}
\caption{Bubble triangulation and curve for linearized $u_{Y_3}$.}
\label{fig:1loop6pt}
\end{center}
\end{figure}

Let us now give a specific example for the case of a loop-level amplitude. We consider a 3-point 1-loop gluon amplitude, which is a 6-point 1-loop scalar amplitude. We are interested in the leading singularity in which one of the legs has a bubble (see figure \ref{fig:1loop6pt}), which corresponds to the triangulation of the punctured disk with chords $\{X_{1,3},X_{3,5},X_{5,1},Y_{1},Y_{5}\}$. Kinematically, $X_{1,5} = X_{5,1}$ but they are different curves on the surface, as we can see in figure \ref{fig:1loop6pt}. We want to find the linearized expression of $u_{Y_3}$ in terms of $\{y_{1,3},y_{3,5},y_{5,1},y_{1},y_{5}\}$. 
As a chord, $Y_3$ only intersects $X_{1,5}$. However, as a lamination, on the side of marked point 3, it intersects $X_{3,5}$; while on the puncture end, it intersects first $Y_1$ and then $Y_5$. Strictly speaking, as a lamination, $Y_3$ intersects $Y_1$ and $Y_5$ an infinite number of times, but since we are only interested in the expansion up to linear order, we can truncate it at intersection 1. This means that the linearized $u_{Y_3}$ is simply:
\begin{equation}
	(u_{Y_3})^{Y_3} = 1 - Y_3 \left[ y_{1,5} - y_{1,5} y_{3,5} - y_{1,5} \left( y_1 + y_1 y_5\right) + y_{1,5} y_{3,5}\left( y_1 + y_1 y_5\right) \right].
\end{equation}

\subsection{Tree level}
Let us study tree-level leading singularities in detail. It suffices to present explicit examples for 3- and 4-point cases, and the general rules for computing any tree-level leading singularities will become clear. The $n=3$ case is trivial in that after scaffolding no more residues are needed thus the only leading singularity corresponds to the amplitudes obtained in~\ref{sec:3pt}. However, we still go through it to illustrate how the same result can be obtained from the linearized $u$'s discussed above. Recall that we simply take $2n=6$ integral and take scaffolding residues in $\{X_{1,3},X_{3,5},X_{1,5}\}$ which already forms a triangulation. The linearized $u$'s we need are simply 
\begin{equation}
    u_{1,4}^{X_{1,4}}=1-X_{1,4}(y_{3,5}-y_{1,3} y_{3,5}),
\end{equation}
\begin{equation}
    u_{2,4}^{X_{2,4}}=1-X_{2,4}\; y_{1,3} y_{3,5},
\end{equation}
and their cyclic images $u_{2,5},u_{3,6}$, and $u_{4,6},u_{2,6}$ respectively. Note that for our purpose we have sent $X_{1,3},X_{3,5},X_{1,5}\to 0$ in these expressions. From here it is easy to read off the terms that are linear in all $y$'s at the leading $\alpha^\prime$ order, which gives the 3-point Yang-Mills amplitude or leading singularity:
\begin{equation}
  X_{1,4} X_{2,6}-X_{1,4} X_{2,5}+ {\rm cyclic},
\end{equation}
and of course in agreement with~\eqref{eq:3ptYM}. In addition, there is one more term that contributes to the next order in $\alpha^\prime$, $-X_{1,4} X_{2,5} X_{3,6}$, which is exactly the $F^3$ amplitude or leading singularity. 
\begin{figure}
    \centering
    \includegraphics[width=\linewidth]{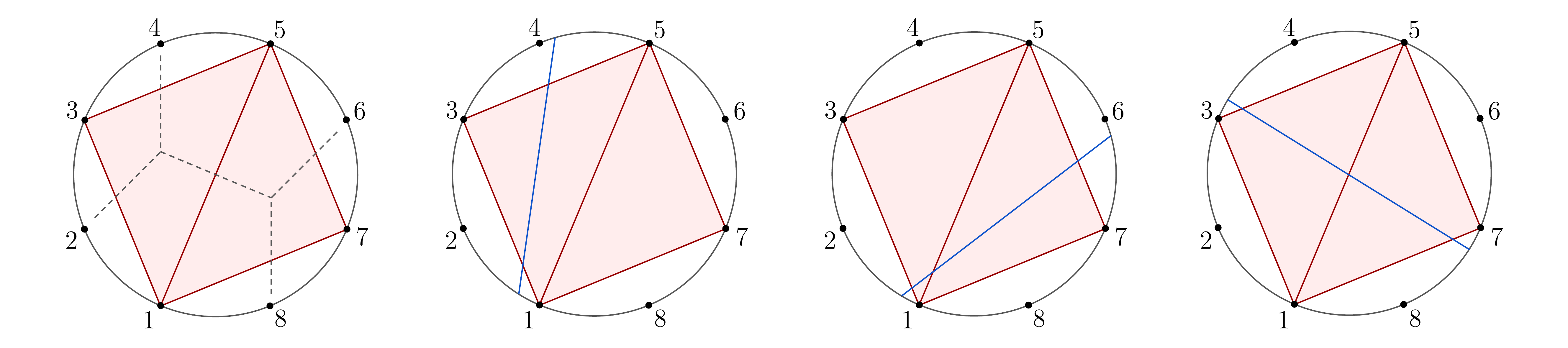}
    \caption{Dual diagram and linearized $u$'s for 4-point leading singularity.}
    \label{fig:LSu4pt}
\end{figure}

Moving on to 4 points and we consider the ``$s$-channel" leading singularity given by the triangulation $\{X_{1,3},X_{3,5},X_{5,7},X_{1,7}$ and $X_{1,5}\}$ (see first of figure~\ref{fig:LSu4pt}). We need to take residue in corresponding $y$'s, which corresponds to taking the coefficient of $y_{1,3} y_{3,5} y_{5,7} y_{1,7}$ of $\prod u^X$. Again it is straightforward to obtain the linearized $u$'s from the rules above, for example (see figure~\ref{fig:LSu4pt}):
\begin{equation}
    u_{1,4}^{X_{1,4}}=1-X_{1,4}(y_{3,5}-y_{1,3} y_{3,5}), 
\end{equation}
\begin{equation}
    u_{1,6}^{X_{1,6}}=1-X_{1,6}(y_{5,7}-y_{1,5} y_{5,7}-y_{1,3} y_{1,5} y_{5,7}), 
\end{equation}
\begin{equation}
    u_{3,7}^{X_{3,7}}=1-X_{3,7}y_{1,5} (-1+y_{1,7}) (-1+y_{3,5}).
\end{equation}

Now the leading singularity obtained from the coefficient above contains three different orders, ${\alpha'}^k$ for $k=3,4,5$, which correspond to gluing 2 three-point amplitudes with $0,1$ or $2$ of them being $F^3$ vertices, respectively. The leading pure YM contribution reads
\begin{equation}\label{eq-ls-4pt-ym}
\begin{aligned}
&c_{5,8} \left(X_{1,4} \left(X_{2,7}-X_{2,5}\right)+X_{2,5} X_{4,7}\right)-c_{1,4} \left(c_{5,8} X_{3,7}-X_{3,6} X_{5,8}+X_{1,6} \left(X_{5,8}-X_{3,8}\right)\right)\\
&-X_{1,4} X_{1,6} \left(X_{2,8}-X_{5,8}\right)-X_{1,4} X_{2,6} X_{5,8}+X_{2,5} \left(X_{1,4}-X_{4,6}\right) X_{5,8}+X_{1,6} X_{2,5} \left(X_{1,4}-X_{4,8}+X_{5,8}\right)\,,
\end{aligned}
\end{equation}
where we have kept the expression directly from the computation, which is written in terms of both $c$'s and $X$'s. The sub-leading contribution with one $F^3$ insertion reads
\begin{equation}
\begin{aligned}
    &X_{1,4} X_{1,6} X_{2,5} X_{3,8}+X_{1,4} X_{1,6} X_{2,7} X_{5,8}+X_{1,6} X_{2,4} X_{5,8} X_{3,7} \\
    -&X_{1,4} X_{1,6} X_{2,5} X_{3,7}-X_{1,4} X_{1,6} X_{2,5} X_{5,8}-X_{1,4} X_{2,5} X_{5,8} X_{3,7}  +(X_{i,j} \to X_{i+4,j+4}).
\end{aligned}
\end{equation}

Finally, the pure $F^3$ term is simply $-X_{1,4} X_{1,6} X_{2,5} X_{3,7} X_{5,8}$. 
\begin{figure}[t]
    \centering
\includegraphics[width=\linewidth]{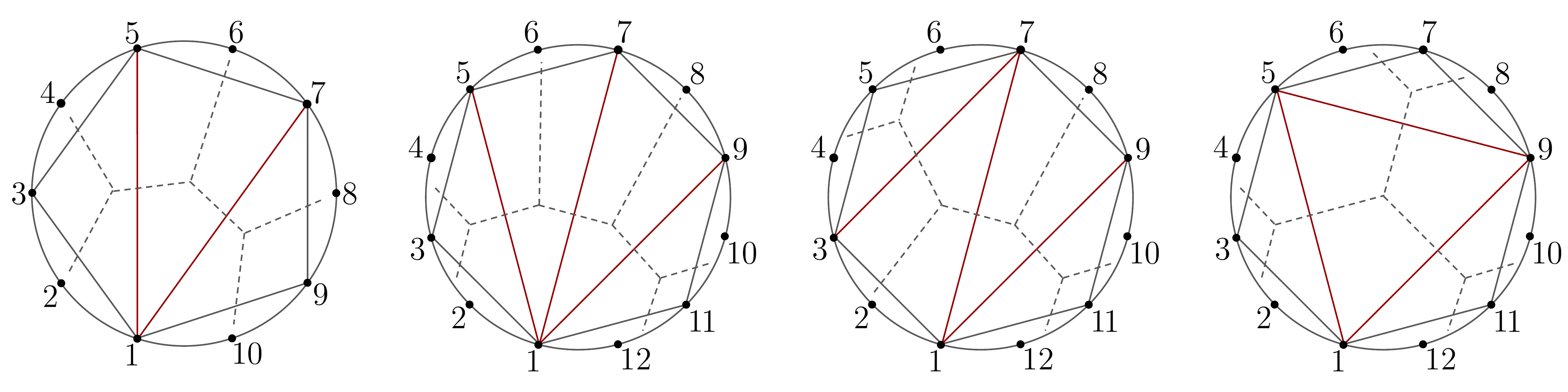}
    \caption{Dual diagrams for 5- and 6-point leading singularities.}
    \label{fig:Tree56ptF3}
\end{figure}
Remarkably, the pure $F^3$ contribution contains only one term in both cases we have considered above. In fact, this is true for any pure $F^3$ leading singularity, or equivalently the highest $\alpha'$ order, to all loops! For now, let us discuss tree-level pure $F^3$ leading singularities for any triangulation to all $n$. According to ~\eqref{eq:linearu}, each $X$ is multiplied with at least one $y$, thus the highest $\alpha^\prime$ order is given by a single term, namely the product of all $X$'s (with a minus sign) that intersects only one of those initial $X$'s. For example, for $n=5$ with scaffolding triangulation $\{X_{1,3},X_{3,5},\ldots,X_{9,1}\}$ and inner triangulation $\{X_{1,5},X_{1,7}\}$ (see the first graph in figure~\ref{fig:Tree56ptF3}), the term with only $F^3$ vertices is
\begin{equation}
   {\rm LS}_{5; X_{15}=X_{17}=0}^{F^3}=-X_{1,4} X_{1,6} X_{1,8} X_{2,5} X_{3,7} X_{5,9} X_{7,10}.
\end{equation}

For $n=6$ with inner triangulations $a=\{X_{1,5},X_{1,7},X_{1,9}\}$, $b=\{X_{3,7},X_{1,7},X_{1,9}\}$ and $c=\{X_{1,5},X_{5,9},X_{1,9}\}$ as presented in figure~\ref{fig:Tree56ptF3}, the LS are
\begin{equation}
\begin{aligned}
&{\rm LS}^{F^3}_{6; a}=-X_{1,4} X_{1,6} X_{1,8} X_{1,10} X_{2,5} X_{3,7} X_{5,9} X_{7,11} X_{9,12}, \\
&{\rm LS}^{F^3}_{6; b}=-X_{1,5} X_{1,8} X_{1,10} X_{2,7} X_{3,6} X_{3,9} X_{4,7} X_{7,11} X_{9,12},\\
&{\rm LS}^{F^3}_{6; c}=-X_{1,4} X_{1,7} X_{1,10} X_{2,5} X_{3,9} X_{5,8} X_{5,11} X_{6,9} X_{9,12}. 
\end{aligned}
\end{equation}

In general, the highest-order contribution of any $n$-point LS is given by a product of $2n{-}3$ $X$'s, which is consistent with the fact that it is obtained from gluing $n{-}2$ $F^3$ vertices. For lower order terms, we replace $k=1, 2, \cdots, n{-}2$ of $F^3$ vertices by three-point Yang-Mills ones, which have mass dimension of $X^{2n{-}3{-}k}$, thus for the lowest order, {\it i.e.} pure YM LS, the mass dimension is the expected $X^{n{-}1}$). Since YM $3$-point amplitude has more terms than $F^3$, the number of terms grows for lower order terms.
\begin{figure}[t]
    \centering
    \includegraphics[width=0.9\linewidth]{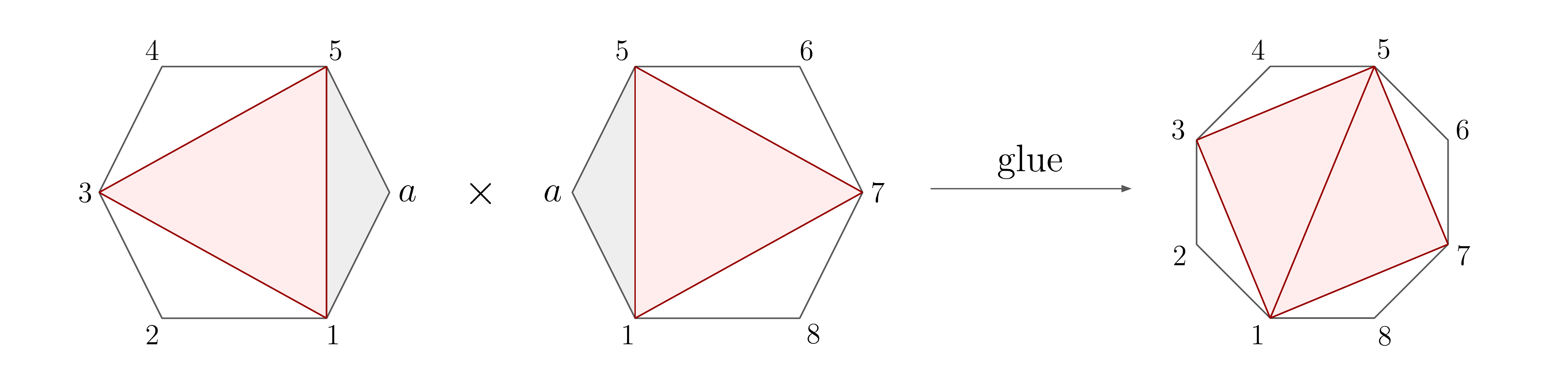}
    \caption{Gluing of 3-point amplitudes.}
    \label{fig:Xgluing}
\end{figure}

Now it is a remarkable fact that any such $(2n{-}3)$-fold residue, computed by extracting the coefficient of $\prod y$ in the expansion, agrees with gluing $n{-}2$ three-point amplitudes (which is the sum of both the YM and $F^3$ pieces) together! Of course, at tree-level this is guaranteed by the fact that the formula is computing $n$-gluon tree amplitude in the bosonic string which must contain all such LS. However, even without such a direct relation with bosonic string at loop level, as we will see shortly residues of loop-level stringy formulas continue to give correct loop-level LS!

Already at tree-level, these residues are compared with the actual leading singularities obtained from gluing three-point amplitudes using the factorization rule~\eqref{eq-fac}. For example, the four-point LS is given by the gluing of two three-point amplitudes, which is illustrated in figure~\ref{fig:Xgluing}. 
\begin{equation}
\begin{aligned}
     \mathrm{LS}^{\rm YM}_{L}(1,2,3,4,5,a)&=\sum_{j=2}^{4}(X_{j,a}-X_{j,5})Q^{L}_{j}=g_{2,a}(-X_{1,4})+g_{3,a}(c_{1,4})+g_{4,a}(-X_{2,5}),\\
      \mathrm{LS}^{\rm YM}_{R}(a,5,6,7,8,1)&=\sum_{J=6}^{8}(X_{a,J}-X_{1,J})Q^{R}_{J}=g_{a,6}(-X_{5,8})+g_{a,7}(c_{5,8})+g_{a,8}(-X_{1,6}),
\end{aligned}
\end{equation}
where $ \mathrm{LS}^{YM}_{L/R}$ denotes three-point YM amplitude/LS on the left/right side, and $g_{j,a}=(X_{j,a}-X_{j,5})$, $g_{a,J}=(X_{a,J}-X_{1,J})$.
The gluing rule directly formulated in $X$ variables is given by the factorization~\eqref{eq-fac}:
\begin{equation}
\begin{aligned}
    \sum_{j,J}(X_{j,a}-X_{j,5})Q^{L}_{j}(X_{a,J}-X_{1,J})Q^{R}_{J}\to  -\sum_{j,J}(X_{j,J}-X_{j,5}-X_{1,J})Q^{L}_{j}Q^{R}_{J},
\end{aligned}
\end{equation}
which is equal to $g_{j,a}g_{a,J}\to g_{j,J}\equiv (X_{j,5}+X_{1,J}-X_{j,J})$.

\begin{equation}
\begin{aligned}
    \mathrm{LS}^{\rm YM}_{4}&=\mathrm{LS}^{\rm YM}_{L}(1,\ldots,5,a) \otimes \mathrm{LS}^{\rm YM}_{R}(a,5,\ldots,1)\\
    &=g_{2,6}(X_{1,4}X_{5,8})+g_{3,6}(-c_{1,4}X_{5,8})+g_{4,6}(X_{2,5}X_{5,8})\\
    &+g_{2,7}(-X_{1,4}c_{5,8})+g_{3,7}(c_{1,4}c_{5,8})+g_{4,7}(-X_{2,5}c_{5,8})\\
    &+g_{2,8}(X_{1,4}X_{1,6})+g_{3,8}(-c_{1,4}X_{1,6})+g_{4,8}(X_{2,5}X_{1,6}),
\end{aligned}
\end{equation}
where of course the gluing result~$\mathrm{LS}^{\rm YM}_{4}$ agrees with the residue~\eqref{eq-ls-4pt-ym}, and $\otimes$ denotes the gluing rule $g_{j,a}g_{a,J}\to g_{j,J}$ .

Recall the $3$-point ${\rm F^3}$ amplitude obtained in section~\ref{sec:3pt} (for both L and R):
\begin{equation}
\begin{aligned}
     \mathrm{LS}^{\rm F^{3}}_{L}(1,2,3,4,5,a)&=g_{3,a}X_{1,4}X_{2,5}\,,\\
      \mathrm{LS}^{\rm F^{3}}_{R}(a,5,6,7,8,1)&=g_{a,7}X_{1,6}X_{5,8}\,.
\end{aligned}
\end{equation}

We can glue ${\rm F^3}$ amplitude with ${\rm YM}$ amplitude to give the $4$-point LS $\mathrm{LS}^{\rm YM+F^{3}}_{4}$, and LS $\mathrm{LS}^{\rm F^{3}}_{4}$ with two $F^3$ vertices:
\begin{equation}
\begin{aligned}
    \mathrm{LS}^{\rm YM+F^{3}}_{4}&= \mathrm{LS}^{\rm F^{3}}_{L}(1,\ldots,5,a) \otimes \mathrm{LS}^{\rm YM}_{R}(a,5,\ldots,1)+ \mathrm{LS}^{\rm YM}_{L}(1,\ldots,5,a)
      \otimes\mathrm{LS}^{\rm F^{3}}_{R}(a,5,\ldots,1)\\
      &=\left(g_{2,7}(-X_{1,4})+g_{3,7}(c_{1,4})+g_{4,7}(-X_{2,5})\right)X_{1,6}X_{5,8}\\
      &+\left(g_{3,6}(-X_{5,8})+g_{3,7}(c_{5,8})+g_{3,8}(-X_{1,6})\right)X_{1,4}X_{2,5},\\
      \\
    \mathrm{LS}^{\rm F^{3}}_{4}&= \mathrm{LS}^{\rm F^{3}}_{L}(1,\ldots,5,a) \otimes \mathrm{LS}^{\rm F^{3}}_{R}(a,5,\ldots,1)\\
      &=g_{3,7}X_{1,6}X_{5,8}X_{1,4}X_{2,5}=-X_{1,4} X_{1,6} X_{2,5} X_{3,7} X_{5,8}.
\end{aligned}
\end{equation}

It is straightforward to see that the gluing of $F^3$ vertices is particularly simple, always giving one term for any $n$ and agrees with the residue discussed above. 
\subsection{Loop level}

\begin{figure}
    \centering
    \includegraphics[width=\linewidth]{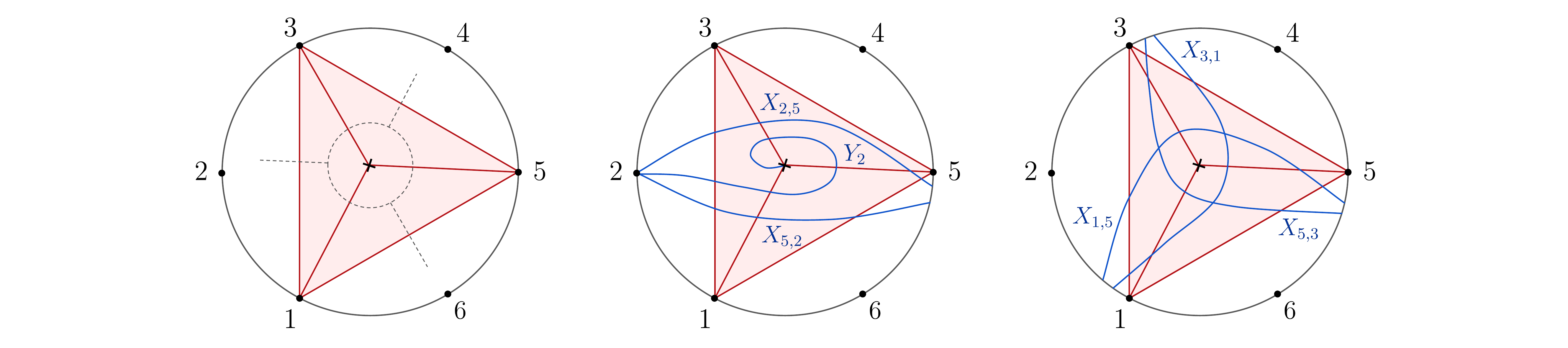}
    \caption{Dual diagram and linearized $u$'s for 1-loop triangle leading singularity.}
    \label{fig:LSutriangle}
\end{figure}

Now we present examples at loop level. Starting with one-loop three-point triangle with triangulation $\{X_{1,3},X_{3,5},X_{5,1},Y_1,Y_3,Y_5\}$, the linearized $u$'s are, for example, (see figure~\ref{fig:LSutriangle})
\begin{equation} \label{eq:triangleLSu1}
    u_{2,5}^{X_{2,5}}=1-X_{2,5} y_{1,3}y_3(1-y_{5}-y_{5}y_{1,5}),
\end{equation}
\begin{equation} \label{eq:triangleLSu2}
    u_{5,2}^{X_{5,2}}=1-X_{5,2} y_{1,3}y_1(1-y_{3,5}),
\end{equation}
\begin{equation} \label{eq:triangleLSu3}
    (u_{Y_2})^{Y_{2}}=1-Y_{2} y_{1,3}(1-y_{1}-y_{1}y_{5}-y_{1}y_{3}y_{5}).
\end{equation}

The remaining $u$'s that give non-vanishing contribution are cyclic of \eqref{eq:triangleLSu1}-\eqref{eq:triangleLSu3} and the one corresponds to the closed curve:
\begin{equation}
    (u_\Delta)^\Delta= 1- \Delta y_1 y_3 y_5.
\end{equation}

Then we can directly obtain the result:
\begin{equation}
    -2 Y_6 X_{1,4} X_{2,5}-2 Y_2 X_{1,4} X_{3,6}-2 Y_4 X_{2,5} X_{3,6}+2 X_{1,4} X_{2,5} X_{3,6}-(\Delta+1) Y_2 Y_4 Y_6.
\end{equation}

It is worth noting that the complete result for this case contains no contribution from the $F^3$ vertices. In fact, if we include the $u$'s corresponding to the curves $\{X_{3,1},X_{5,3},X_{1,5}\}$ (See right of figure~\ref{fig:LSutriangle}), we can compute the $F^3$ term
$Y_2 Y_4 Y_6  X_{3,1} X_{5,3} X_{1,5}$ but it vanishes
since we have set $\{X_{3,1},X_{5,3},X_{1,5}\} \to 0$ (on-shell). Similarly, all the terms with at least one $F^3$ vertex vanish. 

\begin{figure}[t]
    \centering
    \includegraphics[width=\linewidth]{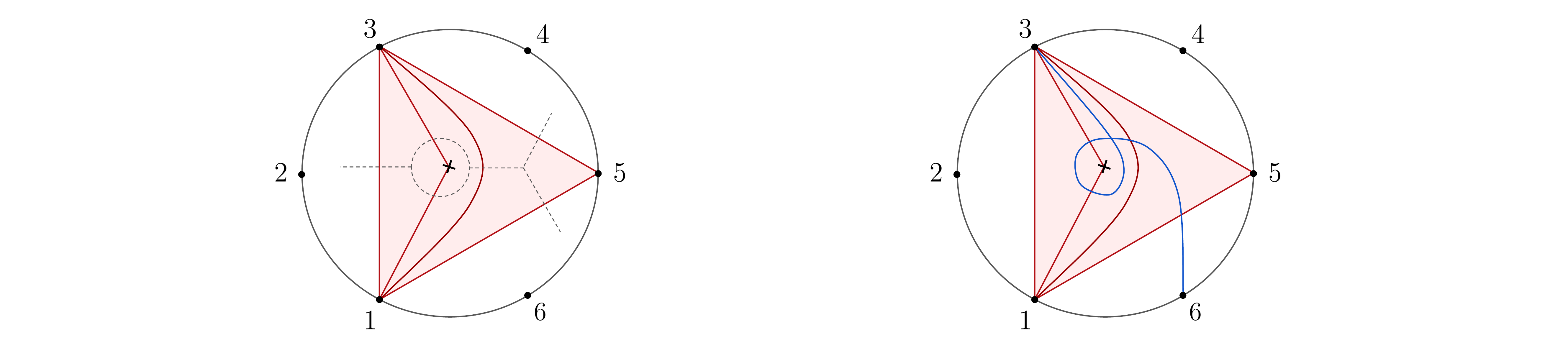}
     \caption{Dual diagram for 1-loop massless bubble and self-intersecting curve needed for the leading singularity computation.}
     \label{fig:LSu3ptbubble}
\end{figure}

For the 3-point massless bubble, as depicted in figure~\ref{fig:LSu3ptbubble} (left), it is crucial to include the curve with a single self-intersection (see figure~\ref{fig:LSu3ptbubble}, right), whose corresponding linearized $u$ reads:
\begin{equation}
    (u_{3,6}^{{\rm S.I.}})^{X_{3,6}}=1-X_{3,6} y_1 y_3 y_{1,3} y_{5,1}.
\end{equation}

Then one can compute the result which contains contributions with zero and one $F^3$ vertex to be
\begin{equation}
    (1+\Delta) Y_2 \left(Y_5 X_{1,4}-Y_6 X_{1,4}-Y_4 X_{3,6}+Y_5 X_{3,6}-Y_5 X_{4,6}+X_{3,6} X_{1,4}+Y_5 X_{3,6} X_{1,4}\right).
\end{equation}

Again curves with higher self-intersection are irrelevant as they begin at quadratic or higher order and don't contribute to the leading singularity. 

For the 4-point box, the result already contains 434 terms which we present in appendix~\ref{sec:boxLScomplete}, however, the highest $\alpha^\prime$ order is again given by a single term:
\begin{equation}
    Y_2 Y_4 Y_6 Y_8 X_{1,5}^2 X_{3,7}^2.
\end{equation}
In fact, using the rule we have previously introduced, we can obtain directly the contribution with only $F^3$ vertices for the 5- and 6-point penta- and hexa-gon:
\begin{equation}
\begin{aligned}
  & \underline{\text{Pentagon}}:  -Y_2 Y_4 Y_6 Y_8 Y_{10} X_{1,5} X_{1,7} X_{3,7} X_{3,9} X_{5,9},\\
  & \underline{\text{Hexagon}}: Y_2 Y_4 Y_6 Y_8 Y_{10} Y_{12} X_{1,5} X_{1,9} X_{3,7} X_{3,11} X_{5,9} X_{7,11}.
\end{aligned}
\end{equation}
\begin{figure}[t]
    \centering
\includegraphics[width=\linewidth]{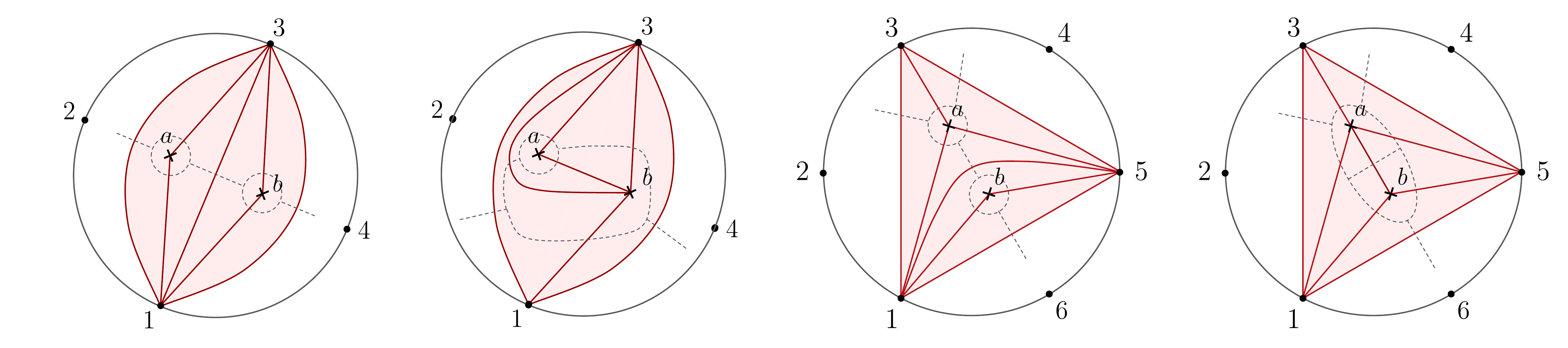}
    \caption{Dual diagrams of 2-loop 2- and 3-point leading singularities.}
    \label{fig:LS2loop}
\end{figure}

It is straightforward to generalize the computation to one-loop higher points, and in particular for $n$-gon LS, one can check that the result agrees with those written in terms of polarization and (external and loop) momenta~\cite{Edison:2022jln}.

Finally, let us present some examples for 2-loop 2- and 3-point. We attach the same $\Delta$ to all closed curves. For simplicity, we will focus on the leading $\alpha^\prime$ order. Now for the gluing of two 1-loop 2-point bubble (first graph in figure~\ref{fig:LS2loop}), the result reads:
\begin{equation}
    -(1+\Delta)^2 Y_{a,2} Y_{b,4} Y_{a,b}+(a \leftrightarrow b),
\end{equation}
where we have the second term since the integrand contains the contribution of the graph from symmetrization of loop punctures $a,b$.
For the second graph in figure~\ref{fig:LS2loop}, the result is 
\begin{equation}
    2(1+\Delta) X_{2,4} Y_{a,1}^2,
\end{equation}
where the factor of 2 comes from the symmetrization $(1 \leftrightarrow 3, a \leftrightarrow b)$ of the graph. For 3-point, the third graph reads:
\begin{equation}
    (1+\Delta) Y_{b,6} \left(-(1+\Delta) Y_{a,2} Y_{a,4} Y_{a,b}+2 (X_{1,4} X_{2,5} Y_{b,3}- X_{1,4} X_{2,5} Y_{a,b}- X_{1,4} Y_{a,2} Y_{b,3}- X_{2,5} Y_{a,4} Y_{b,3})\right),
\end{equation}
and the final graph is:
\begin{align}
    Y_{b,3}&\big(4 X_{1,4} X_{2,5} Y_{a,6}-4 X_{1,4} Y_{a,2} Y_{a,6}-4 X_{2,5} Y_{a,4} Y_{a,6}\\
    -&(1+\Delta) Y_{b,6} (X_{1,4} Y_{a,2}+X_{2,5} Y_{a,4}-X_{1,4} Y_{b,2}+X_{1,4} Y_{b,3}-X_{2,4} Y_{b,3}+X_{2,5} Y_{b,3}-X_{2,5} Y_{b,4})\big). \nonumber
\end{align}

It is important to include the curves that can self-intersect up to once around either of the punctures. Of course the expression again matches what we get from explicit gluing.

\label{sec:LeadSingLoop}

  %\newpage
  \section{Integrand Cuts: Tree-loop cut and loop-cut}
\label{sec:Loopcut}

We finally turn to the most non-trivial checks on our proposal -- the cuts of our integrand. In this section, we will study both the tree-like cut, where we go on residue at  $X_{i,j}=0$ ($i,j$ both odd), that we will call the tree-loop cut, and the loop-cut, given by the residue of our curve integral at loop level, on a ``single cut'' setting a single loop propagator to zero, $Y_i=0$ (i odd). In the first case, this should match the gluing of a tree process with a one-loop integrand, while in the second case, we should match the ``forward limit'' of an amplitude with one loop lower.
Now, in the later case, this match is usually fraught with subtlety in non-supersymmetric theories, due to the fact that both single-cuts and forward limits are naively infected with ``1/0'' divergences. But as we have hinted earlier, dealing with surface kinematics gives us a way out of this impasse, allowing a non-trivial match between ``loop cut'' and ``forward limit'' to be made.

In this section, we will start by deriving the gluing rules in momentum space, and then proceed to explain the generalization of them to surface kinematics. The later part is directly given to us by understanding the residues surface integral defining the surface integrands at low-energies. We will focus on studying both cuts at one-loop. 

We will start with the tree-loop cut where we will see that if we cut a tree-like propagator the final answer matches the generalized gluing of a tree and a loop-integrand, defined in terms of surface kinematics. Next, we will see that if we cut any loop propagator at one loop, it matches the tree-level amplitude in the forward limit, with the cut loop adding two color-adjacent particles,  ``glued'' by summing over polarization states. This will hold not only in the field theory limit but even at finite $\alpha^\prime$.

\subsection{Tree-loop Cut}
In this case, we are considering starting with a one-loop integrand, and cutting it via a given propagator $X_{a,b}$ with $a<b$ and $a,b$ both odd, so that the puncture is on the right of the curve $X_{a,b}$. In this limit, we expect to match the gluing of a tree amplitude, $\mathcal{A}_L(a,a+1,\cdots,b-1,b,x)$, and a loop-integrand on the right, $\mathcal{I}_R(b,b+1,\cdots,a-1,a,x^\prime)$. This is precisely the type of factorization already studied at tree-level in section \ref{sec:factTree}, with the difference that now the lower point object on the right is a loop-integrand instead of a tree amplitude. 

Recall from \ref{sec:factTree}, that in momentum space this cut was written as follows:
\begin{equation}    \mathcal{A}_{\text{glued}} 
     \propto \sum_{j,J} (-X_{b,J} -X_{a,j}+X_{j,J}) \partial_ {X_{x,j}}\mathcal{A}_L \partial_ {X_{x^\prime,J}} \mathcal{A}_R,
     \label{eq:factTree}
\end{equation}
where $j$ ranges over the indices entering $\mathcal{A}_L$ and $J$ those on $\mathcal{A}_R$. Therefore the only difference for the case in which $\mathcal{A}_R \equiv \mathcal{I}_R$ is that the sum over $J$ is now $J\in \{b+1,b+2,\cdots,a-1,p\}$, where $p$ is the puncture. 

In addition, in this formula all the $X$'s entering correspond to momentum invariants that, as explained previously, are read off by homology. This means that in this formula we have $X_{j,J} = X_{J,j}$, as well as the tadpole curves and external bubble that appear in the sum, such as $X_{a+1,a-1}$ and $X_{a,a}$, are set to zero. 

So our goal now is to derive an extension of \eqref{eq:factTree} that is defined for full surface kinematics described in \ref{sec:surfaceKin}. To do this we start with the definition of the surface integrand coming from the surface integral where we assign the kinematics to each curve $X_{\mathcal{C}}$ according to \ref{sec:surfaceKin}, and study what happens on the residue at $X_{a,b}=0$.

\subsubsection{Tree-loop cut from surface integral}

We know that taking the residue at $X_{a,b}=0$ of the full integrands, corresponds to taking the residue of the form in the surface integral at $y_{a,b}=0$, provided we picked an underlying fatgraph containing propagator $X_{a,b}$. So we want to extract the following residue
\begin{equation}
    \text{Res}_{y_{a,b}=0} \left(\frac{1}{y_{a,b}^2} \prod_{\mathcal{C}\in\mathcal{S}} u_\mathcal{C}^{X_{\mathcal{C}}} \right),
\end{equation}
where we are omitting the remaining integration $y$'s as well as the scaffolding residue, required to get the gluon amplitude.

Now the $u$-variables that have a non-zero derivative with respect to $y_{a,b}$, and thus contribute to this residue, correspond to those associated to curves that cross $X_{a,b}$ once or intersect it as lamination. If we read off the kinematics $X_{\mathcal{C}}$ by homology, than these curves are precisely those appearing in the prefactor in \eqref{eq:factTree}. However if we are keeping all the curves, we should be more careful in understanding these residue. 

Let's start by listing all the curves that will contribute to the residue. Of course, as before we have the contribution from curves from $j$ to $J$, where now we distinguish, $u_{j,J}^{(q)}$ and $u_{J,j}^{(q)}$, with $q$ labeling the self-intersection which can be either zero or one, $q=0,1$, and $j\in \textbf{L}=\{a+1,\cdots,b-1\}$ and $J\in \textbf{R}=\{b+1, \cdots, a-1\}$. Now in addition to $u^{(q)}_{a,J}$ and $u^{(q)}_{J,a}$, we have curves $u^{(1)}_{a,j}$ as well as $u^{(q)}_{j,a}$. Similarly, we have curves $u^{(q)}_{j,b}$ and  $u^{(q)}_{b,j}$, as well as $u^{(q)}_{b,a}$. Finally, we have also contribution coming from $u_{a,a}^{(q)}$.

Let's now go through each of these separately and derive the contribution they give on the residue, $X_{a,b}=0$. Starting with $u_{j,J}^{(q)}$ and $u_{J,j}^{(q)}$, it is trivial to show (as shown in an appendix) that in the locus where $y_{a,b}=0$ we have:

\begin{equation}
\begin{aligned}
   & \frac{\partial \log{u_{j,J}^{(q)}}}{\partial y_{a,b}} \bigg \vert_{y_{a,b} =0} = \frac{\partial \log{u_{j,x}}}{\partial y_{a,b}} \bigg \vert_{y_{a,b} =0} \times \frac{\partial \log{u_{x^\prime,J}^{(q)}}}{\partial y_{a,b}} \bigg \vert_{y_{a,b} =0}, \\
    & \frac{\partial \log{u_{J,j}^{(q)}}}{\partial y_{a,b}} \bigg \vert_{y_{a,b} =0} = \frac{\partial \log{u_{j,x}}}{\partial y_{a,b}} \bigg \vert_{y_{a,b} =0} \times \frac{\partial \log{u_{J,x^\prime}^{(q)}}}{\partial y_{a,b}} \bigg \vert_{y_{a,b} =0} .
\end{aligned}
\end{equation}

Now inside the surface integral, these derivatives precisely allow us to interpret the result on the residue in terms of the lower-objects as we can write them as $\partial_{X_{j,x}} \mathcal{A}_L \times \partial_{X_{x^\prime,J}} \mathcal{I}_R$. Therefore the contribution to the residue coming from these curves is simply
\begin{equation}
    \sum_{j,J} X_{j,J} \frac{\partial \mathcal{A}_L}{\partial X_{j,x}} \times \frac{\partial \mathcal{I}_R}{\partial X_{x^\prime,J}} + X_{J,j} \frac{\partial \mathcal{A}_L}{\partial X_{j,x}} \times \frac{\partial \mathcal{I}_R}{\partial X_{J,x^\prime}} + Y_j\frac{\partial \mathcal{A}_L}{\partial X_{j,x}} \times \frac{\partial \mathcal{I}_R}{\partial Y_{x^\prime}},
\end{equation}
where we have added explicitly the case in which $J=p$. Now the remaining curves that cross $X_{a,b}$ are then: $u_{j,b}^{(1)},u_{b,j}^{(q)}$ as well as $u_{j,a}^{(q)},u_{a,j}^{(1)}$, for which we can similarly derive:
\begin{equation}
\begin{aligned}
       & \frac{\partial \log{u_{j,b}^{(1)}}}{\partial y_{a,b}} \bigg \vert_{y_{a,b} =0} = \frac{\partial \log{u_{j,x}}}{\partial y_{a,b}} \bigg \vert_{y_{a,b} =0} \times \frac{\partial \log{u_{x^\prime,b}^{(1)}}}{\partial y_{a,b}} \bigg \vert_{y_{a,b} =0} 
 \quad \Rightarrow \quad X_{j,b} \frac{\partial \mathcal{A}_L}{\partial X_{j,x}} \times \frac{\partial \mathcal{I}_R}{\partial X_{x^\prime,b}}, \\
    & \frac{\partial \log{u_{b,j}^{(q)}}}{\partial y_{a,b}} \bigg \vert_{y_{a,b} =0} = \frac{\partial \log{u_{j,x}}}{\partial y_{a,b}} \bigg \vert_{y_{a,b} =0} \times \frac{\partial \log{u_{b,x^\prime}^{(q)}}}{\partial y_{a,b}} \bigg \vert_{y_{a,b} =0} \quad \Rightarrow \quad X_{b,j} \frac{\partial \mathcal{A}_L}{\partial X_{j,x}} \times \frac{\partial \mathcal{I}_R}{\partial X_{b,x^\prime}}, \\
    & \frac{\partial \log{u_{j,a}^{(q)}}}{\partial y_{a,b}} \bigg \vert_{y_{a,b} =0} = \frac{\partial \log{u_{j,x}}}{\partial y_{a,b}} \bigg \vert_{y_{a,b} =0} \times \frac{\partial \log{u_{x^\prime,a}^{(q)}}}{\partial y_{a,b}} \bigg \vert_{y_{a,b} =0} \quad \Rightarrow \quad X_{j,a} \frac{\partial \mathcal{A}_L}{\partial X_{j,x}} \times \frac{\partial \mathcal{I}_R}{\partial X_{x^\prime,a}},\\
    & \frac{\partial \log{u_{a,j}^{(1)}}}{\partial y_{a,b}} \bigg \vert_{y_{a,b} =0} = \frac{\partial \log{u_{j,x}}}{\partial y_{a,b}} \bigg \vert_{y_{a,b} =0} \times \frac{\partial \log{u_{a,x^\prime}^{(1)}}}{\partial y_{a,b}} \bigg \vert_{y_{a,b} =0} \quad \Rightarrow \quad X_{a,j} \frac{\partial \mathcal{A}_L}{\partial X_{j,x}} \times \frac{\partial \mathcal{I}_R}{\partial X_{a,x^\prime}},
\end{aligned}
\label{eq:tree-loopCutCurves}
\end{equation}
for any $j \in \textbf{L}$. Note that even though $u^{(0)}_{x^\prime,b}$ (similarly for $u^{(0)}_{a,x^\prime}$)  corresponds to a purely boundary curve $u^{(1)}_{x^\prime,b}$ does not, and therefore by keeping curves $X_{x^\prime,b}$ non-zero in the integrand, we can once more interpret the result on the residue as a derivative of the lower-point object. The contributions to the cut of each of these curves are presented on the right in \eqref{eq:tree-loopCutCurves}. 

Now let's proceed to study the curves that intersect $X_{a,b}$ as \textit{laminations}, and as shown in the appendix, in this case the derivative of the $u$-variables still factorize but in a different way. 

To begin we have curves $u^{(0)}_{j,b}$ for all $j\in \textbf{L} \backslash \{b-1\}$, from which we get:
\begin{equation}
     \frac{\partial \log{u_{j,b}^{(0)}}}{\partial y_{a,b}} \bigg \vert_{y_{a,b} =0} = \frac{\partial \log{u_{j,x}}}{\partial y_{a,b}} \bigg \vert_{y_{a,b} =0} \times -\frac{\partial \log{(1-u_{a,b})}}{\partial y_{a,b}} \bigg \vert_{y_{a,b} =0} ,
\end{equation}
from which we get the following contribution to the cut
\begin{equation}
    -X_{j,b} \frac{\partial \mathcal{A}_L}{\partial X_{j,x}}\times \left[ \sum_{J\in \textbf{R}\cup \{a,b\} } \left( \frac{\partial \mathcal{I}_R}{\partial X_{x^\prime,J}} +   \frac{\partial \mathcal{I}_R}{\partial X_{J,x^\prime}}\right) +   \frac{\partial \mathcal{I}_R}{\partial Y_{x^\prime}}\right],
\end{equation}
for all $j \in \textbf{L}\backslash\{b-1\}$. In addition, curves $u^{(q)}_{b,a}$, $u_{a,p}$, $u_{a,p}$ and $u^{(q)}_{a,a}$ also laminate the puncture giving us the following contributions on the cut:
\begin{equation}
    -X_{b,a}  \sum_{j\in \textbf{L}}  \frac{\partial \mathcal{A}_L}{\partial X_{j,x}}  \times \frac{\partial \mathcal{I}_R}{\partial X_{b,x^\prime}} - Y_a\sum_{j\in \textbf{L}}  \frac{\partial \mathcal{A}_L}{\partial X_{j,x}}  \times \frac{\partial \mathcal{I}_R}{\partial Y_{x^\prime}} - X_{a,a} \sum_{j\in \textbf{L}}  \frac{\partial \mathcal{A}_L}{\partial X_{j,x}} \times \left(\frac{\partial \mathcal{I}_R}{\partial X_{x^\prime,a}} + \frac{\partial \mathcal{I}_R}{\partial X_{a,x^\prime}}  \right) ,
\end{equation}
and finally we also have curves $u^{(q)}_{a,J}$ for all $J \in \textbf{R}$ as well as $u^{(q)}_{J,a}$ for $J \in \textbf{R} \backslash \{a-1\}$, since for $J=a-1$, we only have $u^{(1)}_{a-1,a}$ since $u^{(0)}_{a-1,a}$ is a boundary. We will see this last case is a bit more subtle, but for the remaining curves, just like before we get, in the residue
\begin{equation}
    - \sum_{J \in \textbf{R}} X_{a,J}  \sum_{j\in \textbf{L}}  \frac{\partial \mathcal{A}_L}{\partial X_{j,x}}  \times \frac{\partial \mathcal{I}_R}{\partial X_{x^\prime,J}}- \sum_{J \in \textbf{R}\backslash \{a-1\}} X_{J,a}  \sum_{j\in \textbf{L}}  \frac{\partial \mathcal{A}_L}{\partial X_{j,x}}  \times \frac{\partial \mathcal{I}_R}{\partial X_{J,x^\prime}} .
\end{equation}

Now let's look at the contribution of curve $u^{(1)}_{a-1,a}$, just like for the remaining curves laminating curves, the dlog also factorizes to give:
\begin{equation}
    \frac{\partial \log{u_{a-1,a}^{(1)}}}{\partial y_{a,b}} \bigg \vert_{y_{a,b} =0} = -\frac{\partial \log{(1-u_{a,b})}}{\partial y_{a,b}} \bigg \vert_{y_{a,b} =0} \times \frac{\partial \log{u_{a-1,x^\prime}^{(1)}}}{\partial y_{a,b}} \bigg \vert_{y_{a,b} =0} ,
\end{equation}
and once inside the surface integral the first term will produce the $-\sum_{j\in \textbf{L}}  \partial \mathcal{A}_L/\partial X_{j,x} $, also obtained in the previous cases, however now the second term in the product is \textit{not} $\partial \mathcal{I}_R/\partial X_{a-1,x^\prime}$. This would only be true if we had the dlog of the product of $u_{a-1,x^\prime}^{(0)}u_{a-1,x^\prime}^{(1)}$, which both have the same exponent, $X_{a-1,x^\prime}$ in the surface integral, this is:
\begin{equation}
\begin{aligned}
    \frac{\partial \mathcal{I}_R} {\partial X_{a-1,x^\prime}} &= \int_{\mathcal{S}_{\mathcal{I}_R}} \prod_{P\neq (a,b)} \frac{\diff y_P}{y_P^2} \left[\left(\frac{\log{u_{a-1,x^\prime}^{(0)}}}{\partial y_{a,b}}  + \frac{\log{u_{a-1,x^\prime}^{(1)}}}{\partial y_{a,b}}  \right) \prod_{\mathcal{C}\in \mathcal{S}_{\mathcal{I}_R}} u_\mathcal{C} ^{X_\mathcal{C}} \right]\bigg \vert_{y_{a,b} =0} \\
    &= \mathcal{Q}_{a-1}^{(0)} +  \mathcal{Q}_{a-1}^{(1)}
\end{aligned}
\end{equation}
and for the tree-loop cut, we only want the second term in the sum on the r.h.s. This term in precisely the counterpart of the correction term we derived in the gauge-invariance statement for the surface integrand in \ref{sec:gauginvLoop}. So from equation \eqref{eq:GaugeInvLoop}, we have that
\begin{equation}
    \mathcal{Q}_{a-1}^{(1)} = -\frac{\partial   \mathcal{I}_R}{\partial X_{a-1,a}} \bigg \vert_{x^\prime \rightarrow a},
\end{equation}
and thus the contribution from curve $u_{a-1,a}^{(1)}$ to the cut is:
\begin{equation}
    +X_{a-1,a}  \sum_{j\in \textbf{L}}  \frac{\partial \mathcal{A}_L}{\partial X_{j,x}}  \times \frac{\partial \mathcal{I}_R}{\partial X_{a-1,a}}\bigg \vert_{x^\prime \rightarrow a}.
\end{equation}

In summary, collecting all the different contributions derived above, we can write the tree-loop cut as follows:
\begin{equation}
\begin{aligned}    \mathop{\mathrm{Res}}_{X_{a,b}=0}\left(\Omega_{2n}^{\text{loop}}\right) &= \sum_{j \in L, J\in R} \left[\left(X_{j,J}-X_{j,b}\hat{\delta}_{j,b-1} - X_{a,J}\right) \frac{\partial \mathcal{A}_L^\mathcal{S}}{\partial X_{j,x}}\frac{\partial \mathcal{I}_R^\mathcal{S}}{\partial X_{x^\prime,J}} \right. \\
& \quad \quad \quad \quad  \left. + \left(X_{J,j}-X_{j,b}\hat{\delta}_{j,b-1}  - X_{J,a}\hat{\delta}_{J,a-1} \right) \frac{\partial \mathcal{A}_L^\mathcal{S}}{\partial X_{j,x}}\frac{\partial \mathcal{I}_R^\mathcal{S}}{\partial X_{J,x^\prime}}\right]\\
&\quad + \sum_{j\in L} \left[\left(X_{j,p} - X_{a,p} -X_{j,b}\hat{\delta}_{j,b-1}\right)   \frac{\partial \mathcal{A}_L^\mathcal{S}}{\partial X_{j,x}}\frac{\partial \mathcal{I}_R^\mathcal{S}}{\partial X_{x^\prime,p}}\right] \\
& \quad + \sum_{j\in L} X_{a-1,a} \frac{\partial \mathcal{A}_L^\mathcal{S}}{\partial X_{j,x}}\frac{\partial \mathcal{I}_R^\mathcal{S}}{\partial X_{a-1,a}} \bigg\vert_{x^\prime \rightarrow a},
\end{aligned}  
\label{eq:tree-loop}
\end{equation}
with $\hat{\delta}_{i,k} = 1-\delta_{i,k}$.

\subsection{Loop-Cut}
\begin{figure}[t]
\begin{center}
\includegraphics[width=0.9\textwidth]{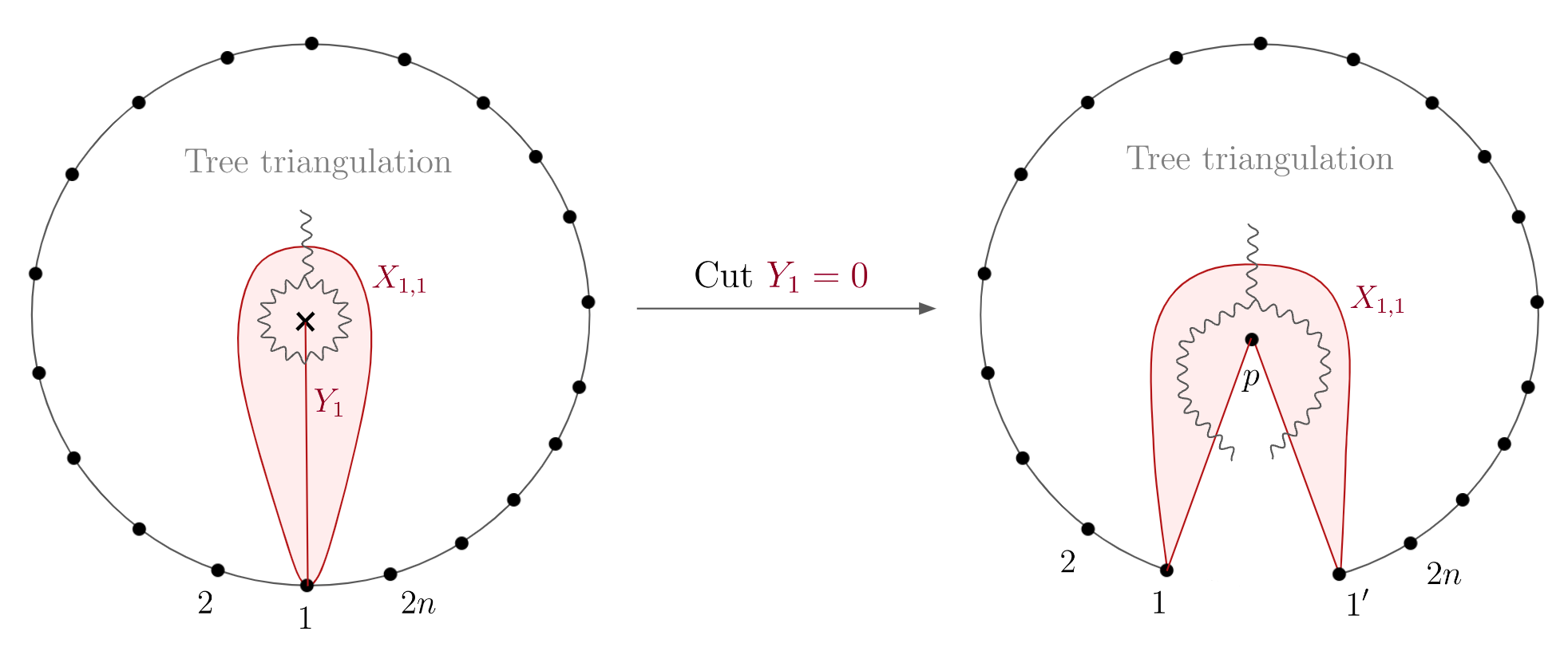}
\caption{Tadpole Triangulation with $Y_{1}$ and $X_{1,1,}$ and loop cut.}
\label{fig:LoopCut}
\end{center}
\end{figure}
On the loop side, we will be considering a $2n$-scalar 1-loop amplitude, from which we get the $n$-point 1-loop gluon amplitude. To define the corresponding surface, we will find it convenient to pick a particular ``tadpole'' triangulation, whose fat-graph includes the loop propagator $Y_1$ as well as the tadpole propagator $X_{1,1}$ (see figure \ref{fig:LoopCut}). Taking the loop cut corresponds to extracting the residue when $Y_1=0$, which is easy to do in this triangulation since $Y_1$ is one of the chords in the triangulation. The resulting tree-level result corresponds to a $(2n+2)$-point amplitude, with the two extra gluon legs glued. We call these extra two legs $1^\prime$ and $p$ (see figure \ref{fig:LoopCut}). 
Regardless of $Y_1=0$ we still keep $X_{1,1}$ as the label of the curve on the cut surface. 

\begin{figure}[t]
\begin{center}
\includegraphics[width=0.9\textwidth]{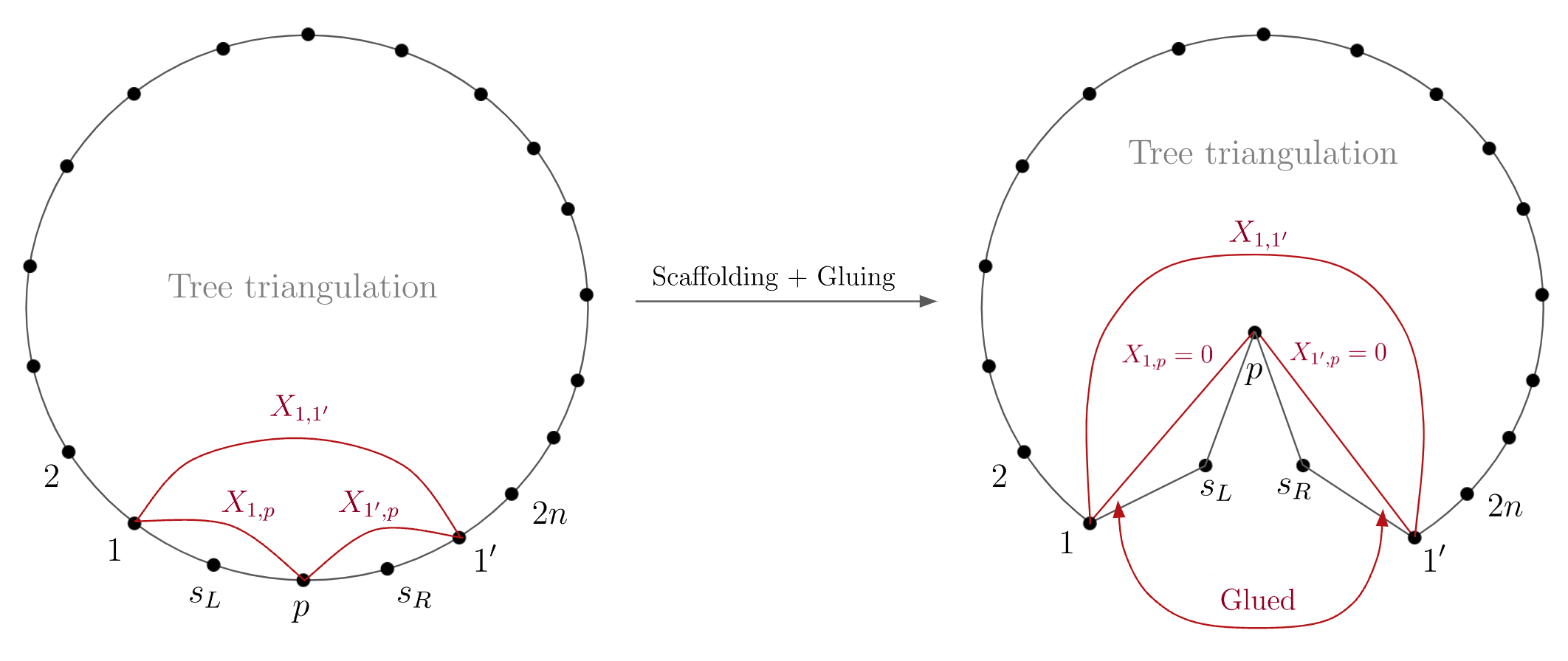}
\caption{Tree triangulation and scaffolding residue.}
\label{fig:TreeGlue}
\end{center}
\end{figure}

\begin{figure}[t]
\begin{center}
\includegraphics[width=\textwidth]{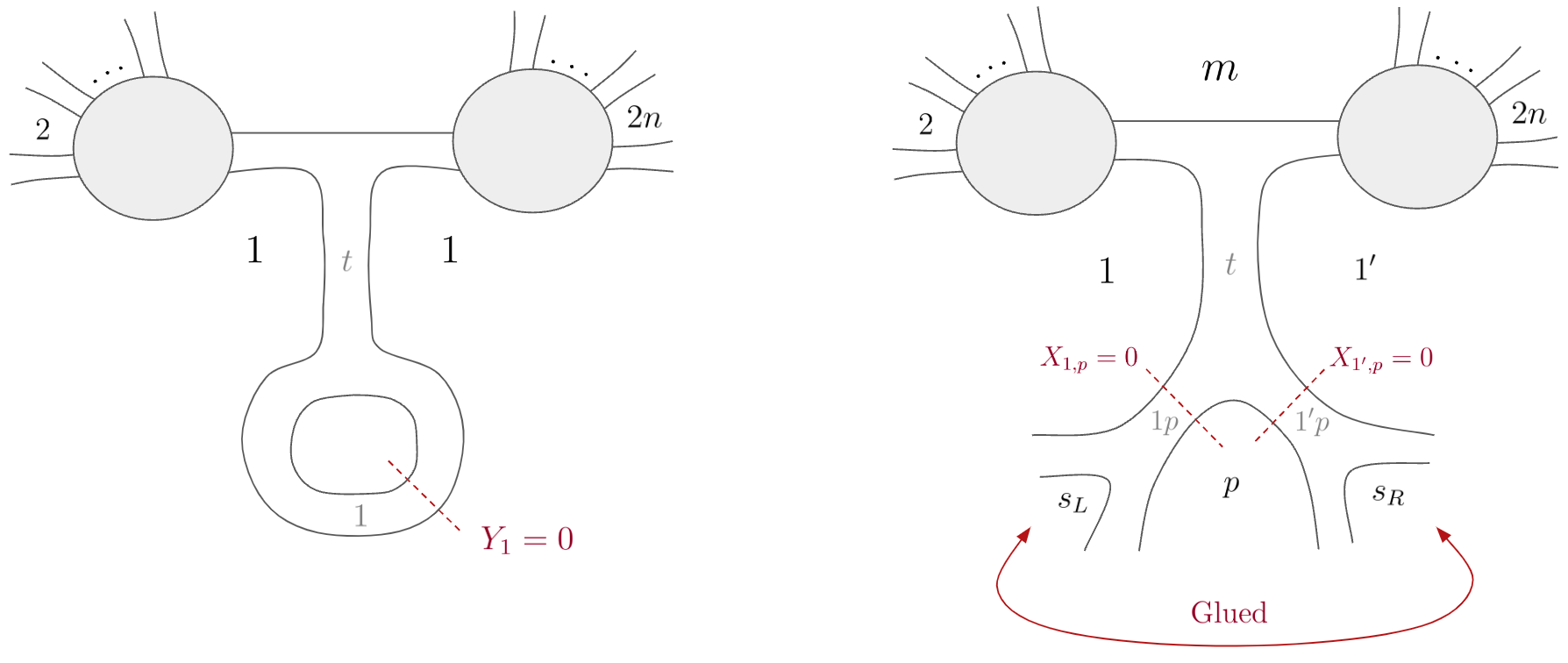}
\caption{1-loop fat graph (left) and tree-level fat graph (right).}
\label{fig:FatGraphs}
\end{center}
\end{figure}

To do the matching, we consider the tree-level amplitude of $2n+4$ scalars, where we label the particles as $(1,2,...,2n,1^\prime,s_R,p,s_L)$ so that the matching with the loop computation becomes clearer.
We make the gluons, $1^\prime$ and $p$, on-shell by taking the scaffolding residue, $X_{1,p}=X_{1^\prime,p}=0$. Therefore we pick a triangulation to define the surface that contains propagators $\{X_{1,p},X_{1,p^\prime},X_{1,1^\prime}\}$, the last propagator is chosen to replicate the triangulation picked in the loop cut part (see figure \ref{fig:TreeGlue}). 
The corresponding fat graphs associated with the 1-loop and tree-level triangulations are presented in figure \ref{fig:FatGraphs}. 

\subsection{Gluing in Momentum Space}

We will now describe the procedure for performing the polarization sum (``gluing'') for the two gluons in the forward limit opened up by cutting a loop. This will be a bit more interesting than the gluing we already encountered in our discussion of tree-level factorization, where we took two separate gauge-invariant objects ${\cal A}_L^\mu(p, \cdots), {\cal A}_R^\nu(-p,\cdots)$ and performed the polarization sum over the intermediate particle which finally yielded (thanks to the gauge-invariance of ${\cal A}_L,{\cal A}_R$) the trivial gluing rule ${\cal A}_L^\mu {\cal A}_R^\nu \eta_{\mu\nu}$. Now instead here we wish to start with a single $n$-point tree-level amplitude, and glue together two of its adjacent legs, say $1$ and $n$. The polarization sum is:
\begin{equation}
	{\cal A}_{\text{glued}}= {\cal A}_n^{\mu \nu} \left( -\eta_{\mu,\nu} + \frac{p_\mu q_\nu + p_\nu q_\mu}{p \cdot q} \right), \quad \text{with} \quad {\cal A}_n^{\mu \nu}= \partial_{\epsilon_{1,\mu}} \partial_{\epsilon_{n,\nu}} {\cal A}_n.
\end{equation} 

Now, there is a crucial difference with the tree-like gluing case. Gauge invariance does {\bf not} imply that ${\cal A}_n^{\mu \nu} p_\mu$ vanishes, and in general for non-Abelian gauge theories we have 
\begin{equation}
	{\cal A}_n^{\mu \nu}p_{\mu} \neq 0.
\end{equation}

This is famously related to the need to introduce ghost fields in the Lagrangian at loop level. However all is not lost, gauge invariance still constrains the form of this object, since we must have:
\begin{equation}
	{\cal A}_n^{\mu \nu}p_{\mu} p_{\nu} = 0 \quad \Rightarrow \quad {\cal A}_n^{\mu \nu}p_{\mu} = \mathcal{N} \, p^{\nu},
\end{equation}
where $\mathcal{N}$ is some function of the remaining kinematics and polarizations. Using this fact, the result after gluing can be written as:
\begin{equation}
	{\cal A}_{\text{glued}} = {\cal A}_n^{\mu \nu} \left( -\eta_{\mu,\nu} + \frac{p_\mu q_\nu + p_\nu q_\mu}{p \cdot q} \right)=- {\cal A}_n^{\mu \nu}\eta_{\mu,\nu} + 2 \mathcal{N},
\end{equation}
where the $q^{\mu}$ dependence dropped out again, as expected. Therefore, to perform the gluing in the forward limit, we need to determine $\mathcal{N}$. 

Let us now rephrase some of these statements in scalar language using the $X_{i,j}$'s from the scaffolded gluon amplitude. Again, let us say we start with an $n$-point tree-level gluon amplitude, so $2n$ scalar amplitude, and we want to glue gluons $1$ and $n$. Then the $X$'s associated with the polarizations of these gluons are, respectively, $X_{2}^{\mu}$ and $X_{2n}^{\nu}$, so we have:
\begin{equation}
	{\cal A}_n^{\mu \nu} = \frac{\partial^2{\cal A}_n}{\partial X_2^\mu \partial X_{2n}^\nu},
\end{equation}
where ${\cal A}_n$ stands for the amplitude obtained after taking the scaffolding residue in gluons $1$ and $n$. Now, gauge invariance and multi-linearity in particles $1$ and $n$, we know we can write:
\begin{equation}
	{\cal A}_n = \sum_j (X_{2,j} - X_{1,j}) \mathcal{Q}^2_j = \sum_j (X_{2n,j} - X_{1,j}) \mathcal{Q}^{2n}_j.
	\label{eq:glueginv}
\end{equation}

Now by linearity in the polarizations the $\mathcal{Q}^2_j$ cannot depend in $X_{2,k}$ for any $k$, and similarly for $\mathcal{Q}^{2n}_j$. Therefore we can rewrite \eqref{eq:glueginv} as:
\begin{equation}
	{\cal A}_n = \sum_j (X_{2,j} - X_{1,j}) \frac{\partial {\cal A}_n}{\partial X_{2,j}} = \sum_j (X_{2n,j} - X_{1,j}) \frac{\partial {\cal A}_n}{\partial X_{2n,j}}.
\end{equation}

Now plugging the second form of ${\cal A}_n$ into the derivative with respect to $X_{2,j}$, we get:
\begin{equation}
	{\cal A}_n = \sum_{j,k} (X_{2,j} - X_{1,j})(X_{2n,k} - X_{1,k})  \frac{\partial^2 {\cal A}_n}{\partial X_{2,j}\partial X_{2n,k}} + X_{2,2n} \frac{\partial {\cal A}_n}{\partial X_{2,2n}}.
\end{equation}
From this expression we can read off ${\cal A}_n^{\mu \nu}$:

\begin{equation}
	{\cal A}_n^{\mu \nu} = 4 \sum_{j,k} (X_2^\mu - X_j^\mu) (X_{2n}^\nu - X_{k}^\nu) \frac{\partial^2 {\cal A}_n}{\partial X_{2,j}\partial X_{2n,k}} -2 \eta^{\mu\nu} \frac{\partial {\cal A}_n}{\partial X_{2,2n}}.
\end{equation}
Now doing a gauge transformation and a rescaling in gluon $2$ and $2n$, let us choose $X_2^\mu = X_1^\mu$ and $X_{2n}^\nu = X_1^\nu$. So we get the following representation of ${\cal A}_n^{\mu \nu}$:

\begin{equation}
	{\cal A}_n^{\mu \nu} \propto 4 \sum_{j,k} (X_1^\mu - X_j^\mu) (X_{1}^\nu - X_{k}^\nu) \frac{\partial^2 {\cal A}_n}{\partial X_{2,j}\partial X_{2n,k}} - 2\eta^{\mu\nu} \frac{\partial {\cal A}_n}{\partial X_{2,2n}}.
\end{equation}
Now this is a particularly nice representation because it is such that $\mathcal{N}$ is zero:
\begin{equation}
\begin{aligned}
	{\cal A}_n^{\mu \nu}p_{\mu} &\propto \left[4 \sum_{j,k} (X_1^\mu - X_j^\mu) (X_{1}^\nu - X_{k}^\nu) \frac{\partial^2 {\cal A}_n}{\partial X_{2,j}\partial X_{2n,k}} -2 \eta^{\mu\nu} \frac{\partial {\cal A}_n}{\partial X_{2,2n}}\right] (X_3-X_1)_{\mu} \\
	&= 2 \sum_{j,k} (X_{3,j} - X_{1,j}) (X_{1}^\nu - X_{k}^\nu) \frac{\partial^2 {\cal A}_n}{\partial X_{2,j}\partial X_{2n,k}} -2 (X_3^\nu-X_1^\nu) \frac{\partial {\cal A}_n}{\partial X_{2,2n}} \\
	& = 2\sum_k (X_{1}^\nu - X_{k}^\nu) \frac{\partial }{\partial X_{2n,k}} \underbrace{\left( \sum_j (X_{3,j} - X_{1,j})\frac{\partial {\cal A}_n}{\partial X_{2,j}} \right)}_{=0} - 2 (X_1^\nu - X_3^\nu)\frac{\partial {\cal A}_n}{\partial X_{2,2n}} \\
	& \quad \quad +2 (X_1^\nu-X_3^\nu) \frac{\partial {\cal A}_n}{\partial X_{2,2n}} \\
	&=0,
\end{aligned}
\end{equation}
where the sum over $j$ vanishes because of gauge invariance. Therefore we found a representation of ${\cal A}_n^{\mu \nu}$ for which $\mathcal{N}=0$, and so in which the gluing becomes simply:
\begin{equation}
\begin{aligned}
	{\cal A}_{\text{glued}} &\propto -{\cal A}_n ^{\mu \nu} \eta_{\mu \nu} \propto 2 \sum_{j,k} (X_1^\mu -X_j^\mu)(X_{1\mu} - X_{k\mu}) \frac{\partial^2 {\cal A}_n}{\partial X_{2,j}\partial X_{2n,k}} -D \frac{\partial {\cal A}_n}{\partial X_{2,2n}} \\
	&=  \sum_{j,k} (X_{1,j} + X_{1,k} - X_{j,k})  \frac{\partial^2 {\cal A}_n}{\partial X_{2,j}\partial X_{2n,k}} -D \frac{\partial {\cal A}_n}{\partial X_{2,2n}} .
\end{aligned}
\label{eq:glue}
\end{equation}

This is finally our expression for gluing in the forward limit, written purely in terms of $X$ variables.

Of course, this holds for gluing any gauge-invariant object in the forward limit, but we now turn to the specific representations we give for $M$ in terms of integrals of forms of $y$ variables, and translate this statement into a gluing rule for these forms. Let's start by defining the two forms we obtain by taking one scaffolding residue after the other:
\begin{equation}
    \Omega_{s_{1,3}} = \text{Res}_{y_{1,3}=0} \Omega_{2n}^{\text{loop}}, \quad \Omega_S = \text{Res}_{y_{1,2n-1}=0} \Omega_{s_{1,3}}.
\end{equation}

Then we have:
\begin{equation}
\begin{aligned}
	\frac{\partial {\cal A}_n}{\partial X_{2,j}} &= \int \frac{\partial \Omega_S}{\partial X_{2,j}} = \int\text{Res}_{y_{1,2n-1}=0} \left( \frac{\partial \Omega_{s_{1,3}}}{\partial X_{2,j}} \right) \\
 &=\int\text{Res}_{y_{1,2n-1}=0} \left(\prod_{s\neq (1,3)} \frac{\diff y_{s}}{y_{s}^2} \, \prod_{i} \frac{\diff y_{i}}{y_{i}^2} \times \frac{\partial \log ( u_{2,j})}{\partial y_{1,3}}\prod_{(k,m) \neq \{(1,j),(2,j)\}} u_{k,m}^{ X_{k,m}}[y] \bigg \vert_{y_{1,3}=0}\right) \\ 
 &=  \int\prod_{\substack{s\neq \{(1,3),\\(1,2n-1)\}}} \frac{\diff y_{s}}{y_{s}^2} \, \prod_{i} \frac{\diff y_{i}}{y_{i}^2}  \left[ \sum_{k} 	(X_{2n,k}-X_{2n-1,k})\frac{\partial \log ( u_{2,j})}{\partial y_{1,3}}  \frac{\partial \log ( u_{2n,k})}{\partial y_{1,2n-1}} \right.\\
 & \quad \quad \quad \quad \quad \quad \quad \quad \quad \quad \quad  \quad \quad \left. + \frac{\partial^2 \log ( u_{2,2n})}{\partial y_{1,3} \partial y_{1,2n-1}}   \right]\times \prod_{\substack{(k,m) \neq \{(1,j),(2,j),\\
 (1,k),(2n-1,k)\}}} u_{k,m}^{ X_{k,m}}[y] \bigg \vert_{y_{1,3}=y_{1,2n-1}=0},
\end{aligned}
\end{equation}
which further taking a derivative on $X_{2n,k}$, gives
\begin{equation}
\begin{aligned}
	\frac{\partial^2 {\cal A}_n}{\partial X_{2n,k}\partial X_{2,j}} =  \int &\underbrace{\prod_{\substack{s\neq \{(1,3),\\(1,2n-1)\}}} \frac{\diff y_{s}}{y_{s}^2} \, \prod_{i} \frac{\diff y_{i}}{y_{i}^2}  \prod_{\substack{(k,m) \neq \{(1,j),(2,j),\\
 (1,k),(2n-1,k)\}}} u_{k,m}^{ X_{k,m}}[y]   }_{\tilde{\text{(KN)}}}\\
 &\quad \quad \quad \quad \quad \quad \quad \quad \quad \quad \times \left(\frac{\partial \log ( u_{2,j})}{\partial y_{1,3}}  \frac{\partial \log ( u_{2n,k})}{\partial y_{1,2n-1}} \right) \bigg \vert_{y_{1,3}=y_{1,2n-1}=0} , 
\end{aligned}
\end{equation}
so we can summarize the gluing rule as follows:
\begin{equation}
	\mathcal{A}_{\text{Glued}} =\int \sum_{j,k}\left(X_{1,j}+X_{1,k}-X_{j,k} \right) \frac{\partial \log{(u_{2,j})}}{\partial y_{1,3}} \frac{\partial \log (u_{2n,k})}{\partial y_{1,2n-1}} \times \tilde{\text{(KN)}} -D \, \cdot \frac{\partial \Omega_S }{\partial X_{2,2n}}.
	\label{eq:glue2}
\end{equation}

We finally translate this result using the variables of the loop-cut we want to compute, and find that the glued form is given by: 
\begin{equation}
\begin{aligned}
	\Omega^{2n}_{\text{Glued}} =& \prod_i \frac{\diff y_i}{y_i^2} \prod u_C^{X_C} \left.\left[\sum_{\substack{j\in\{2,\dots,1^\prime \}, \\ \,k \in \{1,\dots,2n\}} }\left(X_{p,j}+X_{p,k}-X_{j,k} \right) \frac{\partial \log{u_{s_L,j}}}{\partial y_{1,p}} \frac{\partial \log u_{s_R,k}}{\partial y_{1^\prime,p}} \right] \right\vert_{y_{1,p}=y_{1^\prime,p}=0} \\
 & \,-D \, \cdot \frac{\partial \Omega_S}{\partial X_{s_L,s_R}}  ,  
\end{aligned}
\label{eq:FormGlued}
\end{equation}
where $y_i$ are now all remaining variables in the triangulation (other than the two scaffolding ones we extracted the residue) and $C$ stands for the curves living in the smaller surface, obtained after taking the scaffolding residue. $\Omega_S$ is now the form resulting from taking both scaffolding residues that now correspond to $y_{1,p}=y_{1^\prime,p}=0$. As previously we can express this object in terms of derivatives of the lower-point tree amplitude as follows:
\begin{equation}   \Omega_{2n}^{\text{Glued}} = \sum_{j,k} \left(X_{j,p} + X_{k,p} - X_{j,k} \right)\frac{\partial^2 \mathcal{A}_{n+2}^\mathcal{S}}{\partial X_{j,s_L}\partial X_{k,s_R}}  -D \frac{\partial \mathcal{A}_{n+2}^{\text{tree}}}{\partial X_{s_L,s_R}};
\label{eq:Glued_NotSK}
\end{equation}

Now, in this formula we are assuming $X_{j,k}=X_{k,j}$, since in the tree side these variables are indeed the same. Now crucially the glued object is meant to match the result we get from the cut of the one-loop integrand, which if we define in terms of surface kinematics has $X_{j,k} \neq X_{k,j}$. So just like in the tree-loop cut, we want to derive a generalization of \eqref{eq:Glued_NotSK} which allow us to match with the object defined in terms of surface kinematics. 

\subsection{Surface kinematics and the loop-cut }

Manifestly in equation \eqref{eq:FormGlued}, the only part that is sensitive to surface kinematics is the term $X_{j,k}$, as well as the index range for $j,k$ is correct. From the loop-cut side, we know that all the curves $X_{j,k}$ entering must correspond to curves that intersect $Y_1$. Let's assume $j<k$, then $X_{j,k}$ is the curve that goes above the puncture, while $X_{k,j}$ goes below the puncture. So the self-intersection one version of $X_{j,k}$ and the self-intersection zero of $X_{k,j}$ both cross $Y_1$ and thus contribute to the loop-cut. So on the loop side, taking the residue on $y_{1}=0$ we get the contributions from these curves $u_{j,k}^{(1)}$ and $u_{k,j}^{(0)}$. Just like we saw on the tree-loop cut, it's simple to show that the derivatives of these $u$-variables factorize nicely into derivatives of $u$'s of the tree-problem as follows:

\begin{equation}
\begin{aligned}
   & \frac{\partial \log{u_{j,k}^{(1)}}}{\partial y_{1}} \bigg \vert_{y_{1} =0} = \frac{\partial \log{u_{j,s_R}}}{\partial y_{1^\prime,p}} \bigg \vert_{y_{1^\prime,p} =0} \times \frac{\partial \log{u_{k,s_L}}}{\partial y_{1,p}} \bigg \vert_{y_{1,p} =0}, \\
    & \frac{\partial \log{u_{k,j}^{(0)}}}{\partial y_{1}} \bigg \vert_{y_{1} =0} = \frac{\partial \log{u_{k,s_R}}}{\partial y_{1^\prime,p}} \bigg \vert_{y_{1^\prime,p} =0} \times \frac{\partial \log{u_{j,s_L}}}{\partial y_{1,p}} \bigg \vert_{y_{1,p} =0} 
\end{aligned}
\end{equation}

So from here, we see that to follow surface kinematics \eqref{eq:FormGlued} becomes 
\begin{equation}
\begin{aligned}
	\Omega^{2n}_{\text{Glued}} =& \prod_i \frac{\diff y_i}{y_i^2} \prod u_C^{X_C} \left.\left[\sum_{j,k}\left(X_{p,j}+X_{p,k}-X_{j,k} \right) \frac{\partial \log{u_{j,s_R}}}{\partial y_{1^\prime,p}} \frac{\partial \log u_{k,s_L}}{\partial y_{1,p}} \right] \right\vert_{y_{1,p}=y_{1^\prime,p}=0} \\
 & \,-D \, \cdot \frac{\partial \Omega_S}{\partial X_{s_L,s_R}}  ,  
\end{aligned}
\label{eq:FormGlued2}
\end{equation}

where now $j,k \in (1,2,\cdots,2n, 1^\prime)$, and in the end we set $1^\prime \to 1$, and we get $X_{i,i}, X_{i+1,i}, X_{i,i+1}\neq 0$. In terms of derivatives of the tree object we can write:
\begin{equation}
\Omega^{2n}_{\text{Glued}} = \sum_{j,k} \left[\left(X_{j,p} + X_{k,p} - X_{j,k} \right)\frac{\partial^2 \mathcal{A}_{n+2}^\mathcal{S}}{\partial X_{j,s_R}\partial X_{k,s_L}} \right]\bigg\vert_{1^\prime\rightarrow 1} -D \frac{\partial \mathcal{A}_{n+2}^{\mathcal{S}}}{\partial X_{x^\prime,x}}; \quad j,k \in (1,2,\cdots,2n, 1^\prime).
    \label{eq:Glue_SK}
\end{equation}

Now that we have the glued object expressed in terms of the most general kinematics, $X_{i,j} \neq X_{j,i}$, let's proceed to show that the result we get from the residue of the surface loop integrand matches this gluing. 

\subsection{Loop-cut matching with Tree-gluing}

To do this we start by rewriting the tree-level gluing on a way that facilitates the matching of each term separately as follows:
\begin{equation}
\begin{aligned}
\Omega^{2n}_{\text{Glued}} =& - \underbrace{ D \, \cdot \frac{\partial \Omega_S}{\partial X_{s_L,s_R}}}_{(\dagger)} \quad + \quad  \prod_i \frac{\diff y_i}{y_i^2} \prod u_C^{X_C} \left[ - \underbrace{\sum_{j,k\in\{2,\dots,2n \}} X_{j,k} \frac{\partial \log{u_{s_L,k}}}{\partial y_{1,p}} \frac{\partial \log u_{s_R,j}}{\partial y_{1^\prime,p}}}_{(\dagger \dagger)} \right. \\
	&  + \underbrace{\sum_{j\in\{2,\dots,2n \}} X_{p,j} \left( \frac{\partial \log{u_{s_L,j}}}{\partial y_{1,p}} \sum_{k \in \{1,\dots,2n\}} \frac{\partial \log u_{s_R,k}}{\partial y_{1^\prime,p}} + \frac{\partial \log u_{s_R,j}}{\partial y_{1^\prime,p}} \sum_{k \in \{2,\dots,1^\prime\}} \frac{\partial \log u_{s_L,k}}{\partial y_{1,p}} \right)}_{(\dagger \dagger \dagger)}  \\
	&  - \underbrace{\sum_{k \in \{2,\dots,2n\}}X_{1,k} \frac{\partial \log{u_{1,s_R}}}{\partial y_{1^\prime,p}}  \frac{\partial \log u_{k,s_L}}{\partial y_{1,p}} \,+ X_{k,1^\prime} \frac{\partial \log{u_{s_L,1^\prime}}}{\partial y_{1,p}}  \frac{\partial \log u_{k,s_R}}{\partial y_{1^\prime,p}} }_{(\dagger \dagger \dagger \dagger)}\\
	& \left. \left. \underbrace{-X_{1,1^\prime} \left( \frac{\partial \log{u_{s_L,1^\prime}}}{\partial y_{1,p}}  \frac{\partial \log u_{s_R,1}}{\partial y_{1^\prime,p}}  \right)}_{(\dagger \dagger \dagger \dagger \dagger)} \right] \right\vert _{y_{1,p}=y_{1^\prime,p}=0},
\end{aligned}
\end{equation}
where we are setting $X_{1,j}=X_{1\prime,j}$.
Our goal now is to show the different terms agree with the loop cut result, up to a possible total derivative. Let's look at what the loop-cut result looks like before proceeding. 

As explained before, in the loop integrand we include all the curves up to self-intersection one. So by taking the residue on $Y_1$, we are effectively taking the residue of the integrand at $y_1=0$, which leads to:
\begin{equation}
\begin{aligned}
	\text{Res}_{y_1=0} \Omega_{2n}^{\text{loop}} =&  \prod_i \frac{\diff y_i}{y_i^2} \prod u_C^{X_C} \left[ \sum_{j \in \{2,\dots,2n\}} Y_j \frac{\partial \log{u_{Y_j}}}{\partial y_{1}} + \sum_{j\leq k } X_{j,k}  \frac{\partial \log{u_{j,k}^{(1)}}}{\partial y_1} + X_{k,j}  \frac{\partial \log{u_{k,j}^{(0)}}}{\partial y_{1}} \right. \\
	& \left. \left. + \sum_{j \in\{2,\cdots,2n\}} X_{1,j} \frac{\partial \log{u_{1,j}^{(1)}}}{\partial y_{1}} + X_{j,1} \frac{\partial \log{ u_{j,1}^{(0)}u_{j,1}^{(1)}}}{\partial y_{1}}  + X_{1,1} \frac{\partial \log{ u_{1,1}^{(0)}u_{1,1}^{(1)}}}{\partial y_{1}} \quad- \Delta \right] \right\vert _{y_{1,p}=y_{1^\prime,p}=0},
\end{aligned}
\label{eq:loopcut1}
\end{equation}
where once more we are keeping full surface kinematics. So, using $\Delta = 1-D$, we can rewrite \eqref{eq:loopcut1} as follows:
\begin{equation}
\begin{aligned}
	\text{Res}_{y_1=0} \Omega_{2n}^{\text{loop}} =&  \underbrace{D \, \cdot \prod_i \frac{\diff y_i}{y_i^2} \prod u_C^{X_C} }_{(\star)}\quad + \quad  \prod_i \frac{\diff y_i}{y_i^2} \prod u_C^{X_C}\left[ \underbrace{\sum_{j\leq k } X_{j,k}  \frac{\partial \log{u_{j,k}^{(1)}}}{\partial y_1} + X_{k,j}  \frac{\partial \log{u_{k,j}^{(0)}}}{\partial y_{1}}}_{(\star \star)} \right. \\
	& \left. +  \underbrace{\sum_{j \in \{2,\dots,2n\}} Y_j \frac{\partial \log{u_{Y_j}}}{\partial y_{1}}}_{(\star \star \star)} \quad +\quad  \underbrace{\sum_{j \in\{2,\cdots,2n\}} X_{1,j} \frac{\partial \log{u_{1,j}^{(1)}}}{\partial y_{1}} + X_{j,1} \frac{\partial \log{ u_{j,1}^{(0)}u_{j,1}^{(1)}}}{\partial y_{1}}}_{(\star \star \star \star)} \quad \right. \\
	& \left. \left. + \underbrace{X_{1,1} \left(\frac{\partial \log{u_{1,1}}}{\partial y_{1}} + \frac{\partial \log{u^{(1)}_{1,1}}}{\partial y_{1}}\right)}_{(\star \star \star \star \star)} -1 \right]\right\vert _{y_{1,p}=y_{1^\prime,p}=0}.
\end{aligned}
\label{eq:loopcut2}
\end{equation}

From figures \ref{fig:LoopCut} and \ref{fig:TreeGlue}, it's obvious that $Y_j$ on the loop side maps to $X_{p,j}$ on the tree glue side. In addition, the choice of triangulation for the loop/tree problems makes it clear that $\prod_i \frac{\diff y_i}{y_i^2} \prod u_C^{X_C}$ is the same for both of them. 
  
The approach we will follow is to analyze each term separately and understand whether they match directly or differ by a total derivative. The matching works quite intricately, and we devote the appendix~\ref{sec:matchingAll} for more details, here we simply summarize the results. 

Both $(\star)$ and $(\star \star)$ can be shown to match $(\dagger)$ and $(\dagger \dagger)$, respectively. The same is not true for the remaining three terms, however, the difference will be shown to be the following rather non-trivial total derivative:
\begin{equation}
    -\frac{\partial}{\partial y_t} \left(\prod \frac{1}{y_i^2} \frac{\prod_{k\in\{1, \dots, m-1\}} u_{p,k}u_{1^\prime,k}}{u_{1,p}\prod_{j\in\{2, \dots, 2n\}} u_{p,j}^2} \prod_{C} u_C^{X_{C}}\right)\bigg \vert_{y_{1,p}=y_{1^\prime,p}=0}.
\end{equation}

It is very pleasing that the total derivative is with respect to the only variable we could possibly have ``local" to the triangulation. 

This establishes the consistency of our curve integral for scalar-scaffolded gluons, by showing that the loop-cut and the tree-level gluing agree up to a total derivative. We stress again that while this argument is strictly valid at 1-loop, its quite ``local" character makes it likely that it can be extended to any number of loops.

  %\subfile{Sections/7_LoopCut.tex}
  %\newpage
  \section{Outlook}

The first concrete example for reformulating the fundamental physics of scattering amplitudes in a new mathematical language, based on elementary but deep ideas in combinatorics and geometry, arose ten years ago in the content of the positive Grassmannian~\cite{Arkani-Hamed:2012zlh} and the amplituhedron~\cite{Arkani-Hamed:2013jha} for planar ${\cal N}=4$ SYM. The past decade has seen a succession of developments pushing away from this very special theory and closer to the physics of Yang-Mills theory in the real world~\cite{Arkani-Hamed:2017mur, Arkani-Hamed:2019plo, Arkani-Hamed:2023lbd, Arkani-Hamed:2023mvg, Zeros, NLSM}. This program has been motivated by the belief that more magic--not less!--should reveal itself as we move from toy models towards describing the universe we actually live in.

We believe that with the results reported in this paper, we have entered a new phase in this adventure, in which combinatorial geometry in kinematic space is finally seen to directly describe major aspects of the world we see outside our windows. In one fell swoop, we have freed ourselves from the shackles of supersymmetry and integrability and ``dlog forms'', along the way discovering a startling unity between non-supersymmetric gluons, pions, and the simplest theory of colored scalars. 

As such there are a huge number of avenues for exploration, both in better understanding the conceptual origins and underpinnings of our proposal, as well as in seeking to propel us ever closer to describing all aspects of real-world physics from this new point of view. Instead of making a long list of these entirely obvious big questions, we close by mentioning a small number of directions of exploration immediately following from the developments here, which we hope to report on in the near future. 

To begin with, we have left a number of important loose ends that need to be fully understood to precisely define our proposal at all orders in the topological expansion. We have emphasized that the dependence on dimensionality $D$ arises through the product of the closed $\Delta$ curves wrapping around punctures, keeping one self-intersection per puncture. This is a perfectly well-defined proposal for the planar diagrams contributing to single-trace amplitudes, but starting with the torus corrections at two loops there are many kinds of closed $\Delta$ curves that can contribute, and the precise statement about the choice of $\Delta$'s and exponents must be understood. Similarly, the precise statement about $\Delta$'s for multi-trace amplitudes must be determined. We have checked that in the simplest case of 1-loop annulus corrections to double-trace amplitudes, where there is a unique closed $\Delta$ curve separating the two boundaries, the obvious choice of $\Delta$ with the same exponent $(1 - D)$ correctly matches leading singularities, so we expect a consistent choice is possible at all orders in the topological expansion. Also, our proof for the equality of the loop-cut and forward limit was strictly only for the one-loop cut $\to$ tree-level forward limit. The character of the argument was however very ``local" on the surface, and we expect that it can be extended to all orders-- as supported by the successful checks of leading singularities at two loops-- but it is important to establish this definitively. 

Perhaps the most pressing question is learning how to effectively extract the low-energy, field-theory limit of the amplitudes from our stringy integrals. This is more challenging than the analogous problem for the Tr $\phi^3$ theory, where the $dy/y$ singularity makes it easy to extract the low-energy limit of the amplitudes for $\alpha^\prime X \ll 1$, localizing to logarithmic divergences as $|{\rm log} y| \to \infty$. This can be easily computed by ``tropicalizing" the $u$ variables, giving a global Schwinger parametrization for the full amplitude directly in the field theory limit. This procedure does not work straightforwardly for our scaffolded gluons, where the $dy/y^2$ form has power-law divergences and the integral must be defined by analytic continuation, even at tree-level. In~\cite{cexpansion} we will report on an elementary and direct way of extracting the field theory limit, via a ``cone-by-cone" analysis of the singular behavior of the integrand. This will already expose a number of surprising facts about gluon amplitudes. For instance, it will allow us to establish what we call the ``mesh expansion" for tree amplitudes, in terms of interesting sums associated with {\it all} facets of the ABHY associahedron. Extremes of this expansion can be described very efficiently either by closed-form formulae at all multiplicity, or by novel ``tropical" expressions. But we are still seeking an efficient ``tropical" representation of the {\it full} gluon amplitudes, as a  counterpart of the existing ``tropical" formulae for Tr $\phi^3$ theory at all orders in the topological expansion. 

We have emphasized that our surface picture for gluons provides a solution to a long-standing difficulty of making sense of ``nice" integrands for non-supersymmetric Yang-Mills theories, by giving a natural way of treating the infamous ``1/0" difficulties associated with tadpoles and massless external bubbles~\cite{Caron-Huot:2010fvq}. Quite apart from its conceptual importance especially at ``stringy" level, the elementary mechanism by which tadpoles/massless external bubbles are naturally dealt with suggests a recursive procedure for computing integrands for Yang-Mills theory~\cite{YMrec} in any number of dimensions, in the planar limit where integrands are well-defined. We have computed integrands determined by this recursion through a number of one- and two-loop examples, which have passed all obvious consistency checks~\cite{YMrec}. Optimistically an all-loop recursion for non-supersymmetric Yang-Mills theory in the planar limit may be at hand, as a counterpart to the very different recursion based on BCFW recursion for planar ${\cal N}=4$ SYM~\cite{Arkani-Hamed:2010zjl}. 

We have seen that the extremely simple and purely kinematical idea of thinking about particles with spin in the ``scalar-scaffolding'' language allows highly non-trivial hidden features of gluon amplitudes to be brought to light. As such it is clearly interesting to study more general spinning amplitudes in this language. Clearly tree-level gravity amplitudes can be described in exactly the same $2n$ scalar scaffolding language, with an obvious extension of the notions of multilinearity and gauge invariance. Their stringy integrands are again a single term, arising as usual from the mod-square of the scalar-scaffolded gluon integrand. Amplitudes for particles of any mass and spin, including the string resonances, can also be kinematically described using scalar scaffolding. The surprise for gluons is that the ``dynamics" can be captured by a simple kinematic shift/monomial in the $u$ variables. It would therefore be fascinating to look for an analog of our $(\prod u_{e,e}/\prod u_{o,o})$ factor that might describe the scattering of general string states. 

Finally, while our new description for scalar-scaffolded gluons works in any number of dimensions, we know amplitudes in $D=4$ spacetime dimensions are not only particularly important, but should have extra simplicity reflected in the Abelian nature of the little group and the natural decomposition of the amplitude into helicity sectors. In fact we expect the behavior of the amplitude {\it near} four dimensions to also be significant; for instance, it has long been appreciated that the enigmatic ``rational terms" in loop amplitudes are determined by ``$\epsilon/\epsilon$'' effects in dimensional regularization, and hence are sensitive to the behavior of the integrand close to four dimensions. There is a natural way to approach this physics, parametrizing the $X$ kinematic data in $D=4$ using momentum twistor variables and considering general perturbations $\epsilon \delta X$ as an expansion in $\epsilon$. 

This also gives a new avenue for trying to understand the lingering mystery of having an embarrassment of riches in describing the dynamics of gluons using combinatorial geometry. In four dimensions and with maximal supersymmetry, we have the story of the positive Grassmannian~\cite{Arkani-Hamed:2012zlh} and the amplituhedron~\cite{Arkani-Hamed:2013jha}, while now in general dimensions and with no supersymmetry, we have the scalar-scaffolded gluons arising from the combinatorics of surfaces. Obviously, in $D=4$ these two stories must be connected. Perhaps the closest point where a technical connection can be made is at the level of leading singularities, where we have two different formulations as the residues of a master differential form, in the Grassmannian and the binary geometry of surfaces respectively. Of course, the two stories also differ crucially at the very outset, by the way they describe the scattering kinematics. In the positive Grassmannian and amplituhedron, the external data is labelled ``one-particle at a time", by specifying momentum(-twistor) variables for each particle, while in the scaffolding picture it is crucial that the data is specified in terms of variables $X_{ij}$ encoding Lorentz-invariants labelled by {\it pairs} of points/particles. An understanding of the approach to $D=4$ from the scalar-scaffolding story, encoding the kinematic data in momentum-twistor variables, should help bring the connection between these two fascinating worlds closer to light.

\acknowledgments

It is our pleasure to thank Hadleigh Frost, Sebastian Mizera, Pierre-Guy Plamondon, Giulio Salvatori, and Hugh Thomas for numerous inspiring discussions. The work of N.A.H. is supported by the DOE (Grant No. DE-SC0009988), by the Simons Collaboration on Celestial Holography, and further support was made possible by the Carl B. Feinberg cross-disciplinary program in innovation at the IAS. The work of Q.C. is supported by the National Natural Science Foundation of China under Grant No. 123B2075. The work of C.F. is supported by FCT/Portugal
(Grant No. 2023.01221.BD). The work of S.H. has been supported by the National Natural Science Foundation of China under Grant No. 12225510, 11935013, 12047503, 12247103, and by the New Cornerstone Science Foundation through the XPLORER PRIZE.

  %\newpage
  \appendix
\section{Scaffolding residues using worldsheet variables of bosonic string}
\label{sec:bosonicString}
In sec.~\ref{sec:tree} we have given elegant and practical formulas for the $n$-gluon amplitudes obtained from scaffolding $2n$-scalar amplitudes (which are special components of $2n$-gluon amplitudes) in bosonic string theory; we have also shown a consistency check from $4n$ scalars to $2n$ gluons and then $2n$ scalars, which concludes the tree-level story.

However, we can also directly compute scaffolding residue of $2n$-scalar amplitudes using worldsheet variables, which gives again the $n$-gluon correlator of bosonic-string integrand as shown in \eqref{eq:2nScalarUs}. Our starting point is:
\begin{equation}
{\cal I}_{(12)\cdots (2n{-}1 2n)}=\int \frac{d^{2n} z}{{\rm SL}(2,\mathbb{R})}\frac{\prod_{i<j} z_{i,j}^{2p_i \cdot p_j}}{z_{12}^2 z_{34}^2 \cdots z^2_{2n{-}1, 2n}}=\int \frac{d^{2n} z}{{\rm SL}(2,\mathbb{R})}\frac{\left\langle\prod_{a=1}^{2n} e^{i p_a \cdot X\left(z_a\right)}\right\rangle}{z_{12}^2 z_{34}^2 \cdots z^2_{2n{-}1, 2n}}\,,
\end{equation}
where we have rewritten the Koba-Nielsen factor in terms of the OPE of vertex operators. 

It is important to make the special choice $\epsilon_i= p_{2i{-}1}$ (recall $q_i^\mu=p_{2i-1}^\mu+p_{2i}^\mu$): as we have discussed one can check that $\epsilon_a \cdot \epsilon_b\propto c_{2a-1, 2b-1}$ in $W_{a,b}$, and for a given $a$, the collection of $\epsilon_a \cdot q_b$ for $b\neq a$ is equivalent to those $X_{2a,j}-X_{2a-1,j}$ for all $j$. Taking the scaffolding residues corresponds to taking the residues at $z_{i,i+1}=0$ for odd $i$, in terms of the $z$ variables. The residue $z_{12}=0$ is given by taking the derivative of the vertex operator and setting $z_{1}=z_{2}$:
\begin{equation}
\left.{\partial_{z_{1}} }\left\langle\prod_{a=1}^{2n} e^{i p_a \cdot X\left(z_a\right)}\right\rangle\right|_{z_{1}=z_{2}}=\left\langle ip_{1}\cdot{\partial_{z_{2}}}X(z_{2})e^{i p_{1,2} \cdot X\left(z_2\right)}\prod_{a=3}^{2n} e^{i p_a \cdot X\left(z_a\right)}\right\rangle.
\end{equation}

Remember the special choice $\epsilon_a= p_{2a{-}1}$, and $q_a=p_{2a-1}+p_{2a}=p_{2a-1,2a}$, after scaffolding the $2n$-scalar correlator  becomes the $n$-gluon correlator
\begin{equation}
\left\langle \prod_{a=1}^{n}ip_{2a-1}\cdot{\partial_{z_{2a}}}X(z_{2a})e^{i p_{2a-1,2a} \cdot X\left(z_{2a}\right)} \right\rangle=\left\langle \prod_{a=1}^{n}i\epsilon_{a}\cdot{\partial_{z_{2a}}}X(z_{2a})e^{i q_{a} \cdot X\left(z_{2a}\right)} \right\rangle,
\end{equation}
with the relabelling the worldsheet coordinates $z_{2a}\to z_{a}$. Therefore, the $2n$-scalar bosonic string amplitude becomes the $n$-gluon bosonic string amplitude by scaffolding:
\begin{equation}
{\cal I}_{(12)\cdots (2n{-}1 2n)}\xrightarrow{\text{scaffolding}}\int \frac{\prod_{a=1}^{n}dz_{a}}{{\rm vol~SL}(2,\mathbb{R})}\left\langle \prod_{a=1}^{n}i\epsilon_{a}\cdot{\partial_{z_{a}}}X(z_{a})e^{i q_{a} \cdot X\left(z_{a}\right)} \right\rangle\,.
\end{equation}

Having made connections with the more familiar worldsheet variables, we can also repeat such a derivation for closed-string and conclude that the $n$-graviton scattering amplitudes of bosonic string is again given by scaffolding residue of $2n$-scalar amplitudes, which are ``mod-square" closed-stringy integrals over ``complex binary geometries"~\cite{Arkani-Hamed:2019mrd}. Obviously, such a ``mod-square" $2n$-scalar integral for closed-string is related to those for open-string via Kawai-Lewellen-Tye (KLT) relations~\cite{Kawai:1985xq}. In the field-theory limit, they become double copy relations~\cite{Bern:2008qj} between gravity and gluon amplitudes, now written in this scalar-scaffolding language. 

\section{Complete results for $1$-loop $4$-point box leading singularity} \label{sec:boxLScomplete}
We present the complete result of $1$-loop $4$-point box leading singularity (with $Y_1=Y_2=Y_3=Y_4=0$) which contains terms with mass dimension $X^m$ for $m=4,5,6,7,8$. These terms correspond to the pure Yang-Mills leading singularity and those with $1,2,3$ and $4$ insertions of $F^3$ vertices, respectively. The pure Yang-Mills leading singularity reads:
\begin{equation}
\begin{aligned}
&-\frac{(\Delta +1)}{4} Y_2 Y_4 Y_6 Y_8-2 Y_6 Y_8 X_{1,4} X_{2,5}+ \frac{1}{4} X_{3,7}^2 X_{1,5}^2+\frac{1}{2} X_{2,7} X_{3,6} X_{1,5}^2-X_{2,8} X_{3,6} X_{1,5}^2-X_{2,7} X_{3,7} X_{1,5}^2\\
&+X_{2,8} X_{3,7} X_{1,5}^2+\frac{1}{2} X_{2,8} X_{4,6} X_{1,5}^2-X_{1,6} X_{3,7}^2 X_{1,5}+2 Y_6 Y_8 X_{2,4} X_{1,5}-Y_4 Y_8 X_{2,6} X_{1,5}+2 Y_4 Y_8 X_{2,7} X_{1,5}\\
&-Y_8 X_{2,4} X_{3,6} X_{1,5}+Y_4 X_{2,7} X_{3,6} X_{1,5}-X_{1,4} X_{2,7} X_{3,6} X_{1,5}-Y_4 X_{2,8} X_{3,6} X_{1,5}+X_{1,4} X_{2,8} X_{3,6} X_{1,5}\\
&-Y_2 Y_6 X_{3,7} X_{1,5}+Y_6 X_{2,4} X_{3,7} X_{1,5}+Y_8 X_{2,4} X_{3,7} X_{1,5}-Y_4 X_{2,6} X_{3,7} X_{1,5}+X_{1,4} X_{2,7} X_{3,7} X_{1,5}\\
& +X_{1,6} X_{2,7} X_{3,7} X_{1,5}-X_{1,4} X_{2,8} X_{3,7} X_{1,5}-X_{1,6} X_{2,8} X_{3,7} X_{1,5}+Y_4 X_{2,6} X_{3,8} X_{1,5}-X_{2,5} X_{3,6} X_{3,8} X_{1,5}\\
& +X_{1,6} X_{3,7} X_{3,8} X_{1,5}+X_{1,4} X_{1,6} X_{3,7}^2-X_{1,6} X_{2,4} X_{3,7}^2+\frac{1}{2} X_{1,6} X_{2,5} X_{3,7}^2 +Y_4 Y_8 X_{1,6} X_{2,5}-2 Y_4 Y_8 X_{1,6} X_{2,7}\\
&+Y_8 X_{1,4} X_{2,5} X_{3,6}+Y_4 Y_8 X_{2,7} X_{3,6}-Y_8 X_{1,4} X_{2,7} X_{3,6}+2 Y_4 Y_8 X_{1,6} X_{3,7}-Y_8 X_{1,6} X_{2,4} X_{3,7}\\
&-Y_6 X_{1,4} X_{2,5} X_{3,7}-Y_8 X_{1,4} X_{2,5} X_{3,7}+Y_4 X_{1,6} X_{2,5} X_{3,7}-Y_4 Y_8 X_{2,6} X_{3,7}+Y_8 X_{1,4} X_{2,6} X_{3,7}\\
&-X_{1,4} X_{1,6} X_{2,7} X_{3,7}-Y_4 X_{1,6} X_{2,8} X_{3,7}+X_{1,4} X_{1,6} X_{2,8} X_{3,7}+Y_6 X_{1,4} X_{2,5} X_{3,8}-Y_4 X_{1,6} X_{2,5} X_{3,8}\\
&+X_{1,6} X_{2,4} X_{3,7} X_{3,8}-X_{1,6} X_{2,5} X_{3,7} X_{3,8}+\frac{1}{4} X_{1,6} X_{2,5} X_{3,8} X_{4,7} +\frac{1}{4} X_{1,4} X_{2,7} X_{3,6} X_{5,8} \\
&+ (\mathrm{cyclic}, i\to i+2,i+4,i+6)\,,
\end{aligned}
\end{equation}
which, by setting $\Delta=1-D$, agrees with the well-known $D$-dimensional result ({\it c.f.}~\cite{Edison:2022jln}) once we translate it to  polarizations and momenta.

Those with $1,2$ and $3$ insertions of $F^3$ vertices are:
\begin{equation}
\begin{aligned}
&Y_2 X_{3,7}^2 X_{1,5}^2-2 Y_4 Y_6 X_{2,7} X_{1,5}^2+2 Y_4 Y_6 X_{2,8} X_{1,5}^2+Y_4 X_{2,7} X_{3,6} X_{1,5}^2-Y_4 X_{2,8} X_{3,6} X_{1,5}^2+2 Y_4 Y_6 X_{3,7} X_{1,5}^2\\
&-Y_4 X_{2,7} X_{3,7} X_{1,5}^2-Y_6 X_{2,7} X_{3,7} X_{1,5}^2-Y_8 X_{2,7} X_{3,7} X_{1,5}^2+Y_4 X_{2,8} X_{3,7} X_{1,5}^2+Y_6 X_{2,8} X_{3,7} X_{1,5}^2\\
&-Y_2 X_{1,6} X_{3,7}^2 X_{1,5}-Y_4 X_{1,6} X_{3,7}^2 X_{1,5}-Y_8 X_{1,6} X_{3,7}^2 X_{1,5}+2 Y_2 Y_8 X_{1,4} X_{3,6} X_{1,5}+Y_4 Y_8 X_{2,7} X_{3,6} X_{1,5}\\
&-Y_8 X_{1,4} X_{2,7} X_{3,6} X_{1,5}-2 Y_2 Y_8 X_{1,4} X_{3,7} X_{1,5}-2 Y_2 Y_8 X_{1,6} X_{3,7} X_{1,5}+2 Y_6 Y_8 X_{2,4} X_{3,7} X_{1,5}\\
&-Y_4 Y_8 X_{2,6} X_{3,7} X_{1,5}+Y_6 X_{1,4} X_{2,7} X_{3,7} X_{1,5}+Y_8 X_{1,4} X_{2,7} X_{3,7} X_{1,5}+Y_4 X_{1,6} X_{2,7} X_{3,7} X_{1,5}\\
&+Y_8 X_{1,6} X_{2,7} X_{3,7} X_{1,5}-Y_6 X_{1,4} X_{2,8} X_{3,7} X_{1,5}-Y_4 X_{1,6} X_{2,8} X_{3,7} X_{1,5}+2 Y_4 Y_6 X_{2,5} X_{3,8} X_{1,5}\\
&-Y_4 X_{2,5} X_{3,6} X_{3,8} X_{1,5}+Y_2 X_{1,6} X_{3,7} X_{3,8} X_{1,5}+Y_4 X_{1,6} X_{3,7} X_{3,8} X_{1,5}-2 Y_2 Y_4 X_{1,6} X_{3,7}^2\\
&+Y_8 X_{1,4} X_{1,6} X_{3,7}^2-Y_8 X_{1,6} X_{2,4} X_{3,7}^2+Y_4 X_{1,6} X_{2,5} X_{3,7}^2+2 Y_2 Y_8 X_{1,4} X_{1,6} X_{3,7}+Y_2 X_{1,4} X_{1,6} X_{3,7}^2\\
&-2 Y_6 Y_8 X_{1,4} X_{2,5} X_{3,7}+Y_4 Y_8 X_{1,6} X_{2,5} X_{3,7}-Y_8 X_{1,4} X_{1,6} X_{2,7} X_{3,7}-Y_4 X_{1,6} X_{2,5} X_{3,7} X_{3,8}\\
&+ (\mathrm{cyclic}, i\to i+2,i+4,i+6)\,;
\end{aligned}
\end{equation}

\begin{equation}
\begin{aligned}
&Y_2 Y_4 X_{3,7}^2 X_{1,5}^2+2 Y_2 Y_4 Y_6 Y_8 X_{1,5}^2-2 Y_4 Y_6 Y_8 X_{2,7} X_{1,5}^2 +Y_2 Y_4 Y_6 Y_8 X_{3,7} X_{1,5}\\
&+2 Y_2 Y_4 Y_6 X_{3,7} X_{1,5}^2+2 Y_2 Y_4 Y_8 X_{3,7} X_{1,5}^2-Y_4 Y_6 X_{2,7} X_{3,7} X_{1,5}^2-Y_4 Y_8 X_{2,7} X_{3,7} X_{1,5}^2\\
&+Y_4 Y_6 X_{2,8} X_{3,7} X_{1,5}^2-Y_2 Y_4 X_{1,6} X_{3,7}^2 X_{1,5}-Y_2 Y_8 X_{1,6} X_{3,7}^2 X_{1,5}-Y_4 Y_8 X_{1,6} X_{3,7}^2 X_{1,5}\\
&-2 Y_2 Y_6 Y_8 X_{1,4} X_{3,7} X_{1,5}-2 Y_2 Y_4 Y_8 X_{1,6} X_{3,7} X_{1,5}+Y_6 Y_8 X_{1,4} X_{2,7} X_{3,7} X_{1,5} \\
&+Y_2 Y_4 X_{1,6} X_{3,7} X_{3,8} X_{1,5}-2 Y_2 Y_4 Y_8 X_{1,6} X_{3,7}^2+Y_2 Y_8 X_{1,4} X_{1,6} X_{3,7}^2 -Y_6 Y_8 X_{2,7} X_{3,7} X_{1,5}^2\\
&+\frac{1}{2} Y_4 Y_8 X_{2,7} X_{3,6} X_{1,5}^2 +\frac{1}{2} Y_4 Y_8 X_{1,6} X_{2,5} X_{3,7}^2 +\frac{1}{2} Y_2 Y_6 X_{3,7}^2 X_{1,5}^2 \\
&+Y_4 Y_8 X_{1,6} X_{2,7} X_{3,7} X_{1,5} + (\mathrm{cyclic}, i\to i+2,i+4,i+6)\,;
\end{aligned}
\end{equation}
\begin{equation}
\begin{aligned}
    &Y_2 Y_4 Y_6 X_{3,7}^2 X_{1,5}^2+2 Y_2 Y_4 Y_6 Y_8 X_{3,7} X_{1,5}^2-Y_4 Y_6 Y_8 X_{2,7} X_{3,7} X_{1,5}^2\\
    &-Y_2 Y_4 Y_8 X_{1,6} X_{3,7}^2 X_{1,5}+ (\mathrm{cyclic}, i\to i+2,i+4,i+6)\,.
\end{aligned}
\end{equation}

Finally, the one term with $4$ $F^3$ vertices reads:
\begin{equation}
    Y_2 Y_4 Y_6 Y_8 X_{1,5}^2 X_{3,7}^2.
\end{equation}

\section{Factorization for derivatives of $u$ variables}
In this appendix, we show that the derivatives of $u$ variables encountered in computing residues nicely factorize. The statement is true very generally. Suppose we are interested in $u$ variables that depend on a given $y$ variable. Obviously the associated curve $C$ must pass through the road associated with $y$. We will focus on the cases where the curve passes through the road $y$ once. The word for this curve is then of the form $A -{\rm Turn} -y- {\rm Turn}^\prime- B$ where Turn, Turn$^\prime$ are either Left or Right turns into and out of $y$ respectively. 

Let us begin by considering the case where the words $A,B$ are generic; by this we mean that the $2 \times 2$ matrices for $A,B$ are non-zero in all their entries. Geometrically, this corresponds to the case where the curve $C$ intersects $y$ once. The matrix associated with $C$ is then of the form 
\begin{equation}
M_C = \left(\begin{array}{cc} a & b \\ c & d \end{array} \right) M_{L/R}(y) \left(\begin{array}{cc} a^\prime & b^\prime \\ c^\prime & d^\prime \end{array} \right)
\end{equation}
Let's consider the case with $M_L(y)$, then 
\begin{equation}
u_C = \frac{M_C^{1,2} M_C^{2,1}}{M_C^{1,1} M_C^{2,2}} = \frac{(a b^\prime y + d^\prime (b + a y)) (a^\prime c y + c^\prime (d + c y))}{(a a^\prime y + 
   c^\prime (b + a y)) (b^\prime c y + d^\prime (d + c y))}
\end{equation}
and we can compute 
\begin{equation}
\partial_y {\rm log} u_C\vert_{y=0} = - \left(\frac{a d - b c}{b d} \right) \times \left(\frac{a^\prime d^\prime - b^\prime c^\prime}{c^\prime d^\prime} \right)
\end{equation}
This expression beautifully factorizes into a part that only depends on the part of the words on either side of $y$.
\footnote{Note that since $C$ intersects $y$, $u_C \to 1$ as $y \to 0$, so that $\partial_y {\rm log} u_C$ is actually equal to $\partial_y u_C$; we work with $\partial_y {\rm log} u_C$ simply because the logarithmic derivative is most natural for manipulations with the surface integral.}. 

This also means that we can interpret the two factors as $\partial {\rm log} u$'s for curves on new surfaces, which we get by cutting on $y$ and adding ``scaffolding'' to them. Consider a curve $C_{{\rm left}}$ whose word is $A y L \alpha$ where $\alpha$ is a boundary. Since the matrix for this word is exactly the same as the one we looked at above putting the right matrix to the identity, we have that  $\partial_y {\rm log} u_{C_{\rm left}}|_{y = 0} = - (a d - b c)/(b d)$. Similarly, we can define the curve $C_{{\rm right}}$ whose word is $\beta R y B$ where $\beta$ is a boundary. Then the matrix for this word is exactly the same as for $C$ with the choice that the ``left" matrix is given by $M_R(\alpha \to 1)$, so that again $\partial_y {\rm log} u_{C_{{\rm right}}} \vert_{y=0} = - (a^\prime d^\prime - b^\prime c^\prime)/(c^\prime d^\prime)$. 

Thus we have discovered that the $\partial_y {\rm log} u_C \vert_{y=0}$ factorizes. We can write this explicitly in terms of words as 
\begin{equation}
\partial_y{\rm log} u[X y Z]\vert_{y = 0} = - \partial_y{\rm log} u[X y L \alpha]\vert_{y = 0} \times  \partial_y{\rm log} u[\beta R y Z]\vert_{y = 0}
\end{equation}
where $X=A$ and $Z = L B$. We can recognize this in terms of the left and right curves on the two surfaces we get by cutting on $y$ and adding scaffolding as in the above:
\begin{equation}
\partial_y{\rm log} u_C\vert_{y = 0} = \partial_y{\rm log} u_{{C_{{\rm left}}}} \vert_{y=0}  \times \partial_y {\rm log} u_{{C_{{\rm right}}}} \vert_{y=0}
\end{equation}

Exactly the same manipulations show precisely the same factorization with a right turn at $y$. Note that in defining the left and right curves, we decided to turn left at the end for $C_{\rm {left}}$ and right at the beginning for $C_{{\rm right}}$. Flipping the choice for each one gives the same factorizing expression with an extra minus sign. 

Note that in this argument, it was important that the left and right matrices $X, Z$ were generic, so that the $u$ variables for the left and right curves were well-defined. This will not be the case if the left or right part of the curves correspond to pure boundaries, that is, if $X$ has all right turns or $Z$ has all left turns. This corresponds to a case where the curve $C$ does pass through $y$ but does not intersect $y$ as a curve; if we draw $y$ as a chord, then we can say this is a case where the chord associated with $C$ doesn't intersect the chord for $y$, but the lamination for $C$ does intersect the chord for $y$. 

Let's consider the case where the left part of the word for $C$ is of the form $(\rm All Right) y L B$. The result will end up being exactly the same if $y$ is followed by a right turn instead. Now a general all right matrix is of the form 
\begin{equation}
M_{{\rm all right}} = \left(\begin{array}{cc} a & 0 \\ b & 1 \end{array} \right) 
\end{equation}
and hence 
\begin{equation}
M_C = \left(\begin{array}{cc} a & 0 \\ c & 1 \end{array} \right) M_L(y) \left(\begin{array}{cc} a^\prime & b^\prime \\ c^\prime & d^\prime \end{array} \right)
\end{equation}
From here we can compute that the derivative of $u$ again factorizes \footnote{Note this expression does not coincide with what we'd get from our previous expressions simply putting $c \to 0$, this is why we have to treat the case where the words are ``all right" to the left of $y$ or ``all left" to the right of $y$ separately}. 
\begin{equation}
\partial_y u_C = - c \times \frac{(b^\prime + d^\prime) (a^\prime d^\prime - b^\prime c^\prime)}{(a^\prime + c^\prime) d^{\prime 2}}.
\end{equation}

Just as before we can interpret each factor in terms of curves on the cut surfaces that are further scaffolded along the cut curve $y$. In terms of words we have that 
\begin{equation}
\partial_y u[({\rm All Right}) y Z] \vert_{y=0}= \partial_y u[({\rm All Right}) y L \alpha] \vert_{y=0} \times \partial_y u[\beta L y Z]\vert_{y=0}
\end{equation}
Note that as in our previous example, we could also choose $\beta R Z$ above, at the cost of an overall sign. We write it using $L$, since the curve with word $(\beta L y Z)$ intersect the curve for $y$ which has word $(\beta R y ({\rm All Left}))$, and hence it's $\partial_y u[\beta L y Z]$ can be replaced by $\partial_u {\rm log} u[\beta L y Z]$ at $y=0$.  Meanwhile, the word for the first factor $(({\rm All Right}) y L \alpha)$ is just that for the curve $y$ itself on the left surface, so we have $u_{{\rm curve}y}$ for that factor.  
It is again useful to write all derivatives as logarithmic derivatives, but clearly $\partial_y u_{{\rm curve y}} = - \partial_y ( 1- u_{{\rm curve} y})$ so $\partial_y u_{{\rm curve y}}\vert_{y=0} = - \partial_y {\rm log} ( 1 - u_{{\rm curve y}})\vert_{y=0}$ since $u_{{\rm curve y}} \to 0$ as $y \to 0$. Thus expressed in terms of logarithmic derivatives we have that for curves $C$ with words of the form $(({\rm All Right}) y Z)$, we have the factorization for the derivatives as 
\begin{equation}
\partial_y u_C \vert_{y=0}= - \partial_y {\rm log} (1 - u_{{\rm curve} y})\vert_{y=0} \times \partial_y {\rm log} u_{C_{{\rm right}}} \vert_{y=0}
\end{equation}
where $C_{{\rm right}}$ has the word $(\beta L y Z)$.

\section{Matching of the loop-cut with gluing of tree in the forward limit}
\label{sec:matchingAll}
In this appendix, we show the highly non-trivial matching of the loop-cut and the gluing of tree amplitude. We separate the matching of each term into different parts, and in the end, we show the difference is actually a total derivative.

\subsection{Matching of the $D$-dependent part}

We will start by showing that $(\dagger)$=$(\star)$, $i.e.$ that the $D$-dependent part of both answers agree. 
  
To do this, we look at the tree-level side and understand how we can write $\partial_{X_{s_L,s_R}}\mathcal{A}$, in terms of $u$'s. After taking the scaffolding residue on $s_L$, we know we can write the answer as: 

\begin{equation}
\begin{aligned}
	\text{Res}_{y_{1,p}=0} \mathcal A &= \prod_i \frac{\diff y_i}{y_i^2}  \sum_j (X_{s_L,j} - X_{p,j}) \frac{\partial \log(u_{s_L,j})}{\partial y_{1,p}} \times  \prod_{X}  u_X^X \bigg \vert_{y_{1,p}=0} \\
 &= \prod_i \frac{\diff y_i}{y_i^2}  \sum_j (X_{s_L,j} - X_{p,j}) \frac{\partial u_{s_L,j}}{\partial y_{1,p}} \times  \prod_{X}  u_X^X \bigg \vert_{y_{1,p}=0},
\end{aligned}
\end{equation}
where in the last line we used the fact that $u_{s_L,j} \to 1$ when $y_{1,p}=0$, which follows directly from the $u$-equations.  Therefore, the part proportional to $X_{s_L,s_R}$ after taking the residue in $y_{1^\prime,p}$, is simply:
\begin{equation}
	\text{Res}_{y_{1^\prime,p}=0}\left(\text{Res}_{y_{1,p}=0} \mathcal A \right)= \prod_i \frac{\diff y_i}{y_i^2}\left( X_{s_L,s_R} \frac{\partial^2 u_{s_L,s_R}}{\partial y_{1^\prime,p} \partial y_{1,p}} \times  \prod_{X}  u_X^X \bigg \vert_{y_{1,p}=y_{1^\prime,p}=0} + ...\right).
\end{equation}

Now using the word for curve $X_{s_L,s_R}$, we can write  $u_{s_L,s_R}$ in terms of $y_{1,p}$ and $y_{1^\prime,p}$:
\begin{equation}
	u_{s_L,s_R} = \frac{1+y_{1^\prime,p}}{1+y_{1^\prime,p}+y_{1^\prime,p}y_{1,p}} \quad \Rightarrow  \quad \frac{\partial^2 u_{s_L,s_R}}{\partial y_{1^\prime,p} \partial y_{1,p}} \bigg \vert_{y_{1,p}=y_{1^\prime,p}=0} = -1.
\end{equation}

Therefore we have:
\begin{equation}
	-D \, \cdot \frac{\partial \mathcal{A}}{\partial X_{s_L,s_R}} =  D \, \cdot \prod_i \frac{\diff y_i}{y_i^2} \prod_{X}  u_X^X \bigg \vert_{y_{1,p}=y_{1^\prime,p}=0},
\end{equation}
which is exactly what we get on the loop-cut side. So indeed we have  $(\dagger)$=$(\star)$, and the D-dependent parts agree without any need for total derivative identities.  

\subsection{Matching of variables of tree triangulation}
\label{sec:match2ndterm}

Let's proceed to the matching of ($\dagger \dagger $) with ($\star \star $), $i.e.$ the parts proportional to $X_{j,k}$. Let's start by looking at the loop-side. In this case, we have contributions from two curves, $X_{k,j}$, $u_{k,j}^{(0)}$, intersecting $Y_1$ and $X_{j,k}$ that self-intersects once, $u_{j,k}^{(1)}$. Let's now understand how we can write the derivatives $ \partial_{y_1} \log{u_{k,j}}$ in terms of the $u$'s of the cut surface. Let's start by looking at $u_{k,j}$, using the matrix method to get this $u$-variable we have:

\begin{equation}
\begin{aligned}
	\underline{kj}: &\quad 
	 M_1  M_{R}(y_{1})  M_2 = \begin{bmatrix}
		x & y  \\
		z & w 
	\end{bmatrix} \begin{bmatrix}
		y_1 & 0  \\
		1 & 1 
	\end{bmatrix} \begin{bmatrix}
		a & b  \\
		c & d 
	\end{bmatrix} = \begin{bmatrix}
		c y + a (y+x y_1) & d y + b (y+x y_1)  \\
		c w + a (w+z y_1)& d w + b (w+z y_1)
	\end{bmatrix}. \\
	\\
	& \quad \Rightarrow u_{k,j}^{(0)} = \frac{(b y + d y + b x y_1) (a w + c w + a y_1 z)}{(a y + c y + a x y_1) (b w + d w + b y_1 z)} \quad \Rightarrow \frac{\partial \log{u_{k,j}}}{\partial y_{1}} \bigg \vert_{y_1=0} = \frac{(b c - a d) (w x - y z)}{(a + c) (b + d) w y},
\end{aligned}
\end{equation}
\begin{figure}[t]
    \centering
    \includegraphics[width=0.6\linewidth]{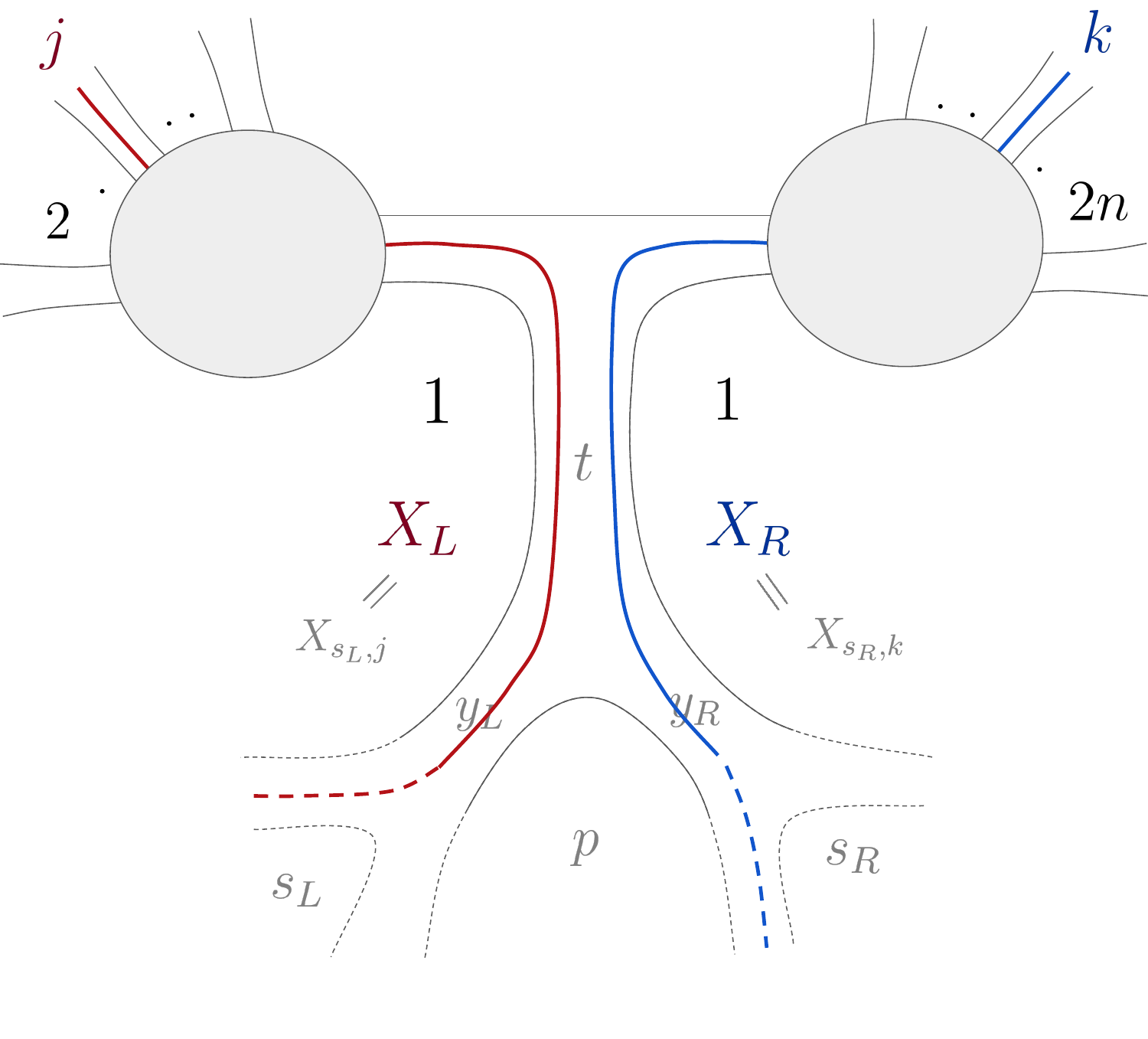}
    \caption{Curve $X_{k,j}$ after loop cut, and corresponding extension to the scaffolded surface (dashed).}
    \label{fig:curvext}
\end{figure}
where $M_1$ and $M_2$ are the matrices corresponding to the word to the left/right of $y_1$, respectively. After cutting the loop surface, this curve breaks into two, one on the left and one on the right, that we call $X_L$ and $X_R$ (see figure \ref{fig:curvext}). Now we consider their extensions into the scaffolded surface which turns them into curves that are defined on the tree-level side (represented in dashed in figure \ref{fig:curvext}). This way of extending $X_R$ and $X_L$ is exactly such that the corresponding $u$'s are
\begin{equation}
\begin{aligned}
	\underline{u_{X_L} = u_{j,s_L}}:& \quad M_1 M_{R}(y_{L}) =  \begin{bmatrix}
		y + x y_L & y  \\
		w + z y_L & w
	\end{bmatrix} \\
 &\Rightarrow u_{X_L} = \frac{y (w + y_L z)}{w (y + x y_L)} \quad \Rightarrow \frac{\partial \log{u_{X_L}}}{\partial y_{L}} \bigg \vert_{y_L=0} = -\frac{(w x - y z)}{(w y)}.\\
	\underline{u_{X_R} = u_{k,s_R}}:& \quad M_L(1)M_{R}(y_R) M_2 =  \begin{bmatrix}
		c + a (1 + y_R) &  d + b (1 + y_R)  \\
		a + c & b + d
	\end{bmatrix} \\
 &\Rightarrow u_{X_R} = \frac{(a + c) (b + d + b y_R)}{(b + d) (a + c + a y_R)}  \Rightarrow \frac{\partial \log{u_{X_R}}}{\partial y_{R}} \bigg \vert_{y_R=0} = \frac{(b c - a d)}{(a + c) (b + d)}.
\end{aligned}
\end{equation}

Therefore we have:
\begin{equation}
	\frac{\partial \log{u_{k,j}}}{\partial y_{1}} \bigg \vert_{y_1=0} = -  \frac{\partial \log{u_{X_R}}}{\partial y_{R}} \bigg \vert_{y_R=0} \times \frac{\partial \log{u_{X_L}}}{\partial y_{L}} \bigg \vert_{y_L=0},
\end{equation}
from figure \ref{fig:curvext}, we can read off directly the extensions in the scaffolded problem, which gives us exactly the tree-level result:
\begin{equation}
	\frac{\partial \log{u_{k,j}}}{\partial y_{1}} \bigg \vert_{y_1=0} = - \frac{\partial \log{u_{s_L,j}}}{\partial y_{1,p}} \frac{\partial \log u_{s_R,k}}{\partial y_{1^\prime,p}}.
\end{equation}

Proceeding exactly in the same way for the self-intersecting curve, we find:
\begin{equation}
	\frac{\partial \log{u^{(1)}_{j,k}}}{\partial y_{1}} \bigg \vert_{y_1=0} = - \frac{\partial \log{u_{s_L,k}}}{\partial y_{1,p}} \frac{\partial \log u_{s_R,j}}{\partial y_{1^\prime,p}}.
\end{equation}

\subsection{Matching of loop variables}

We now want to relate ($\dagger \dagger \dagger$) with ($\star \star \star$), $i.e.$ the parts proportional to $X_{p,j}$/$Y_{j}$ in the tree and loop answers, respectively. In this case, it turns out that these two factors will not match at integrand level, but we will show that they differ by a total derivative.
  
Let's start by looking at the tree-level glue term ($\dagger \dagger \dagger$), in particular to the piece proportional to $X_{p,J}$ with $J\in\{2,\dots,2n \}$:

\begin{equation}
	 X_{p,J} \left.\left( \underbrace{\frac{\partial \log{u_{s_L,J}}}{\partial y_{1,p}} \sum_{k \in \{1,\dots,2n\}} \frac{\partial \log u_{s_R,k}}{\partial y_{1^\prime,p}}}_{\textbf{(1)}} + \underbrace{\frac{\partial \log u_{s_R,J}}{\partial y_{1^\prime,p}} \sum_{k \in \{2,\dots,1^\prime\}} \frac{\partial \log u_{s_L,k}}{\partial y_{1,p}}}_{\textbf{(2)}} \right)\right\vert_{y_{1,p}=y_{1^\prime,p}=0}.
\label{eq:term3}
\end{equation}

Let's start by analyzing term \textbf{(1)}:
\begin{equation}
	 X_{p,J}\frac{\partial \log{u_{s_L,J}}}{\partial y_{1,p}} \sum_{k \in \{1,\dots,2n\}} \frac{\partial \log u_{s_R,k}}{\partial y_{1^\prime,p}} \bigg \vert_{y_{1,p}=y_{1^\prime,p}=0}=  X_{p,J}\frac{\partial \log{u_{s_L,J}}}{\partial y_{1,p}} \times \left. \frac{\partial \log\left( \prod_{k \in \{1,\dots,2n\}}u_{s_R,k}\right)}{\partial y_{1^\prime,p}} \right\vert_{y_{1,p}=y_{1^\prime,p}=0}.
\end{equation}

Now inside the second logarithm we have the product of all the u-variables corresponding to curves that start somewhere on the tree part and eventually end in the following path: from $t$ to $1^\prime p$ and exit in road $p,s_R$. This means that this product is effectively the $u$-variable of the curve of the smaller surface obtained by cutting through road $t$, $U_{s_R,t}$, that goes from road $p,s_R$ to road $t$. In particular, this means that this product is \textbf{independent} of $y_t$. Now the $u$-variable for this curve on this smaller surface is really simple:
\begin{equation}
	\prod_{k \in \{1,\dots,2n\}}u_{s_R,k} \equiv U_{s_R,t} = \frac{1}{(1+y_{1^\prime,p})}  \quad \Rightarrow \quad \frac{\partial \log\left( \prod_{k \in \{1,\dots,2n\}}u_{s_R,k}\right)}{\partial y_{1^\prime,p}} \bigg \vert_{y_{1,p}=y_{1^\prime,p}=0} = -1.
	\label{eq:effcurve}
\end{equation} 

So we can rewrite \eqref{eq:term3} as follows:
\begin{equation}
	X_{p,J}\left. \left( \underbrace{-\frac{\partial \log{u_{s_L,J}}}{\partial y_{1,p}} }_{\textbf{(1)}} + \underbrace{\frac{\partial \log u_{s_R,J}}{\partial y_{1^\prime,p}} \sum_{k \in \{2,\dots,1^\prime\}} \frac{\partial \log u_{s_L,k}}{\partial y_{1,p}}}_{\textbf{(2)}} \right) \right\vert_{y_{1,p}=y_{1^\prime,p}=0}.
\end{equation}

Before looking at term \textbf{(2)}, let's look at what we have on the loop-side. From \eqref{eq:loopcut2}, we have that the term in $(\star \star \star)$ proportional to $Y_J$ is:
\begin{equation}
	Y_J \frac{\partial \log{u_{Y_J}}}{\partial y_{1}} \bigg \vert_{y_1=0}.
\end{equation}

Since $Y_J$ does not intersect $Y_1$, we can't use the $u$-equations to compute this derivative. Instead, we consider all curves that, just like $Y_J$, enter the fat graph in $J,J+1$ and go until road $1$, and from there continue to anywhere. Let's denote this set of curves by $\mathcal{C}$. This is of course an infinite family of curves since they can self-intersect infinitely often. Now, exactly for the same reason as before, have that:

\begin{equation}
	\partial_{y_1} \left(u_{Y_J} \times \prod_{X\in \mathcal{C}} u_X  \right) = 0 \quad \Rightarrow \quad \partial_{y_1} \log u_{Y_J} = - \partial_{y_1}  \sum_{X\in \mathcal{C}} \log u_X.
\end{equation}

\begin{figure}[t]
\begin{center}
\includegraphics[width=\linewidth]{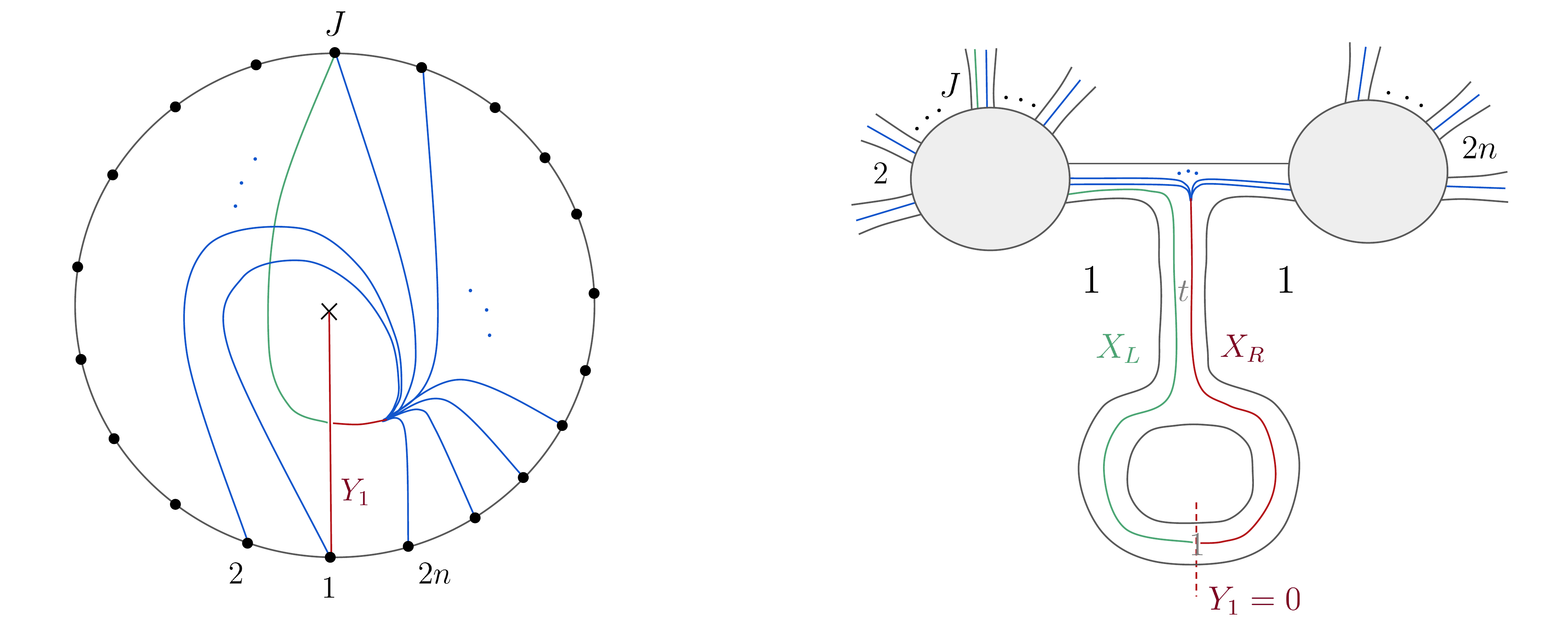}
\caption{Extending curve $Y_J$  -- curves that intersect $Y_1$ only once.}
\label{fig:ExtensionYJ}
\end{center}
\end{figure}

Now since we want to evaluate the derivative at $y_1=0$, the contribution coming from $X\in \mathcal{C}$ that intersects $Y_1$ more than once is zero. Therefore we have:

\begin{equation}
	\partial_{y_1} \log u_{Y_J} \bigg \vert_{y_1=0} = - \partial_{y_1}  \log \prod_{k\in\{1,\dots,2n\}} u_{J,k}\bigg \vert_{y_1=0},
\end{equation}
where for $k\in \{1,\dots,J-1\}$ these correspond to curves that self-intersect once; while for $k\in \{J,\dots,2n\}$ these are curves that do not intersect but still go around the loop (see figure \ref{fig:ExtensionYJ}, left). So this product of curves can be interpreted as a curve that starts on road $(J,J+1)$, goes around the loop once, and ends in $t$ (union of curve green and red). In particular, after cutting the loop surface, this curve breaks into $X_L$ and $X_R$ (see figure \ref{fig:ExtensionYJ}, right). $X_R$ is effectively a curve of the smaller surface that starts in the tadpole road.
Therefore, using the result from section \ref{sec:match2ndterm}, we can write the derivative of the logarithm as follows:
\begin{equation}
	\frac{\partial \log{u_{Y_J}}}{\partial y_{1}} \bigg \vert_{y_1=0}=  \frac{\partial \log{u_{X_{J,1}}}}{\partial y_{1}} \cdot \frac{\partial \log{u_{X_{1,t}}}}{\partial y_{1}} \bigg \vert_{y_1=0} = \frac{\partial \log{u_{X_L}}}{\partial y_{1}} \cdot \frac{\partial \log{u_{X_R}}}{\partial y_{1}} \bigg \vert_{y_1=0} .
\end{equation}

Once again, to relate to the tree-level variables, we consider the extension of curves $X_L$ and $X_R$, to the scaffolded cut surface in the same way we did previously. While $u_{X_{1,t}}$ becomes $U_{s_R,t}$ giving a trivial contribution, as shown in \eqref{eq:effcurve}, $u_{X_{J,1}}$ becomes $u_{s_L,J}$,  allowing us to write  
\begin{equation}
	\frac{\partial \log{u_{Y_J}}}{\partial y_{1}} \bigg \vert_{y_1=0}= - \frac{\partial \log{u_{s_L,J}}}{\partial y_{1}}\bigg \vert_{y_1=0},
\end{equation}
which precisely matches part of the tree-level contribution. So currently the difference between the tree-gluing and the loop cut for ($\dagger \dagger \dagger$) is simply term \textbf{(2)}:

\begin{equation}
	X_{p,J} \, \left.\left(\frac{\partial \log u_{s_R,J}}{\partial y_{1^\prime,p}} \sum_{k \in \{2,\dots,1^\prime\}} \frac{\partial \log u_{s_L,k}}{\partial y_{1,p}}\right)\right\vert_{y_{1,p}=y_{1^\prime,p}=0} .
\end{equation}

Let's now try to use the $u$-equations to simplify \textbf{(2)}. We start by once more turning the sum into a product inside the logarithm:

\begin{equation}
	X_{p,J} \, \frac{\partial \log u_{s_R,J}}{\partial y_{1^\prime,p}} \cdot \frac{\partial \log \prod_{k \in \{2,\dots,1^\prime\}}u_{s_L,k}}{\partial y_{1,p}} \bigg \vert_{y_{1,p}=y_{1^\prime,p}=0} .
\label{eq:totalDer1}
\end{equation}

As opposed to \textbf{(1)}, this product of $u$'s \textbf{cannot} be identified as the $u$ of a smaller surface. So instead, in order to simplify this result we use the extended $u$-equations:
\begin{equation}
	\prod_{k \in \{2,\dots,1^\prime\}}u_{s_L,k} + u_{1,p}u_{1,s_R} =1,
\end{equation}
which allows us to write:
\begin{equation}
	  \frac{\partial \log \prod_{k \in \{2,\dots,1^\prime\}}u_{s_L,k}}{\partial y_{1,p}}\bigg \vert_{y_{1,p}=y_{1^\prime,p}=0} = - u_{1,s_R} \cdot 
\frac{\partial u_{1,p}}{\partial y_{1,p}}\bigg \vert_{y_{1,p}=y_{1^\prime,p}=0} = -\frac{\partial u_{1,p}}{\partial y_{1,p}}\bigg \vert_{y_{1,p}=y_{1^\prime,p}=0}.
\end{equation}

Finally to compute $\partial_{y_{1,p}} u_{1,p}$, we use \eqref{eq:yi} which allows us to write $y_{1,p}$ in terms of $u$'s in the following way:
\begin{equation}
	y_{1,p} = \frac{u_{1,p} \prod_{k\in\{2,\dots,2n\}}u_{p,k}}{u_{s_L,s_R}u_{s_L,1^\prime}}  \quad \Rightarrow \quad 1 = \frac{\partial u_{1,p}}{\partial y_{1,p}} \prod_{k\in\{2,\dots,2n\}}u_{p,k} \bigg \vert_{y_{1,p}=y_{1^\prime,p}=0},
\end{equation} 
and so we can rewrite \eqref{eq:totalDer1} as
\begin{equation}
	X_{p,J} \, \frac{\partial \log u_{s_R,J}}{\partial y_{1^\prime,p}} \cdot \frac{1}{\prod_{k\in\{2,\dots,2n\}}u_{p,k}} \bigg \vert_{y_{1,p}=y_{1^\prime,p}=0}.
\end{equation}

Using, once again, the $u$-equations we can find an expression for $\partial_{y_{1^\prime,p}}\log u_{s_R,J}$ in terms of the remaining $u$ variables as follows:
\begin{equation}
\begin{aligned}
	&u_{s_R,J} + u_{1^\prime,p} \prod_{\substack{a \in\{1, \dots , J-1\} \\ b \in\{J+1,\dots,1^\prime\}} } u_{a,b} \times \prod_{k\in \{J+1,\dots 2n \}} u_{p,k} =1\\
	&\Rightarrow \frac{\partial \log u_{s_R,J}}{\partial y_{1^\prime,p}} = -  \underbrace{\frac{\partial \log u_{1^\prime,p}}{\partial y_{1^\prime,p}}}_{=1} \times  \prod_{\substack{a \in\{1, \dots , J-1\} \\ b \in\{J+1,\dots,1^\prime\}} } u_{a,b} \times \prod_{k\in \{J+1,\dots 2n \}} u_{p,k},
\end{aligned} 
\end{equation}
which we can further simplify with the help of the generalized $u$-equations that tell us 
\begin{equation}
	\prod_{\substack{a \in\{1, \dots , J-1\} \\ b \in\{J+1,\dots,1^\prime\}} } u_{a,b} = 1 - u_{p,J}u_{s_R,J} \quad \Rightarrow \quad \prod_{\substack{a \in\{1, \dots , J-1\} \\ b \in\{J+1,\dots,1^\prime\}} } u_{a,b} \bigg \vert_{y_{1,p}=y_{1^\prime,p}=0} = 1 - u_{p,J}\bigg \vert_{y_{1,p}=y_{1^\prime,p}=0} .
\end{equation}

In summary, we have that the tree-level contribution that is not present in the loop side is:
\begin{equation}
	X_{p,J} \, \left(\frac{1 - u_{p,J}}{u_{p,J}} \cdot \frac{1}{\prod_{k\in\{2,\dots,J-1\}}u_{p,k}} \right)\bigg \vert_{y_{1,p}=y_{1^\prime,p}=0} .
	\label{eq:totalDer2}
\end{equation}
In subsection~\ref{sec:totalD}, we show that the above difference between loop integrand and the forward limit of the tree is actually a total derivative and thus vanishes after integration.

\subsection{Matching of curves $X_{1,j}$}

Now we want to see how ($\dagger \dagger \dagger \dagger$) matches $(\star\star\star\star)$. 
Let's start with the tree side. Since $X_{1,2}=X_{1^\prime,2n}=0$ we have that $(\dagger \dagger \dagger \dagger)$ is :
\begin{equation}
\sum_{j\in\{2,\dots,2n \}} -X_{j,1^\prime}   \underbrace{\frac{\partial \log{u_{s_L,1^\prime}}}{\partial y_{1,p}} }_{=-1}\frac{\partial \log u_{s_R,j}}{\partial y_{1^\prime,p}} \bigg\vert_{y_{1,p}=y_{1^\prime,p}=0} \,- X_{1,j} \frac{\partial \log{u_{s_L,j}}}{\partial y_{1,p}}  \frac{\partial \log u_{s_R,1}}{\partial y_{1^\prime,p}}  \bigg\vert_{y_{1,p}=y_{1^\prime,p}=0}.
\end{equation}

\begin{figure}[t]
\begin{center}
\includegraphics[width=0.4\textwidth]{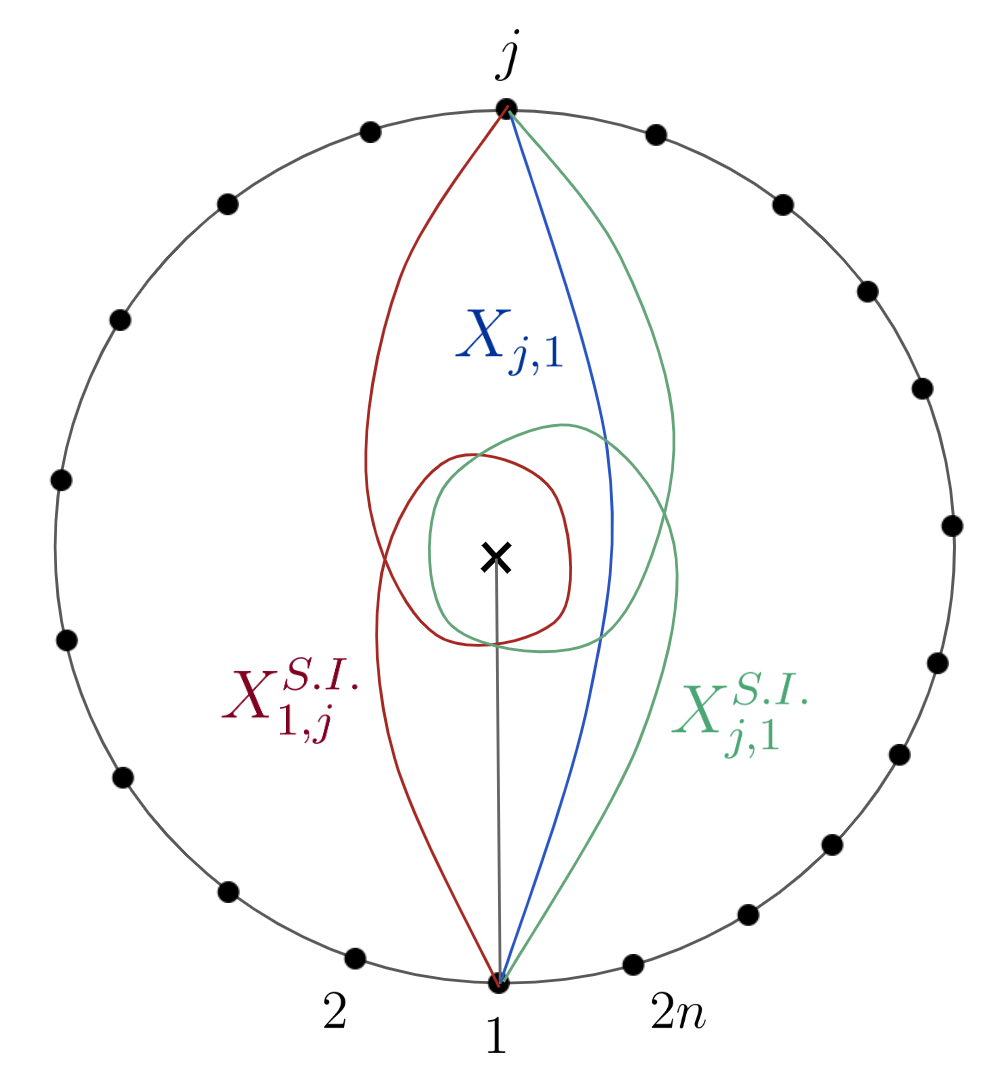}
\caption{Curves contributing to $X_{1,j}$ term.}
\label{fig:X1j}
\end{center}
\end{figure}

Now on the loop side, we have to consider the contributions of three different curves, see figure \ref{fig:X1j}. 
Let's start by looking at the contribution coming from the two self-intersecting curves. From the word for curve $X_{1,j}^{S.I.}$, we get the following $u$-variable:
\begin{equation}
\begin{aligned}
	\underline{u_{1,j}^{(1)}}: &\quad 
	 M_1 M_{L}(y_{t}) M_{L}(y_{1})  M_2 = \begin{bmatrix}
		x & 0  \\
		z & 1 
	\end{bmatrix} \begin{bmatrix}
		y_t & y_t  \\
		0 & 1 
	\end{bmatrix} \begin{bmatrix}
		y_1 & y_1  \\
		0 & 1 
	\end{bmatrix} \begin{bmatrix}
		a & b  \\
		c & d 
	\end{bmatrix}  \\
	& \Rightarrow u_{1,j}^{(1)} = \frac{(d + b y_1 + d y_1) (c + c y_t z + a y_1 y_t z + c y_1 y_t z)}{(c + a y_1 + 
   c y_1) (d + d y_t z + b y_1 y_t z + d y_1 y_t z)} \Rightarrow \frac{\partial \log{u_{1,j}^{(1)}}}{\partial y_{1}} \bigg \vert_{y_1=0} = \frac{(b c - a d)}{cd(1 + y_t z)}.
\end{aligned}
\end{equation}

Now proceeding like we did in section \ref{sec:match2ndterm}, we can show that:
\begin{equation}
	\frac{\partial \log{u_{1,j}^{(1)}}}{\partial y_{1}} \bigg \vert_{y_1=0} = -\frac{\partial \log u_{s_R,1}}{\partial y_{1^\prime,p}}  \frac{\partial \log{u_{s_L,j}}}{\partial y_{1,p}}. 
\end{equation}

Doing exactly the same thing for $X_{j,1}^{S.I.}$ we get:

\begin{equation}
	\frac{\partial \log{u_{j,1}^{(1)}}}{\partial y_{1}} \bigg \vert_{y_1=0} = \frac{\partial \log u_{s_R,j}}{\partial y_{1^\prime,p}} .
	\label{eq:X1jSI}
\end{equation}

This means that the contribution coming from the two self-intersecting curves perfectly matches the tree-level gluing term. However, on the loop side, we still have a contribution from the non-self-intersecting curve $X_{j,1}$ (see figure \ref{fig:X1j}), for which the $u$-variable is:
\begin{equation}
\begin{aligned}
	\underline{j1}: &\quad 
	 M_1  M_{R}(y_{1})  M_2 = \begin{bmatrix}
		x & 0  \\
		z & 1 
	\end{bmatrix} \begin{bmatrix}
		y_1 & 0  \\
		1 & 1 
	\end{bmatrix} \begin{bmatrix}
		a & b  \\
		c & d 
	\end{bmatrix} = \begin{bmatrix}
		a x y_1 &  b x y_1  \\
		c  + a (1+z y_1)& d  + b (1+z y_1)
	\end{bmatrix} \\
	& \quad \Rightarrow u_{j,1} = \frac{b  (a  + c  + a y_1 z)}{a  (b  + d  + b y_1 z)} \quad \Rightarrow \frac{\partial \log{u_{j,1}}}{\partial y_{1}} \bigg \vert_{y_1=0} = \frac{b z (ad -bc)}{a (b + d)^2}, 
\end{aligned}
\end{equation}
which can be expressed in terms of the scaffolded-tree variables as follows:
\begin{equation}
	\frac{\partial \log{u_{j,1}}}{\partial y_{1}} \bigg \vert_{y_1=0} = - z \frac{\partial \log u_{s_R,j}}{\partial y_{1^\prime,p}},
\end{equation}
with $z$ being the $F$-polynomial appearing in $M_1$. So on the loop side, we have an extra term:
\begin{equation}
	\sum_{j=3,\dots,2n-1}- z X_{1,j} \frac{\partial \log u_{s_R,j}}{\partial y_{1^\prime,p}}.
\end{equation} 

Now, one can easily check that $z$ is exactly the $F$-polynomial corresponding to the curve that turns consistently right until $y_t$, and thus has the following simple form:
\begin{equation}
	z = (1 + y_t (1 + y_{1,m}(1+ \dots))).
\end{equation}

Finally, we note that this $F$-polynomial appears in the expression of $u_{Y_1}$, allowing us to write:
\begin{equation}
	\frac{1}{z} = \frac{y_1}{u_{Y_1}} \bigg \vert_{y_1=0} = \prod_{k\in\{2,\dots,2n\}} u_{Y_{k}} \bigg \vert_{y_1=0} =  \prod_{k\in\{2,\dots,2n\}} u_{p,k} \bigg \vert_{y_1=0},
\end{equation}
where we use \eqref{eq:yi} in the first equality. In summary, the extra term in the loop side is:
 
\begin{equation}
	\sum_{j=3,\dots,2n-1}-  X_{1,j} \frac{\partial \log u_{s_R,j}}{\partial y_{1^\prime,p}} \frac{1}{\prod_{k\in\{2,\dots,2n\}} u_{p,k}}.
\end{equation}

\subsection{Matching the tadpole part}

From the tree-level side $(\dagger \dagger \dagger \dagger \dagger)$, we have the following 
\begin{equation}
	-X_{1,1^\prime}\left.\frac{\partial \log{u_{s_L,1^\prime}}}{\partial y_{1,p}} \frac{\partial \log u_{s_R,1}}{\partial y_{1^\prime,p}}  \right\vert_{y_{1,p}=y_{1^\prime,p}=0} = X_{1,1^\prime}\left.\frac{\partial \log u_{s_R,1}}{\partial y_{1^\prime,p}}  \right\vert_{y_{1,p}=y_{1^\prime,p}=0}.
\end{equation} 

From the loop side, we have two contributions, one from the regular tadpole curve, and one from the tadpole that self-intersects once. To obtain the contribution from the latter we can use directly \eqref{eq:X1jSI}, setting $j=1$ to get:
\begin{equation}
	X_{1,1} \frac{\partial \log{u_{1,1}^{(1)}}}{\partial y_{1}} \bigg \vert_{y_1=0} = X_{1,1} \frac{\partial \log u_{s_R,1}}{\partial y_{1^\prime,p}},
\end{equation}
which precisely matches the tree glue result. As for the contribution from the non-self-intersecting curve, proceeding exactly in the same way as we did for $X_{j,1}$, we get:
\begin{equation}
	-X_{1,1}  \frac{\partial \log u_{s_R,1}}{\partial y_{1^\prime,p}} = -  X_{1,1} \frac{\partial \log u_{s_R,1}}{\partial y_{1^\prime,p}} \frac{1}{\prod_{k\in\{2,\dots,2n\}}u_{p,k}}.
\end{equation}

\subsection{Difference between loop-cut and tree-gluing as a total derivative}
\label{sec:totalD}
So in summary the difference between the loop cut and the tree-level glue is:

\begin{equation}\label{eq:totaldiff}
\begin{aligned}
	&\prod \frac{1}{y_i^2} \left( \sum_{j=1,\dots,2n} (X_{1,j} - X_{p,j}) \frac{\partial \log u_{s_R,j}}{\partial y_{1^\prime,p}} \frac{1}{\prod_{k\in\{2,\dots,2n\}}u_{p,k}}+1\right) \prod_{C} u_C^{X_{C}}  \\
	=	&\prod \frac{1}{y_i^2} \left( -\sum_{j=1,\dots,2n} (X_{1,j} - X_{p,j}) \frac{1-u_{p,j}}{\prod_{k\in\{2,\dots,j\}}u_{p,k}}+1\right) \prod_{C} u_C^{X_{C}}.
\end{aligned}
\end{equation}
\\
The claim is that \eqref{eq:totaldiff} is equal to the following total derivative:

\begin{equation}\label{eq:total d}
	-\frac{\partial}{\partial y_t} \left.\left(\prod \frac{1}{y_i^2} \frac{\prod_{k\in\{1, \dots, m-1\}} u_{p,k}u_{1^\prime,k}}{u_{p,1} \prod_{j\in\{2, \dots, 2n\}} u_{p,j}^2} \prod_{C} u_C^{X_{C}}\right)\right\vert_{y_{1,p}=y_{1^\prime,p}=0} ,
\end{equation}
where $m$ is the region indicated in figure \ref{fig:FatGraphs}. 

Let's now proceed to the proof this result. We start by using~\eqref{eq:yi}, $y_i = \prod_{X} \, u_X^{g_X^{(i)}}$, to write $y_{t}$ in terms of $u$'s as follows
\begin{equation}
	y_{t}=\left(\frac{\prod_{k\in\{1, \dots, m-1\}} u_{1^\prime,k}}{ \prod_{j\in\{m, \dots, 2n\}} u_{p,j}} \right),
\end{equation}
and so we can simplify expression~\eqref{eq:total d} by defining $Q$ as
\begin{equation}
    Q=\frac{\prod_{k\in\{1, \dots, m-1\}} u_{p,k}u_{1^\prime,k}}{u_{p,1} \prod_{j\in\{2, \dots, 2n\}} u_{p,j}^2}=\frac{y_{t}}{\prod_{j\in\{2, \dots, 2n\}} u_{p,j}}=\frac{y_{t}}{1-u_{1,1^\prime}},
\end{equation}
where we use the $u$-equations in the last equality. Therefore the claim is equivalent to
\begin{equation}\label{eq:target total d}
    -Q \frac{\partial \log u_{p,J}}{\partial y_{t}}\bigg \vert_{y_{1,p}=y_{1^\prime,p}=0} =\frac{1 - u_{p,J}}{u_{p,J}} \cdot \frac{1}{\prod_{k\in\{2,\dots,J-1\}}u_{p,k}} \bigg \vert_{y_{1,p}=y_{1^\prime,p}=0}.
\end{equation}

It is sufficient to prove~\eqref{eq:target total d} under a specific triangulation since the integrated result is triangulation-independent. For example, taking the scaffolding triangulation with a ray-like triangulation inside the gluon $n$-gon: $\{X_{1,p^\prime},X_{1,p},X_{1,3},X_{3,5},\ldots,X_{2n-1,1^\prime},X_{1,1^\prime},X_{3,1^\prime},\ldots,$ $X_{2n-1,1^\prime}\}$, allow us to write $Q$ as 
\begin{equation}
    Q=\frac{y_t(1+y_t)}{u_{p,2}}.
\end{equation}

Then one can prove~\eqref{eq:target total d} indeed holds after some tedious algebra involving the $2\times 2$ matrices defining the $u$ variables.

  \newpage
  \bibliographystyle{JHEP}
  \bibliography{Refs}
\end{document}